\documentclass[11pt,a4paper]{article}
\pdfoutput=1

\usepackage{jheppub}
\usepackage{hyperref} 
\usepackage[dvipsnames]{xcolor}
\usepackage{amssymb}
\usepackage{amsfonts}
\usepackage{graphicx}
\usepackage{epstopdf}
\usepackage{dcolumn}
\usepackage{amsmath}
\usepackage{latexsym,bm}
\usepackage{amsthm}
\usepackage{slashed}
\usepackage{float}
\usepackage{color}
\usepackage{url}
\usepackage{longtable}
\usepackage{tikz}
\usepackage[all,cmtip]{xy}
\usepackage{multirow}
\usepackage{longtable}
\usepackage{extarrows}

\usepackage[titletoc]{appendix}
\makeatletter
\newtheorem*{rep@theorem}{\rep@title}
\newcommand{\newreptheorem}[2]{%
\newenvironment{rep#1}[1]{%
 \def\rep@title{#2 \ref{##1}}%
 \begin{rep@theorem}}%
 {\end{rep@theorem}}}
\makeatother

\newreptheorem{lemma}{Lemma}

\newreptheorem{conj}{Conjecture}

\theoremstyle{definition}

\newcommand \xoverline[2][0.75]{
    \sbox{\myboxA}{$\m@th#2$}
    \setbox\myboxB\null
    \ht\myboxB=\ht\myboxA
    \dp\myboxB=\dp\myboxA
    \wd\myboxB=#1\wd\myboxA
    \sbox\myboxB{$\m@th\overline{\copy\myboxB}$}
    \setlength\mylenA{\the\wd\myboxA}
    \addtolength\mylenA{-\the\wd\myboxB}
    \ifdim\wd\myboxB<\wd\myboxA
       \rlap{\hskip 0.5\mylenA\usebox\myboxB}{\usebox\myboxA}%
    \else
        \hskip -0.5\mylenA\rlap{\usebox\myboxA}{\hskip 0.5\mylenA\usebox\myboxB}%
    \fi}
\makeatother

\newcommand{\ba}{\begin{aligned}}
\newcommand{\ea}{\end{aligned}}
\def\be{\begin{equation}}
\def\ee{\end{equation}}
\def\bsp{\begin{split}}
\def\esp{\end{split}}
\def\bea{\begin{eqnarray}}
\def\eea{\end{eqnarray}}

\def \bp{\begin{pmatrix}}
\def\ep{\end{pmatrix}}

\def\R{\mathbb{R}}
\def\D{\mathbb{D}}
\def\T{\mathbb{T}}
\def\I{\mathbb{I}}
\def\N{\mathcal{N}}

\def\C{\mathbb{C}}
\def\Z{\mathbb{Z}}

\def\Q{\mathbb{Q}}
\def\O{\mathbb{O}}

\def\dim{\mathrm{dim}}
\def\mk{\mathfrak}

\def\br{\breve}
\newcommand{\vol}{\mathrm{vol}}

\usepackage{tikz}
\usepackage{diagbox}
\usetikzlibrary{positioning}
\usetikzlibrary{calc}
\usetikzlibrary{decorations.pathreplacing,calligraphy}

\usepackage{xstring}
\usetikzlibrary{decorations.pathmorphing} 
\usetikzlibrary{decorations.markings} 
\usetikzlibrary{arrows} 
\usetikzlibrary{shapes} 
\usetikzlibrary{matrix} 
\usetikzlibrary{positioning} 
\usepackage[english]{babel} 
\usepackage[autostyle]{csquotes}

\usepackage{tikz}
\usetikzlibrary{arrows}
\usetikzlibrary{arrows.meta}
\usetikzlibrary{shapes.geometric,calc,arrows, positioning,shapes.misc,decorations.markings}
\tikzset{
  big arrow/.style={
    decoration={markings,mark=at position 1 with {\arrow[scale=2,#1]{>}}},
    postaction={decorate},
    shorten >=0.4pt},
  big arrow/.default=black}
  
\pgfdeclarelayer{edgelayer}
\pgfdeclarelayer{nodelayer}
\pgfsetlayers{edgelayer,nodelayer,main} 
\tikzstyle{none}=[inner sep=0pt] 

\tikzstyle{NodeCross}=[draw, shape=circle, cross out, inner sep=0pt, minimum size=6pt,line width=0.25mm]
\tikzstyle{Circle}=[draw, shape=circle, black,  fill=black, inner sep=0pt, minimum size=6pt]
\tikzstyle{Star}=[draw, shape=star, fill=black, star points=8, inner sep=0pt, minimum size=8pt]

\tikzstyle{DashedLine}=[-, densely dashed, line width=0.25mm]
\tikzstyle{DottedLine}=[-, dotted, line width=0.25mm]
\tikzstyle{ThickLine}=[-, line width=0.25mm]
\tikzstyle{ArrowLineRight}=[-, -{Stealth[scale=1.75]}, line width=0.1mm, scale=5]
\tikzstyle{RedLine}=[-, draw={rgb,255: red,191; green,0; blue,0}, fill=none, line width=0.25mm]
\tikzstyle{DottedRed}=[-, dotted, draw={rgb,255: red,191; green,0; blue,0}, fill=none, line width=0.25mm]
\tikzstyle{DashedLineThin}=[-, densely dashed, line width=0.125mm, fill=none, draw=black]
\tikzstyle{ArrowLineRed}=[-, -{Stealth[scale=1.75]}, draw={rgb,255: red,191; green,0; blue,0}, line width=0.1mm, scale=5]
\tikzstyle{brane}=[draw]
\tikzset{D7/.style={circle, draw=black, inner sep=0pt, fill=white, minimum size=3mm}}
\tikzset{hasse/.style={circle, fill,inner sep=2pt}}
\tikzset{flavor/.style={regular polygon,fill=white,regular polygon sides=4,inner sep=2.5pt, draw}}
\tikzset{gauge/.style={circle, draw,inner sep=2.5pt}}
\tikzset{gaugeb/.style={circle, draw,fill=black,inner sep=2.5pt}}
\tikzset{gauger/.style={circle, draw,fill=cyan,inner sep=2.5pt}}
\tikzset{gaugeg/.style={circle, draw,fill=red,inner sep=2.5pt}}
\tikzset{SUd/.style={circle, draw=black, inner sep=0pt, fill=yellow, minimum size=2mm}}
\tikzset{bd/.style={circle, draw=black, inner sep=0pt, fill=black, minimum size=2mm}}
\tikzset{wd/.style={circle, draw=black, inner sep=0pt, fill=white, minimum size=2mm}}
\tikzset{Dynkin/.style={circle, draw=black, inner sep=0pt, fill=white, minimum size=2mm}}
\tikzstyle{ligne}=[draw, thick] 
\tikzset{doublearrow/.style={ draw=black!75, color=black!75, thick, double distance=3pt, }} 

\def\bbS{\mathbb{S}}

\usepackage{dsfont}

\usepackage{bclogo}

\newcommand\restr[2]{{
  \left.\kern-\nulldelimiterspace 
  #1 
  \littletaller 
  \right|_{#2} 
  }}

\newcommand{\littletaller}{\mathchoice{\vphantom{\big|}}{}{}{}}

\usepackage{subfigure}

\usepackage{pst-node}
\usepackage{tikz-cd}

\usepackage{etoolbox}
\usepackage{dynkin-diagrams}
\def\row#1/#2!{#1_{\IfStrEq{#2}{}{n}{#2}} & \dynkin{#1}{#2}\\}

\usepackage{mathrsfs}  

\usepackage{color} 

\usepackage{xcolor}

\usetikzlibrary{positioning}

\usepackage{tikz-cd}

\usepackage{tikz}
\usepackage{tikz-3dplot}

\usepackage{enumitem} 

\usepackage{physics}

\usepackage{chemfig}
\usepackage{import}

\usepackage{booktabs}

\usepackage{braket}

\graphicspath{ {figs/} }

\title{Aspects of 4d $\mathcal{N}=1$ $ADE$ gauge theories from M-theory: decomposition, automorphisms, and generalised symmetries}

\author[\,\spadesuit]{Osama Khlaif}
\emailAdd{osama.khlaif@icloud.com}

\author[\,\clubsuit]{Marwan Najjar}
\emailAdd{marwan.najjar@pku.edu.cn}

\affiliation[\spadesuit]{School of Mathematics, University of Birmingham,\protect\\ Watson Building, Edgbaston, Birmingham B15 2TT, UK}

\affiliation[\clubsuit]{Center for High Energy Physics, Peking University,\protect\\ Beijing 100871, China}

\abstract{
We study the decomposition of 4d $\mathcal{N}=1$ gauge theories with Lie algebras of type $\mathfrak{su}(N)$, $\mathfrak{so}(2N)$, and $\mathfrak{e}_{6}$, realized via M-theory geometric engineering. These theories, together with their novel decomposition structure, arise from quotienting the Bryant--Salamon spin bundle over the 3-sphere by special finite subgroups acting simultaneously on both the fiber and base. We show that these gauge theories admit both inner and outer automorphisms, enabling sequences of gauge theory breaking. In particular, outer automorphisms extend the decomposition structure to theories with $\mathfrak{so}(2N+1)$, $\mathfrak{sp}(2N)$, $\mathfrak{f}_{4}$, and $\mathfrak{g}_{2}$ gauge algebras. For these theories, including both simply-laced and non-simply-laced cases, we analyze their $p$-form symmetries, including $(-1)$-form symmetries, derive the corresponding SymTFTs, and identify the M-theoretic origin of their symmetry topological operators and defects. Finally, we demonstrate that these gauge theories exhibit modified instanton sums and higher 4-group structures, and we derive the associated topological sector directly from M-theory.
}

\begin{document}

\maketitle

\section{Introduction and summary}

Perhaps the most powerful framework for analyzing and characterizing quantum field theories (QFTs), including non-abelian gauge theories and their dynamics, is that of symmetries and their anomalies. Early work on distinguishing the phases of non-abelian gauge theories \cite{Wilson:1974sk,tHooft:1977nqb,Alford:1990fc,Alford:1991vr,Lo:1994jp,Bucher:1991bc,Alford:1992yx} and subsequent developments in related directions \cite{Witten:1998wy,Gukov:2006jk,Hellerman:2006zs,Kapustin:2013uxa,Kapustin:2014gua} culminated in the seminal formulation of \emph{generalized global symmetries} in \cite{Gaiotto:2014kfa}. This framework treats symmetries as topological operators acting via linking on $p$-dimensional defects in a QFT. Since then, the notion of symmetry has been significantly expanded to encompass higher-form symmetries \cite{Gaiotto:2014kfa}, higher-group structures, non-invertible symmetries, and $(-1)$-form symmetries\footnote{The notion of $(-1)$-form symmetry is closely tied to the concept of \emph{decomposition}, first explored in \cite{Pantev:2005rh,Pantev:2005wj,Pantev:2005zs,Hellerman:2006zs,Sharpe:2014tca} and further developed in subsequent works, including \cite{Robbins:2020msp,Sharpe:2021srf,Pantev:2022kpl,Sharpe:2022ene,Tanizaki:2019rbk,Najjar:2024vmm,Najjar:2025htp,Najjar:2025rgt}.}. To illustrate the rapid growth of this field |both from field theory and string theory perspectives| over the past decade, we refer the reader to the following sample of representative works \cite{Sharpe:2015mja,Tachikawa:2017gyf,Cordova:2018cvg,Benini:2018reh,Tanizaki:2019rbk,Cordova:2019jnf,Cordova:2019uob,Gaiotto:2020iye,Brennan:2020ehu,Hidaka:2020izy,McNamara:2020uza,Yu:2020twi,Gukov:2020btk,Hidaka:2021mml,Kaidi:2021xfk,Heidenreich:2021xpr,Choi:2021kmx,Nguyen:2021naa,Acharya:2021jsp,Apruzzi:2022rei,Apruzzi:2021nmk,Lin:2022xod,Bhardwaj:2022lsg,Antinucci:2022eat,Bashmakov:2022jtl,Vandermeulen:2022edk,Choi:2022jqy,Choi:2022zal,Hayashi:2022fkw,Bhardwaj:2022yxj,Hubner:2022kxr,DelZotto:2022joo,vanBeest:2022fss,Heckman:2022xgu,Roumpedakis:2022aik,Chatterjee:2022kxb,Damia:2022seq,Antinucci:2022cdi,Apruzzi:2023uma,Closset:2023pmc,Baume:2023kkf,Chen:2023qnv,Bashmakov:2023kwo,Najjar:2023hee,Wang:2023iqt,DelZotto:2024tae,Bharadwaj:2024gpj,GarciaEtxebarria:2024fuk,Franco:2024mxa,Cvetic:2024dzu,Tian:2024dgl,Heckman:2024oot,Furrer:2024zzu,Closset:2024sle,Gagliano:2024off,Aloni:2024jpb,Santilli:2024dyz,Garcia-Valdecasas:2024cqn,KNBalasubramanian:2024bcr,Duan:2024xbb,Brennan:2024tlw,Chen:2024fno,Cui:2024cav,Najjar:2024vmm,Najjar:2025htp,Najjar:2025rgt,Yu:2024jtk,Braeger:2025rov,Robbins:2025apg,Closset:2025lqt,Jia:2025jmn,Tian:2025yrj}.

Since the mid-1990s, string theory geometric engineering has emerged as a powerful framework for studying supersymmetric QFTs such as non-abelian gauge theories. The essential degrees of freedom for the realization of such gauge theories localize near specific type of singularities in the extra spatial dimensions  \cite{Witten:1995ex,Katz:1996th,Katz:1996fh,Bershadsky:1996nh,Ooguri:1997ih,Sen:1997kz,Vafa:1998vs,Acharya:1998pm,Acharya:2000gb,Atiyah:2001qf}. This motivates formulating string theory on non-compact spaces with cone-like geometries, thereby isolating these singularities and their associated gauge dynamics from the bulk gravitational sector.

Remarkably, the geometric engineering philosophy captures key physical features of the resulting QFTs, most notably their generalized symmetries and associated ’t Hooft anomalies, which remain invariant under RG flows. These symmetries can be systematically encoded using the framework of Symmetry Topological Field Theory (SymTFT) \cite{Freed:2012bs,Gaiotto:2020iye,Apruzzi:2021nmk}. For a QFT, denoted by $\mathcal{T}$, defined on a spacetime manifold $M_{d}$, the SymTFT approach isolates the intrinsic dynamics of $\mathcal{T}$ from its symmetries by embedding it into a bulk $(d+1)$-dimensional topological field theory on a space $Y_{d+1}$, taken as a radial extension of $M_{d}$. The physical degrees of freedom reside on the physical boundary $M_{d}\times \{0\}$, while the symmetries of $\mathcal{T}$ are encoded on the opposite symmetry boundary through appropriate topological boundary conditions.

It turns out that the cone-like geometries used in geometric engineering naturally realize the setup of the SymTFT framework \cite{Apruzzi:2021nmk}. Specifically, the physical boundary $M_{d}\times\{0\}$ corresponds to the low-energy QFT localized at the apex of the cone, while the symmetry boundary is identified with the boundary of the cone, commonly referred to as the link space.

The SymTFT degrees of freedom—namely, the background gauge fields for finite symmetries and the field strengths associated with continuous $p$-form symmetries—are extracted from the topological sector of the supergravity limit of string theory \cite{Apruzzi:2021nmk,Najjar:2024vmm}. This involves interpreting the kinetic terms of the supergravity field strengths as topological BF couplings, together with cubic Chern–Simons interactions. Finite $p$-form symmetries require a formulation in differential cohomology to correctly capture their discrete background fields \cite{Apruzzi:2021nmk}. In contrast, continuous $p$-form symmetries are obtained by reducing the topological action over free cocycles, supplemented by correction terms induced by Chern–Simons couplings \cite{Najjar:2024vmm,Najjar:2025htp}.

In this work, we focus on novel quotients of the Bryant--Salamon spin bundle over the 3-sphere within the framework of M-theory geometric engineering. We analyze their implications for the resulting low-energy gauge theories and their associated symmetry structures, as summarized in the subsection below.

In M-theory, $\N=1$ supersymmetric Minkowski vacua arise when the seven extra spatial dimensions are modeled by manifolds of $G_{2}$ holonomy \cite{Papadopoulos:1995da}. Supersymmetric non-abelian gauge theories with $ADE$ Lie algebras are engineered from codimension-four singularities of $ADE$ type, as explicitly demonstrated in \cite{Acharya:2000gb,Atiyah:2001qf}.

It is worth noting that constructing new examples of compact or non-compact 7-dimensional manifolds with $G_{2}$ structure remains challenging, owing to the absence of an existence theorem analogous to that for Calabi--Yau manifolds. Nonetheless, substantial progress has been made in recent years, as reviewed in \cite{Braun:2017ryx,Braun:2017uku,Braun:2018vhk,Foscolo:2018mfs,Acharya:2020vmg,Xu:2020nlh,DelZotto:2021ydd,Sabag:2022hyw,McOrist:2025sdy,Farakos:2025shl}.

In the rest of this section, we give an overview of the topics we will be interested in in this work, and we summarize our results.


\subsection{Summary of results}

\paragraph{Non-splittable quotients of the Bryant--Salamon space.}

In M-theory geometric engineering, the Bryant--Salamon spin bundle over the 3-sphere $\bbS^{3}$ plays a central role, as its finite $ADE$ quotients furnish the co-dimension four singularities required to engineer 4d $\N=1$ $ADE$ gauge theories. In this work, we focus on a class of quotients that act simultaneously on both the spin bundle fiber and its $\bbS^{3}_{\rm{b}}$ base, modeled via \textit{Goursat's theorem}. We refer to such quotient groups as \emph{non-splittable} and denote them by $\Gamma_{\rm{ns}}$, defined as
\begin{equation}
    \Gamma_{\rm ns}\,=\,\{\, (c,b)\,\in\,C\times B \,|\, \varphi(c)\,=\,b \,\}~,
\end{equation}
where $C$ and $B$ are finite $ADE$ subgroups of the isometry group of the Bryant--Salamon space, and $\varphi$ is a homomorphism relating their actions. We discuss explicitly Goursat's theorem and its implications in \ref{sec:non-split-section}. We note that the quotient space of interest inherits the $G_{2}$ structure from the Bryant--Salamon space, as the group $\Gamma_{\rm{ns}}$ is a subgroup of its isometry group.

In fact, our interest lies in the cases where the group $C$ can be expressed as a $B$-extension of another finite $ADE$ subgroup $C_{0}$. Remarkably, such extensions are highly constrained: as classified by Reid in \cite{Reid:1985}, there exist only six possible cases, namely:
\begin{equation}\label{eq:intro-6-cases}
    \begin{split}
       \mathrm{Case}-1:& \qquad   1\,\to\,\Z_{N}\,\to\,\Z_{pN}\,\to\, \Z_{p}\,\to\,1~,
        \\
       \mathrm{Case}-2:& \qquad  1\,\to\,\Z_{2N+1}\,\to\,2\D_{2(2N+1)}\,\to\, \Z_{4}\,\to\,1~,
        \\
        \mathrm{Case}-3:& \qquad  1\,\to\,\Z_{2N}\,\to\,2\D_{2N}\,\to\, \Z_{2}\,\to\,1~,
        \\
        \mathrm{Case}-4:& \qquad 1\,\to\,2\D_{4}\,\to\,2\T\,\to\, \Z_{3}\,\to\,1~,
        \\
        \mathrm{Case}-5:& \qquad 1\,\to\,2\D_{2N}\,\to\,2\D_{4N}\,\to\, \Z_{2}\,\to\,1~, 
        \\
        \mathrm{Case}-6:&  \qquad 1\,\to\,2\T\,\to\,2\O\,\to\, \Z_{2}\,\to\,1 ~.
    \end{split}
\end{equation}
Our notation for the finite $ADE$ subgroups is clarified after equation (\ref{eq:the-6-cases}).

Additionally, we require that these six cases act freely on the link space of the Bryant–Salamon space, following the conditions derived by Cort\'es and V\'asquez in \cite{cortés2014locallyhomogeneousnearlykahler}. In some instances, this necessitates an extra extension by an independent $ADE$ subgroup $A$. These conditions are summarized in Table \ref{Table:list-AA0-dihedral}.

Finally, the action of $\Gamma_{\rm{ns}}$ on the Bryant–Salamon space can be expressed topologically as:
\begin{equation}\label{X7 defn}
X_{7} \ = \ \frac{\,\mathbb{R}^{4}\times\mathbb{S}^{3}_{\mathrm{b}}\,}{\Gamma_{\rm ns}}\ \cong \   \frac{\,(\,\mathbb{R}^{4} /C_{0}\,)\ \times\ \mathbb{S}^{3}_{\mathrm{b}}\,}{B}~,
\end{equation}
where the simultaneous action of $B$ on both components motivates the terminology \emph{non-splittable quotient}.

\paragraph{Physics of the non-splittable quotients.}

The structure of the non-splittable quotient group $\Gamma_{\rm{ns}}$ exerts a profound influence on the associated physics in M-theory geometric engineering. As demonstrated in section \ref{subsec: physics dictionary}, the physical implications of the action of $\Gamma_{\rm{ns}}$ on the Bryant–Salamon space can be systematically classified into three categories:
\begin{itemize}
\item \textbf{Decomposition and many `universes':} The local $ADE$ gauge algebra $\mk{g}$ is determined by the corresponding finite $ADE$ subgroup $C_{0}$. This is because the codimension four singularity, which engineers the gauge theory, is given by $C_{0}$, while $B$ acts freely on the base 3-sphere.

However, this is not the complete picture. As we discuss in section \ref{sec:non-split-section}, the quotient group $\Gamma_{\rm{ns}}$ decomposes into a disjoint union:
\begin{equation}\label{eq:intro-gamma-ns}
    \Gamma_{\rm ns}\,=\,  \, (C_{0},\,1)\, \sqcup\, (\widetilde{\varphi}(b_{1})\cdot C_{0},\, b_{1})\,\sqcup\,\cdots\,\sqcup\, (\widetilde{\varphi}(b_{m-1})\cdot C_{0},\,b_{m-1})~,
\end{equation}
with $\widetilde{\varphi}$ is a homomorphism from $B$ to $C/C_{0}$. This structure implies the existence of $m$ distinct vacua (or `universes'), each characterized by the same local gauge algebra $\mk{g}$. Consequently, the field theory associated with $\Gamma_{\rm{ns}}$ decomposes into a disjoint union of $m$ gauge theories associated with $C_{0}$:
\begin{equation}\label{eq:intro-Theory-decomposition}
\mathcal{T}_{\Gamma_{\mathrm{ns}}}\,=\, \bigsqcup_{j=0}^{m-1}\, \mathcal{T}^{(j)}_{C_{0}}~.
\end{equation}

\item \textbf{Inner automorphisms:} Within this framework, each local gauge theory is naturally accompanied by a massive group-valued scalar field that carries charges under $\pi_{1}(\bbS^{3}_{\mathrm{b}})$. These massive fields, often referred to as flat connections, define homomorphisms from $\pi_{1}(\bbS^{3}_{\mathrm{b}})$ to the gauge group $G$, which are identified with inner automorphisms of $G$. Physically, these flat connections break the gauge algebra to a proper sub-algebra while preserving its rank \cite{Witten:1985xc,Acharya:2020vmg,Najjar:2022eci}.

\item \textbf{Outer automorphisms:} Monodromies of the crepant resolution of the $ADE$ singularities around $\pi_{1}(\bbS^{3}_{\mathrm{b}})$ define outer automorphisms of the Lie algebra $\mk{g}$ \cite{Aspinwall:1996nk,Bershadsky:1996nh,Witten:1997kz,Vafa:1997mh}. Specifically, these monodromies act by permuting the vanishing 2-cycles of the crepant resolution, and, via the McKay correspondence, induce permutations of the nodes of the associated Dynkin diagram (DD), thereby realizing DD automorphisms:
\begin{equation}\label{mon on DDADE}
    \mathrm{Mon}\,:\, \pi_{1}(\mathbb{S}^{3}_{\mathrm{b}}/B) \ \to \  \mathrm{Aut}(\mathrm{DD}_{\scriptscriptstyle ADE})\,.
\end{equation}
Quotienting by such outer automorphisms |also referred to as \emph{folding} or \emph{twisting}|reduces the $ADE$ gauge Lie algebra $\mk{g}$ to a non-simply laced Lie algebra $BCFG$ of lower rank. 
\end{itemize}

\paragraph{Non-splittable quotients and (co)homology of the link.} For each one of the non-splittable cases, one is interested in computing the (co)homology ring of the corresponding link space $L_6$ given by:
\begin{equation}
  L_6 \,  = \, \frac{\mathbb{S}^{3}_{\mathrm{f}}\times\mathbb{S}^{3}_{\mathrm{b}}}{\Gamma_{\rm ns}}\ \cong \   \frac{\,(\,\mathbb{S}^{3}_{\mathrm{f}} /C_{0}\,)\,\times\,\mathbb{S}^{3}_{\mathrm{b}}\,}{B}~.
\end{equation}
As we will see in more detail in later sections, the calculation of these (co)homology groups is not straightforward. This is due to the synchronized diagonal action of the group $B$ on both 3-sphere factors, as can be seen from \eqref{eq:intro-gamma-ns}.

Yet, motivated by the decomposition of $\Gamma_{\rm ns}$ that we discussed above, along with some physical intuition, in section \ref{sec:SymTFT}, we give a proposal for how to deal with this obstacle, and then we review how to use K\"unneth theorem to work out the full $H_\bullet(L_6, \Z)$. The final result is:
\begin{equation}\label{hom of L6}
    H_{\bullet}(L_{6},\Z) \,=\,  \Z \,\oplus\, \begin{matrix}B^{\rm ab}\\\oplus \\\widetilde{\Gamma}^{\mathrm{ab}}_{\mathrm{ext}}\end{matrix}\,\oplus\,  {\rm Tor}_1^{\Z}(B^{\rm ab}, \widetilde{\Gamma}^{\mathrm{ab}}_{\mathrm{ext}})  \oplus\, \begin{matrix}\Z^2\\\oplus\\ {\rm Tor}_1^{\Z}(B^{\rm ab}, \widetilde{\Gamma}^{\mathrm{ab}}_{\mathrm{ext}}) \end{matrix}\,\oplus\, \begin{matrix}B^{\rm ab}\\\oplus\\ \widetilde{\Gamma}^{\mathrm{ab}}_{\mathrm{ext}}\end{matrix}\, \oplus\, 0\,\oplus\, \Z~.
\end{equation}
Here, $\widetilde{\Gamma}^{\rm ab}$ is the abelianisation of the group $\Gamma$, and $\widetilde{\Gamma}_{\rm ext}$ is the actual group that acts of the fibered factor $\mathbb{S}^3_{\rm f}$. This is defined as an extension of the group $C_0$ by $B$ that is dictated by Goursat's theorem. As for the Tor$_1^{\Z}(\bullet, \bullet)$, this is the Tor-functor whose properties we review in subsection \ref{sec:co-homology}

\paragraph{Reducing M-theory action and 5d SymTFTs.} Knowing the explicit form of the homology ring of the link space, the next step is to study the reduction of the 11d M-theory action on $L_6$. This action is of the decomposed as:
\begin{equation}
            S_{\rm tot}^{\rm M}\, = \,S_{\rm CS}^{\rm M}\,+\,S_{\rm kin}^{\rm M}~,
\end{equation}
where the first term contains topological information and the second one governs the dynamics. When reducing this full action, and to keep track of the torsional information appearing in the explicit form of the homology groups of $L_6$ (as shown above), we use the tools of differential cohomology, which we also review in section \ref{sec:SymTFT}. 

As argued earlier in the introduction, the resulting theory is the 5d SymTFT of the geometrically engineered 4d $\mathcal{N}=1$ gauge theory.  In general, this action decomposes into the following form:
\begin{equation}
		S_{\rm SymTFT}^{L_6} = S_{\rm twist}^{L_6} \,+\,S_{\rm BF}^{L_6}~.
\end{equation}
The first term comes from the reduction of $S_{\rm CS}^{\rm M}$ and detects any potential (mixed) 't Hooft anomalies that can be exhibited in the 4d theory. As for the second term, this contains BF terms that couple all possible higher-form symmetries in the 4d theory with their duals in the sense that will be more clear later in section \ref{sec:charged top defects}.

As immediate applications of these reduction results, we will explore the following two aspects:

\begin{itemize}
\item \textbf{Universe decomposition of 5d SymTFT.} As we will discuss in detail in subsection \ref{sec:local-observer}, in terms of our universe decomposition proposal, the full 5d SymTFT theory can be decomposed as follows:
\begin{equation}
		{\rm SymTFT}_{\rm full}\,=\, \bigoplus_{j=0}^{m-1} {\rm SymTFT}^{(j)} \, \oplus\, {\rm SymTFT}_{\rm global}~.
\end{equation}
Here, the LHS is the full SymTFT whose action is $S_{\rm SymTFT}^{L_6}$. This decomposes into contributions SymTFT$^{(j)}$ from each one of the $m$ universes (here we are taking $B\cong\Z_m$), and SymTFT$_{\rm global}$ is the collection of global information that the local observers cannot perceive.

One can define a projection operator $\pi_j$ such that:
\begin{equation}
\pi^{(j)}({\rm SymTFT}_{\rm full}) = {\rm SymTFT}^{(j)}~.
\end{equation}
These maps forget the global information contained in SymTFT$_{\rm global}$. More explicit details on the action of these projection operators on the different differential characters appearing in the SymTFT action are given in subsection \ref{sec:local-observer}.

\item \textbf{Charged defects and symmetry operators.} For all the possible (discrete and continuous) higher-form symmetries that can be present in the 4d theory, in section \ref{sec:charged top defects}, we construct the topological operators generating these symmetries. Geometrically, this is done by wrapping branes on homology cycles along the link space. For discrete symmetries, we wrap M2 or M5 branes on torsional cycles in $L_6$. Meanwhile, for the continuous cases, we wrap $P_4$ or $P_7$ fluxbranes (Page charges) along free cycles in $L_6$. In this construction, we closely follow section 4 of \cite{Najjar:2024vmm} where they study 4d gauge theories with gauge algebra $\mathfrak{su}(N)$.

For instance, codimension one defects (or interfaces) arise from M5-branes wrapping torsional 2-cycles of order $m$ in the link space and extend to the zero-section. These torsional cycles emerge naturally as a direct consequence of the non-splittable quotients considered in this work. Physically, such interfaces admit a natural interpretation as domain walls with infinite tension separating the $m$ distinct vacua (or `universes') described around (\ref{eq:intro-Theory-decomposition}).

\end{itemize}

\paragraph{Outer automorphisms, foldings, and SymTFTs.} Resolving the codimension-4 singularity in \eqref{X7 defn} amounts to adding vanishing 2-cycles whose intersection numbers are dictated by the Cartan matrix of the resulting $ADE$ Lie algebra $\mathfrak{g}$. This is none other than the McKay correspondence \cite{mckay} (which we review in more detail in section \ref{subsec: physics dictionary}), which asserts that:
\begin{equation}
\mathrm{DD}_{\scriptscriptstyle ADE} \,\cong\, H_2(\widetilde{X}_{7},\mathbb{Z})~,
\end{equation}
where $\widetilde{X}_7$ is the resolved space. As a result, the monodromy action \eqref{mon on DDADE} can be lifted to an action on these 2-cycles. 

On the other hand, one can consider the following exact sequence in relative homology:
\begin{equation}
H_{2}(\widetilde{X}_{7},\Z)\,\xrightarrow[]{g}\,H_{2}(\widetilde{X}_{7},L_{6},\Z)\,\xrightarrow[]{f}\, \mathrm{Tor}H_{1}(L_{6},\Z)\,\to\,0~,
\end{equation}
where the link space $L_6$ is viewed as the boundary of $\widetilde{X}_7$ and the torsional part of $H_1(L_6,\Z)$ can be read off from \eqref{hom of L6} above. In subsection \ref{sec:Tor-outer-folding}, we argue that, as a result of this sequence and the McKay correspondence statement above, the monodromy action on $H_2(\widetilde{X}_7, \Z)$ can be extended to a particular subgroup of Tor$H_1(L_6,\Z)$. Namely, it extends to the $C_0^{\rm ab}$-valued torsional $1$-cycles of the link space. Our argument uses the geometric engineering fact that the center of the gauge group $\mathcal{Z}(G)\cong C_0^{\rm ab}$. 

With this in mind, one can perform the folding we discussed above around \eqref{mon on DDADE} at the level of this subsector of torsional 1-cycles. We do this in detail in subsection \ref{sec:folding-center-torH1}. We find that the center of the $ADE$ gauge groups are folded into the centers of their non-simply laced counterparts as expected. Moreover, we find that the universe decomposition discussed above is preserved upon folding.

As an application for this, in subsection \ref{subsec: new 5d SymTFT}, we discuss aspects of the 5d SymTFTs of the resulting 4d gauge theories with non-simply laced gauge algebras, which we claim to be one of the novel results of our work. We discuss the modifications to the higher-form symmetries in the new theories and the different coupling constants that explicitly appear in their 5d actions.

\paragraph{Physical implications of non-splittable quotients.} Based on our results in sections \ref{sec:SymTFT}--\ref{subsec: symtft outer}, in section \ref{sec:phy-implications}, we discuss some physical aspects of the 4d $ADE$ gauge theories we obtained in section \ref{sec:ADE-gauge-theories} and their corresponding non-simply laced obtained by foldings. Our discussion focuses on three main themes:
\begin{itemize}
\item \textbf{Mixed 't Hooft anomalies.} In subsection \ref{subsec: mixed t hooft}, after reviewing the periodicity of the Yang--Mills theta angle $\theta_{\rm YM}$ and its couplings to the 4d action, we interpret $\theta_{\rm YM}$ as a background $0$-form gauge field of the Chern--Weil $(-1)$-form symmetry $U(1)^{[-1]}$ following \cite{Cordova:2019uob,Vandermeulen:2022edk,Santilli:2024dyz,Aloni:2024jpb,Najjar:2024vmm,Najjar:2025htp}.

Coupling the 4d theory to a $2$-form background gauge field $B_2^{\rm e}$ of the electric $1$-form symmetry $\mathcal{Z}(G)^{[1]}$, the periodicity of $\theta_{\rm YM}$ gets modified. At the level of the partition function, this can be detected as:
\begin{equation}
    \frac{Z^{G}[\theta_{\text{YM}}+2\pi,B_{2}^{\mathrm{e}}]}{Z^{G}[\theta_{\text{YM}},B_{2}^{\mathrm{e}}]}   \, = \, \exp(2\pi i \,{\Phi_{G}}\,\int_{M_4}\frac{\mathcal{P}(w_2)}{2})~.
\end{equation}
Here, $\mathcal{P}(w_2)$ is the Pontryagin square of the second Stiefel--Whitney class $w_2\sim B_2^{\rm e}$ of the $G/\mathcal{Z}(G)$-bundle. This indicates a mixed $U(1)^{[-1]}$--$\mathcal{Z}(G)^{[1]}$ 't Hooft anomaly measured by the factor $\Phi_G$. For earlier results on this, see \cite{Cordova:2019uob} and references therein.

In our work here, following \cite{Najjar:2024vmm,Najjar:2025htp}, we re-derive these anomaly factors by identifying the anomaly-inform action with the following particular term that we get in our 5d SymTFT action:
\begin{equation}\label{eq:intro-anomaly-inflow}
    \mathrm{CS}_{C_{0}}\, \int_{Y_5} \frac{F_{1}^{\rm{b}}}{2\pi}\smile B_{2}^{\rm{e}}\smile B_{2}^{\rm{e}}~.
\end{equation}
The coupling constant is the 3d Chern--Simons invariant associated with the finite $ADE$ subgroup $C_0$--See around \eqref{CS invariants: def} for its definition and computation. Moreover, $F_1^{\rm b}\in H^1(Y_5, \Z)$ is a 1-form background field strength for $U(1)^{[-1]}$. More explicitly, we have the following identification:
\begin{equation}
    F_1^{\rm b}\quad \longleftrightarrow\quad \dd\theta_{\rm YM}~.
\end{equation}

Using the results we obtained in table \ref{tab: CS invariants for folded g} for the effect of folding on the CS invariants, we also compute the anomaly factor in the non-simply laced 4d gauge theories. We find that it also matches the results summarized in table of \cite{Cordova:2019uob}.

As we will also discuss in section \ref{sec:phy-implications}, another term of interest that appears in the 5d SymTFT action is:
\begin{equation}
            \int_{Y_5}  \frac{F_1^{\rm b}}{2\pi} \, \wedge\, \frac{\widetilde{h}^{\rm f}_4}{2\pi}~.
\end{equation}
Similar to the interpretation we gave for the 1-form $F_1^{\rm b}$, the 4-form $\widetilde{h}_4^{\rm f}$ is interpreted as the field strength of the continuous 2-form symmetry $U(1)^{[2]}$ dual to $U(1)^{[-1]}$ discussed above. With this in mind, we identify the above term of the action with the standard topological term for $\theta_{\rm YM}$ upon which we have:
\begin{equation}
        \widetilde{h}_4^{\rm f}\quad \longleftrightarrow \quad \tr\,\mathsf{F}\,\wedge\,\mathsf{F}~.
\end{equation}
Here, $\mathsf{F}$ is the field strength of the $G$ Yang--Mills theory.

\item \textbf{Modified instanton sums.} As we discuss in section \ref{sec:charged top defects}, wrapping $P_7$ fluxbrane around the link space $L_{6}$ gives us a `universal' continuous 2-form symmetry $U(1)^{[2]}$. Inspired by the earlier results of \cite{Tanizaki:2019rbk, Najjar:2024vmm,Najjar:2025htp}, in subsection \ref{sec:mod-inst-sum}, we consider gauging a finite subgroup $\Z_K^{[2]}\subset U(1)^{[2]}$ of this symmetry. We find that, upon projecting on the 4d topological boundary, we get the following equation of motion:
\begin{equation}\label{eq:intro-mod-inst-sum}
    \frac{1}{8\pi^{2}} \tr(\mathsf{F}\,\wedge\,\mathsf{F}) \,=\, \frac{K}{2\pi}\,\dd c_{3} ~.
\end{equation}
Here, $c_3$ is the 3-form $\Z_K^{[2]}$ gauge field. Note that the LHS is none other than the instanton number density. Integrating both sides over the 4-manifold, we find a modification on the instanton sectors of the theory: Instead of being $\Z$-valued, they become $(K\,\Z)$-valued. Furthermore, the periodicity of $\theta_{\rm YM}$ gets modified as well, such that, instead of being $2\pi$ periodic, it becomes $\frac{2\pi}{K}$.

At the end of the subsection, we argue that this process can be applied as well after folding to get a modification for the instanton counting in the non-simply laced gauge theories. Our main argument is that the universal $2$-form symmetry $U(1)^{[2]}$ is preserved after the folding.

\item \textbf{Higher 4-group structure.} As observed in \cite{Tanizaki:2019rbk}, for $\mk{su}(N)$ gauge theories, further gauging of the electric 1-form symmetry $\Z_{N}^{[1]}$ modifies the EOM in (\ref{eq:intro-mod-inst-sum}) to include the anomaly inflow in (\ref{eq:intro-anomaly-inflow}). This modification introduces fractional instantons, rendering the equation of motion inconsistent. The inconsistency is resolved by introducing a gauged finite 3-form symmetry, which naturally leads to the emergence of a 4-group structure.  

Motivated by the M-theory derivation of the 4-group structure in \cite{Najjar:2024vmm}, in section \ref{subsec:4group} we extend this framework to encompass $\mk{su}(N)$, $\mk{so}(8)$, $\mk{so}(2N)$, and $\mk{e}_{6}$ gauge theories arising from the cases listed in (\ref{eq:intro-6-cases}), as well as gauge theories with Lie algebras $\mk{so}(2N+1)$ and $\mk{sp}(2N)$ obtained by the folding procedure. This is possible because the class of non-splittable quotients considered in this work naturally admits a finite 3-form symmetry, denoted by $\Z_{m}^{[3]}$.

Specifically, we demonstrated that one can consistently derive |from M-theory| the following TQFT sector:
\begin{equation}
\begin{split}
S_{\text{TQFT}}^{\,G/\Z_{m}^{[2]}\times\Z_{n}^{[1]}\widetilde{\times} \Z_{m}^{[3]}} \,&=\, \int_{M_{4}}\,\,\left[\, \frac{\theta_{\text{YM}}}{8\pi^{2}}\,\tr((\widetilde{\mathsf{F}}-b_{2})^{2})\,+\, \frac{\theta^{(2)}}{2\pi}\,(\dd c_{3}\,-\,a_{4})\,\right]\\
&-\int_{M_4}\chi\,\wedge\,\left[\,\frac{1}{8\pi^{2}}\tr(\widetilde{\mathsf{F}}\,\wedge \,\widetilde{\mathsf{F}}) -\frac{\dd a_{3}}{2\pi}\,-\, \frac{m}{2\pi}\,\dd c_{3}\,\right]\,,
\end{split}
\end{equation}
which applies universally to all $\mk{a}_{N}$, $\mk{b}_{N}$,$\mk{c}_{N}$, $\mk{d}_{N}$ and $\mk{e}_{6}$ gauge theories we obtained in section \ref{sec:geo-eng-4d}. Here, $\widetilde{\mathsf{F}}$ is the extended field strength of the $G$ Yang--Mills theories, which we introduce around (\ref{eq:tr-tilde-F=nB2e}). $b_{2}$ is the gauge field for the gauged electric 1-form symmetry $\Z_{n}^{[1]}$, $(a_{4},a_{3})$ are a pair of $U(1)$-valued gauge fields associated with the finite 3-form symmetry $\Z_{m}^{[3]}$, and $\chi$ is a 0-form background field seen as a Lagrange multiplier.

\end{itemize}

\section{Geometric engineering of 4d \texorpdfstring{$\N = 1$}{N=1} SYM theories}\label{sec:geo-eng-4d}

In this section, we review key properties and constructions of the Bryant--Salamon (BS) $G_{2}$-manifold, which is the spin bundle over $\mathbb{S}^{3}$ with link space $\mathbb{S}^{3}\times\mathbb{S}^{3}$. We then turn to the study of its quotients by non-splittable or semi-splittable $ADE$ finite subgroups of $SU(2)$. In particular, we restrict our attention to finite groups acting freely on the link space. As already discussed in the introduction, we will focus on six distinct cases of such quotients.

We further examine the physical interpretation of these quotient spaces in M-theory, including the associated automorphism structures. The physics of the six distinct cases is discussed in detail.  We conclude the section with remarks on the M-theory physics of general non-splittable and semi-splittable quotients of $G_{2}$ spaces of B7-type. 

\subsection{The Bryant--Salamon space and its symmetries}\label{subsec:BS and symmetries}

The  Bryant--Salamon (BS) space is a non-compact, cohomogeneity-one manifold with $G_{2}$-holonomy, exhibiting a cone-like asymptotic structure. Before delving into the specific details of the BS space, we begin by reviewing some general properties of cohomogeneity-one spaces.

\paragraph{Cohomogeneity-one manifold.} 

In geometric engineering, we are often interested in conical Riemannian spaces of the form $X\simeq[0,\infty)\times L$, where $L$ denotes the link space (or principal orbit) of the cone, situated at infinity. Mathematically, such a space $X$ is a cohomogeneity-one space: There exists an isometry group $G$ whose action on $X$ produces codimension-one orbits so that the quotient $X/G$ reduces to the half-line $[0,\infty)$ \cite{Cvetic:2001zx,Acharya_2004}.

A cohomogeneity-one space $X$ is generally characterized by group data consisting of an isometry group $G$ and a pair of isotropy subgroups $H$ and $K$ that satisfy the inclusion chain \cite{Cvetic:2001zx,Acharya_2004,Foscolo:2018mfs}:
\begin{equation}
    K\,\subset\, H\,\subset\, G\,.
\end{equation}
The codimension-one orbits over the open interval $(0,\infty)$ are diffeomorphic to the link space $L=G/K$. In contrast, the orbit over $0$ has codimension greater than one and is referred to as the zero-section (or singular orbit), which is given by $S:=G/H$. Furthermore, the quotient $H/K$ must be diffeomorphic to a sphere $\mathbb{S}^{n}$. Consequently, the link space $L=G/K$ admits the structure of an $\mathbb{S}^{n}$-bundle over the zero-section $S$. This implies the total space $X$ is topologically an $\R^{n+1}$-bundle over $S$.

\paragraph{The BS space and its symmetries.}

Topologically, the BS space is expressed as $\R^{4}\times \mathbb{S}^{3}$ with a link $\mathbb{S}^{3}\times \mathbb{S}^{3}$ \cite{Bryant1989OnTC,Gibbons:1989er}. The space can be expressed effectively by the following group data (see e.g. \cite{Gibbons:1989er,Atiyah:2001qf,Acharya_2004}):
\begin{equation}
    \Delta SU(2) \,\subset\, SU(2)\,\times\, SU(2) \,\subset\, SU(2)\,\times\, SU(2)\,\times\, SU(2)\,.
\end{equation}
Here, $\Delta SU(2)$ is the diagonally embedded subgroup of $G=SU(2)\times SU(2)\times SU(2)$. The link space $\mathbb{S}^{3}\times \mathbb{S}^{3}$ can be defined through unit quaternions $\mathsf{g}_{i},\, i=1,2,3$. Explicitly, it is given by the following identification:
\begin{equation}\label{eq:def-link-g123h}
    L\,=\, \frac{(SU(2))^{3}}{\Delta SU(2)}\,\,:\, \quad (\mathsf{g}_{1},\mathsf{g}_{2},\mathsf{g}_{3})\,\sim\, (\mathsf{g}_{1}\mathsf{h},\mathsf{g}_{2}\mathsf{h},\mathsf{g}_{3}\mathsf{h})~,
\end{equation}
with $\mathsf{h}$ being a unite quaternion that corresponds to $\Delta SU(2)$.

The spin bundle over $\mathbb{S}^{3}$ with topology $\R^{4}\times \mathbb{S}^{3}$ can be found in the following way \cite{Atiyah:2001qf,Foscolo:2018mfs}:
\begin{itemize}
    \item \textbf{Gauge a particular $\mathsf{g}_{i}$.} Without loss of generality, we gauge $\mathsf{g}_{3}$. The gauging is performed by taking $\mathsf{h}=\mathsf{g}^{-1}_{3}=\Bar{\mathsf{g}}_{3}$. Therefore, the link space can be effectively described by two unit quaternions $(\mathsf{q}_{1},\mathsf{q}_{2})$ defined simply by:
\begin{equation}\label{eq:q1q2-g1g2}
     L\,\,:\,\, (\mathsf{q}_{1},\mathsf{q}_{2})\,\, \mapsto\,\, (\mathsf{g}_1\overline{\mathsf{g}}_3, \mathsf{g}_2\overline{\mathsf{g}_3},1)\, \sim\, (\mathsf{g}_1, \mathsf{g}_2, \mathsf{g}_3)~. 
\end{equation}

    \item \textbf{Filling-in a particular $\mathbb{S}^{3}$.} The cone space, $[0,\infty)\times \mathbb{S}^{3}_{1}\times \mathbb{S}^{3}_{2}$ can be mapped to $\R^{4}\times \mathbb{S}^{3}_{\rm b}$ via \cite{Foscolo:2018mfs}:
    \begin{equation}\label{eq:def-space-R4S3}
\begin{aligned}
        \left[ 0,\infty \right) \times \mathbb{S}^{3}_{1} \times \mathbb{S}^{3}_{2}  \ & \rightarrow \ \R^4 \times \mathbb{S}_{\mathrm{b}}^3 \\
        (t, \mathsf{q}_1, \mathsf{q}_2 ) \ & \mapsto \ (t \mathsf{q}_1, \mathsf{q}_1 \bar{\mathsf{q}}_2 )\,.
\end{aligned}
\end{equation}
The above map accounts for filling in the 3-sphere parametrized by $\mathsf{q}_{1}$. The zero-section of the 7d geometry is given by $\mathbb{S}^{3}_{\mathrm{b}}$ parametrized by $\mathsf{q}_{1}\Bar{\mathsf{q}}_{2}$. We denote the above 7-dimensional $G_{2}$-holonomy manifold by $X_{(1;3)}$. This notation indicates that the first 3-sphere, associated with $\mathsf{g}_{1}$, is filled in, while the third 3-sphere, corresponding to $\mathsf{g}_{3}$, is subject to gauging.

The above map can be taken in two steps: First, we have a ``twist" map on the 3-spheres:
\begin{equation}\label{eq:twist-psi}
    \begin{split}
        \psi\ &:\   \mathbb{S}^{3}_{1}\,\times\,\mathbb{S}^{3}_{2}\ \to\ \,   \mathbb{S}^{3}_{\mathrm{f}}\,\times\,\mathbb{S}^{3}_{\mathrm{b}}
        \\
        \ &:\  (\mathsf{q}_1, \mathsf{q}_2 ) \ \, \  \mapsto \ (\mathsf{q}_1, \mathsf{q}_1 \bar{\mathsf{q}}_2 )\,.
    \end{split}
\end{equation}
Here, $\mathbb{S}^{3}_{\mathrm{f}}$ is the fibered 3-sphere over the base 3-sphere $\mathbb{S}^{3}_{\mathrm{b}}$. Second, we fill-in $\mathbb{S}^{3}_{\mathrm{f}}$ as:
\begin{equation}
     (t,  \mathsf{q}_1, \mathsf{q}_{1} \bar{\mathsf{q}}_{2}  ) \ \mapsto \ (t \mathsf{q}_1, \mathsf{q}_1 \bar{\mathsf{q}}_2 ) \,.
\end{equation}
Therefore, one can regard $\mathbb{S}^{3}_{\mathrm{f}}\times \mathbb{S}^{3}_{\mathrm{b}}$ as the link space of $X_{(1;3)}$.
\end{itemize}

Note that, with the above process, one obtains 6 equivalent copies of $\R^{4}\times \mathbb{S}^{3}$ denoted by $X_{(i;j)}$ with $i\neq j$ and $i,j=1,2,3$. The notation $X_{(i;j)}$ means filling-in $\mathsf{g}_{i}$ and gauging $\mathsf{g}_{j}$.  

The isometry of the BS spin bundle over $\mathbb{S}^{3}_{\mathrm{b}}$ is \cite{Foscolo:2018mfs,Acharya:2021jsp}:
\begin{equation}\label{eq:BS-isometry-(1)}
    G^{\mathrm{BS}}\,=\, \frac{SU(2) \times SU(2) \times SU(2)}{\Delta\Z_{2}} \,,
\end{equation}
with $\Delta\Z_{2}$ is the diagonally embedded center in $(SU(2))^{3}$. A generic element $(a,b,c)\in (SU(2))^{3}$ acts on a point $(\mathsf{g}_{1},\mathsf{g}_{2},\mathsf{g}_{3})\in L$ via:
\begin{equation}\label{eq:action-abc-ggg}
  (a,b,c)\cdot  (\mathsf{g}_{1},\mathsf{g}_{2},\mathsf{g}_{3})\,=\, (a\,\mathsf{g}_{1},b\,\mathsf{g}_{2},c\,\mathsf{g}_{3})~.
\end{equation}
On the $\mathbb{S}_{1}^{3}\times\mathbb{S}^{3}_{2}$, the above action is given through (\ref{eq:q1q2-g1g2}):
\begin{equation}
    (a,b,c)\cdot (\mathsf{q}_{1},\mathsf{q}_{2})\,=\,(a\,\mathsf{q}_{1}\,c^{-1},\,b\,\mathsf{q}_{2}\,c^{-1})~.
\end{equation}

As for the twisted link space of $X_{(1;3)}=\R^{4}\times\mathbb{S}^{3}_{\mathrm{b}}$, the action of the isometry by an element $(a,b,c)$ is given by:
\begin{equation}\label{eq:action-abc-S3fS3b}
    (a,b,c)\cdot (\mathsf{q}_1, \mathsf{q}_1 \bar{\mathsf{q}}_2 )\,=\, (a\,\mathsf{q}_{1}\,c^{-1},a\, \mathsf{q}_1 \bar{\mathsf{q}}_{2}\,b^{-1} )~.
\end{equation}
In particular, this defines the isometry action on the zero-section $\mathbb{S}^{3}_{\mathrm{b}}$. The isometry group for the space $X_{(1;3)}$ can be labeled as:
\begin{equation}\label{eq:BS-isometry}
    G^{\mathrm{BS}}\,=\, \frac{SU(2)_{\ell,1}\times SU(2)_{\ell,2}\times \Delta SU(2)_{r}}{\Delta\Z_{2}} ~.
\end{equation}
The above isometry group can be understood from the structure of (\ref{eq:twist-psi}) (and (\ref{eq:action-abc-S3fS3b})). In particular, we know that the isometry of each $\mathbb{S}_{i}^{3}$ is given by $SU(2)_{\ell\,,i}\times SU(2)_{r\,,i}$. However, one can verify that only the diagonal subgroup $ \triangle SU(2)_{r} \subset  SU(2)_{r,1} \times SU(2)_{r,2}$ preserves (\ref{eq:twist-psi}); hence, one arrives at (\ref{eq:BS-isometry}).

\paragraph{Triality symmetry and classical branches.}

The metric on the link space $\mathbb{S}_{1}^{3}\times\mathbb{S}^{3}_{2}$ has finite isometry known as the triality symmetry $\Sigma_{3}$ \cite{Atiyah:2001qf}. Its action on the $(\mathsf{g}_{1},\mathsf{g}_{2},\mathsf{g}_{3})$ quaternions is given by permutation:
\begin{equation}
    \begin{split}
        \alpha\,&:\, (\mathsf{g}_{1},\mathsf{g}_{2},\mathsf{g}_{3})\ \to \  (\mathsf{g}_{2},\mathsf{g}_{3},\mathsf{g}_{1})\quad \mathrm{with}\quad \alpha^{3}=1\,,
        \\
        \beta_{(1)}\,&:\, (\mathsf{g}_{1},\mathsf{g}_{2},\mathsf{g}_{3})\ \to \ ({\mathsf{g}}_{3},{\mathsf{g}}_{2},{\mathsf{g}}_{1}) \quad \mathrm{with}\quad \beta_{(1)}^{2}=1\,.
    \end{split}
\end{equation}
The generators $\alpha$ and $\beta_{(1)}$ satisfy $\beta_{(1)}\alpha\beta_{(1)}=\alpha^{-1}$. The other two order-2 elements of $\Sigma_{3}$ can be obtained by conjugation of $\beta_{(1)}$ by $\alpha$:
\begin{equation}
    \alpha\beta_{(i)}\alpha^{-1}\,=\, \beta_{(i+1)}~,\qquad  \text{for}~~ i=1,2~.
\end{equation}
It follows that $\Sigma_{3}$ acts by permutations on the 7d spaces $X_{(i;j)}$. On top of this, one can also perform a complex conjugation:
\begin{equation}\label{complex cojugation}
    \mathrm{C}\ :\  (\mathsf{g}_{1},\mathsf{g}_{2},\mathsf{g}_{3})\ \to (\bar{\mathsf{g}}_{1},\bar{\mathsf{g}}_{2},\bar{\mathsf{g}}_{3})\,.
\end{equation}

Recall that, the $(\mathsf{q}_{1},\mathsf{q}_{2})$ unit quaternions that describe $\mathbb{S}_{1}^{3}\times\mathbb{S}_{2}^{3}$ can be used after gauging one of the $\mathsf{g}_{i}$. The triality symmetry $\Sigma_{3}$ action on these quaternions is given by \cite{Foscolo:2018mfs}:
\begin{equation}
\begin{split}
        \gamma \ : \ (\mathsf{q}_{1}, \mathsf{q}_{2})\ \to \ ( \bar{\mathsf{q}}_{1}, \mathsf{q}_{2}\bar{\mathsf{q}}_{1})\,, \qquad 
        \delta \ : \  (\mathsf{q}_{1}, \mathsf{q}_{2})\ \to \ (\mathsf{q}_{1}\bar{\mathsf{q}}_{2}, \bar{\mathsf{q}}_{2})\,.
\end{split}
\end{equation}
One can verify that $(\gamma\delta)^{3}=(\delta\gamma)^{3}=1$, and the operators $\mathrm{f} \equiv \delta\gamma\delta =\gamma\delta\gamma$ exchanges the pair $(\mathsf{q}_{1}, \mathsf{q}_{2})$:
\begin{equation}
    \mathrm{f}\ :\ (\mathsf{q}_{1}, \mathsf{q}_{2}) \ \to \ (\mathsf{q}_{2}, \mathsf{q}_{1})~.
\end{equation}
In terms of the quaternions $\mathsf{q}_{1,2}$, the complex conjugation \eqref{complex cojugation} becomes:
\begin{equation}
    \mathrm{C}\ : \  (\mathsf{q}_{1}, \mathsf{q}_{2}) \ \to \  (\bar{\mathsf{q}}_{1}, \bar{\mathsf{q}}_{2})\,.
\end{equation}
Moreover, we can define the following ``twist'' map:\footnote{One can also define another twist map:
\protect{\begin{equation}
    \widetilde{\psi}\,\equiv\, \mathrm{f}\cdot\mathrm{C} \cdot \gamma~.
\end{equation}}
But this does not play a role in our work here.
}
\begin{equation}
    \psi \,\equiv\,\mathrm{C} \cdot \gamma~.
\end{equation}
And indeed, this is the one we used in (\ref{eq:twist-psi}).

Note that taking quotients of the link space (and the BS geometry), by finite subgroups of the isometry, breaks the triality symmetry to subgroup(s) of $\Sigma_{3}$. For a generic quotient, this symmetry is completely broken, leaving only the trivial subgroup.

Moreover, a $G_{2}$ flop on the $X_{(1;3)}$ space can be obtained by filling in the second entry of (\ref{eq:def-space-R4S3}) instead of the first. This flop geometry is connected to $X_{(1;3)}$ and other phases via a moduli space (see, e.g., \cite{Atiyah:2001qf}). However, we emphasize that in this work, we do not explore the full moduli space associated with the BS space. Instead, we focus on specific components that give rise to non-abelian gauge theories, as will be demonstrated below.

\subsubsection{Non-splittable and semi-splittable quotients}\label{sec:non-split-semi-split}

We consider quotients of the BS space by freely acting finite subgroups $\Gamma_{\rm tot}\subset G^{\mathrm{BS}}$. Generically speaking, one may consider non-freely acting finite subgroups $\Gamma_{\rm tot}$ on the link space. However, the physical interpretation of the fixed point sets in this case is not clear. For further discussion, the reader is invited to consult \cite{Acharya:2023bth}. To obtain freely acting subgroups $\Gamma_{\rm tot}$'s, we follow the partial classification of Cort\'es and V\'asquez \cite{cortés2014locallyhomogeneousnearlykahler}. Since the finite groups are subgroups of the isometry $G^{\mathrm{BS}}$, then the $G_2$-structure of the BS descends to the quotient space $(\R^{4}\times\mathbb{S}^{3})/\Gamma_{\rm tot}$.\footnote{We thank Lorenzo Foscolo for a discussion on this point.}

\paragraph{(Non-)Freely acting subgroup.} Before proceeding to the details of the class of freely acting subgroup $\Gamma_{\rm tot}$ of interest, we discuss the case of having non-free action of $\Gamma_{\rm tot}$ following \cite{cortés2014locallyhomogeneousnearlykahler}.    

Consider the action of $(a,b,c)\in\Gamma_{\rm tot}$ on $(\mathsf{g}_{1},\mathsf{g}_{2},\mathsf{g}_{3})$ subject to the equivalence relation in (\ref{eq:def-link-g123h}):
\begin{equation}
  (a,b,c)\cdot  (\mathsf{g}_{1},\mathsf{g}_{2},\mathsf{g}_{3})\,=\,(a\,\mathsf{g}_{1},b\,\mathsf{g}_{2},c\,\mathsf{g}_{3})\,\sim\,(\mathsf{g}_{1}\,\mathsf{h},\mathsf{g}_{2}\,\mathsf{h},\mathsf{g}_{3}\,\mathsf{h})~.
\end{equation}
Applying $(\mathsf{g}_{1},\mathsf{g}_{2},\mathsf{g}_{3})^{-1}$ from the left, we get:
\begin{equation}
 (\mathsf{g}_{1},\mathsf{g}_{2},\mathsf{g}_{3})^{-1}\, (a,b,c)\,  (\mathsf{g}_{1},\mathsf{g}_{2},\mathsf{g}_{3})\,=\,(\mathsf{h},\mathsf{h},\mathsf{h}) \, \in\, K\,\subset G^{\mathrm{BS}}~.
\end{equation}
As quaternions, this conjugation leaves the real part of $a,b$, and $c$ invariant. Therefore, the subgroup $\Gamma_{\rm tot}$ with non-trivial elements $(a,b,c)$ satisfying:
\begin{equation}\label{equal real parts}
{\rm Re}(a)\,=\,{\rm Re}(b)\,=\,{\rm Re}(c)\,=\,{\rm Re}(\mathsf{h})~,
\end{equation}
acts non-freely on the link space. 

So, to sum up, to have a freely acting group $\Gamma_{\rm tot}$, the set
\begin{equation}\label{eq:the-set-W(Gamma)}
    \mathcal{W}(\Gamma_{\rm tot}) :=\,\{\,{\rm Re(a)}\,\,|\,\, {\rm Re}(a)\,=\,{\rm Re}(b)\,=\,{\rm Re}(c)\,\}~,
\end{equation}
for all elements $(a,b,c)\in \Gamma_{\rm tot}$ should satisfy:
\begin{equation}\label{eq:free-action-condition}
    \mathcal{W}(\Gamma_{\rm tot})\,=\,\{\,1\,\}~.
\end{equation}

Let us now look at the twisted link $\mathbb{S}^{3}_{\mathrm{f}}\times\mathbb{S}^{3}_{\mathrm{b}}$. First, we consider the twist map $\psi$ on the elements $(\mathsf{g}_{1},\mathsf{g}_{2},\mathsf{g}_{3})$ as:
\begin{equation}
    \psi((\mathsf{g}_{1},\mathsf{g}_{2},\mathsf{g}_{3})) \,=\, (\mathsf{g}_{1},\mathsf{g}_{1}\bar{\mathsf{g}}_{2},\mathsf{g}_{3}) \,\sim\, (\mathsf{g}_{1}\,\mathsf{h},\mathsf{g}_{1}\bar{\mathsf{g}}_{2},\mathsf{g}_{3}\,\mathsf{h})\,.
\end{equation}
Proceeding similarly to the above discussion, one finds:
\begin{equation}
\begin{split}
      (a,b,c)\cdot  \psi((\mathsf{g}_{1},\mathsf{g}_{2},\mathsf{g}_{3}))\,&=\,  \psi((a,b,c)\cdot(\mathsf{g}_{1},\mathsf{g}_{2},\mathsf{g}_{3}))
      \\
      \,&=\,(a\,\mathsf{g}_{1},a\,\mathsf{g}_{1}\bar{\mathsf{g}}_{2}\,\bar{b},c\,\mathsf{g}_{3})
      \\
      \,&\sim\, (\mathsf{g}_{1}\,\mathsf{h},\mathsf{g}_{1}\bar{\mathsf{g}}_{2},\mathsf{g}_{3}\,\mathsf{h})\,.
\end{split}
\end{equation}
Applying $(\mathsf{g}_{1},\mathsf{g}_{1}\bar{\mathsf{g}}_{2},\mathsf{g}_{3})^{-1}$ from the left, one gets:
\begin{equation}
(\mathsf{g}_{1},\mathsf{g}_{1}\bar{\mathsf{g}}_{2},\mathsf{g}_{3})^{-1}\, (a,b,c)\,  (\mathsf{g}_{1},\mathsf{g}_{1}\bar{\mathsf{g}}_{2},\mathsf{g}_{3})\,=\,(\mathsf{h},1,\mathsf{h})\,.
\end{equation}
Comparing both sides, we obtain the relations \eqref{equal real parts}. Therefore, we showed that any subgroup $\Gamma_{\rm tot}$ acting freely on the link $\mathbb{S}^{3}_{1}\times \mathbb{S}^{3}_{2}$ survive the twist and act freely on $\mathbb{S}^{3}_{\mathrm{f}}\times\mathbb{S}^{3}_{\mathrm{b}}$.

\subsection{Semi-splittable freely acting finite \texorpdfstring{$ADE$}{ADE} subgroups}\label{sec:non-split-section} 

Let us consider finite subgroup $\Gamma_{\rm tot}$ acting on the link space $\mathbb{S}^{3}_{\mathrm{f}}\times \mathbb{S}^{3}_{\mathrm{b}}$ as given in (\ref{eq:action-abc-S3fS3b}). We also take $\Gamma_{\rm tot} \subseteq A\times B\times C$, with $A$, $B$, and $C$ are non-trivial finite subgroups of $(SU(2))^{3}\cong G^{\rm{BS}}$,\footnote{Here we are ignoring the $\Z_2$ factor appearing in \protect\eqref{eq:BS-isometry-(1)}.} such that a generic element of $\Gamma_{\rm tot}$ is of the form $(a,b,c)\in A\times B\times C$. As discussed in \cite{cortés2014locallyhomogeneousnearlykahler}, the finite group $\Gamma_{\rm tot}$ can be one of the following: 
\begin{itemize}
    \item \textbf{Splittable.} In this case, $\Gamma_{\rm tot} \equiv \Gamma_{\rm ss}= A\times B\times C$, i.e., the three subgroups act independently on the link space. Some examples of this type exist in the physics literature, such as \cite{Friedmann:2002ct,Friedmann:2002ty}.
    \item \textbf{Semi-splittable.} In this case, we consider $\Gamma_{\rm tot} = A\times \Gamma_{\rm ns}$ with $\Gamma_{\rm ns}\subset B\times C$. The embedding of $\Gamma_{\rm ns}$ into the product is constrained by non-trivial relations (homomorphisms) given by \textit{Goursat's theorem}--which we review momentarily. This leads to a correlated structure that prevents a direct product decomposition. Abusing terminology, we will refer to the semi-splittable cases where $A$ can be taken to be trivial as \textbf{non-splittable}. We will denote this case by $\Gamma_{\rm ns}$.
\end{itemize}
In this work, we focus on the latter two cases—namely, the non-splittable finite subgroups and their semi-splittable extensions\footnote{In certain cases, freely acting finite group actions on the link space can only be realized through semi-splittable quotients.}—which yield non-trivial codimension-four $ADE$-type singularities in the BS space and act freely on the associated link space. An example of this is given in \cite{Acharya:2021jsp,Najjar:2022eci}. It is important to note that the general classification of non-splittable finite subgroups lies beyond the scope of this work. Rather, here we will focus on a partial classification that we will review below.

Before proceeding to discuss the finite subgroups of interest, let us recall the definition of Goursat’s theorem following \cite{deMedeiros:2010dn,cortés2014locallyhomogeneousnearlykahler}.

\paragraph{Goursat’s theorem.} Let $G_{1}$ and $G_{2}$ be two groups, then there is a one-to-one correspondence between a subgroup $H\subset G_{1}\times G_{2}$ and a quintuplet:
\begin{equation}\label{eq:goursat-Q(H)}
    \mathcal{Q}(H)\,=\, \{\,A,\,A_{0},\,B,\,B_{0},\,\varphi\,\}~,
\end{equation}
with $A\subseteq G_{1}$ and $B\subseteq G_{2}$, and $A_{0}\subseteq A$ and $B_{0}\subseteq B$ are normal subgroups. Furthermore, $\varphi$ is an isomorphism:
\begin{equation}
    \varphi\,:\, A/A_{0} \,\to\, B/B_{0}~.  
\end{equation}
In terms of this data, the subgroup $H$ can be written as:
\begin{equation}
    H\,=\, \{\,(a,b)\,\in A\times B \, \, |\, \, \varphi(aA_{0})=bB_{0}\,\}\,.
\end{equation}
In particular, one can consider this theorem in modeling a non-splittable subgroup $H$ of $G_{1}\times G_{2}$ via the quintuplet $\mathcal{Q}(H)$.

The subgroups $A$ and $B$ are determined through the projection maps $\pi_{i}:G_{1}\times G_{2}\to G_{i}$ for $i=1,2$. Specifically, the subgroup $A=\pi_{1}(H)\subset G_{1}$: 
\begin{equation}
    A \,=\, \{a\,\in\, G_{1}   \,\,|\,\, \exists \, b\in G_{2} \, \,\text{with}\,\, (a,b)\,\in \, H    \}~.
\end{equation}
The normal subgroup $A_{0}\triangleleft A$ can be written explicitly as:
\begin{equation}\label{eq:def-A0}
    A_{0} \,=\, \{a\,\in\, G_{1}   \,\,\,|\,\, \, (a,1)\,\in \, H \}~,
\end{equation}
where $1$ is the identity in $G_{2}$. Similarly, one can define $B=\pi_{2}(H)\subset G_{2}$ and its normal subgroup $B_{0}$.

\paragraph{Non-splittable quotient subgroup $\Gamma_{\rm ns}$.} As advertised earlier, a class of interesting non-splittable and semi-splittable groups can be obtained through Goursat's theorem of $\Gamma_{\rm ns}\subset B\times C$. We will first consider the non-splittable groups and then discuss the semi-splittable extension.  

Let us consider the action of $\Gamma_{\rm ns}\subset B\times C$ on the link space $\mathbb{S}^{3}_{\mathrm{f}}\times\mathbb{S}^{3}_{\mathrm{b}}$  (\ref{eq:action-abc-S3fS3b}):
\begin{equation}
    (1,b,c)\cdot (\mathsf{q}_1, \mathsf{q}_1 \bar{\mathsf{q}}_2 )\,=\, (\mathsf{q}_{1}c^{-1}, \mathsf{q}_1 \bar{\mathsf{q}}_{2}\,b^{-1} )\,.
\end{equation}
Following the discussion around (\ref{eq:BS-isometry}), recall that $\Gamma_{\rm ns}\subset SU(2)_{\ell,2}\times\Delta SU(2)_{r}$. We specialize to non-splittable finite quotient group where $\Gamma_{\rm ns}$ is modelled by:
\begin{equation}\label{eq:goursat-B0=1}
    \mathcal{Q}(\Gamma_{\rm ns})\,=\, \{\, C,\, C_{0},\, B,\, \{1\} ,\, \varphi   \,\}~.
\end{equation}
That is, we take $B_0 = \{1\}$, the trivial normal subgroup. Hence, we can write the subgroup $\Gamma_{\rm ns}$ as:
\begin{equation}\label{eq:our-general-Gamma-Goursats-defn}
    \Gamma_{\rm ns}\,=\,\{\, (c,b)\,\in\,C\times B \,|\, \varphi(c)\,=\,b \,\}~,
\end{equation}
with $\varphi(c_{0})=1\in B$ for $c_{0}\in C_{0}$.  

In all the cases we will study in this work, and which we will list shortly, $B\cong \Z_m$ for some positive integer $m$. Suppose that $B = \{1,b_{1},\cdots,b_{m-1}\}$, then, it follows from Goursat's theorem that there exists $\widetilde{c}_{j}\in C/C_{0}$ with $j=0,\cdots m-1$ such that: 
\begin{equation}
    \varphi(\widetilde{c}_{j})\,=\, b_{j}~.
\end{equation}
Therefore, the group $\Gamma_{\rm ns}$ can be decomposed as follows:
\begin{equation}\label{eq:our-general-Gamma-Goursats-(1)}
    \Gamma_{\rm ns}\,=\, \, ( C_{0},\,1)\, \sqcup\, (\widetilde{c}_{1}\cdot C_{0},\, b_{1})\,\sqcup\,\cdots\,\sqcup\,  (\widetilde{c}_{m-1}\cdot\,C_{0},\,b_{m-1})~,  
\end{equation}
where here $(\widetilde{c}_j\cdot C_0, b_j) := \{(\widetilde{c}_j\cdot c_0, b_j)~|~~c_0\in C_0\}$.

Since $\varphi$ is an isomorphism, then there exist a homomorphism $\widetilde{\varphi} : B \longrightarrow C$ such that $\varphi\cdot \widetilde{\varphi} = {\rm id}_C$ and $\widetilde{\varphi}\cdot \varphi={\rm id}_{B}$. In particular, we have:
\begin{equation}
   \widetilde{\varphi}(b_{j})\,=\,\widetilde{c}_{j} \,.
\end{equation}
Therefore, the structure of $\Gamma_{\rm ns}$ \eqref{eq:our-general-Gamma-Goursats-(1)} can be equivalently expressed as:
\begin{equation}\label{eq:our-general-Gamma-Goursats}
    \Gamma_{\rm ns}\,=\,  \, (C_{0},\,1)\, \sqcup\, (\widetilde{\varphi}(b_{1})\cdot C_{0},\, b_{1})\,\sqcup\,\cdots\,\sqcup\, (\widetilde{\varphi}(b_{m-1})\cdot C_{0},\,b_{m-1})~.  
\end{equation}

Moreover, it acts on the link space $\mathbb{S}^{3}_{\mathrm{f}}\times\mathbb{S}^{3}_{\mathrm{b}}$ as:
\begin{equation}\label{eq:quotient-Gamma-link-non-splittable}
  \Gamma_{\rm ns} \ :\   (\mathsf{q}_1, \mathsf{q}_1 \bar{\mathsf{q}}_2 ) \  \sim\  (\mathsf{q}_{1}\,(\widetilde{\varphi}(b)c_{0})^{-1} , \mathsf{q}_1 \bar{\mathsf{q}}_{2}\,b^{-1})\,.
\end{equation}
Here, the notion of non-splittable action is understood through the action of $B$ on the link space as: 
\begin{equation}
  B \ :\   (\mathsf{q}_1, \mathsf{q}_1 \bar{\mathsf{q}}_2 ) \  \sim\  (\mathsf{q}_{1}\,(\widetilde{\varphi}(b))^{-1} , \mathsf{q}_1 \bar{\mathsf{q}}_{2}\,b^{-1})\,.
\end{equation}
In other words, the action of $B$, or $\Gamma_{\rm ns}$, is synchronized on the two 3-spheres of the link space. Topologically, we have:
\begin{equation}\label{eq:non-split-on-link}
  L_6 = \frac{\mathbb{S}^{3}_{\mathrm{f}}\times\mathbb{S}^{3}_{\mathrm{b}}}{\Gamma_{\rm ns}}\ \cong \   \frac{\,(\,\mathbb{S}^{3}_{\mathrm{f}} /C_{0}\,)\,\times\,\mathbb{S}^{3}_{\mathrm{b}}\,}{B}~.
\end{equation}

As we will see in the next section, these observations will become essential for us when computing the homology ring of the total link space $L_6$. For that purpose, let us also state here that the sector of $\Gamma_{\rm ns}$ that acts on the fibered 3-sphere, which we will denote as $\widetilde{\Gamma}_{\rm ext}$, can be viewed as an extension:
\begin{equation}
         1\,\to\,C_{0}\,\to\,\widetilde{\Gamma}_{\rm ext}\,\to\,B\,\to\,1~,
\end{equation}
where the extension is dictated by the isomorphism $\varphi$. One can further view this group as the semi-direct product $ \widetilde{\Gamma}_{\rm ext}\,\cong\,C_{0} \,\rtimes_{\widetilde{\varphi}} B$.

\paragraph{Semi-splittable extensions $\Gamma_{\rm tot}$ of quotient groups.}

To construct semi-splittable finite groups, we extend a non-splittable group $\Gamma$ by an additional finite group $A$, 
\begin{equation}
    \Gamma_{\rm ns}\,\longmapsto\, \Gamma_{\rm ss} := A\,\times\, \Gamma_{\rm ns}\,,
\end{equation}
 The group $A$ acts independently on the link space, as described in \eqref{eq:action-abc-S3fS3b}, and there are no constraints or relations between $A$ and $\Gamma_{\rm ns}$. Consequently, generalizing the structure \eqref{eq:our-general-Gamma-Goursats}, the structure of $\Gamma_{\rm ss}$ takes the form:
\begin{equation}\label{eq:general-Gamma-Goursats-semi-split}
   \Gamma_{\rm ss}\,=\, (A,C_{0},\,1)\, \sqcup\, (A,\,\widetilde{\varphi}(b_{1})\cdot\,C_{0},\, b_{1})\,\sqcup\,\cdots\,\sqcup\, (A\,,\widetilde{\varphi}(b_{m-1})\cdot C_{0},\,b_{m-1})~.  
\end{equation}
Here, $(A, \widetilde{\varphi}(b_j)\cdot C_0, b_j) := \{(a,\widetilde{\varphi}(b_j)\cdot c_0, b_j )~|~~a\in A~, ~c_0\in C_0\}$.

\begin{table}[t]
\centering
\begin{tabular}{|c|c|c|c|c|}
\hline
 Case number & $B \cong C/C_{0}$ & $(C,C_{0})$ & Required $A$ & Conditions  \\
\hline
$\mathrm{Case}-1$ & $\Z_{p}$ & $(\Z_{pN},\Z_{N})$  &  --- & $\mathrm{gcd}(1\pm rN,pN)=1$  \\
$\mathrm{Case}-2$ & $\Z_{4}$ & $(2\D_{2(2N+1)},\Z_{2N+1})$ &  $\Z_{m}$  & $m$ is odd     \\
$\mathrm{Case}-3$ & $\Z_{2}$ & $(2\D_{4N},\Z_{2N})$ & --- & ---  \\
$\mathrm{Case}-4a$ & $\Z_{3}$ & $(2\T,2\D_{4})$ & $\Z_{k}$  & $3 \nmid k  $\\
$\mathrm{Case}-4b$ & $\Z_{3}$ & $(2\T,2\D_{4})$ & $2\D_{2l}$  & $3 \nmid 2l  $\\
$\mathrm{Case}-5$ & $\Z_{2}$ & $(2\D_{4N},2\D_{2N})$ & ---   &  \textbf{\textemdash} \\
$\mathrm{Case}-6$ & $\Z_{2}$ & $(2\O,2\T)$ &  ---  &  \textbf{\textemdash}  \\
\hline
\end{tabular}
\caption{The table lists the six cases of (\ref{eq:the-6-cases}), along with the corresponding finite group $A$ required to ensure a free action, as well as further necessary conditions. We note that, in the cases where a non-splittable group already acts freely, the group $A$ can be taken to be any finite $ADE$ subgroup, yielding a valid semi-splittable extension.}
\label{Table:list-AA0-dihedral}
\end{table}

\paragraph{The list of non-splittable and semi-splittable $ADE$ finite groups.}

In fact, the possible choices for the $ADE$ triplets $(C,C_{0},B)$ that can be modeled through (\ref{eq:goursat-B0=1}), have already been classified by Reid in \cite[page 376]{Reid:1985}--see also \cite{Witten:1997kz} for an earlier discussion from the physics perspective. There are 6 different cases:
\begin{equation}\label{eq:the-6-cases}
    \begin{split}
       \mathrm{Case}-1:& \qquad   1\,\to\,\Z_{N}\,\to\,\Z_{pN}\,\to\, \Z_{p}\,\to\,1~,
        \\
       \mathrm{Case}-2:& \qquad  1\,\to\,\Z_{2N+1}\,\to\,2\D_{2(2N+1)}\,\to\, \Z_{4}\,\to\,1~,
        \\
        \mathrm{Case}-3:& \qquad  1\,\to\,\Z_{2N}\,\to\,2\D_{2N}\,\to\, \Z_{2}\,\to\,1~,
        \\
        \mathrm{Case}-4:& \qquad 1\,\to\,2\D_{4}\,\to\,2\T\,\to\, \Z_{3}\,\to\,1~,
        \\
        \mathrm{Case}-5:& \qquad 1\,\to\,2\D_{2N}\,\to\,2\D_{4N}\,\to\, \Z_{2}\,\to\,1~, 
        \\
        \mathrm{Case}-6:&  \qquad 1\,\to\,2\T\,\to\,2\O\,\to\, \Z_{2}\,\to\,1 ~.
    \end{split}
\end{equation}
Here, $2\D_{2\bullet}$ is the binary dihedral group, $2\T$ is the binary tetrahedral group, and $2\O$ is the binary octahedral group. For the explicit definition of each one of these groups, see subsection \ref{sec:ADE-gauge-theories} below.

The above list classifies the possible non-splittable finite groups $\Gamma$'s acting on the link $\mathbb{S}^{3}\times\mathbb{S}^{3}$ when modeled by Goursat's theorem. However, it does not guarantee that the groups $\Gamma$'s are acting freely on the link space. To determine that, we shall impose the condition(s) given in (\ref{eq:free-action-condition}). Following the discussion in \cite[Sec.4.2]{cortés2014locallyhomogeneousnearlykahler}, there are only four non-splittable cases that are acting freely, whereas the other two cases are semi-splittable. These cases are listed in table \ref{Table:list-AA0-dihedral}.

\subsection{The M-theory physical dictionary: decomposition and automorphisms}\label{subsec: physics dictionary}

In this section, we discuss the general dictionary for obtaining gauge theory from quotients of the Bryant--Salamon space within M-theory geometric engineering. We discuss the realization of inner and outer automorphisms. Further, we consider the physics of non-splittable and semi-splittable quotients of the BS space. 

\subsubsection{General philosophy}

The conventional wisdom for realizing $4d$ $\N=1$ $ADE$ gauge theories via M-theory geometric engineering involves the quotient space $\R^{4}/\Gamma_{ADE}\times\mathbb{S}^{3}$, where $\R^{4}\times\mathbb{S}^{3}$ is the spin bundle over $\mathbb{S}^{3}$ and $\Gamma_{ADE}$ is a finite $ADE$ subgroup of $SU(2)$. In what follows, we will review some key features of this construction following \cite{Acharya:2000gb,Atiyah:2001qf,Witten:2001bf,Acharya_2004}.

\medskip
\noindent
\textbf{Reducing to a 7d gauge theory.}
It is well known that placing M-theory on a non-compact Calabi–Yau twofold (CY2) of the form $\R^{4}/\Gamma_{ADE}$ breaks the 11d spacetime symmetry $\mk{so}(1,10)$ to:
\begin{equation}\label{7dglobalsymm}
 \mk{g}_{\scriptscriptstyle \text{global}}^{\scriptscriptstyle 7d}\,  = \, \mk{su}(2)_{\scriptscriptstyle\text{R}} \, \times\, \mk{so}(1,6) \,. 
\end{equation}
Here, $\mk{so}(1,6)$ is the Lorentz symmetry for the effective 7d theory and $\mk{su}(2)_{\scriptscriptstyle\text{R}}$ is its R-symmetry.    

Moreover, compactification on $\R^{4}/\Gamma_{ADE}$ gives a 7d $ADE$ effective gauge theory with 16 supercharges \cite{Sen:1997kz}. The field content and their transformation properties under the global symmetry algebra (\ref{7dglobalsymm}) are given by:
\begin{equation}\label{7dfields}
        {A}: (\textbf{1},\textbf{7}), \ \ \ \phi: (\textbf{3}, \textbf{1}),
        \ \ \ \ \psi: (\textbf{2}, \textbf{8}), \ \ \ Q: (\textbf{2}, \textbf{8})\,.
\end{equation}
Here, $A$ is the gauge field, $\phi$ denotes the scalar field(s), and the fermions are denoted by $\psi$ and $Q$. 

To determine the $ADE$ gauge group, we resolve the $\Gamma_{ADE}$ singularity by a small (crepant) resolution $\widetilde{\R^4/\Gamma_{ADE}}$. This amounts to introducing $\mathrm{rank}(\Gamma_{ADE})$ independent vanishing 2-cycles. The vanishing 2-cycles allow us to obtain $\mathrm{rank}(\Gamma_{ADE})$ photons, which correspond to the Coulomb branch . This is due to the fact that, for each one of the vanishing 2-cycles, one can associate an L$^2$-normalisable 2-form via Poincaré duality \cite{Hausel:2002xg}. To get the photons we expand the M-theory $C_{3}$ field along the L$^{2}$-normalizable harmonic two-forms $\omega_{2}$:
\begin{equation}
    C_{3}\,=\, \sum_{i=1}^{\mathrm{rank}(\Gamma_{ADE})}\, A^{i}\,\wedge\,\omega_{2}^{i}~.
\end{equation}

The structure of the 2-cycles is determined through the 2d McKay correspondence \cite{mckay} (see also \cite{Ito1994mckay,ReidMcKay1999}). In particular, the second homology of $\widetilde{\R^{4}/\Gamma_{ADE}}$ coincides with the root lattice of the $ADE$ algebra. Moreover, the intersection numbers between the 2-cycles coincide with the corresponding elements of the $ADE$ Cartan matrix.

Let us now introduce $\mathrm{rank}(\Gamma_{ADE})$ M2-branes wrapping—in both orientations—the vanishing 2-cycles. From the perspective of the 7d theory, these configurations correspond to massive charged particles, with their charges determined by the intersection matrix of the vanishing 2-cycles (i.e., the Cartan matrix of the $ADE$ Lie algebra). In other words, they behave as massive ‘gluons’ with masses proportional to the volumes of the vanishing 2-cycles. In the limit where we shrink these vanishing 2-cycles (the singular limit), we have all the required degrees of freedom to get an $ADE$ gauge theory.

\medskip
\noindent
\textbf{Reducing to a 4d gauge theory.}
Reducing M-theory on the $G_{2}$-space $\R^{4}/\Gamma_{ADE}\times\mathbb{S}^{3}$ amounts to further compactifying the 7d theory further on $\mathbb{S}^{3}$ \cite{Acharya:2000gb}. The $G_{2}$-structure on the total space $\R^{4}\times\mathbb{S}^{3}$ requires the global symmetry to be given by \cite{Acharya:2000gb,Acharya_2004}:
\begin{equation}\label{4dglobalsymm}
 \mk{g}_{\scriptscriptstyle \text{global}}^{\scriptscriptstyle 4d} \,  = \, \mk{su}(2) \, \times\, \mk{so}(1,3) ~. 
\end{equation}
The 7d fields \eqref{7dfields} reduce as: 
\begin{equation}\label{4Dfieldsfrom7D}
    \begin{aligned}
        &(\textbf{1},\textbf{7})\,\rightarrow \, (\textbf{3},\textbf{1})\, \oplus \,(\textbf{1}, \textbf{4})~, \ && (\textbf{3}, \textbf{1})\,\rightarrow\, (\textbf{3}, \textbf{1})~,
        \\ 
        &(\textbf{2}, \textbf{8})\,\rightarrow \,(\textbf{1}, \textbf{2})\, \oplus \,(\textbf{3}, \textbf{2})\, \oplus \, \text{c.c.}~.
    \end{aligned}
\end{equation}
Since the first Betti number of $\mathbb{S}^{3}$ is zero, then the massless degrees of freedom are given by \cite[Section 5]{Acharya_2004}
\begin{equation}
    (\textbf{1},\textbf{4})\,\oplus\,(\textbf{1},\textbf{2})\,\oplus\, \text{c.c}\,,
\end{equation}
corresponding to the vector multiplet of 4d $\N=1$ theory with $ADE$ gauge algebra.

The above discussion establishes that the codimension-four singularity is localized at the fixed-point set
\begin{equation}\label{eq:fixed-poin-set(1)}
        \{0\}\,\times\,\mathbb{S}^{3}\,\subset\, \R^{4}/\Gamma_{ADE}\times\mathbb{S}^{3}~. 
    \end{equation}
gives rise to a $4d$ $\N=1$ gauge theory with an $ADE$ Lie algebra.

\paragraph{On line defects: Groups vs. Algebras.}

Generally speaking, geometric engineering techniques determine the Lie algebra $\mk{g}$ of the gauge theory, rather than the global structure of the corresponding gauge group. The latter is determined by the set of line operators \cite{Kapustin:2005py,Gaiotto:2014kfa}. 

In 4d field theory, the global structure of the gauge group is characterized by discrete electric and magnetic 1-form symmetries, whose charged operators are Wilson and ’t Hooft lines, respectively. These line defects serve as order parameters that distinguish between confining and deconfining phases in non-abelian gauge theories \cite{Wilson:1974sk,tHooft:1977nqb}. The spectrum of such line operators is classified by the center of the gauge group
\begin{equation}\label{eq:ze-zm-in-center}
    (z_{e},\,z_{m})\,\in\, \mathcal{Z}(G)\,\times\, \mathcal{Z}(G^{\ast})~.
\end{equation}
Here, $G^{\ast}$ is the Goddard--Nuyts--Olive (GNO) dual Lie group \cite{Goddard:1976qe,Kapustin:2005py}.  Notably, its center coincides with $\mathcal{Z}(G)$, the center of the original gauge group. Consequently, specifying the allowed spectrum of line defects amounts to selecting a particular global structure of the gauge group $G$. For an in-depth discussion of these ideas, we refer the reader to \cite{Kapustin:2005py,Gaiotto:2014kfa}.

From the perspective of geometric engineering, the spectrum of allowed line defects and the associated topological symmetry operators are determined by homological data of the link space (see, e.g., \cite{Apruzzi:2021nmk, Najjar:2024vmm, Najjar:2025htp}). Consistency with field theory results establishes a duality between the homology of the link space and the Lie-algebraic data of the gauge theory, as we will elaborate in sections~\ref{sec:SymTFT} and \ref{sec:charged top defects} below.

\subsubsection{Physics of non-splittable and semi-splittable quotients }\label{sec:physics-non=split}

Let us apply the geometric dictionary discussed above to the non-splittable quotient of the BS space, $(\R^{4}\times\mathbb{S}^{3}_{\mathrm{b}})/\Gamma_{\mathrm{ns}}$, given as \eqref{eq:non-split-on-link}:
\begin{equation}
 \frac{\R^{4}\times\mathbb{S}^{3}_{\mathrm{b}}}{\Gamma_{\mathrm{ns}}} \ = \  \frac{(\R^{4}/C_{0})\,\times\, \mathbb{S}^{3}_{\mathrm{b}}\,}{B}~,
\end{equation}
with $\Gamma_{\mathrm{ns}}$ has the structure given in (\ref{eq:our-general-Gamma-Goursats}). To arrive at the above $G_{2}$-space, we fill-in the $\mathbb{S}^{3}_{\mathrm{f}}$ factor of (\ref{eq:non-split-on-link}) to get the non-compact $\R^{4}$ space as we reviewed in subsection \ref{subsec:BS and symmetries} above. 

The finite group $B$ acts freely on the base 3-sphere $\mathbb{S}^{3}_{\mathrm{b}}$--It acts on its Hopf fiber. This fact can be traced back to the action of $\Gamma_{\mathrm{ns}}$ on the link space discussed in the previous subsection. Therefore, the group $B$ acts freely on the total space $\R^{4}/C_{0}\times\mathbb{S}^{3}_{\mathrm{b}}$ \cite{Witten:1997kz,Acharya:2020vmg}. With this being said, the codimension-4 singularity does not get enhanced any further by this quotient. Hence, the codimension-four singularities are determined only by the normal subgroup $C_{0}$.

More precisely, the codimension-4 singularities originate from the group-theoretic structure of $\Gamma$ as defined in (\ref{eq:goursat-B0=1}). Since the normal subgroup $B_{0}$ is trivial, i.e., $B_{0}=1$, then Goursat's theorem ensures that for every $c\in C_{0}$ the image under the map $\varphi$ satisfies $\varphi(c)=1\in B$, as given in (\ref{eq:def-A0}). Therefore, the singular locus is generated precisely by the subgroup
\begin{equation}\label{eq:codim-4-sing-C0}
    C_{0}\,\cong\, \{\,(c,1)\,\,|\,\,\varphi(c)\,=\,1\,\}\,.
\end{equation} 
The restriction of $C_{0}$ to $ADE$-type subgroups (see the discussion around (\ref{eq:the-6-cases}) and table \ref{Table:list-AA0-dihedral}) directly determines that the low-energy gauge algebra must be of the corresponding $ADE$-type. We will illustrate this explicitly in the next subsection \ref{sec:ADE-gauge-theories}.

\paragraph{Many ``universes".}

Given the structure of the non-splittable group $\Gamma_{\mathrm{ns}}$ as written in (\ref{eq:our-general-Gamma-Goursats}), with $B\cong\Z_{m}$, one observes that the codimension-4 singularities identified (\ref{eq:codim-4-sing-C0}) correspond to the subgroup $(C_{0},1)$, i.e., the first factor above.  However, this does not imply that the remaining components are inert or physically irrelevant.

In particular, consider acting on $\Gamma_{\mathrm{ns}}$ by an element $(\widetilde{\varphi}(b_{j}^{-1}),b_{j}^{-1})\in \Gamma_{\mathrm{ns}}$ with $j=0,1,\cdots,m-1$. This operation cyclically permutes the decomposition of $\Gamma_{\mathrm{ns}}$, effectively shifting the codimension-4 singular locus to the $j^{\mathrm{th}}$ sector. Physically, this structure can be understood by associating a local observer with each component of $\Gamma_{\mathrm{ns}}$ as presented in \eqref{eq:our-general-Gamma-Goursats}. The $j^{\mathrm{th}}$ observer, equipped with the frame defined by $(\widetilde{\varphi}(b_{j}^{-1}),b_{j}^{-1})$ perceives an $ADE$ gauge theory determined by the subgroup $C_{0}$.

This conclusion holds for all $m$ components, each corresponding to a distinct observer. However, due to the non-splittable structure of the quotient, these observers are not simultaneously accessible within a single local frame. As a result, the geometry gives rise to $m$ physically distinct but locally indistinguishable $ADE$ gauge theories. A similar observation in the context of $\mk{su}(n)$ gauge theories was made in \cite{Witten:1997kz,Acharya:2020vmg,Najjar:2022eci}. However, our treatment presents a more general, formal, and physically motivated discussion. We interpret these as a collection of separate vacua—or, metaphorically, as distinct `universes'—each governed by an identical gauge algebra but realized in a different sector of the global orbifold. The terminology `universes' is borrowed from \cite{Tanizaki:2019rbk,Cherman:2020cvw}, for reasons we now elaborate.

In particular, let $\mathcal{T}_{\Gamma_{\mathrm{ns}}}$ denote the field theory associated with the non-splittable finite quotient group $\Gamma_{\mathrm{ns}}$. Based on the preceding discussion, this theory naturally decomposes into a disjoint union of $m$ distinct sectors:
\begin{equation}\label{eq:ns-theory-decompose}
\mathcal{T}_{\Gamma_{\mathrm{ns}}}\,=\, \bigsqcup_{j=0}^{m-1}\, \mathcal{T}^{(j)}_{C_{0}}~.
\end{equation}
For each $j$, $\mathcal{T}^{(j)}_{C_{0}}$ is the 4d gauge theory determined by the normal subgroup $C_{0}$. Each of these sectors is dynamically isolated from the others due to the non-trivial structure imposed by $\Gamma_{\mathrm{ns}}$. As we will demonstrate in sections \ref{sec:SymTFT} and \ref{sec:phy-implications}, these vacua exhibit a $\Z_{m}^{\scriptscriptstyle[3]}$ 3-form symmetry, with charged defects corresponding to non-dynamical UV domain walls (or interfaces). These interfaces separate the different vacua, thus providing a physical realization of the argument above. 

The coexistence of multiple separate universes, is intimately related to the concept of decomposition first explored in \cite{Pantev:2005rh, Pantev:2005wj, Pantev:2005zs, Hellerman:2006zs} and subsequently applied in a range of contexts including \cite{Seiberg:2010qd, Tachikawa:2013hya, Sharpe:2014tca, Tanizaki:2019rbk, Cherman:2020cvw, Cherman:2021nox, Nguyen:2021naa, Sharpe:2022ene, Closset:2024sle, Yu:2024jtk, Closset:2025lqt,Najjar:2025rgt, Najjar:2025htp}.

\paragraph{The physical dictionary of semi-splittable quotients.} 

Recall that the semi-splittable is built as the direct product of the $ADE$ group $A$ with $\Gamma_{\scriptscriptstyle\mathrm{ns}}$:
\begin{equation}\label{eq:semi-splittable-Gamma-ss}
    \Gamma_{\mathrm{ss}} \ = \ A\,\times\, \Gamma_{\mathrm{ns}}~.
\end{equation}
The group $A$ acts diagonally on the two components of the link space $L_6$ as in \eqref{eq:action-abc-S3fS3b}. Therefore, the quotient BS space can be expressed as:
\begin{equation}\label{semi-splittable-on-link}
 \frac{\R^{4}\times\mathbb{S}^{3}_{\mathrm{b}}}{\Gamma_{\mathrm{ss}}} \ = \  \frac{(\R^{4}/C_{0})\,\times\, \mathbb{S}^{3}_{\mathrm{b}}\,}{A\,\times\,B}\,.
\end{equation}

The introduction of the group $A$ as described above does not alter the codimension-four singularities since $A$ acts freely on $\mathbb{S}_{\mathrm{b}}^{3}$ and hence on the whole $G_2$-manifold. Similar to the non-splittable case \eqref{eq:our-general-Gamma-Goursats}, the structure of the finite group $\Gamma_{\mathrm{ss}}$ can be written as in \eqref{eq:general-Gamma-Goursats-semi-split}. Therefore, the low-energy physics again consists of $m$ distinct but identical universes, each supporting an $ADE$ gauge theory determined by the normal subgroup $C_{0}$.

\subsubsection{Flat connections: Inner automorphisms}

The massive 4d scalar field  $(\textbf{3},\textbf{1})$, which we will denote by $\Phi$, accompanying the 4d vector field $(\textbf{1},\textbf{4})$ in \eqref{4Dfieldsfrom7D} plays a distinguished role when a finite quotient acts on $\mathbb{S}^{3}$, as we now demonstrate. 

\paragraph{Getting flat connections.} Consider a splittable quotient group of the form $\Gamma_{ADE}=\Gamma_{1}\times\Gamma_{2}$, with $\Gamma_{1}$ acting on $\R^{4}$ and $\Gamma_{2}$ acting on $\mathbb{S}^{3}$.  Such splittable structures have been considered in \cite{Friedmann:2002ty,Friedmann:2002ct,Friedmann:2012uf}, where explicit constraints are derived to ensure that the combined action of $\Gamma_{1}\times\Gamma_{2}$ acts freely on the link space $\mathbb{S}^{3}\times\mathbb{S}^{3}$. The constraints on such splittable quotient can also be found in section 3 of \cite{cortés2014locallyhomogeneousnearlykahler}.

The triplet-scalar field $\Phi$ arises as the holonomy of the 7d gauge field $A$ along 1-cycles of $\mathbb{S}^{3}/\Gamma_{2}$:
\begin{equation}\label{eq:Wilson-line}
\Phi_{\gamma}\,=\,\mathcal{P}\,\exp(i\oint_{\gamma}\,\widetilde{A})\,, \qquad \gamma\in \pi_{1}(\mathbb{S}^3/\Gamma_2,\Z)\cong \Gamma_2~.
\end{equation}
Here, $\widetilde{A}$ denotes the components of  $A$ along $\mathbb{S}^3/\Gamma_{2}$. $\mathcal{P}$ denotes the path order operator for the $ADE$ gauge theory, as $\widetilde{A}$ is adjoint-valued in the gauge group. Consequently, the holonomy $\Phi_{\gamma}$ is group-valued and may be labeled by a representation $R$ of the gauge group $G$, i.e., $\Phi_{\gamma}:=\Phi_{\gamma}^{R}$. One may, for instance, choose $R$ to be the fundamental representation. We identify the scalar fields $\Phi_{\gamma}$ as Wilson loops -- flat connections that preserve supersymmetry\footnote{The supersymmetry transformation of $\widetilde{A}$ necessarily involves its field strength. Supersymmetric invariance therefore requires the existence of spinors on the three-sphere $\mathbb{S}^{3}$. However, the field content resulting from the decomposition in equation (\protect\ref{4Dfieldsfrom7D}) lacks such spinors. Consequently, in order to preserve supersymmetry, the connection $\widetilde{A}$ must be flat. In that sense, $\widetilde{A}$ can be taken as a background field.}. For further discussion on such flat connections one may consider \cite{Breit:1985ud,Witten:1985xc,delAguila:1985hkb,Hosotani:1988bm,Gopakumar:1997dv,Witten:1997kz,Hebecker:2001jb,Hall:2001tn,Witten:2001bf,Friedmann:2002ct,Cachazo:2002zk,Hosomichi:2005ja,Acharya:2020vmg,Najjar:2022eci}.

\paragraph{Flat connections as inner automorphism.}

The action of $\Gamma_{2}$ on $\mathbb{S}^{3}$ introduces non-trivial first homotopy group $\pi_{1}(\mathbb{S}^{3}/\Gamma_{2})$. Consequently, the flat connections $\Phi_{\gamma}^{R}$ carry charge under $\pi_{1}(\mathbb{S}^{3}/\Gamma_{2})$. Specifically, under a gauge transformation associated with $\xi\in\pi_{1}(\mathbb{S}^{3}/\Gamma_{2})$, the holonomy transforms as \cite{Witten:1985xc}:
\begin{equation}
    \Phi_{\gamma}^{R} \ \longmapsto \ U_{\xi}\,\Phi_{\gamma}^{R}\,,\quad \mathrm{for}  \quad  \ \ \ U_{\xi}\in G\,,
\end{equation}
with $U_{\xi}U_{\xi'}=U_{\xi\xi'}$. The gauge transformation $U_{\xi}$ is independent of the spacetime coordinates. The above transformation law for the group-valued flat connections $\Phi_{\gamma}^{R}$ defines the following homomorphisms \cite{Hall:2001tn}:
\begin{equation}\label{eq:hom-map}
    \text{Hom$(\pi_{1}(\mathbb{S}^{3}/\Gamma_{2})$ $\to G)$}/G\,,
\end{equation}
where the quotient by $G$ accounts for gauge equivalence. The homomorphisms define the moduli space $\mathcal{M}_{\Phi}$ that classifies all inequivalent flat connections.

By choosing the background field $\widetilde{A}$ in (\ref{eq:Wilson-line}) such that $\Phi^{R}_{\gamma}=U_{\xi}$, we identify the holonomies with the gauge transformations themselves. Since these flat connections are given as exponential of the adjoint-valued field, the set $\{\Phi_{\gamma}^{R}\}$ generates inner automorphisms of the gauge group $G$ through (\ref{eq:hom-map}).

\paragraph{Breaking gauge theories via flat connections.}

Let us specialize to case with $\Gamma_{2}=\Z_{m}$. By the above homomorphism(s) we associate $U_{\xi}\in G$ for the generator $\xi$ of $\Z_{m}$. As $\Z_{m}$ is abelian, then, one can always write \cite{Hebecker:2001jb}:
\begin{equation}
  U_{\xi} \,=\, \exp(-2\pi i \,\, V_\xi^{j} H_{j})\,\,\in \,\, G~.
\end{equation}
Here, the set $H_{j}$ corresponds to the Cartan generators of the gauge group $G$, and $V_\xi^{j}$ is the \textit{twist vector} that specifies the action of $\xi\in\Z_{m}$. Recall, the above can also be taken as the Wilson loops $\Phi_{\xi}^{R}$.

Consider the generators of the gauge group $G$  in the Cartan--Weyl basis:
\begin{equation}\label{[Hi, Ealpha]}
    [\,H_{i}\,,\,E_{\alpha}\,]\,=\,\alpha_{i}\,E_{\alpha}~,
\end{equation}
where
\begin{itemize}
    \item $H_{i}$ are the Cartan generators with $i=1,2,\cdots,\mathrm{rank}(G)$,
    \item $E_\alpha$ are the generators associated with the roots $\alpha \in\Delta$.
\end{itemize}

The action of $\Z_{m}$ on the generators $\{H_{i},E_{\alpha}\}$ is given by conjugation:
\begin{equation}\label{eq:Uxi-action-on-H-E}
    U_{\xi}\,H_{k}\,U_{\xi}^{-1}\,=\, H_{k}\,,\qquad U_{\xi}\,E_{\alpha}\,U_{\xi}^{-1}\,=\,  \exp(-2\pi i \,\, V_\xi^{j} \alpha_{j})\,\, E_{\alpha}\,.
\end{equation}
From the above, we observe that the Cartan generators are invariant under $\Z_{m}$, ensuring the rank of $G$ is preserved. This corresponds to an inner automorphism of the algebra. Furthermore, only root vectors $E_{\alpha}$ satisfying $V^{j}_\xi\alpha_{j}\in \Z$ are invariant. Therefore, the described $\Z_{m}$ flat connections break the gauge group by keeping $\Z_{m}$-invariant generators. Further discussion can be found in, e.g., \cite{SLANSKY19811,Dienes:1996yh,Hebecker:2001jb}.

\paragraph{{An example: $\mathfrak{e}_6\rightarrow \mathfrak{su}(6)\oplus \mathfrak{su}(2)$.}} Let us consider the following example where we start with algebra $\mathfrak{e}_6$. In this case, we have six simple roots that are given as the rows of the corresponding Cartan matrix:
\begin{equation}\label{rows of e6 Cartan}
\begin{split}
        &\alpha_1 = (\,2\,,\,-1\,,\,0\,,\,0\,,\,0\,,\,0\,)~,\\
        &\alpha_2 = (\,-1\,,\,2\,,\,-1\,,\,0\,,\,0\,,\,0\,)~,\\
        &\alpha_3 = (\,0\,,\,-1\,,\,2\,,\,-1\,,\,0\,,\,-1\,)~,\\
        &\alpha_4 = (\,0\,,\,0\,,\,-1\,,\,2\,,\,-1\,,\,0\,)~,\\
        &\alpha_5 = (\,0\,,\,0\,,\,0\,,\,-1\,,\,2\,,\,0\,)~,\\
        &\alpha_6 = (\,0\,,\,0\,,\,-1\,,\,0\,,\,0\,,\,2\,)~.
\end{split}
\end{equation}

Consider the twist vector $V = (1/2,1,1,1,0,1)$. One can show that this is a $\Z_2$ twist vector in the sense that $g = \exp(-2\pi i \sum_{a=1}^6 V^aH_a)$ has the property that $g^2 = \mathbb{I}_{27}$. Here, $H_a$ are the 6 Cartan generators of $\mathfrak{e}_6$.

Extending the Dynkin diagram of $\mathfrak{e}_6$ by adding an extra node associated with $-\alpha_0 \equiv (0,0,0,0,0,-1)$\footnote{$\alpha_0$ is the referred to as the \textit{highest root}. It is defined to be the root such that any $\alpha\in \Delta$ can be obtained by adding or subtracting simple roots from it. For ABCDEFG Lie algebras, the associated highest roots are listed in table 8 of \protect\cite{SLANSKY19811}}. One can see that the twist vector $V$ preserves all the nodes of the extended diagram except for the second one. That is, 
\begin{equation}\label{E6 example inner auto}
    (V\cdot \alpha_{i\neq 2}) \in \Z~, \quad {\rm but, } \quad (V\cdot \alpha_2) \notin \Z~.
\end{equation}
Therefore, this $\Z_2$ inner automorphism breaks the gauge algebra according to the pattern $\mathfrak{e}_6 \rightarrow \mathfrak{su}(6)\oplus \mathfrak{su}(2)$. This is the last case in table \ref{Table:Z2-Wilson-loops}.

Note that, similarly, one can pick a $\Z_2$ twist vector that removes the node 4 or node 6 instead of node 2. These two cases will still give us the same breaking pattern of the $\mathfrak{e}_6$ algebra. This observation is related to the $\Z_3$ outer automorphism group of the extended Dynkin diagram.

\medskip
\noindent
\textbf{An equivalent approach.} From (\ref{eq:Uxi-action-on-H-E}), we learn that the reduced gauge group $K\subset G$ can be determined as the centralizer of the subgroup $\Z_{m}\subset G$. That is (see e.g. \cite{delAguila:1985hkb}):
\begin{equation}
    K\,=\, \{\, g\,\in\, G \,\,\ |\ \,\, [g,U_{\xi}]\,=\,0\,,  \ \forall \,\xi\,\in\,\Z_{m} \,\}\,.
\end{equation}
Equivalently, $K$ is determined by the elements $g\in G$ satisfying $[U_{g},\Phi_{\gamma}^{R}] = 0$.

For the fundamental representation $F$, with dimension $\mathrm{dim}(F)$, $U_{\xi}$ takes a block-diagonal form with eigenvalues $\xi^{k}$:
\begin{equation}
 U_{\xi}\,=\,   \mathrm{diag}\left(\,\xi^{0}\,\mathds{1}_{n_{0}}\,,\, \xi^{1}\, \mathds{1}_{n_{1}}\,,\,\cdots,\, \xi^{m-1}\,\mathds{1}_{n_{m-1}}\,\right)_{\scriptscriptstyle\mathrm{dim}(F)\times\mathrm{dim}(F)}\,.
\end{equation}

\begin{table}[t]
\centering
\begin{tabular}{|c|c|c|}
\hline
$\mk{g}$ & $\mk{k}$ & Conditions   \\
\hline
$\mk{su}(N)$ & $\mk{su}(n_{1})\oplus \mk{su}(n_{2})\oplus \mk{u}(1)$  & $N=n_{1}+n_{2}$  \\
$\mk{so}(2N)$ & $\mk{su}(N)\oplus \mk{u}(1)$  & \\
$\mk{so}(n_{1}+n_{2})$ & $\mk{so}(n_{1})\oplus \mk{so}(n_{2})$  & $n_{1}$ or $n_{2}$ even \\
$\mk{e}_{6}$ & $\mk{so}(10)\oplus \mk{u}(1)$  &   \\
$\mk{e}_{6}$ & $\mk{su}(6)\oplus \mk{su}(2)$  &   \\
\hline
\end{tabular}
\caption{Possible breaking patterns of $\Z_{2}$ Wilson loops (flat connections) identified with $\Z_{2}$ inner automorphisms \cite{SLANSKY19811,Hebecker:2001jb}.}
\label{Table:Z2-Wilson-loops}
\end{table}

For example, applying this $\Z_{m}$  inner automorphism  to an $SU(N)$ gauge theory leads to the following symmetry breaking pattern:
\begin{equation}\label{eq:break-SU(N)-Zp}
    SU(N) \, \rightarrow\, SU(n_{0})\,\times\, SU(n_{1}) \,\times\, \cdots \,\times \,SU(n_{m-1})\,\times\, (U(1))^{s-1}.
\end{equation}
Here, we have $N=\sum_{j=0}^{m-1} n_{j}$ and the integer $s$ counts the number of non-trivial $SU$ factors, which ensures rank preservation. 

The general results for $\Z_{2}$ inner automorphisms in the gauge theories of interest are summarized in table \ref{Table:Z2-Wilson-loops} above.

\paragraph{Inner automorphisms and the non-splittable quotients.}

The existence of the above flat connections applies to the non-splittable quotients \eqref{eq:non-split-on-link} that we are interested in. As there are $m$ universes, each with local $ADE$ gauge theory, then there are $m$ different sets of flat connections. Each set introduces inner automorphisms into the local $ADE$ gauge theory as discussed above. Similar observation for $\mk{su}(N)$ gauge theories were made in section 4 of \cite{Acharya:2020vmg} and chapter 4 of \cite{Najjar:2022eci}. 

Since flat connections are elements of the gauge group $G$, their determinants are well-defined within each local frame. However, since there is no single local frame, then the different sets of the flat connections would have different determinants. Specifically, the determinant of a flat connection in the $j^{\mathrm{th}}$ universe is related to that in the $0^{\mathrm{th}}$ universe via 
\begin{equation}
    \det(U^{j}_{\xi})\,=\, \xi^{j}\,  \det(U^{0}_{\xi})\,,\qquad \mathrm{for} \,\ \xi\,\in\, \Z_{m}~.
\end{equation}
 This relation arises from the structure of the non-splittable group $\Gamma_{\mathrm{ns}}$ given in \eqref{eq:our-general-Gamma-Goursats}, reflecting the group-theoretic twisting induced by the nontrivial quotient.

\paragraph{Inner automorphisms and the semi-splittable quotients.} Let us briefly here comment on the semi-splittable extension case. Recall that the semi-splittable quotient group is given by (\ref{eq:general-Gamma-Goursats-semi-split}), and its action on the $G_{2}$ space takes the form:
\begin{equation}
    \frac{\,(\R^{4}/C_{0})\,\times\,\mathbb{S}^{3}_{\mathrm{b}}\,}{A \times B}\,.
\end{equation}
In this case, the group $A\times B$ acts on the base 3-sphere $\mathbb{S}^{3}_{\mathrm{b}}$, and hence the flat connections
are classified by:
\begin{equation}\label{eq:hom-A-B-to-G(1)}
     \text{Hom$\,\left(\,\pi_{1}\left(\frac{\mathbb{S}^{3}}{A\,\times\, B}\right) \, \to\,  G\right)$}/G\,.
\end{equation}
Since the group action is a direct product $A\times B$, the associated flat connections may be considered independently, leading to the decomposition  
\begin{equation}\label{eq:hom-A-B-to-G(2)}
     \text{Hom$\,\left(\, A\, \times\, B \, \to\,  G\right)$}/G  \ \cong  \    \text{Hom$\,\left(\, A \, \to\,  G\right)$}/G \,\times \, \text{Hom$\,\left(\,  B \, \to\,  G\right)$}/G\,.
\end{equation}

\subsubsection{Monodromies: Outer 
automorphisms}\label{sec:outer-auto}

Let us go back to the non-splittable quotient of the BS space, given in \eqref{eq:non-split-on-link}. As reviewed earlier, upon resolving the $\R^{4}/C_{0}$ singularities, one obtains a collection of vanishing 2-cycles whose intersection structure is captured by the $ADE$ Dynkin diagrams $\mathrm{DD}{\scriptscriptstyle ADE}$ associated with the gauge Lie algebra $\mk{g}\equiv \mk{g}_{C_0}$. These Dynkin diagrams are fibered over the 3-sphere $\mathbb{S}^{3}_{\mathrm{b}}/B$.

Remarkably, the Dynkin diagrams can be transported along loops in the base space $\mathbb{S}^{3}_{\mathrm{b}}/B$, with the transport governed by the fundamental group $\pi_{1}(\mathbb{S}^{3}_{\mathrm{b}}/B)$. This process realizes a monodromy action on the $ADE$ Dynkin diagrams\footnote{Each local observer associated with the $m$ distinct universes can independently define an identical monodromy action.}, leading to automorphisms of the gauge algebra $\mk{g}$:
\begin{equation}\label{Mon action as Aut}
    \mathrm{Mon}\,:\, \pi_{1}(\mathbb{S}^{3}_{\mathrm{b}}/B)\,\to\, \mathrm{Aut}(\mk{g})~.
\end{equation}
The effects of such monodromies on gauge symmetry and the associated Dynkin diagrams have been previously investigated in \cite{Aspinwall:1996nk,Bershadsky:1996nh,Witten:1997kz,Vafa:1997mh}.

For a Lie algebra $\mathfrak{g}$, there are two types of automorphisms, inner and outer automorphisms. The group of inner automorphisms, denoted by Inn$(\mathfrak{g})$ consists of those automorphisms that can be written as the exponential of some adjoint-valued operators as in (\ref{eq:Wilson-line}). Meanwhile, the group of outer automorphisms, denoted by Out$(\mathfrak{g})$ consists of the rest, i.e., Out$(\mathfrak{g}) = {\rm Aut}(\mathfrak{g})/{\rm Inn}(\mathfrak{g})$. For further details, the reader may consult \cite{samelson1969notes, Kac:1990gs, fuchs2003symmetries} on which our discussion here is based.

\begin{figure}[t]
\begin{minipage}{0.49\linewidth}
\subfigure{\includegraphics[scale=0.85]{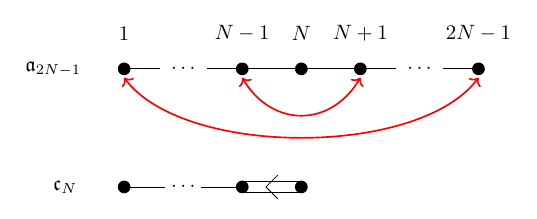}
}\\[10pt]
\subfigure{
\includegraphics[scale=0.85]{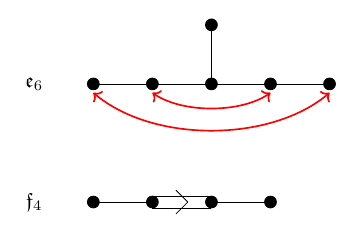}
}
\end{minipage}
\hfill\hfill
\begin{minipage}{0.49\linewidth}
\subfigure{\includegraphics[scale=0.85]{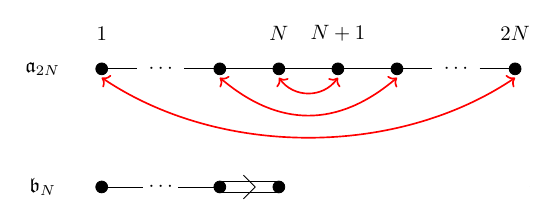}
}\\[10pt]
\subfigure{
\includegraphics[scale=0.85]{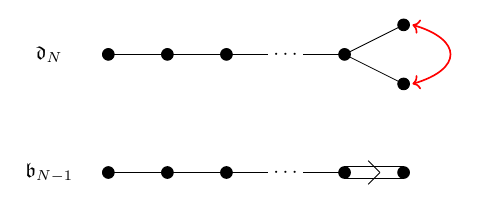}
}
\end{minipage}
\\[10pt]
\begin{minipage}{0.49\linewidth}
\subfigure{
\includegraphics[scale=0.85]{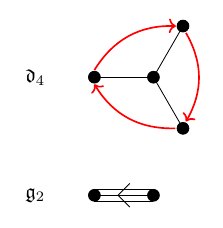}
}
\end{minipage}
\caption{The figure illustrates the outer automorphisms of simply-laced $ADE$ Dynkin diagrams—i.e., automorphisms of the Dynkin graphs themselves—together with their folding actions and the resulting non-simply-laced Dynkin diagrams.}
\label{Figure:ADE-folding}
\end{figure}

The monodromy $\mathrm{Mon}_{\gamma}$, with $\gamma\in\pi_{1}(\mathbb{S}^3_{\rm b}/B)$, is an inner automorphism of $\mk{g}$ precisely when it is realized as an element of the Weyl group $\mathrm{W}\subset \mathrm{Aut}(\mk{g})$, and lifts to the group $G$. In fact, for simply connected groups, all Weyl actions are inner automorphisms.

The distinction between inner and outer automorphisms is reflected in the structure of the automorphism group $\mathrm{Aut}(\mk{g})$. This is given by the semidirect product:\footnote{Throughout this work, simply-laced Lie algebras will be denoted by $\mathfrak{g}$. We will denote the corresponding Dynkin diagram by DD$_{ADE}$ or DD$_\mathfrak{g}$ interchangeably depending on the context.}
\begin{equation}
    \mathrm{Aut}(\mk{g})\,=\, \mathrm{Aut}(\mathrm{DD}_{\scriptscriptstyle ADE}) \,\rtimes\, \mathcal{W}\,,
\end{equation}
with $\mathrm{Aut}(\mathrm{DD}_{\scriptscriptstyle ADE})$ denotes the automorphism group of the Dynkin diagram $\mathrm{DD}_{\scriptscriptstyle ADE}$. The group $\mathrm{Aut}(\mathrm{DD}_{\scriptscriptstyle ADE})$ is naturally identified with the group of outer automorphisms of $\mk{g}$ denoted by $\mathrm{Out}(\mk{g})$.

In general, the set of all outer automorphisms may not form a group. However, the subset that permutes the nodes of the Dynkin diagram forms a finite group which is identified with $\mathrm{Aut}(\mathrm{DD}_{\scriptscriptstyle ADE})$ \cite{samelson1969notes,gilmore1974,fuchs2003symmetries,Aspinwall:1996nk,Bershadsky:1996nh,Witten:1997kz,Vafa:1997mh}.

Therefore, modulo a Weyl group $\mathcal{W}$ action, we focus on the monodromy action defined by:
\begin{equation}\label{eq:Mon-def-on-DD-ADE}
    \mathrm{Mon}\,:\, \pi_{1}(\mathbb{S}^{3}_{\mathrm{b}}/B) \ \to \  \mathrm{Aut}(\mathrm{DD}_{\scriptscriptstyle ADE})\,.
\end{equation}
That is, for a loop $\gamma:[0,1]\to\mathbb{S}^{3}_{\mathrm{b}}/B$ based at the point $x_{0}\in\mathbb{S}^{3}_{\mathrm{b}}/B$, transporting the Dynkin diagram $\mathrm{DD}_{\scriptscriptstyle ADE}$ along $\gamma$ and return to the base point, we have a permuted version of $\mathrm{DD}_{\scriptscriptstyle ADE}$ denoted by $\mathrm{Mon}_{\gamma}(\mathrm{DD}_{\scriptscriptstyle ADE})$. The action of these outer automorphisms is depicted in figure \ref{Figure:ADE-folding} for Lie algebras of $ADE$ type. 

When we consider semi-splittable cases with an $ADE$ group $A$ (as in (\ref{eq:semi-splittable-Gamma-ss})), then the monodromy is defined as:
\begin{equation}\label{eq:monodromy-SS-AxB}
    \mathrm{Mon}\,:\, \pi_{1}\left(\frac{\mathbb{S}^{3}_{\mathrm{b}}}{A\,\times\,B}\right) \ \to \  \mathrm{Aut}(\mathrm{DD}_{\scriptscriptstyle ADE})\,.
\end{equation}

\paragraph{Foldings and non-simply laced Lie algebras.}

Taking a quotient by an outer automorphism $\sigma$ is equivalent to folding the $ADE$ Dynkin diagram associated with the normal subgroup associated with $C_{0}$. This folding procedure yields a non-simply laced Lie algebra $\widetilde{\mk{g}}$. Similar construction for $\widetilde{\mk{g}}$ has been discussed in \cite{Aspinwall:1996nk,Bershadsky:1996nh,Witten:1997kz,Vafa:1997mh}. We emphasize that breaking the gauge theory by folding (also known as the twist of $ADE$ Lie algebra or outer automorphism twist) generally reduces the rank of the original $ADE$ gauge group.

\begin{table}[t]
\centering
\begin{tabular}{|c|c|c|}
\hline
$\mk{g}$ & $\mk{k}$ & permutation   \\
\hline
$\mk{su}(2N)$ & $\mk{sp}(2N)$  & $\Z_{2}$  \\
$\mk{su}(2N+1)$ & $\mk{so}(2N+1)$  & $\Z_{2}$  \\
$\mk{so}(2N)$ & $\mk{so}(2N-1)$  & $\Z_{2}$  \\
$\mk{so}(8)$ & $\mk{g}_{2}$  & $\Z_{3}$  \\
$\mk{e}_{6}$ & $\mk{f}_{4}$  & $\Z_{2}$  \\
\hline
\end{tabular}
\caption{Possible breaking patterns by permutation (outer automorphisms) \cite{gilmore1974,SLANSKY19811,Kac:1990gs,Fuchs:1995zr,Bershadsky:1996nh,Witten:1997kz,Hebecker:2001jb,fuchs2003symmetries}. The folding $\mk{su}(2N+1)$ can be found in \cite[\S 8.3]{Kac:1990gs} and \cite[\S2]{Fuchs:1995zr}. The $\Z_{3}$ permutation is a subgroup of $\mathcal{S}_{3}$. Note that for the $\mk{so}(8)$ case we have another permutation by the three $\Z_{2}\subset \mathcal{S}_{3}$ which are related by a $\Z_{3}$-transformation. After folding, we get three $\mk{so}(7)$ gauge theories. We note that we can have $\mk{e}_{6}\to\mk{sp}(8)$ which is achieved by acting exclusively on the two distal simple roots (e.g., \cite{Sammani:2025wtx}).}
\label{Table:outer-automorphisms}
\end{table}

\begin{table}[t]
\centering
\begin{tabular}{|c|c|c|}
\hline
$\mk{g}$ & $\mk{k}$ & Conditions   \\
\hline
$\mk{g}_{2}$ & $\mk{su}(2)\oplus \mk{su}(2)$  & \\
$\mk{f}_{4}$ & $\mk{su}(2)\oplus \mk{sp}(6)$  & \\
$\mk{f}_{4}$ & $\mk{so}(9)$  & \\
$\mk{so}(n_{1}+n_{2})$ & $\mk{so}(n_{1})\oplus \mk{so}(n_{2})$  & $n_{1}$ or $n_{2}$ even \\
$\mk{sp}(2N)$ & $\mk{su}(N)\oplus \mk{u}(1)$  & \\
$\mk{sp}(2N)$ & $\mk{sp}(2n_{1})\oplus \mk{sp}(2n_{2})$  & $n_{1}+n_{2}=N$ \\
\hline
\end{tabular}
\caption{Possible breaking patterns of non-simply-laced Lie algebra by $\Z_{2}$ flat connections \cite{gilmore1974,SLANSKY19811}.}
\label{Table:Z2-Wilson-non-simply-laced}
\end{table}

The folding of $ADE$ Dynkin diagrams according to the monodromy action determined by $B$ (i.e., permutation by $B$) can be systematically characterized through its action on simple roots \cite[chapter 7]{Kac:1990gs}. For the Lie algebras $\mk{su}(2N)$, $\mk{so}(2N)$, $\mk{so}(8)$, and $\mk{e}_{6}$, which we denote generically by $\mk{g}$, let $\Pi = \{\alpha_{1},\cdots,\alpha_{r}\}$ be the set of simple roots, where $r$ is the rank. The monodromy $B\cong\Z_{m}$ induces a decomposition on $\Pi$ as:
\begin{equation}\label{eq:Pi-decompos-inv-non-inv}
\Pi \,=\, \Pi_{\mathrm{inv}} \,\oplus\, \Pi_{\mathrm{non-inv}}~.
\end{equation}
Here, $\Pi_{\mathrm{inv}}\subset \Pi$ comprises the $\Z_{m}$-invariant roots: $\sigma(\alpha_{i})=\alpha_{i}$, and $\Pi_{\mathrm{non-inv}}\subset\Pi$ consist of orbits of roots related by the $\Z_{m}$ action. Taking a linear combination for all the elements in a $\Z_{m}$-orbit gives a new invariant element. The set of such new invariant elements will be donated by $\Pi_{\mathrm{orb-inv}}$. A similar decomposition holds for the set of all roots $\Delta$, which can be reconstructed from $\Pi$.

Following the discussion in chapter 7 of \cite{Kac:1990gs}, the $\Z_{m}$-invariant simple roots of $\mk{g}$ coincide with the long simple roots of $\widetilde{\mk{g}}$:
\begin{equation}\label{eq:Pi-long}
    \Pi_{\ell}\,=\,\{\, \alpha_{i}\,\in \Pi_{\mathrm{inv}} \,\,|\,\,\sigma(\alpha_{i})=\alpha_{i}    \,\}\,.
\end{equation}
On the other hand, the set $\Pi_{\mathrm{orb-inv}}$ defines the short roots for $\widetilde{\mk{g}}$: 
\begin{equation}\label{eq:Pi-short}
    \Pi_{s}\,=\,\Pi_{\mathrm{orb-inv}}\,=\,\left\{\, \frac{1}{n}(\sum_{j}\alpha_{j})\,\,|\,\,\sigma(\alpha_{j})\neq\alpha_{j}   \,\right\}~.
\end{equation}

For the case of the Lie algebra $\mk{su}(2n+1)$, the decomposition of the simple root system $\Pi$, as described in \eqref{eq:Pi-decompos-inv-non-inv}, does not apply. This is evident from the folding patterns illustrated in figure \ref{Figure:ADE-folding}. Nevertheless, it is still possible to define long and short roots of the folded Lie algebra $\mk{so}(2n+1)$ as done in chapter 7 of \cite{Kac:1990gs}. Explicitly, we have,
\begin{equation}\label{eq:long-short-su-odd}
   \begin{split}
        \Pi_{\ell}\,&=\, \left\{\, \frac{1}{2}\left(\alpha_{i} + \sigma(\alpha_{i})  \right)\ |\  \langle \alpha_{i} | \sigma(\alpha_{i}) \rangle \,=\, 0 \,\right\}~,
        \\
        \Pi_{s}\,&=\, \left\{\, \frac{1}{2}\left(\alpha_{i} + \sigma(\alpha_{i})  \right)\ |\  \langle \alpha_{i} | \sigma(\alpha_{i}) \rangle \,\neq\, 0 \,\right\}~.
   \end{split}
\end{equation}
Here, $\langle \alpha_{i} | \alpha_{j} \rangle$ stands for the inner product (i.e., Killing form) on the root space of $\mk{su}(2n+1)$. 

\paragraph{{An example: $\mathfrak{e}_6\rightarrow \mk{f}_{4}$.}} Consider the folding of the Lie algebra $\mathfrak{e}_{6}$ under its order-two outer automorphism. The simple root system of $\mathfrak{e}_{6}$ is given by $\{\alpha_{1},\alpha_{2},\cdots,\alpha_{6}\}$ with inner products specified by the corresponding Cartan matrix. The $\Z_{2}$ outer automorphism, as evident from  figure \ref{Figure:ADE-folding}, acts by exchanging:
\begin{equation}
    \alpha_{1}\,\longleftrightarrow\,\alpha_{5}~,\quad \alpha_{2}\,\longleftrightarrow\,\alpha_{4}~,
\end{equation}
while leaving $\alpha_{3}$ and $\alpha_{6}$ invariant.

Under this folding, the simple roots of the resulting algebra $\mk{f}_{4}$ are constructed through (\ref{eq:Pi-long}) and (\ref{eq:Pi-short}). Explicitly, the folded simple roots are:
\begin{equation}
     \widetilde{\alpha}_{1}\,:=\,\frac{1}{2}(\alpha_{1}+\alpha_{5})~,\quad \widetilde{\alpha}_{2}\,:=\,\frac{1}{2}(\alpha_{2} + \alpha_{4})~,\quad \widetilde{\alpha}_{3}\,:=\, \alpha_{3}~,\quad \widetilde{\alpha}_{4}\,:=\, \alpha_{6}~.
\end{equation}

Using the known Cartan matrix of $\mk{e}_{6}$,\footnote{The rows of this matrix are given by \protect\eqref{rows of e6 Cartan}, in order.} one can explicitly compute the inner products among $\{\widetilde{\alpha}_{1},\cdots,\widetilde{\alpha}_{4}\}$ and verify that they reproduce the Cartan matrix of the Lie algebra $\mk{f}_{4}$\footnote{See table 6 of \cite{SLANSKY19811} for the explicit form of this matrix.}, confirming the consistency of the folding procedure.

The general result of the above procedure in summarized in table \ref{Table:outer-automorphisms}. We will revisit these in more details in section \ref{subsec: symtft outer} from a slightly different perspective.

Let us conclude this subsection with the following comments:
\begin{itemize}
    \item \textbf{Semi-splittable folding:} For the semi-splittable cases with a cyclic group $A$ (as in (\ref{eq:semi-splittable-Gamma-ss})), we fold the $ADE$ Dynkin digram according to the monodromy in (\ref{eq:monodromy-SS-AxB}). However, in practice, we fold the diagram either according to the group $A$ or the group $B$. 

    \item \textbf{Interplay:} We can have interplay between flat connections (inner automorphisms) and monodromies (outer automorphisms). A similar observation was made in \cite{Witten:1997kz}. This interplay enables us to consider flat connections for the non-simply-laced Lie algebra $\widetilde{\mk{g}}$. We collect the results of $\Z_{2}$ inner automorphism in table \ref{Table:Z2-Wilson-non-simply-laced}.

\end{itemize}

\subsection{ADE gauge theories with automorphisms}\label{sec:ADE-gauge-theories}

In this subsection, we present a systematic analysis of the low-energy physics (within M-theory geometric engineering) arising from non-splittable and semi-splittable quotients of the spin bundle over $\mathbb{S}^{3}_{\mathrm{b}}$, as classified by the six distinct cases in (\ref{eq:the-6-cases}).

\subsubsection{An \texorpdfstring{$\mk{su}(N)$}{su(N)} gauge theory}\label{sec:case-1-su(N)theory}

Let us start with the first case in \eqref{eq:the-6-cases}. We have the triplet $(C, C_0, B) = (\Z_{pN},\Z_{N},\Z_{p})$. This case is non-splittable and the group $\Gamma$ \eqref{eq:our-general-Gamma-Goursats-defn} is given by:
\begin{equation}
    \Gamma^{(1)}\,=\, \{\, (\alpha,\beta)\,\in\,\Z_{pN}\,\times\,\Z_{p} \,\,|\,\, \varphi(\alpha) = \alpha^{rNk}=\beta^{rk};\, k\in\Z,\, r\in\Z_{p}^{\times} \,\}\,.
\end{equation}
Here, $\Z_{p}^{\times}$ is the multiplicative group of units in the ring $\Z_{p}$\footnote{This is a subgroup of $\Z_p$ consisting of the invertible elements of $\Z_p$ under multiplication mod $p$. For example, $\Z_4^\times = \{1,3\} \cong \Z_2$}. The elements of the finite group $\Gamma$ can be written as:
\begin{equation}
    \Gamma^{(1)}\,:=\,\Gamma_{N,p,r}\,=\,\{\, (\alpha^{k},\beta^{rk}) \,\,|\,\, k\in\Z,\,r\in\Z_{p}^{\times} \,\}\,.
\end{equation}

Let us apply the (non-)freely acting condition as discussed around (\ref{eq:the-set-W(Gamma)}). In this case, the condition gives
\begin{equation}
    \mathrm{Re}(\alpha^{k})\,=\,\mathrm{Re}(\beta^{rk})\quad \iff  \quad \cos\left(\frac{2\pi k}{pN}\right)\,=\,\cos\left(\frac{2\pi rk}{p}\right)\,.
\end{equation}
This condition can be simplified into:
\begin{equation}
    (1\pm rN)k\,=\,0\,\,\mathrm{mod}\,\,pN \quad \Leftrightarrow \quad \gcd(1\pm rN,pN)\,\neq\,1\,.
\end{equation}

When satisfied, the above condition implies $\Gamma^{(1)}$ is acting non-freely on the link space. Following the discussion around (\ref{eq:free-action-condition}), to get a free acting finite group $\Gamma^{(1)}$, we shall impose
\begin{equation}\label{eq:condition-on-gamma-Npr}
    \gcd(1\pm rN,pN)\,=\,1\,.
\end{equation}
For instance, the case with $p=2$ gives a freely acting $\Gamma^{(1)}$ only when $N$ is even. Furthermore, the above condition is satisfied whenever $N$ is an integer multiple of $p$. In our work here, we will be assuming this case. 

\medskip
\noindent
\textbf{An example.} To get a feeling for the group $\Gamma_{N,p,r}$, we consider the case $N=4$ and $p=2$. The value or $r$ is determined through the condition $\gcd{(2,r)}=1$, which gives $r=1$. The list of all pairs $(\alpha,\beta)\in \Z_{8}\times\Z_{2}$ is given as
\begin{equation}
 \left\{  \begin{aligned}  
        &(1,1),\quad &&(\alpha,\beta),\quad &&&(\alpha^{2},1), \quad &&&& (\alpha^{3},\beta),
        \\
        &(\alpha^{4},1),\quad &&(\alpha^{5},\beta),\quad &&&(\alpha^{6},1),\quad &&&&(\alpha^{7},\beta)
    \end{aligned}
\right\}\,.
\end{equation}
The subgroup of $\Gamma_{4,2,1}$ that acts trivially on the base $\mathbb{S}^{3}_{\mathrm{b}}$ is given by:
\begin{equation}
    \Gamma_{4,2,1}^{0}\,=\,\{\,(1,1),\  (\alpha^{2},1),\ (\alpha^{4},1) ,\  (\alpha^{6},1)  \,\}\,,
\end{equation}
which is isomorphic to $C_{0}=\Z_{4}$.

As another example, let us take $r=1$. In this case, the non-splittable quotient is expressed as
\begin{equation}
  \Gamma_{N,p,1}\,=\, \{\,(\Z_{N}\,,\,1)\,,\,(e^{\frac{2\pi i}{ pN}}\,\cdot\,\Z_{N},e^{\frac{2\pi i }{p}})\,,\,\cdots\,,\,  ( e^{\frac{2\pi i(p-1)}{ pN}}  \,\cdot\,\Z_{N}\,,\,e^{\frac{2\pi i (p-1) }{p}})  \,\}~. 
\end{equation}
In particular, we observe that the group $C_0 = \Z_N$ acts trivially on the base 3-sphere $\mathbb{S}^3_{\rm b}$. This remains to be the case for any $r$.

Based on the above discussion, the action of $\Gamma_{N,p,r}$ on the total $G_{2}$ space can thus be written as:
\begin{equation}
    \frac{(\,\R^{4}\,\times\,\mathbb{S}^{3}_{\mathrm{b}}\,)}{\Gamma_{N,p,r}}\,=\, \frac{(\R^{4}/\Z_{N}\,)\,\times\,\mathbb{S}^{3}_{\mathrm{b}}\,}{\Z_{p}}\,.
\end{equation}
Following the general physical dictionary in subsection \ref{sec:physics-non=split}, the low-energy physics associated with the $G_{2}$ space $(\R^{4}\times \mathbb{S}^{3}_{\mathrm{b}})/\Gamma_{N,p,r}$ is given as
\begin{equation}\label{eq:phys-case-1}
   \begin{split}
        &\textit{$4$d $\N=1$ $\mk{su}(N)$ gauge theory with $\Z_{p}$ inner automorphisms}
        \\
        &\qquad\qquad \textit{and $\Z_{p}$ outer automorphisms}\,.
   \end{split}
\end{equation}
Furthermore, we expect, at least, the existence of $\Z_{N}$-valued line defects correspond to Wilson and 't Hooft lines associated with the Lie algebra $\mk{su}(N)$ as reviewed around (\ref{eq:ze-zm-in-center}). Moreover, the structure of the quotient group suggests the existence of $p$ distinct $\mk{su}(N)$ theories with $\Z_{p}$-valued interfaces (non-dynamical domain walls) separating them as discussed around \eqref{eq:our-general-Gamma-Goursats}. 

To connect this with the discussion in subsection \ref{sec:outer-auto}, we want to have a $\Z_2$ outer automorphism for our theory. For our case here, this happens for even values of $p$--hence $N$ needs to be even as well. In this case, the $\Z_2$ outer automorphism induces the fundamental breaking pattern:
\begin{equation}
    \mk{su}(2M) \,\xrightarrow{\Z_{2}\,\mathrm{outer}}\, \mk{sp}(2M)\,.
\end{equation}
For the case $p=2$, the effect of $\Z_{2}$ flat connections is given in Table \ref{Table:Z2-Wilson-loops}. The interplay between outer and inner automorphisms yields rich breaking patterns. For instance, we have the following possibility
\begin{equation}\label{eq:su-sp-su+u}
    \mk{su}(2M) \,\xrightarrow{\Z_{2}\,\mathrm{outer}}\, \mk{sp}(2M) \,\xrightarrow{\Z_{2}\,\mathrm{inner}}\, \mk{su}(M)\,\oplus\,\mk{u}(1)\,.
\end{equation}

For the cases where $p$ is odd, although the resulting $\Z_p$ outer automorphism does not have a $\Z_2$ subgroup, there is a way out. What we need to do is to simply `extend' the non-splittable group $\Gamma$ to become semi-splittable, as we now discuss.

\paragraph{Semi-splittable extension.} 

The group $\Gamma_{N,p,r}$, satisfying the condition in (\ref{eq:condition-on-gamma-Npr}), can be extended by any finite $ADE$ subgroup \cite{cortés2014locallyhomogeneousnearlykahler}. In particular, we consider an extension by the finite group $\Z_{2}$, acting as the group $A$ introduced in \eqref{semi-splittable-on-link}. The resulting singular geometry is given by:
\begin{equation}
    \frac{(\,\R^{4}\,\times\,\mathbb{S}^{3}_{\mathrm{b}}\,)}{\Z_{2}\,\times\,\Gamma_{N,p,r}}\,=\, \frac{(\R^{4}/\Z_{N}\,)\,\times\,\mathbb{S}^{3}_{\mathrm{b}}\,}{\Z_{2}\,\times\,\Z_{p}}~.
\end{equation}
In M-theory, the physics associated with this space resembles that of \eqref{eq:phys-case-1}, with the additional inclusion of $\Z_{2}$ inner and outer automorphisms as discussed in \eqref{eq:hom-A-B-to-G(1)}.

The semi-splittable extension framework allows us to incorporate the case where $N$ is odd and the $\Z_{2}$ acts as an inner automorphism--a scenario that is inaccessible in the non-splittable discussion above. In particular, by specializing to $N=5$, we can realize an $\mk{su}(5)$ gauge theory and study its symmetry breaking to a Standard Model–like gauge algebra:
\begin{equation}
    \mk{su}(5) \,\xrightarrow{\Z_{2}\,\mathrm{inner}}\, \mk{su}(3)\oplus\mk{su}(2)\oplus\mk{u}(1)\,.
\end{equation}
At the gauge theory level, this setup was considered in the original work for $\mk{su}(5)$ GUT theory \cite{Georgi:1974sy}. 

\subsubsection{An \texorpdfstring{$\mk{su}(2N+1)$}{su(2N+1)} gauge theory}
For the second case in \eqref{eq:the-6-cases}, we have the triplet $(C, C_0, B) = (2\D_{2(2N+1)},\Z_{2N+1},\Z_{4})$. As discussed around table \ref{Table:list-AA0-dihedral}, to have a freely acting group involving this triplet, we should consider the semi-splittable group with $A=\Z_{m}$ for odd integer $m$. 

In this case, the acting group $\Gamma$ is given by:
\begin{equation}\label{Gamma 2}
    \Gamma^{(2)}\,=\,\{\, (\eta,\delta)\, \in\, (2\D_{2(2N+1)},\Z_{4})\,\,|\,\, \varphi:2\D_{2(2N+1)}/\Z_{2N+1}\,\longrightarrow\,\Z_{4}   \,\}\,.
\end{equation}
where $\varphi$ is an isomorphism that we determine momentarily. Let us recall that the  binary dihedral group $2\D_{2(2N+1)}$: 
\begin{equation}\label{2D22N+1}
    2\mathbb{D}_{2(2N+1)} = \langle t, s \mid {t}^{2(2N+1)} = 1, \,  t^{(2N+1)} =  s^{2}, \, {s}{ t}{ s}^{-1} = {t}^{-1} \rangle~.
\end{equation}
This can be decomposed as follows:
\begin{equation}
    2\D_{2(2N+1)}\,=\, \Z_{2N+1}\,\sqcup\, s\cdot \Z_{2N+1}\,\sqcup\, t\cdot{\Z_{2N+1}}\,\sqcup\, st\cdot \Z_{2N+1}~,
\end{equation}
where, 
\begin{equation}
    \begin{split}
            &\Z_{2N+1}\cong \{1,t^{2}, t^{4},\cdots, t^{2(2N+1)-2}\} ~.\\
    \end{split}
\end{equation}

Following \cite{cortés2014locallyhomogeneousnearlykahler}, one can assign the following  homomorphism:
\begin{equation}
        {\varphi}: 2\D_{2(2N+1)}\longrightarrow \Z_4~,
\end{equation}
such that, for $\eta\in2\D_{2(2N+1)}$:

\begin{equation}
   {\varphi}(\eta)\,:=\,
   \begin{cases}
        \begin{aligned}
           &1~,  \qquad &&\text{for}\quad \eta\,\in\, \Z_{2N+1}~,
           \\
           &e^{\frac{\pi i r}{2}}~, \qquad  &&\text{for}\quad \eta \,\in\, s\cdot \Z_{2N+1}~, 
           \\
           &e^{\pi i r}~,  \qquad &&\text{for}\quad \eta\,\in\, t\cdot{\Z_{2N+1}}~,
           \\
           &e^{\frac{3\pi i r}{2}}~, \qquad  &&\text{for}\quad \eta \,\in\, st\cdot {\Z_{2N+1}}~, 
        \end{aligned}
    \end{cases}
\end{equation}
for $r\in\Z_{4}^{\times}$. It is clear how one can define the isomorphism $\varphi$ introduced in \eqref{Gamma 2} in terms of this homomorphism. Therefore, we denoted them both by the same symbol. 

The choice of $\varphi$ is independent of the subgroup $A=\Z_{m}$. From the above homomorphism, we note that the subgroup of $\Z_{m}\times \Gamma^{(2)}$ acting trivially on $\mathbb{S}^{3}_{\mathrm{b}}$ is $\Gamma_{0}^{(2)}\,\cong\,\Z_{2N+1}\,.$ Similar to \eqref{semi-splittable-on-link}, the semi-splittable quotient in this case can be written in a similar fashion.

From this, we deduce that the total $G_{2}$-space is expressed as:
\begin{equation}
   \frac{(\R^{4}/\Z_{2N+1})\,\times\,\mathbb{S}^{3}_{\mathrm{b}}\,\,}{\Z_{4}\,\times\,\Z_{m}}\,, 
\end{equation}
Therefore, from our physical dictionary, we obtain the following effective theory:\footnote{While a $\Z_{m}$ outer automorphism might be anticipated from geometric considerations, the $\mk{su}(2N+1)$ Dynkin diagram admits only a $\Z_{2}$ outer automorphism (diagram reflection), with no higher-order automorphisms. We do not know the interpretation of these extra outer automorphisms.}
\begin{equation}
   \begin{split}
        &\textit{$4$d $\N=1$ $\mk{su}(2N+1)$ gauge theory with $\Z_{4}\times\Z_{m}$ inner automorphisms}
        \\
        &\qquad\qquad \textit{and $\Z_{4}$ outer automorphisms}\,.
   \end{split}
\end{equation}
Following (\ref{eq:ze-zm-in-center}), the spectrum of line defects is valued in $\Z_{2N+1}$. Moreover, we expect the existence of four distinct vacua (universes) corresponding to the structure of the quotient group $\Gamma^{(2)}$. These universes are separated by $\Z_{4}$-valued codimension-one defects.   

Note that we can consider the special case where the flat connections and outer automorphisms are given by $\Z_{2}\subset \Z_{4}$. The effects of such Wilson loops and monodromies are given in table \ref{Table:Z2-Wilson-loops}-\ref{Table:Z2-Wilson-non-simply-laced}. The case $N=2$ provides us with the $\mk{su}(5)$ theory, where the effect of the $\Z_{2}$ flat connections gives the doublet-triplet splitting:
\begin{equation}
    \mk{su}(5) \,\xrightarrow{\Z_{2}\,\mathrm{inner}}\, \mk{su}(3)\oplus\mk{su}(2)\oplus\mk{u}(1)\,.
\end{equation}
This is a reminiscent of the $\mk{su}(5)$ GUT theory \cite{Georgi:1974sy}. An analogous M-theory realization of the doublet-triplet splitting can be found in \cite{Witten:2001bf}.

\subsubsection{An \texorpdfstring{$\mk{su}(2N)$}{su(2N)} gauge theory}

For our third case in \eqref{eq:the-6-cases}, we have the triplet $(C, C_0, B) = (2\D_{2N},\Z_{2N},\Z_{2})$. This case is strictly non-splittable. The finite group $\Gamma^{(3)}$ is given by:
\begin{equation}\label{Gamma 3}
 \Gamma^{(3)}\,=\,\{(\eta,\delta)\,\in \, 2\D_{2N} \times\Z_{2}\,\,|\,\,\varphi:2\D_{2N}/\Z_{2N}\longrightarrow \Z_{2}\}\,.
\end{equation}
To determine the homomorphism $\varphi$, recall the definition of the group $2\D_{2N}$--see \eqref{2D22N+1} above:
\begin{equation}\label{2D2N}
    2\mathbb{D}_{2N} = \langle t, s \mid {t}^{2N} = 1, \,  t^{N} =  s^{2}, \, {s}{ t}{ s}^{-1} = {t}^{-1} \rangle~.
\end{equation}
This can be decomposed into:
\begin{equation}
    2\D_{2N}\,=\,\{1,t, t^{2},\cdots, t^{2N-1}\}\,\sqcup\,\{s, st,st^{2},\cdots,st^{2N-1}\}~.
\end{equation}
The first set here gives the normal subgroup $C_0 \cong \Z_{2N}$. While the second set is the coset $2\D_{2N}/\Z_{2N}$ represented by the generator $s$. In other words, this set corresponds to $s\cdot \Z_{2N}$. Therefore, the group $2\D_{2N}$ can be expressed as:
\begin{equation}
    2\D_{2N}\,=\, 1\cdot\Z_{2N}\,\sqcup\,s\cdot\Z_{2N}\,,
\end{equation}
which reflects the fact $2\D_{2N}/\Z_{2N}\cong\Z_{2}$.

We define the homomorphism ${\varphi}: 2\D_{2N}\longrightarrow \Z_2$ as:
\begin{equation}
   {\varphi}(\eta)\,=\,
   \begin{cases}
        \begin{aligned}
           &+1~,  \qquad &&\text{for}\quad \eta\,\in\, 1\cdot \Z_{2N}~,
           \\
           &-1~, \qquad  &&\text{for}\quad \eta \,\in\, s\cdot \Z_{2N}~. 
        \end{aligned}
    \end{cases}
\end{equation}
Extending this to an isomorphism $\varphi:2\D_{2N}/\Z_{2N} \longrightarrow \Z_2$, it follows that the group $\Gamma^{(3)}$ \eqref{Gamma 3} can be expressed as:
\begin{equation}
    \Gamma^{(3)}\,=\, \{(t,\,1)\,,\,\,(st\,,\,-1)\,   \}~.
\end{equation}

According to the physical dictionary discussed in the previous subsection, one observes that only elements of the form:
\begin{equation}
    \Gamma^{(3)}_{0}\,=\, \{\,(t,1)\,\,|\,\,t^{2N}\,=\,1\,\} \cong \Z_{2N}~,
\end{equation}
develop a codimension-four singularity as in (\ref{eq:codim-4-sing-C0}). 

In particular, the quotient space $\text{B7}/\Gamma^{(3)}$ is given as:
\begin{equation}
    \frac{(\R^{4}/\Z_{2N})\,\times\, \mathbb{S}^{3}_{\mathrm{b}}\,}{\Z_{2}}~.
\end{equation}
Moreover, the low-energy theory is
\begin{equation}
   \begin{split}
        &\textit{$4$d $\N=1$ $\mk{su}(2N)$ gauge theory with $\Z_{2}$ inner automorphisms}
        \\
        &\qquad\qquad \textit{and $\Z_{2}$ outer automorphisms.}
   \end{split}
\end{equation}
Furthermore, we expect the presence of $\Z_{2N}$-valued Wilson and 't Hooft line defects, as well as two distinct universes separated by $\Z_{2}$-valued non-dynamical domain walls.

The effect of the $\Z_{2}$ inner automorphism and the $\Z_{2}$ outer automorphism are listed in Table \ref{Table:Z2-Wilson-loops} and Table \ref{Table:outer-automorphisms}, respectively. Similar to subsection \ref{sec:case-1-su(N)theory}, one can have interesting sequences of gauge theory breaking as the example given in (\ref{eq:su-sp-su+u}).

We would like to point out, in the current case (and the case of section \ref{sec:case-1-su(N)theory}), non-trivial $\Z_{2}$ flat connections break the $\mk{su}(4)$ gauge theory to:
\begin{equation}
    \mk{su}(4) \,\xrightarrow{\Z_{2}\,\mathrm{inner}}\, \mk{su}(3)\oplus\mk{u}(1)~,
\end{equation}
which is a reminiscent of the Pati--Salam theory \cite{Pati:1974yy}.

\subsubsection{An \texorpdfstring{$\mk{so}(8)$}{so(8))} gauge theory}
Let us now consider the fourth case in \eqref{eq:the-6-cases}. We have the triplet $(C, C_0, B) = (2\T,2\D_{4},\Z_{3})$. As discussed around table \ref{Table:list-AA0-dihedral}, this case gives a free action on the link space only when treated as a semi-splittable quotient. The additional splittable group component $A$ admits two possible realizations given in table \ref{Table:list-AA0-dihedral}, namely, the cyclic quotient with $\Z_{k}$ such that $3\nmid k$  or the dihedral with $2\D_{2l}$ such that $3\nmid 2l$. In both cases, the additional quotient by the group $A$ does not enhance the codimension-four singularity--which in this case is a $2\D_{4}$ singularity as we now show. In the following, we will only focus on the case $A=\Z_{k}$.

The non-splittable part for this case is given by:
\begin{equation}\label{Gamma 4}
    \Gamma^{(4)}\,=\, \{\,(\eta,\omega)\,\in\,2\T\,\times\,\Z_{3}\,\,|\,\,\varphi:2\T/2\D_{4}\longrightarrow\Z_{3}\}~.
\end{equation}
To determine the isomorphism $\varphi$, let us recall here that the binary tetrahedral group $2\T$ can be written as follows:
\begin{equation}\label{eq:2T-group}
    2\T = 2\D_{4} \,\sqcup\, \frac{1}{2}\begin{pmatrix}
        -1-i&-1-i\\1-i&-1+i
    \end{pmatrix}\cdot2\D_{4} \,\sqcup\, \frac{1}{2}\begin{pmatrix}
        -1+i&1+i\\-1+i&-1-i
    \end{pmatrix}\cdot2\D_{4}\,.
\end{equation}
Here, we are representing the quaternions as matrices.\footnote{For a quaternion $\mathsf{q} = a+b \,i + c\, j+d\,k$, we can write it in a matrix form as $\mathsf{q} = \begin{pmatrix}
    a+b\,i& c+d\,i\\
    -c+d\,i&a-b\,i
\end{pmatrix}$.} Inspired by this decomposition, we define the homomorphism ${\varphi}: 2\T \longrightarrow \Z_3$ via \cite{cortés2014locallyhomogeneousnearlykahler}:
\begin{equation}
   {\varphi}(\eta)\,=\,
   \begin{cases}
        \begin{aligned}
           &+1~,  \qquad &&\text{for}\quad \eta\,\in\, 2\D_{4}~,
           \\
           &e^{\frac{2\pi i r}{3}}~, \qquad  &&\text{for}\quad \eta \,\in\,\frac{1}{2}\begin{pmatrix}
        -1-i&-1-i\\1-i&-1+i
    \end{pmatrix}\,\cdot\, 2\D_{4}~, 
           \\
           &e^{\frac{4\pi i r}{3}}~, \qquad  &&\text{for}\quad \eta \,\in\, \frac{1}{2}\begin{pmatrix}
        -1+i&1+i\\-1+i&-1-i
    \end{pmatrix}\,\cdot\,2\D_{4}~,
        \end{aligned}
    \end{cases}
\end{equation}
with $r\in\Z_{3}^{\times}$. It is clear how to define the isomorphism $\varphi$ appearing in \eqref{Gamma 4} in terms of ${\varphi}$.

The subgroup acting trivially on $\mathbb{S}_{\mathrm{b}}^{3}$ consists of elements $\eta \in 2\T$ with $\varphi(\eta)=1$, which precisely yields a $2\D_{4}$ singularity:
\begin{equation}
    \Gamma^{(4)}_{0}\,=\,2\D_{4}~.
\end{equation}
Moreover, without delving into details, one finds that the general structure of the quotient group $\Gamma^{(4)}$ can be expressed in a form analogous to \eqref{eq:general-Gamma-Goursats-semi-split}, with three components inherited from the construction in (\ref{eq:2T-group}).

Following our general physical dictionary, the low-energy physics of the quotient $G_{2}$-space:
\begin{equation}
   \frac{(\R^{4}/2\D_{4})\,\times\,\mathbb{S}^{3}_{\mathrm{b}}}{\Z_{3}\,\times\,\Z_{k}}~, 
\end{equation}
is described by:
\begin{equation}
   \begin{split}
        &\textit{$4$d $\N=1$ $\mk{so}(8)$ gauge theory with $\Z_{3}\times\Z_{k}$ inner automorphisms}
        \\
        &\qquad\qquad \textit{and $\Z_{3}\times\Z_{k}$ outer automorphisms.}
   \end{split}
\end{equation}

Due to the structure of the semi-splittable quotient groups, there are three different universes as discussed around \eqref{eq:general-Gamma-Goursats-semi-split} separated by $\Z_{3}$-valued co-dimension one defects. As we have $\mk{so}(8)$ gauge theory in each universe, the set of Wilson and 't Hooft line defects take values in $\Z_{2}\oplus\Z_{2}$.

We consider inner automorphisms and outer automorphisms separately (either $\Z_{3}$ or $\Z_{k}$ valued). In general, the $\mk{so}(8)$ gauge theory undergoes sequential symmetry breaking. For the special case $k=2$, three distinct scenarios emerge (see tables \ref{Table:outer-automorphisms} and \ref{Table:Z2-Wilson-non-simply-laced}):
\begin{itemize}
    \item \textbf{The first scenario:} The $\mk{so}(8)$ gauge algebra is broken first by a $\Z_{3}$ outer automorphism and then by a $\Z_{2}$ flat connection giving:
    \begin{equation}
      \mk{so}(8)\ \xrightarrow[]{\Z_{3}\text{\, outer}} \ \mk{g}_{2} \xrightarrow[]{\Z_{2}\text{\, inner}}  \ \mk{su}(2)\,\oplus\,\mk{su}(2)~.
    \end{equation}

    \item \textbf{The second scenario:} The $\mk{so}(8)$ gauge algebra is broken first by a $\Z_{2}$ outer automorphism and then by a $\Z_{2}$ flat connection giving:
     \begin{equation}
      \mk{so}(8)\ \xrightarrow[]{\Z_{2}\text{\, outer}} \ \mk{so}(7) \xrightarrow[]{\Z_{2}\text{\, inner}}  \ \mk{so}(3)\,\oplus\,\mk{so}(4)~.
    \end{equation}
    
    \item \textbf{The third scenario:} The $\mk{so}(8)$ gauge algebra is broken first by a $\Z_{2}$ flat connection and then by a $\Z_{2}$ outer automorphism giving:
    \begin{equation}
      \mk{so}(8)\ \xrightarrow[]{\Z_{2}\text{\, inner}} \ \mk{su}(4)\,\oplus\,\mk{u}(1) \xrightarrow[]{\Z_{2}\text{\, outer}}  \ \mk{sp}(4)\,\oplus\,\mk{u}(1)~.
    \end{equation}
\end{itemize}


\subsubsection{An \texorpdfstring{$\mk{so}(2(N+2))$}{so(2(N+2))} gauge theory}
Next, we look at the fifth case in \eqref{eq:the-6-cases}. Here, we have the defining triplet $(C, C_0, B) = (2\D_{4N},2\D_{2N},\Z_{2})$. This case is non-splittable. Let us start by recalling the explicit form of the group $2\D_{4N}$:
\begin{equation}
    2\mathbb{D}_{4N} = \langle r, s \mid {t}^{4N} = 1, \,  t^{2N} =  s^{2}, \, {s}{ t}{ s}^{-1} = {t}^{-1} \rangle~.
\end{equation}
One can verify that the above group can be decomposed as:
\begin{equation}\label{eq:all-elements-2D4N}
    2\D_{4N}\,=\,1\cdot2\D_{2N} \, \sqcup\,t\cdot2\D_{2N}~,
\end{equation}
where \eqref{2D2N}, 
\begin{equation}
        2\D_{2N} =    \{ 1,s,t^{2},st^{2}, \cdots, t^{2n},st^{2n},\cdots \} ~.
\end{equation}

Therefore, one can define the isomorphism $\varphi : 2\D_{4N}/2\D_{2N}\longrightarrow \Z_2$ in terms of the homomorphism ${\varphi}:2\D_{4N}\longrightarrow \Z_2$:
\begin{equation}
   {\varphi}(\eta)\,=\,
   \begin{cases}
        \begin{aligned}
           &+1~,  \qquad &&\text{for}\quad \eta\,\in\, 1\cdot 2\D_{2N}~,
           \\
           &-1~, \qquad  &&\text{for}\quad \eta \,\in\, t\cdot 2\D_{2N}~, 
        \end{aligned}
    \end{cases}
\end{equation}
with $\eta$ represents a generic element in $2\D_{4N}$. Building on these observations, the quotient group for this case, $\Gamma^{(5)}$, can be written explicitly as:
\begin{equation}\label{eq:Gamma(5)}
    \Gamma^{(5)}\,=\, \{ \, (t^{2k},1)\,,\,(st^{2k},1)\,,\,(t^{2k+1},-1)\,,\,(st^{2k+1},-1)~, \quad k\in \Z_{4N}\,\}~,
\end{equation}

The subgroup that acts trivially on $\mathbb{S}_{\mathrm{b}}^{3}$ is isomorphic to the normal subgroup $2\D_{2N}$:
\begin{equation}
    \Gamma_{0}^{(5)}=\,\{\,(t^{2k},1)\,,\,(st^{2k},1) ~, \quad k\in \Z_{4N}\,\}\,\cong\, 2\D_{2N}~.
\end{equation}
Therefore, the quotient $G_{2}$ geometry can be expressed as
\begin{equation}
   \frac{(\R^{4}/2\D_{2N})\,\times\,\mathbb{S}^{3}_{\mathrm{b}}\,}{\Z_{2}}\,.
\end{equation}
Moreover, the associated low-energy field theory is given by:
\begin{equation}
   \begin{split}
        &\textit{$4$d $\N=1$ $\mk{so}(2(N+2))$ gauge theory with $\Z_{2}$ inner automorphisms}
        \\
        &\qquad\qquad \textit{and $\Z_{2}$ outer automorphisms.}
   \end{split}
\end{equation}

From the structure of the quotient group in (\ref{eq:Gamma(5)}), we deduce the existence of two distinct vacua, separated by $\Z_{2}$ non-dynamical domain walls. Within each universe, there exist (at least) line defects, namely, Wilson and 't Hooft lines, whose charges take values in the center of the corresponding gauge group.

The action of these $\Z_{2}$ Wilson loops and $\Z_{2}$ monodromies are given in tables \ref{Table:Z2-Wilson-loops}--\ref{Table:Z2-Wilson-non-simply-laced}. In this case, we assume that $N\geq 2$.\footnote{For the case $N=1$, the group $C_0 \cong 2\D_{2}\cong\Z_4$. This gives us $\mathfrak{su}(4)$ theory, which is of the form $\mathfrak{su}(2N)$ that we studied earlier.} For example, the case $N=3$ is a reminiscent of the $SO(10)$ GUT theory. According to table \ref{Table:Z2-Wilson-loops}, $\mk{so}(10)$ can be broken in two different ways  \cite{Georgi:1974my,Fritzsch:1974nn,Mohapatra:1986uf}:
\begin{equation}\label{eq:inner-on-so-10}
    \begin{split}
        &\mk{so}(10)\ \to \ \mk{so}(6)\,\oplus\,\mk{so}(4)\ \to \ \mk{su}(4)\oplus \mk{su}(2)\oplus \mk{su}(2)~,
        \\
        &\mk{so}(10)\ \to \ \mk{su}(5)\,\oplus\,\mk{u}(1)~.
    \end{split}
\end{equation}

The interplay between inner and outer automorphisms yields the following sequences:
\begin{itemize}
    \item \textbf{The first scenario:} The $\mk{so}(2(N+2))$ gauge algebra is broken first by a $\Z_{2}$ outer automorphism and then by a $\Z_{2}$ flat connection giving:
    \begin{equation}
      \mk{so}(2(N+2))\ \xrightarrow[]{\Z_{2}\text{\, outer}} \ \mk{so}(2(N+2)-1) \xrightarrow[]{\Z_{2}\text{\, inner}}  \ \mk{so}(n_{1})\,\oplus\,\mk{so}(n_{2})~,
    \end{equation}
with $n_{1}+n_{2}=2(N+2)-1$.
    
    \item \textbf{The second scenario:} The $\mk{so}(2(N+2))$ gauge algebra is broken first by a $\Z_{2}$ flat connection and then by a $\Z_{2}$ outer automorphism. This scenario splits into two cases depending on $N$ being even or odd. For the even case, we have:
    \begin{equation}
      \mk{so}(2(N+2))\ \xrightarrow[]{\Z_{2}\text{\, inner}} \ \mk{su}(N+2)\,\oplus\,\mk{u}(1) \xrightarrow[]{\Z_{2}\text{\, outer}}  \ \mk{sp}(N+2)\,\oplus\,\mk{u}(1)~.
    \end{equation}
Meanwhile, for the odd case, we have: 
    \begin{equation}
      \mk{so}(2(N+2))\ \xrightarrow[]{\Z_{2}\text{\, inner}} \ \mk{su}(N+2)\,\oplus\,\mk{u}(1) \xrightarrow[]{\Z_{2}\text{\, outer}}  \ \mk{so}(N+2)\,\oplus\,\mk{u}(1)~.
    \end{equation}
Note that in both cases above, we assumed that the $\mk{u}(1)$ factor survive the breaking by the $\Z_{2}$ outer automorphism. 
\end{itemize}

\medskip
\noindent
\underline{\textit{A side note:}} Following the discussion in section 6 of \cite{Atiyah:2001qf}, the $2\D_{2N}$ singularity described above can give rise to $\mk{sp}(N-2)$ gauge theory. From the type IIA string theory perspective, this corresponds to $(N-2)$ D6-branes and an O6$^{+}$ wrapping $\mathbb{S}^{3}$ in the deformed conifold $T^{\ast}\mathbb{S}^{3}$. While this scenario provides an interesting alternative realization, we will not explore it further here. For additional details, we refer the reader to the original reference.


\subsubsection{An \texorpdfstring{$\mk{e}_{6}$}{e6} gauge theory}
Let us now move on to the last case in our list \eqref{eq:the-6-cases}. This case consists of the triplet $(C, C_0,B) = (2\O,2\T,\Z_{2})$. This case is non-splittable. The finite quotient group is given by: 
\begin{equation}\label{Gamma 6}
   \Gamma^{(6)}\,=\,\{(\eta,\delta)\,\in \, 2\O \times\Z_{2}\,\,|\,\,\varphi:2\O/2\T\longrightarrow\Z_{2}\}~,
\end{equation}
with $\varphi$ an isomorphism that we construct momentarily. Let us recall here that the binary octahedral group $2\O$ can be written in terms of $2\T$ \eqref{eq:2T-group} as:
\begin{equation}
    2\O\,=\, 1\cdot  2\T\,\sqcup\,e^{\frac{i\pi}{4}}\cdot2\T~.
\end{equation}
Therefore, we can associate the homomorphism ${\varphi}$ for $\eta\in2\O$ as:  
\begin{equation}
   {\varphi}(\eta)\,:=\,
   \begin{cases}
        \begin{aligned}
           &+1~,  \qquad &&\text{for}\quad \eta\,\in\, 1\cdot 2\T~,
           \\
           &-1~, \qquad  &&\text{for}\quad \eta \,\in\, e^{\frac{i\pi}{4}}\cdot 2\T~.
        \end{aligned}
    \end{cases}
\end{equation}
This can be uplifted to an isomorphism $\varphi:2\O/2\T\longrightarrow\Z_2$. 

The elements of the group $\Gamma^{(6)}$ are given as:
\begin{equation}
    \Gamma^{(6)}\,=\, \{\, (\zeta,1)\,,\,(e^{\frac{i\pi}{4}}\zeta,-1)   \,\,|\,\, \zeta\in 2\T\,\}\,.
\end{equation}
From this, we see that the subgroup $\Gamma_0^{(6)}$ that acts trivially on the base 3-sphere is isomorphic to $2\T$. Therefore, the quotient $G_{2}$ space can be expressed as
\begin{equation}
    \frac{\,(\R^{4}/2\T)\,\times\, \mathbb{S}^{3}_{\mathrm{b}}\,}{\Z_{2}}~.
\end{equation}
The interpretation of such a quotient space is given by the following low-energy gauge theory:
\begin{equation}
  \begin{split}
       &\textit{$4$d $\N=1$  $\mk{e}_{6}$  gauge theory with $\Z_{2}$ inner automorphisms}
        \\
        &\qquad\qquad \textit{and $\Z_{2}$ outer automorphisms.}
   \end{split}
\end{equation}

We find two distinct universes with the above physical interpretation, separated by codimension-one $\mathbb{Z}_2$ defects. The presence of the $\mathfrak{e}_6$ Lie algebra dictates the existence of $\mathbb{Z}_3$-valued electric and magnetic line defects, consistent with the discussion around (\ref{eq:ze-zm-in-center}).

The effects of the $\Z_{2}$ flat connections and monodromies are given in tables \ref{Table:Z2-Wilson-loops}--\ref{Table:Z2-Wilson-non-simply-laced}. The interplay between the inner and outer automorphisms leads to the following possible cases:
\begin{itemize}
    \item \textbf{The first scenario:} breaking the $\mk{e}_{6}$ gauge algebra first by am outer $\Z_{2}$ and then by an inner $\Z_{2}$ flat connection. This scenario splits into two possibilities: The first gives
    \begin{equation}
    \begin{split}
         &\mk{e}_{6}\ \xrightarrow[]{\Z_{2}\text{\, outer}} \ \mk{f}_{4} \xrightarrow[]{\Z_{2}\text{\, inner}}  \ \mk{su}(2)\,\oplus\,\mk{sp}(6)~.\\
        &\mk{e}_{6}\ \xrightarrow[]{\Z_{2}\text{\, outer}} \ \mk{f}_{4} \xrightarrow[]{\Z_{2}\text{\, inner}}  \ \mk{so}(9)~.
     \end{split}
    \end{equation}
    
\item \textbf{The second scenario:} breaking the $\mk{e}_{6}$ gauge algebra first by an inner $\Z_{2}$ and then by an outer $\Z_{2}$. Again, this scenario splits into two cases:

The first gives
      \begin{equation}
         \mk{e}_{6}\ \xrightarrow[]{\Z_{2}\text{\, inner}} \ \mk{so}(10)\oplus\mk{u}(1) \xrightarrow[]{\Z_{2}\text{\, outer}}  \ \mk{so}(4)\,\oplus\,\mk{u}(1)\,,
    \end{equation}
The intermediate step above can be found in \cite{Breit:1985ud}. The last step is equivalent to having $\mk{su}(2)_{L}\,\oplus\,\mk{su}(2)_{R}\,\oplus\,\mk{u}(1)$.

Let us comment here that the intermediate theory $\mk{so}(10) \oplus \mk{u}(1)$ inherits both inner and outer $\mathbb{Z}_2$ automorphisms from the original $\mk{e}_6$ theory. This provides an alternative perspective on the action of the $\Z_{2}$ outer automorphism in the second step of the symmetry-breaking chain. Rather than invoking the outer automorphism, one could instead activate a $\Z_{2}$ inner automorphism, realizing the symmetry-breaking pattern described in (\ref{eq:inner-on-so-10}). Therefore, in this way, one is able to reconstruct the phenomenologically relevant chain of Lie algebra embeddings:
\begin{equation}
    \mk{su}(5)\,\subset\,\mk{so}(10)\,\subset\,\mk{e}_{6}\,.
\end{equation}
As for the  second  case gives
      \begin{equation}
         \mk{e}_{6}\ \xrightarrow[]{\Z_{2}\text{\, inner}} \ \mk{su}(6)\oplus\mk{su}(2) \xrightarrow[]{\Z_{2}\text{\, outer}}  \ \mk{sp}(6)\,\oplus\,\mk{sp}(2)\,.
    \end{equation}

\item \textbf{The third scenario:} this is built on the observation made in  table \ref{Table:outer-automorphisms}. That is, the $\Z_{2}$ outer automorphism of the $\mk{e}_{6}$ algebra can be taken to act only on the two distal simple roots. Again, this scenario gives two cases: 
    \begin{equation}
    \begin{split}
         &\mk{e}_{6}\ \xrightarrow[]{\Z_{2}\text{\, outer}} \ \mk{sp}(8) \xrightarrow[]{\Z_{2}\text{\, inner}}  \ \mk{su}(4)\,\oplus\,\mk{u}(1)~.\\
         &\mk{e}_{6}\ \xrightarrow[]{\Z_{2}\text{\, outer}} \ \mk{sp}(8) \xrightarrow[]{\Z_{2}\text{\, inner}}  \ \mk{sp}(4)\,\oplus\,\mk{sp}(4)~.
         \end{split}
    \end{equation}
\end{itemize}
The embedding of the above intermediate steps in the $\mk{e}_{6}$ Lie algebra can be found, e.g., in \cite{Kephart:1981gf} and appendix A of \cite{Babu:2023zsm}.

\paragraph{Semi-splittable extension.} As the group $\Gamma^{(6)}$ is freely acting non-splittable finite group (see table \ref{Table:list-AA0-dihedral}), then it can be extended by any $ADE$ group $A$ such that the resulting semi-splittable finite group $A\times\Gamma^{(6)}$ still acts freely. The structure of the subgroup $\Gamma^{(6)}$ and its homomorphism $\varphi$ are independent of the choice of the group $A$. Here, we would like to consider the case $A=\Z_{3}$.

In this case, the quotient BS space is given as:
\begin{equation}
    \frac{\,(\R^{4}/2\T)\,\times\, \mathbb{S}^{3}_{\mathrm{b}}\,}{\Z_{3}\,\times\,\Z_{2}}\,,
\end{equation}
The associated low-energy gauge theory is given as:
\begin{equation}
   \begin{split}
        &\textit{$4$d $\N=1$  $\mk{e}_{6}$  gauge theory with $\Z_{3}\times \Z_{2}$ inner automorphisms}
        \\
        &\qquad\qquad \textit{and $\Z_{3}\times \Z_{2}$ outer automorphisms.}
   \end{split}
\end{equation}
Taking all automorphisms to be trivial but the $\Z_{3}$ flat connections, then, the $\mk{e}_{6}$ gauge algebra breaks according to:
\begin{equation}
         \mk{e}_{6}\ \xrightarrow[]{\Z_{3}\text{\, inner}} \ \mk{su}(3)\,\oplus\,\mk{su}(3)\,\oplus\,\mk{su}(3) \,.
\end{equation}
Such a breaking was considered in \cite{Gursey:1975ki} and further analyzed, e.g., in \cite{Nanopoulos:1980kk,Gursey:1981kf,Mohapatra:1985xm}. From a string theory perspective, it was also studied in \cite{delAguila:1985hkb}.

\subsection{Further comments on the B7 space}

The 7-dimensional B7 space is a $G_{2}$-manifold admitting a 1-parameter family of cohomogeneity-one $G_{2}$-metrics on $\R^{4}\times \mathbb{S}^{3}$ \cite{Brandhuber:2001yi,bazaikin2013complete}. The space is asymptotically locally conical (ALC), smooth, and simply connected. At a special value in the parameter space, we obtain a unique metric of the family with asymptotically conical (AC) geometry. This limit corresponds precisely to the BS metric \cite{Bryant1989OnTC}, which has been discussed earlier in this work.

The key distinction between an AC and an ALC metric is that the latter maintains a finite size circle ($U(1)$ fiber) at the link space--see e.g. \cite{Brandhuber:2001yi}. Therefore, one of the $SU(2)$'s in $G^{\mathrm{BS}}$ should be broken to the normalizer of the finite circle, identified with the Hopf fiber of the broken $SU(2)$. By definition, the normalizer of a $U(1)$ inside an $SU(2)$ is given by all elements $\omega$ that conjugate the $U(1)$ subgroup back to itself in $SU(2)$. This is given by: 
\begin{equation}
    N\,=\, U(1)\,\rtimes\, \Z_{2}~.
\end{equation}
Here, the $2\times 2$ representation of the $\Z_{2}\cong N/U(1)$ can be taken as:
\begin{equation}
   j\,=\, \begin{pmatrix}
        \, \, \, \, 0   & \,1\,\\
        -1 & \,0\,
    \end{pmatrix}~.
\end{equation}

For the semi-classical branch $X_{(1;3)}=\R^{4}\times\mathbb{S}^{3}_{\rm{b}}$, whose link space and quotient is given in (\ref{eq:action-abc-S3fS3b}). The isometry group for $X_{(1;3)}$ is given by:
\begin{equation}\label{eq:B7-isometry}
    G^{\mathrm{B7}} \,=\,\frac{SU(2)_{\ell,1}\times SU(2)_{\ell,2}\times (\, \Delta U(1)_{r}\rtimes \Z_{2}\,)}{\Delta\Z_{2,\ell}}~.
\end{equation}
Here, the finite circle is identified with the Hopf fiber of $\Delta SU(2)_{r}$.

Since the finite quotient groups $\Gamma$ considered in the cases \eqref{eq:the-6-cases} are subgroups of $C\times B$--where $C\subset (\Delta U(1)_{r}\rtimes \Z_{2})$ and $B\subset SU(2)_{\ell,2}$--we are restricted by the reduced isometry of the B7 space. Specifically, the full diagonal $\Delta SU(2)_{r}$ symmetry of the Bryant--Salamon limit is broken to $(\Delta U(1)_{r}\rtimes \Z_{2})$ in the generic ALC $G_{2}$-metrics on B7. Consequently, only cases 1, 2, 3, and 5 among the six cases listed in (\ref{eq:the-6-cases}) can be defined. Therefore, in a generic B7 geometry (away from the BS limit), and in particular on the semi-classical branch $X_{(1;3)}$, the gauge theories associated with the algebras $\mk{so}(8)$ and $\mk{e}_{6}$ cannot be realized.

\section{SymTFT of 4d \texorpdfstring{$\N=1$}{N=1} SYM due to non-splittable quotients}\label{sec:SymTFT}

In this section, we study the 5d SymTFT for the geometrically engineered 4d $\mathcal{N}=1$ gauge theory from M-theory. After giving a lightning review of some of the essential tools from differential cohomology, we give a proposal for the computation of the homology ring of the link space $L_6$ coming from the six cases of interest \eqref{eq:the-6-cases}. Using these results, we work out the full 5d SymTFT action and calculate the different coefficients (couplings) that appear in it. Motivating the work in the following sections, we make some comments on the higher-form symmetries of these 4d gauge theories.

\subsection{SymTFT and M-theory geometric engineering: general discussion}

In this subsection, we give some general discussion on the geometric engineering of 5d SymTFTs from 11d M-theory. We will also give a lightning review of some of the essential ingredients of differential cohomology that we will use later on.

\subsubsection{Reducing the M-theory action: general philosophy}

We are interested in reducing the action of the 11d M-theory along the link space $L_6$ and studying aspects of the 5d SymTFTs that we obtain from it. Recall that the bosonic part of the 11d supergravity action involving the 3-form potential $C_{3}$ is given by \cite{Cremmer:1978km}:
\begin{equation}\label{S M tot}
    S^{\rm M}_{\rm tot} = S_{\rm CS}^{\rm M} + S^{\rm M}_{\rm kin} ~, 
\end{equation}
where the first term is the topological part given as the Chern--Simons functional of the 11d supergravity 4-form $G_{4}=\dd C_{3}\in H^4(M_{11},\Z)$. This can be uplifted to ${\br G}_4\in\br{H}^4(M_{11})$ in the differential cohomology of the 11-manifold $M_{11}$. In terms of this Cheegar--Simons character, the topological action can be written as:
\begin{equation}\label{S M CS}
    S_{\rm CS}^{\rm M} \,=\, -\frac{1}{6} \,\int_{M_{11}}^{\br H} \,{\br G}_4\, \star\,{\br G}_4\,\star\,{\br G}_4 \,\in\, \R/\Z~.
\end{equation}

As for the second term, this is the kinetic part, which is given by:
\begin{equation}\label{S M kin}
    S_{\rm kin}^{\rm M} \,=\, \int_{M_{11}}^{\br H}\, {\br G}_4 \star \br{dG}_7 \,\in\, \R/\Z~,
\end{equation}
where here $\br{dG}_7\in {\br H}^8(M_{11})$ is the CS character uplift of the globally defined 7-form $G_7\in \Omega^7(M_{11})$.

For our purposes in this work, let us take the 11-dimensional manifold $M_{11}$ to be of the form $M_{11} = Y_5\times L_6$. Here $L_6$ is the link space we have been discussing so far, and $Y_5$ is a smooth 5-manifold of the form $Y_5 = M_4\times [0,\infty)$.\footnote{To simplify the calculations here, we will assume that the (co)homology ring of $Y_5$ is torsion free. With this in mind, we will not have to worry about usual contributions from such terms to the K\"unneth formula.}

With this in mind, our goal in this section is to explicitly work out the form of the 5d SymTFT action that we obtain from all the possible Link spaces defined by the cases listed in \eqref{eq:the-6-cases}. To do so, we will decompose the characters $\br{G}_4$ and $\br{dG}_7$ into products of modes along $Y_5$ and $L_6$ and then perform the differential cohomology integral along the link space. For this, let us review how to perform these integrals and list some results that will be utilised later on.

    \begin{figure}[t]
    \centering
    \begin{tikzpicture}
    \node (brH) at (0,0) {$\br{H}^{p} (M_{11})$};
    \node (OZ) at (3,1) {$\Omega^{p}_{\Z} (M_{11})$};
    \node (ohr) at (6,2) {$0$};
    \node (ohl) at (-6,-2) {$0$};
    \node (Hhl) at (-3,-1) {$H^{p-1} (M_{11}, \R/\Z)$};
    \draw[->] (ohl) -- (Hhl);
    \draw[->] (Hhl) edge node[anchor=south] {${\scriptstyle i}$} (brH);
    \path[->] (brH) edge node[anchor=south] {${\scriptstyle \mathscr{F}}$} (OZ);
     \draw[->] (OZ) -- (ohr);
     
    \node (HZ) at (3,-1) {$H^{p} (M_{11},\Z)$};
    \node (ovb) at (6,-2) {$0$};
    \node (ovt) at (-6,2) {$0$};
    \node (Oc) at (-3,1) {$\Omega^{p-1} (M_{11}) / \Omega^{p-1}_{\Z} (M_{11})$};
    
    \path[->] (brH) edge node[anchor=north] {$\scriptstyle c$} (HZ);
    \draw[->] (ovt) -- (Oc);
    \draw[->] (Oc) -- (brH);
    \draw[->] (HZ) -- (ovb);
    \end{tikzpicture}    
    \caption{The two diagonal sequences are exact, connecting the differential cohomology with the standard types of cohomologies. The map $\mathscr{F}$ is the field strength map, and the map $c$ is the topological class map.}
    \label{2exactseqbrH}
    \end{figure}    

\subsubsection{A lightning review of integration in differential cohomology}
Before we dive into the details of the reduction along the link space $L_6$, let us take a pause and review some of the essential ingredients and properties concerning integration $\int^{\br{H}}$ in differential cohomology. The important components that we will need for what follows are exhibited in the two exact sequences in Figure \ref{2exactseqbrH}, where we refer to $\mathscr{F}$ as the field strength map and $c$ as the topological class map.

With these maps in mind, for a CS character $\br{a}\in \br{H}^{11}(M_{11})$ we define:
\begin{equation}\label{primary inv}
    \int^{\br H}_{M_{11}} \br{a} \,=\, \int_{M_{11}} c(\br{a}) \,\in\, \Z~,
\end{equation}
where the RHS is the usual integral over $H^p(M,\Z)$. This differential cohomology integral is usually referred to as \textit{the primary invariant} of the differential cohomology class $\br{a}$.

Let us comment here on the difference between the topological class and the field strength maps. Note that, for any differential character $\br{a}\in \br{H}^p(M_{11})$, we can consider the two corresponding images $c(\br{a})$ and $\mathscr{F}(\br{a})$. In general, the topological class contains more information about the differential character than does the field strength map. This follows from the fact that $c$ lands in the singular cohomology, which can be torsional with the torsion being detectable by $c$. Meanwhile, $\mathscr{F}$ lands in the de Rham cohomology of $M_{11}$, which is insensitive to torsion. Therefore, throughout the paper, we will use the topological class rather than field strength to compute integrals of the form above since the former reduces to the latter in the free case. For more discussion on this point, see section 2 of \cite{Apruzzi:2021nmk}.

The injective map $i$ in figure \ref{2exactseqbrH} associates a character to each `flat connection' on $M_{11}$. For dimensional reasons, it is clear that any character $\br{A}\in \br{H}^{12}(M_{11})$ is flat (i.e., the associated field strength is trivial). Therefore, one can find a unique class $[u]\in H^{11}(M_{11}, \mathbb{R}/\Z)$ such that $i(u) = \br{A}$. With this in mind, we define the \textit{secondary invariant} associated with $\br{A}$ as follows:
\begin{equation}\label{second inv}
    \int_{M_{11}}^{\br H} \br{A} \,=\, \int_{M_{11}} u \,\in\, \R/\Z~.
\end{equation}

As mentioned earlier, let us take the case $M_{11} = Y_5\times L_6$. When decomposing the $\br{G}_4$ and $\br{G}_7$ characters and plugging them into the actions \eqref{S M tot}, we need to perform the integral along each component separately. For this, one needs to decompose the integral $\int^{\br{H}}$ over the product space $Y_5\times L_6$ and perform the integral over the link space to obtain the 5d SymTFT action on $Y_5$:

\begin{equation}\label{int splitting}
    \int_{Y_5\times L_6}^{\br H} \br{F}_p \star \br{G}_q = (-1)^{q} \left(\int_{Y_5}^{\br H} \br{F}_p\right)\left(\int_{L_6}^{\br H} \br{G}_q\right)~.
\end{equation}
where $\br{F}_p\in \br{H}^{p}(Y_5)$ and $\br{G}_q\in \br{H}^q(L_6)$. Then, one can use the primary and secondary invariants defined in \eqref{primary inv} and \eqref{second inv} respectively to work out these integrals.

The product operation that appears on the LHS above is usually referred to as the internal product in the following sense:
\begin{equation}
        \star \,:\, \br{H}^p(M_{11}) \,\otimes\, \br{H}^q(M_{11}) \,\longrightarrow\, \br{H}^{p+q}(M_{11})~,
\end{equation}
 with the commutativity property: $\br{F}_p\star \br{G}_q = (-1)^{pq} \br{G}_{q} \star \br{F}_p$\footnote{Due to this commutation relation, the ordering of the characters appearing in the integrand \eqref{int splitting} matters. In particular, for $\br{F}_p\in {\br H}^{p}(X)$ and $\br{G}_q\in \br{H}^q(Y)$, we have--see \cite[Theorem 59]{bar2014differential}:
\protect\begin{equation}
    \int^{\br H}_{X\times Y} {\br F}_p\star{\br G}_q = (-1)^{(q-\dim(Y))\dim(X)} \left(\int^{\br H}_{X}{\br F}_{p}\right)\star \left(\int_Y^{\br H} {\br G}_{q}\right)~.
\end{equation}}.
We also remark here that, upon using the field strength map $\mathscr{F}$, this reduces to the wedge product in the cohomology ring $\Omega^\bullet(M_{11})$. Meanwhile, via the topological class map, this becomes the cup product between differential forms in $H^{\bullet}(M_{11},\Z)$.

 Another type of product operation we have in differential cohomology is the external product, usually denoted by $\times$. This is usually in cases similar to our situation, where $M_{11} = Y_5\times L_6$. In this case, we have:
 \begin{equation}   
    \times \,:\, \br{H}^p(Y_5) \,\otimes\, \br{H}^q(L_6) \,\longrightarrow\, \br{H}^{p+q}(M_{11})~,
 \end{equation}
such that
\begin{equation}
        \br{F}_p\times \br{G}_{q} \,:=\, (\pi_{5}^* \br{F}_p) \star (\pi_{6}^* \br{G}_{q})~.
\end{equation}
Here, we used the projection maps:
\begin{equation}
    Y_5 \,\xlongleftarrow{\pi_{5}}\, M_{11}\, \xlongrightarrow{\pi_{6}}\,L_6~.
\end{equation}
Although we will be abusing the notation in what follows by dropping $\pi^*_{\bullet}$, which differential cohomology ring we are working on should be understood from context and the type of product we are using.

\subsection{(Co)homology groups of the link space \texorpdfstring{$L_{6}$}{L6}}\label{sec:co-homology}

In this subsection, we discuss the details of the cohomology ring of the link spaces $L_6$ that we obtain from quotients with respect to the groups listed in \eqref{eq:the-6-cases}. In this part of the discussion, we will go through all six cases, non-splittable and semi-splittable, where for the latter case, we will postpone the discussion of the quotient by the finite group $A$ till the end.

Naively, one might go on with computing the homology of the quotient by applying K\"unneth formula to the product space:
\begin{equation}\label{link as product}
     L^{\rm naive}_6 := \frac{\mathbb{S}^3_{\rm f}}{C_0\times B} \times \frac{\mathbb{S}^3_{\rm b}}{B}~,
\end{equation}
using the homology of the two separate components. But, as one would anticipate from the discussion of the previous section, this does not give the correct homology groups. Motivated by the decomposition structure in \eqref{eq:our-general-Gamma-Goursats}, we will propose an algorithm to compute the actual homology of the link space by an extension trick that will be clear below. We will also give some supporting arguments for our proposal.

But before we discuss this proposal, let us give a brief review for the homology ring of the 3-sphere quotiented by a finite $SU(2)$ subgroup $G$ of $ADE$ type.

\begin{table}[t]
    \centering
    \begin{tabular}{|c|c|c|c|c|c|}
    \hline
        $\Gamma$ & $\mathfrak{g}_\Gamma$&$G_{\Gamma}$ & $\Gamma^{\rm ab} = \mathcal{Z}(G_{\Gamma})$ \\
        \hline
        \hline
        $\Z_N$ &$\mathfrak{su}(N)$&$SU(N)$& $\mathbb{Z}_N$ \\
        $2\D_{4N}$ &$\mathfrak{so}(4N)$& Spin$(4N)$ & $\Z_2\oplus\Z_2$ \\
        $2\D_{4N+2}$ &$\mathfrak{so}(4N+2)$& Spin$(4N+2)$ & $\Z_4$\\
        $2\T$ &$\mathfrak{e}_6$& $E_6$ & $\Z_3$\\
        $2\O$  &$\mathfrak{e}_7$&$E_7$ & $\Z_2$ \\
        $2\I$ &$\mathfrak{e}_8$& $E_8$ & $\{1\}$\\
        \hline
    \end{tabular}
    \caption{A list of all possible $SU(2)$ subgroups of $ADE$ type, the corresponding gauge algebra and gauge group. Moreover, in the fourth column, we exhibit their abelianisation for each case.}
    \label{tab:Gamma ab and Linking}
\end{table}

\medskip
\noindent
\textbf{Comments on the homology of $\mathbb{S}^3/\Gamma$.}
Recall that the homology ring of the 3-sphere is trivial except for the degree $0$ case, which is spanned by the point class and the top degree component, which is spanned by the 3-sphere class. Viewing the 3-sphere as a Hopf fibration over $\mathbb{S}^2$, the non-trivial action of a finite subgroup $\Gamma$ on the $\mathbb{S}^1$ fibers leads to torsional elements in the first homology ring $H_1(\mathbb{S}^3/\Gamma,\Z)$. This group is isomorphic to the abelianisation of the subgroup $\Gamma$: $\Gamma^{\rm ab} = \frac{\Gamma}{[\Gamma,\Gamma]}$. In table \ref{tab:Gamma ab and Linking}, we list the abelianisations for all possible finite subgroups of $SU(2)$ of $ADE$ type. To sum up, the (co)homology ring of $\mathbb{S}^3/\Gamma$ is given by:
\begin{equation}\label{CoH S3/Gamma}
    H_{3-\bullet}(\mathbb{S}^3/\Gamma,\Z) = H^{\bullet}(\mathbb{S}^3/\Gamma,\Z) = \Z\,\oplus\,0\,\oplus\,\Gamma^{\rm ab}\,\oplus\, \Z~.
\end{equation}

Before moving on, let us introduce the notation we will follow for the different cohomology classes of $\mathbb{S}^3/\Gamma$:
\begin{itemize}
    \item We take $\mathsf{1}\in H^0(\mathbb{S}^3/\Gamma,\Z)$ to denote the Poincaré dual class to $[\mathbb{S}^3/\Gamma]\in H_3(\mathbb{S}^3/\Gamma,\Z)$.
    \item Since $\Gamma^{\rm ab}$ is an abelian finite group, then it has the following general form:
    \begin{equation}\label{ab gamma general}
        \Gamma^{\rm ab} = \Z_{n_1}\,\oplus\,\cdots\,\oplus\,\Z_{n_{k}}~.
    \end{equation}
    We denote the generator of $a$-th torsional component by $t_{2,a}$ which satisfies $n_a \cdot t_{2,a} = 0$.
    \item For the top form in cohomology, this is none other than the volume form, which we denote by vol$_3\in H^3(\mathbb{S}^3/\Gamma,\Z)$.
\end{itemize}

\subsubsection{The computation algorithm: a proposal}
Following the discussion in the previous section, the link space for the non-splittable quotient group $\Gamma$ is given as:
\begin{equation}\label{eq:def-of-L6}
L_{6}\ =\  \frac{\,(\,\mathbb{S}^{3}_{\mathrm{f}} /C_{0}\,)\,\times\,\mathbb{S}^{3}_{\mathrm{b}}\,}{B}~.
\end{equation}
Our proposal for computing the homology of this space goes as follows: First, we compute the homology of the fibered factor ignoring the extension of $C_{0}$ by $B$. That is, we compute the homology of $\mathbb{S}_{\rm f}^3/(C_0\times B)$ treating the product appearing in the denominator as a direct product. More explicitly, the first homology of this space is torsional of the form:
\begin{equation} \label{tor H1 naive}
\begin{split}
      \mathrm{Tor}\,  H_{1}\left(\mathbb{S}^{3}_{\mathrm{f}}/(C_{0}\times B)\right)\,&\,=\,  \mathrm{Tor}\,H_{1}(\mathbb{S}^{3}_{\mathrm{f}}/C_{0})\,\oplus\,   \mathrm{Tor}\,H_{1}(\mathbb{S}^{3}_{\mathrm{f}}/B)\,
      \\
      \,&\,=\, C_{0}^{\mathrm{ab}}\,\oplus\,B^{\mathrm{ab}}\,.
\end{split}
\end{equation}
Here, $C_{0}^{\mathrm{ab}}$ and $B^{\mathrm{ab}}$ refer to the abelianization of $C_{0}$ and $B$, respectively -- In table \ref{tab:Gamma ab and Linking} we list the abelianization form for all possible subgroups of $ADE$ type. The above decomposition follows from the fact that $(G_1\times G_2)^{\rm ab} = G_1^{\rm ab}\times G_2^{\rm ab}$. As we will see later, to work out the (co)homology ring of the full link space using K\"unneth formula, it will be enough to work out the first homology group of this space.

Second, we bring back the effect of the extension on the quotient $\mathbb{S}^3_{\rm f}/(C_0 \times B)$. 
From table \ref{tab:Gamma ab and Linking}, we see that the result of abelianisation is always a cyclic group. Therefore, let us take $C_{0}^{\mathrm{ab}}\cong\Z_{n}$ and $B^{\mathrm{ab}}\cong\Z_{m}$. The extension is classified by the $\mathrm{Ext}^{1}$-group (functor) (for definitions and further properties, see \cite{78c6565d-147a-31db-a39a-874596950677,dummit2003abstract}). Specifically, we have: 
\begin{equation}\label{eq:Ext-functor}
    \mathrm{Ext}^{1}(\Z_{n},\Z_{m})\,=\,\Z_{\gcd(n,m)}~.
\end{equation}
Thus, the distinct (inequivalent) extensions are in one-to-one correspondence with the elements of $\Z_{\gcd(n,m)}$. Throughout the examples, when available, we pick the non-trivial extension such that we obtain:
\begin{equation}\label{non-trivial ext}
    1\,\to\,\Z_{n}\,\to\,\Z_{nm}\,\to\,\Z_{m}\,\to\,1~.
\end{equation}

In fact, in most cases $B=B^{\mathrm{ab}}\cong\Z_{2}$. From \eqref{eq:Ext-functor}, there would be only one non-trivial extension for even $n$. Meanwhile, for the odd $n$ cases, we already can write $\Z_{2}\times\Z_{n} \cong\Z_{2n}$.

Knowing the actual form of the first homology group of the fibered sphere, third, we find the homology of (\ref{eq:def-of-L6}) using K\"unneth formula, which we review momentarily. We do so by taking the link space $L_6$ to be of the product form:
\begin{equation}\label{eq:def-L6-with-Gamma-ext}
    L_6 = \frac{\mathbb{S}^3_{\rm f}}{\widetilde{\Gamma}_{\rm{ext}}} \times \frac{\mathbb{S}^3_{\rm b}}{B}~.
\end{equation}
Here, $\widetilde{\Gamma}_{\rm ext}$ is the actual group acting on $\mathbb{S}^3_{\rm f}$ \eqref{eq:our-general-Gamma-Goursats}:
\begin{equation}
        \widetilde{\Gamma}_{\rm ext} \,=\, C_0 \,\sqcup\, \widetilde{\varphi}(b_1)\cdot C_0\,\sqcup\,\cdots\,\sqcup\,\widetilde{\varphi}(b_{m-1})\cdot C_0~.
\end{equation}
The extension discussed above gives us the homology group:
\begin{equation}
H_1(\mathbb{S}^3_{\rm f}/\widetilde{\Gamma}_{\rm ext},\Z)\, =\, \widetilde{\Gamma}_{\rm ext}^{\rm ab}\, \cong\, \Z_{mn}~.
\end{equation}

This algorithm for computing the homology is justified by meeting the physics presented in subsection \ref{sec:ADE-gauge-theories}. We note that exchanging the second and the third steps would not affect the final results.

\subsubsection{Further comments on the proposal}

Before moving on to the details of calculating the homology ring of the link space using the above proposal, let us here make some comments supporting it.

As we have seen in (\ref{CoH S3/Gamma}), the torsional cycles are given by the abelianisation of the quotient group $\Gamma$ for the $\bbS^{3}/\Gamma$ space. Therefore, to understand the structure of homology groups of (\ref{eq:def-of-L6}), one may aim to take the abelianisation of the non-splittable quotient group acting on $\bbS^{3}_{\rm{f}}\times\bbS^{3}_{\rm{b}}$. Taking this point of view gives a supporting argument and a deeper physical understanding for our proposal for computing the (co)homology of the link space $L_{6}$.

From the general structure of $\Gamma_{\rm ns}$ \eqref{eq:our-general-Gamma-Goursats}, its abelianisation is given by:
\begin{equation}\label{eq:Gamma-ns-abelianisation}
    \Gamma_{\mathrm{ns}}^{\rm{ab}}\,=\, \, (C_{0}^{\rm{ab}},\,1)\, \sqcup\, (\widetilde{\varphi}(b_{1})_{\rm{ab}}\,\cdot\,C_{0}^{\rm{ab}},\, b_{1})\,\sqcup\,\cdots\,\sqcup\, (\widetilde{\varphi}(b_{m-1})_{\rm{ab}}\,\cdot\,C_{0}^{\rm{ab}},\,b_{m-1})~,
\end{equation}
keeping in mind that $B\cong \Z_{m}$. The map $\widetilde{\varphi}(b_{j})_{\rm{ab}}$, with $j=0,1,\cdots,m-1$, is inherited from \eqref{eq:our-general-Gamma-Goursats}. The above structure suggests that we may understand the effect of $\widetilde{\varphi}(b_{j})_{\rm{ab}}$ as extending $C_{0}^{\rm{ab}}$ by $B^{\rm ab}\cong\Z_{m}$. This is in agreement with our proposal given in (\ref{non-trivial ext}). In particular, taking $C_{0}^{\rm{ab}}\cong\Z_{n}$, we can write: 
\begin{equation}\label{eq:Znm-disjoint-unioun}
         \Z_{nm}\,=\, \bigsqcup_{b_{j}\in\Z_{m}}\,\Z_{n}^{(b_{j})}~,\qquad   \quad\Z_{n}^{(b_{j})}\,:=\, \widetilde{\varphi}(b_{j})_{\rm{ab}}\,\cdot\,\Z_{n}~.
\end{equation}

Therefore, the above observations along with \eqref{CoH S3/Gamma} suggest that the torsional first homology of the link space $L_{6}$ can be written as:
\begin{equation}\label{eq:TorH1-Znm-Zm}
    \mathrm{Tor}H_{1}(L_{6},\Z)\,=\,\Z_{nm}\,\oplus\,\Z_{m}\,.
\end{equation}
where the first factor comes from the fiber part, meanwhile, the second part comes from the base 3-sphere $\mathbb{S}^3_{\rm b}/B$.

\subsubsection{Computing \texorpdfstring{$H_\bullet(L_6, \Z)$}{HL6Z}: applying the proposal}

Let us now apply the proposal above for the six cases we are interested in in this work \eqref{eq:the-6-cases}. For the first step, we compute the `naive' first homology of the fibered component $\mathbb{S}^3/(C_0\times B)$. This is given by \eqref{tor H1 naive}. For each one of the cases, we compute this homology by using columns 4 and 5 of table \ref{tab:H1 for S3 fiber}.

Next, we work out the extension group of $C_0$ by $B$ for each of the cases. As we mentioned earlier, for the cases where the extension is trivial, we keep the homology group to be the naive one. Meanwhile, for the cases where there can be non-trivial extensions, then we take the homology group to be the one as in \eqref{non-trivial ext}. In the last two columns of table \ref{tab:H1 for S3 fiber}, we respectively list the extension groups and the actual first homology for each one of the six cases.

\begin{table}[t]
    \centering
    \begin{tabular}{|c|c|c||c|c||c||c|}
    \hline
        Case &$B$ & $C_0$ &$B^{\rm ab}$ & $C_0^{\rm ab}$& Ext$^1(C_0^{\rm ab}, B^{\rm ab})$&$H_1(\mathbb{S}_{\rm f}^3/\widetilde{\Gamma}_{\rm ext},\Z)$\\
        \hline
        \hline
         1&$\Z_p$& $\Z_{N}$ & $\Z_p$ & $\Z_{N}$ &$\Z_p$ &$\Z_{pN}$\\
         \hline
          2&$\Z_4$ & $\Z_{2N+1}$ & $\Z_4$ & $\Z_{2N+1}$&$0$ & $\Z_{4(2N+1)}$ \\
          \hline
        3&$\Z_{2}$ & $\Z_{2N}$ &$\Z_2$ &$\Z_{2N}$ &$\Z_2$ &$\Z_{4N}$\\
          \hline
         4&$\Z_3$& $2\D_4$ & $\Z_3$ & $\Z_2\oplus \Z_2$ & $0$ & $\Z_2\oplus \Z_6$ \\
         \hline
        5(a)&$\Z_2$ & $2\D_{4N}$ & $\Z_2$ & $\Z_2\oplus \Z_2$& $\Z_2\oplus \Z_2$ &$\Z_2\oplus \Z_4$\\
         \hline
          5(b)&$\Z_2$ & $2\D_{4N+2}$ & $\Z_2$ & $\Z_4$& $\Z_2$ &$\Z_8$\\
         \hline
         6&$\Z_2$ & $2\T$ & $\Z_2$ & $\Z_3$ & $0$ &$\Z_6$ \\
         \hline
    \end{tabular}
    \caption{For each one of the six cases \eqref{eq:the-6-cases}, we consider the possible extensions of $C_0^{\rm ab}$ by $B^{\rm ab}$ using the Ext-functor. For the cases where the Ext is non-zero, we consider the non-trivial extension of the homology of $\mathbb{S}^3_{\rm f}/(C_0\times B)$.}
    \label{tab:H1 for S3 fiber}
\end{table}

Now that we have the actual first homology groups of $\mathbb{S}_{\rm f}^3/\widetilde{\Gamma}_{\rm ext}$, we come to the last step of our algorithm: computing the homology of the total link space using K\"unneth formula. Recall that, K\"unneth decomposition theorem simply relates the homology rings $H_\bullet(\mathbb{S}^3_{\rm f}/\widetilde{\Gamma}_{\rm ext}\times {\mathbb{S}^3_{\rm b}}{/B},\Z)$ with the homology rings $H_\bullet(\mathbb{S}^3_{\rm b}/B,\Z)$ and $H_{\bullet}(\mathbb{S}^3_{\rm f}/\widetilde{\Gamma}_{\rm ext},\Z)$ up to torsional terms. More explicitly, we have the following exact sequence:
\begin{multline}\label{Kunneth theorem}
        0\longrightarrow \bigoplus_{i+j=k} H_i({\mathbb{S}_{\rm f}^3}/{\widetilde \Gamma}_{\rm ext},\Z)\otimes H_j(\mathbb{S}_{\rm b}^3/B,\Z) \longrightarrow H_k(\mathbb{S}_{\rm f}^3/\widetilde{\Gamma}_{\rm ext}\times \mathbb{S}_{\rm b}^3/B,\Z)\\ \longrightarrow \bigoplus_{i+j=k-1} {\rm Tor}^\Z_1(H_i(\mathbb{S}_{\rm f}^3/\widetilde{\Gamma}_{\rm ext},\Z), H_j(\mathbb{S}_{\rm b}^3/B,\Z))\longrightarrow 0~.
\end{multline}
 Here, Tor$_1^\Z(A, B)$ is the first Tor-functor. For definition, see the review in appendix A of \cite{Najjar:2024vmm}. What is important for us at this stage is that this functor has the following properties:
\begin{itemize}
    \item Tor$_1^\Z(A,B) = 0 $ whenever $A$ or $B$ is a free $\Z$-module. In particular, 
    \begin{equation}\label{tor Z Zn}
        {\rm Tor}_1^\Z(\Z, B) = 0~.
    \end{equation}
    \item For the case $A = \Z_m$ and $B = \Z_n$ we have:
    \begin{equation}\label{tor Zm Zn}
        {\rm Tor}_1^{\Z}(\Z_m, \Z_n) = \Z_{\gcd(m,n)}~.
    \end{equation}
    \item The Tor functor distributes over direct sums:
    \begin{equation}
        {\rm Tor}_1^{\Z}(\bigoplus_\alpha \Z_{m_\alpha}, \Z_n) = \bigoplus_\alpha {\rm Tor}_1^\Z(\Z_{n_\alpha}, \Z_m) = \bigoplus_\alpha \Z_{\gcd(n_\alpha,m)}
    \end{equation}
\end{itemize}

\begin{table}[t]
    \centering
    \begin{tabular}{|c|c|c||c|c|}
    \hline
       {\rm Case} &$B$ & $C_0$ &$H_1(L_6,\Z)$&$H_2(L_6,\Z)$\\
        \hline
        \hline
         1&$\Z_p$& $\Z_{N}$ & $\Z_{pN}\oplus \Z_{p}$ & $\Z_p$ \\
         \hline
          2&$\Z_4$ & $\Z_{2N+1}$ &$\Z_{4(2N+1)} \oplus \Z_4$& $\Z_4 $\\
          \hline
        3&$\Z_{2}$ & $\Z_{2N}$ & $\Z_{4N} \oplus \Z_2$ & $\Z_2$\\
           \hline
         4&$\Z_3$& $2\D_4$ & $ \Z_2 \oplus \Z_6 \oplus \Z_3$& $\Z_3$\\
         \hline
        5(a)&$\Z_2$ & $2\D_{4N}$ & $\Z_2 \oplus \Z_4 \oplus \Z_2$ & $\Z_2 \oplus \Z_2$\\
         \hline
          5(b)&$\Z_2$ & $2\D_{4N+2}$ & $\Z_8 \oplus \Z_2$ &$\Z_2$ \\
         \hline
         6&$\Z_2$ & $2\T$ & $\Z_6\oplus \Z_2$ & $\Z_2$ \\
         \hline
    \end{tabular}
    \caption{For each of the six cases in \eqref{eq:the-6-cases}, this table shows the first and the second homologies of the corresponding link space $L_6$. }
    \label{tab:H1 H2 for L6}
\end{table}

\medskip
\noindent
\textbf{The final result for $H_\bullet(L_6,\Z)$.} Using the exact sequence \eqref{Kunneth theorem}, it is straightforward to compute the homology of the link space $L_6$:

\begin{equation}\label{CoH of L6}
    H_{\bullet}(L_{6},\Z) \,=\,  \Z \,\oplus\, \begin{matrix}B^{\rm ab}\\\oplus \\\widetilde{\Gamma}^{\mathrm{ab}}_{\mathrm{ext}}\end{matrix}\,\oplus\,  {\rm Tor}_1^{\Z}(B^{\rm ab}, \widetilde{\Gamma}^{\mathrm{ab}}_{\mathrm{ext}})  \oplus\, \begin{matrix}\Z^2\\\oplus\\ {\rm Tor}_1^{\Z}(B^{\rm ab}, \widetilde{\Gamma}^{\mathrm{ab}}_{\mathrm{ext}}) \end{matrix}\,\oplus\, \begin{matrix}B^{\rm ab}\\\oplus\\ \widetilde{\Gamma}^{\mathrm{ab}}_{\mathrm{ext}}\end{matrix}\, \oplus\, 0\,\oplus\, \Z~.
\end{equation}
where we denoted the extended first homology group by $\widetilde{\Gamma}^{\mathrm{ab}}_{\mathrm{ext}}$.

As we pointed out earlier, it is enough to work out the extension of the first homology of $\mathbb{S}^3_{\rm f}/(C_0\times B)$. This follows from the observation that the K\"unneth formula \eqref{Kunneth theorem} can recursively work out the torsional components of the higher homology groups of $L_6$, keeping in mind that the homology of $\mathbb{S}^3_{\rm b}/B$ is given in \eqref{CoH S3/Gamma}. For instance, take $k=2$ in \eqref{Kunneth theorem}, we have:
\begin{equation}
    H_2(L_6,\Z) = {\rm Tor}_1^{\Z} \left(H_1(\mathbb{S}^3_{\rm f}/\Gamma_{\rm ext}^{\rm ab},\Z), H_1(\mathbb{S}^3_{\rm b}/B,\Z)\right) = {\rm Tor}_1^{\Z}(\widetilde{\Gamma}_{\rm ext}^{\rm ab}, B^{\rm ab})~.
\end{equation}
Moreover, knowing the torsions of these two homology groups of $L_6$, one can use the following duality to  work out the rest:
\begin{equation}\label{torsion duality}
        {\rm Tor}\, H_{k}(L_6,\Z) \cong {\rm Tor}\, H_{5-k}(L_6,\Z)~.
\end{equation}
This duality follows from Poincar\'e duality and the universal coefficient theorem. By the first, we can relate Tor $H_k(L_6,\Z)$ with Tor $H^{6-k}(L_6,\Z)$. Meanwhile, the second theorem related Tor $H^{6-k}(L_6,\Z)$ with Tor $H_{6-k+1}(L_6,\Z)$--For instance, see corollary 3.3 of \cite{Hatcher:478079}.

Using the general result \eqref{CoH of L6} for each one of the six quotients introduced in \eqref{eq:the-6-cases}, one gets the last two columns in table \ref{tab:H1 H2 for L6}.

\medskip
\noindent
\textbf{A comment for the semi-splittable extensions.} As we discussed in the second part of subsection \ref{subsec:BS and symmetries}, one can have a semi-splittable extension of $\Gamma_{\rm ns}$ of the form $\Gamma_{\rm ss} = \Gamma_{\rm ns} \times A$ which still acts freely on the total fiberation $\mathbb{S}^3_{\rm f}\times \mathbb{S}^3_{\rm b}$. From the decomposition in \eqref{eq:general-Gamma-Goursats-semi-split} and since $A$ acts diagonally on both factors, one can apply the algorithm proposed above a second time to compute the homology ring of the link space:
\begin{equation}
    \frac{\mathbb{S}^3_{\rm f}/C_0\times \mathbb{S}^3_{\rm b}}{A\times B}~.
\end{equation}
But we will not do this here. Rather, we focus our discussion on the non-splittable factors.

\subsection{Reducing the M-theory action: 5d SymTFT}\label{sec:reduce-action-SymTFT}

At this stage, we have all the necessary machinery and ingredients to go forward with the reduction of the M-theory action \eqref{S M tot} along the link space $L_6$. But first, we will uplift all the homology classes we computed in the previous section to differential cohomology.

\medskip
\noindent
\textbf{The classes of $\br{H}^\bullet(\mathbb{S}^3/\Gamma)$.}
In the previous section, we gave a review of the homology ring of the quotient of the 3-sphere $\mathbb{S}^3/\Gamma$. Building on that discussion, let us now introduce the classes of differential cohomology ring of this space. From the diagram \eqref{2exactseqbrH}, we see that the exactness of these sequences implies that the topological class map $c$ is surjective. Thus, for cohomology class $\alpha \in H^p(\mathbb{S}^3/\Gamma,\Z)$ can be `uplifted' to a character $\br{\alpha}\in \br{H}^p(\mathbb{S}^3/\Gamma)$. Following the notation we introduced for the cohomology classes below \eqref{CoH S3/Gamma}, we have:
\begin{itemize}
    \item The uplift of the identity class $\mathsf{1}\in H^0(\mathbb{S}^3/\Gamma,\Z)$ is denoted by ${\br {\sf 1}}\in {\br H}^0(\mathbb{S}^3/\Gamma)$.
    \item The uplift of the torsional class $t_{2,a}\in H^2(\mathbb{S}^3/\Gamma,\Z) \cong \Gamma^{\rm}$ is denoted by ${\br t}_{2,a}\in {\br H}^2(\mathbb{S}^3/\Gamma)$. Here $a$ runs over all torsional $\Z_n$ components in $\Gamma^{\rm ab}$.
    \item As for the volume element vol$_3\in H^3(\mathbb{S}^3/\Gamma,\Z)$, we will denote its uplift by ${\br {\rm vol}}_3\in {\br H}^3(\mathbb{S}^3/\Gamma)$.
\end{itemize}

\medskip
\noindent
\textbf{The classes of $H^\bullet(L_6)$.} Using the differential cohomology classes and the two $\mathbb{S}^3/\Gamma$ factors defining $L_6$ and the projection maps: 
\begin{equation}\label{L6 projections}
     \mathbb{S}_{\rm b}^3/\widetilde{\Gamma}\,\xlongleftarrow{\pi_{\rm f}}\, L_6\, \xlongrightarrow{\pi_{\rm b}}\,\mathbb{S}_{\rm f}^3/B~,
\end{equation}
we can work out elements of the differential cohomology ring of $L_6$:
\begin{itemize}
    \item For ${\br H}^{0}(L_6)$ we have one generator ${\br {\sf 1}}$. This is associated with the fact that we have only one $\Z$ factor in $H^0(L_6,\Z)$  as shown in \eqref{CoH of L6}.
    \item For ${\br H}^2(L_6)$, we have the generators $\{{\br t}_{2,a}^{\rm f}, \, {\br t}_{2,b}^{\rm b}\}$ where the indices $a$ and $b_{\rm b}$ depend on the explicit form of the two groups $\widetilde{\Gamma}_{\rm ext}^{\rm ab}$ and $B^{\rm ab}$ respectively -- as we discussed around \eqref{ab gamma general}\footnote{ Let us insist here that we are abusing the notation where, instead of writing $\pi^*_{\rm f}({\br t}_{2,a}^{\rm f})\in {\br H}^2(L_6)$ and $\pi^*_{\rm b}({\br t}^{\rm b}_{2,b_{\rm b}})\in {\br H}^2(L_6)$ for the pull-back of the torsional classes, we are denoting them with their original notation.}. Since, in the six cases we are considering in this work \eqref{eq:the-6-cases}, $B \cong \Z_m$, for some $m$, and to make the notation lighter, we will have only one differential character coming from the base sphere which we denote by $\br{t}_{2}^{\rm b}$.
    \item For ${\br H}^3(L_6)$, we have the generators ${\br {\rm vol}}_3^{\rm f}, {\br {\rm vol}}_3^{\rm b}$ and ${\br t}_{3,a}$. The first two are associated with the $\Z^2$ part of $H_3(L_6,\Z)$ in \eqref{CoH of L6} and the last one is associated with Tor$_1^\Z(B^{\rm ab}, C^{\rm ab}) = B^{\rm ab}\otimes\widetilde{\Gamma}_{\rm ext}^{\rm ab}$ where, as above, the index $a$ depends on the number of the torsional factors in $\widetilde{\Gamma}_{\rm ext}^{\rm ab}$.  
    \item For ${\br H}^4(L_6)$, we have the following generators $\{{\br t}_{4, a} := {\br t}_{2,a}^{\rm f}\times {\br t}_{2}^{\rm b}\}$. 
    \item For ${\br H}^5(L_6)$ we have the following two sets of generators $\{{\br {\rm vol}_3^{\rm f}\times {\br t}_{2}^{\rm b}},{\br {\rm vol}_3^{\rm b}\times {\br t}_{2,a}^{\rm f}} \}$.
    \item For ${\br H}^6(L_6)$ is generated by the element ${\br {\rm vol}}_3^{\rm f}\times {\br {\rm vol}_3^{\rm b}}$.
\end{itemize}

\subsubsection{Reduction of the M-theory topological action}

Let us now reduce the topological term in the M-theory action given in \eqref{S M CS} to a 5d action on $Y_5$. For this, let us decompose the uplift of the 4-form field strength $\br{G}_4 \in \br{H}^{4}(M_{11})$:

\begin{multline}\label{G4 decomp}
    {\br G}_4 \,=\, {\br F}_4\star {\br 1} +{\br B}_{2}^{\rm b} \star {\br t}_{2}^{\rm b}+\sum_{a} {\br B}_{2,a}^{\rm f} \star {\br t}_{2,a}^{\rm f}+\sum_{\alpha={\rm f},{\rm b}} {\br F}^{\alpha}_1 \star {\br {\rm vol}}_3^{\alpha} \\+ \sum_{a}{\br B}_{1,a}\star {\br t}_{3,a}+ \sum_{a}{\br B}_{0,a}\star \br{t}_{4,a}~.
\end{multline}

The above expansion encodes the topological information carried by the 4-form flux $\br{G}_{4}$. In particular, the differential characters $\br{F}_{4-p},\br{B}_{4-p}\in \br{H}^{4-p}(Y_5)$ are interpreted, respectively, as background field strengths and gauge fields for the higher-form symmetries (HFSs) of the geometrically engineered $\N=1$ gauge theory. For more details, see Section 2 of \cite{Apruzzi:2021nmk}. More explicitly, we have:
\begin{itemize}
    \item \textbf{Continuous HFSs}. Each differential character $\br{F}_{4-p}$ appearing in the decomposition of the $\br{G}_4$ field \eqref{G4 decomp} corresponds to a $(3-p)$-form background gauge field for a $(2-p)$-form symmetry $U(1)^{[2-p]}$ symmetry. From this perspective, $F_{4-p}$ defined above is the field strength for this symmetry.
    \item \textbf{Discrete HFSs}. Meanwhile, the character $\br{B}_{4-p}$ corresponds to a $(4-p)$-form background gauge field for a $(3-p)$-form $\Z_{n}^{[3-p]}$ symmetry. Here $n$ is the torsional order of ${B}_{4-p}\in {\rm Tor}\, H^{4-p}(Y_5,\Z)$.
\end{itemize}

In Table \ref{tab: HFS from G4}, we summarize all possible HFSs we get from the different differential characters coming from the decomposition of $\br{G}_4$ \eqref{G4 decomp}. The corresponding symmetry topological operators associated with these HFSs will be discussed in the next section. For now, our focus is on deriving the SymTFT actions as outlined above.

\begin{table}[t]
    \renewcommand{\arraystretch}{1.7}
    \centering
    \begin{equation*}
        \begin{array}{|c|c|c|c|c|}
        \hline
        \br{B}_{2, \bullet}^{\bullet} & \br{B}_{1, \bullet}& \br{B}_{0, \bullet} & \br{F}_4 & \br{F}_{1}^\bullet\\
        \hline
        \Z_n^{[1]}&\Z_m^{[0]}&\Z_k^{[-1]}&U(1)^{[2]}& U(1)^{[-1]}\\
        \hline
        \end{array}
    \end{equation*}   
    \caption{\textsc{First row:} all possible differential characters appearing in twist action \eqref{S L6 twist}. \textsc{Second row:} the corresponding HFS associated with these background gauge fields. Here $n, m$, and $k$ are the torsional degrees of the corresponding background gauge fields. The explicit form of these degrees depends on $C_0$ and $B$ of each one of the six cases \eqref{eq:the-6-cases}.}
    \label{tab: HFS from G4}
\end{table}

\paragraph{Mixed anomalies from M-theory.} The potential anomalies for the HFSs , whose background fields were discussed above, can be extracted from the topological Chern--Simons action of M-theory. The resulting expression is referred to as the twist action. In particular, by expanding the topological action \eqref{S M CS} using the reduction \eqref{G4 decomp}. We find the following final form:
\begin{align}\label{S L6 twist}
\begin{split}
    S_{\rm twist}^{L_6} &= \sum_{a,b} {\sf K}_{a,b} \left(\int_{Y_5}^{\br H} \br{B}_{2, a}^{\rm f} \star \br{B}_{2}^{\rm b}\star \br{B}_{1,b}+\int_{Y_5}^{\br H} \br{F}_4 \star \br{B}_{1,a} \star \br{B}_{0,b}\right)\\
    &+ {\rm CS}_{B} \int_{Y_5}^{\br H} \br{B}_{2}^{\rm b} \star \br{B}^{\rm b}_{2} \star \br{F}_1^{\rm f} +\sum_{a,b}\left[{\rm CS}_{\widetilde{\Gamma}_{\rm ext}}\right]_{a,b} \int_{Y_5}^{\br H} \br{B}_{2,a}^{\rm f} \star \br{B}^{\rm f}_{2,b} \star \br{F}_1^{\rm b}\\
    &+\int^{\br H}_{Y_5} \br{F}_4\star \br{F}_1^{\rm f} \star\br{F}_1^{\rm b}~.
\end{split}
\end{align}

Before fixing the coefficients in the above action, let us write the action \eqref{S L6 twist} more explicitly as a standard integral over $Y_5$. For this, one needs to use invariants \eqref{primary inv} and \eqref{second inv} along with the following properties:
\begin{equation}
    c(\br{B}\star \br{C}) = c(\br{B}) \smile c(\br{C})~,  \qquad \br{B}, \br{C}\in \br{H}^\bullet(X)~,
\end{equation}
where $\smile$ is the cup product in $H^\bullet(X,\Z)$. Using these, we get:
\begin{align}\label{S L6 twist SymTFT}
\begin{split}
    S_{\rm twist}^{L_6} &=\sum_{a, b} {\sf K}_{a,b} \left(\int_{Y_5} {B}_{2, a}^{\rm f} \smile {B}_{2}^{\rm b}\smile {B}_{1,b}+ \int_{Y_5} \frac{{F}_4}{2\pi} \smile {B}_{1,a} \smile {B}_{0,b}\right)\\
    &+{\rm CS}_B \int_{Y_5}{B}_{2}^{\rm b} \smile {B}^{\rm b}_{2} \smile \frac{{F}_1^{\rm f}}{2\pi}+ \sum_{a,b} \left[{\rm CS}_{\widetilde{\Gamma}_{\rm ext}}\right]_{a,b} \int_{Y_5} {B}_{2,a}^{\rm f} \smile {B}^{\rm f}_{2,b} \smile \frac{{F}_1^{\rm b}}{2\pi}\\
    &+\int^{\br H}_{Y_5} \br{F}_4\star \br{F}_1^{\rm f} \star\br{F}_1^{\rm b}~.
    \end{split}
\end{align}
Here, we have used:
\begin{equation}\label{top class for dG4 modes}
B^{\bullet}_{\bullet, \bullet} = c(\br{B}^{\bullet}_{\bullet, \bullet})~,\qquad \frac{F_\bullet}{2\pi} = c(\br{F}_{\bullet})~.
\end{equation}

The coefficients appearing in the first term in the action \eqref{S L6 twist SymTFT} have the following explicit form:
\begin{equation}\label{mathsf K}
{\sf K}_{a, b}\,\equiv\,  \int_{L_{6}}^{\br{H}}\, \br{t}_{4, a}\,\star\, \br{t}_{3,b}\,=\,-\, \frac{\delta_{a,b}}{n_a}~,
\end{equation}
where $n_a$ here is the torsional order of $t_{3,a} = c(\br{t}_{3,a})$ and $t_{4,a}= c(\br{t}_{4,a})$. 

As for the other coefficients appearing in the second and third lines of \eqref{S L6 twist SymTFT}, these are referred to as the Chern--Simons invariants of the corresponding 3-manifold. Let us now make some comments on how to compute the CS-invariants. Further, we propose a framework for dealing with CS invariants on a quotient of the 3-sphere $\bbS^{3}$ when the quotient group admits a disjoint union decomposition as given in \eqref{eq:our-general-Gamma-Goursats}.

\medskip
\noindent
\textbf{3d Chern--Simons invariants.} For a finite subgroup $\Gamma$, the (matrix) Chern--Simons invariants associated with the 3-manifold $\mathbb{S}^3/\Gamma$ are defined as follows:
\begin{equation}\label{CS invariants: def}
        \left[{\rm CS}_{\Gamma}\right]_{a,b} := \frac{1}{2} \int_{{\mathbb{S}^3}/{\Gamma}} \br{t}_{2,a}\, \star\,\br{t}_{2,b}~,
\end{equation}
where $\br{t}_{2,a}\in \br{H}^\bullet(\mathbb{S}^3/\Gamma)$. The size of the matrix depends on the number of torsional components in $\Gamma^{\rm ab}$.

To compute these invariants explicitly, one needs to use the fact that for each torsional element $t_{2,a} \in H^2({\mathbb{S}^3}/{\Gamma},\Z)$ one can associate a 2-cycle $Z_a\in H_2(\widetilde{X}_{4}, \Z)$. Here, $X_{4}$ is an auxiliary singular space with a corresponding $ADE$-type singularity. We donate the resolved space by $\widetilde{X}_{4}$. In this context, the auxiliary space $\widetilde{X}_{4}$ can always be taken to be the minimal resolution of $\C^{2}/\Gamma$. With this in mind, we have \cite[(3.25)]{Apruzzi:2021nmk}:
\begin{equation}\label{CS-integral-intersection}
    \left[{\rm CS}_{\Gamma}^{(3)}\right]_{a,b} = \frac{1}{2} \int_{{\mathbb{S}^3}/{\Gamma}} \br{t}_{2,a}\, \star\,\br{t}_{2,b}~ = \frac{1}{2} \left(\frac{1}{n_a}Z_a\right)\cdot_{\widetilde{X}_4}\left(\frac{1}{n_b}Z_b\right) \mod 1~,
\end{equation}
where $\cdot_{\widetilde{X}_{4}}$ refers to the intersection number in the homology ring $H_2(\widetilde{X}_{4}, \Z)$ and $n_a, n_b$ are the torsional order of $t_{2,a}$ and $t_{2,b}$ respectively. In \cite[section 3]{Bhardwaj:2020phs}, the authors worked out the central divisors that one should consider to obtain the elements of the center $\mathcal{Z}(G_\Gamma)$ for the resulting simply-connected gauge group $G_{\Gamma}$. We will revisit this point in more detail in subsection \ref{subsec: symtft outer} below. We list the explicit form of these divisors in table \ref{tab: ZGamma and CSGamma}.

\begin{table}[t]
    \renewcommand{\arraystretch}{1.7}
    \centering
    \begin{equation*}
        \begin{array}{|c|c|c|}
        \hline
         \Gamma & Z_{\Gamma}  & -{\rm CS}_\Gamma^{(3)}  \\
        \hline
        \hline
        \Z_N & \sum_{i=1}^{N-1} iS_i  & \frac{N-1}{2N} \\
        \hline
        \multirow{2}{*}{$2\D_{4N}$} &  \frac{1}{2}\sum_{i=1}^{2N-1} (1-(-1)^i)S_i & \multirow{2}{*}{$\frac{1}{4}\begin{pmatrix}
            N&N-1\\N-1&N
        \end{pmatrix}$}\\ &\frac{1}{2} \sum_{i=1}^{2N-2} (1-(-1)^i)S_i + S_{2N}&\\
        \hline
        2\D_{4N+2} & \sum_{i=1}^{2N-1} (1-(-1)^i) S_i + S_{2N} + 3S_{2N+1}&\frac{2N+1}{8}\\
        \hline
        2\T & S_1 + 2\,S_2 + 3\,S_3 + 4\,S_4 + 5\,S_5  &\frac{5}{3}\\
        \hline
        2\O & S_1 + S_3 + S_7 &\frac{3}{4}\\
        \hline
        \end{array}
    \end{equation*}   
    \caption{\textsc{Second column:} the central divisors associated with the $SU(2)$ subgroups $\Gamma_{\scriptscriptstyle ADE}$ of $ADE$ type \cite[\S 3]{Bhardwaj:2020phs}. \textsc{third column:} the Chern--Simons invariants associated with each possible $\Gamma_{\scriptscriptstyle ADE}$ \cite[\S2]{GarciaEtxebarria:2019caf}.}
    \label{tab: ZGamma and CSGamma}
\end{table}

In terms of these surface divisors, the intersection number $\cdot_{\widetilde{X}_{4}}$ is given by the inverse of the Cartan matrix $A({\mathfrak{g}_\Gamma})$ of the associated Lie algebra $\mathfrak{g}_{\Gamma}$ --  For an explicit form of these Cartan matrices see table 6 of \cite{SLANSKY19811}. For each possible case for $\Gamma$, we give the explicit form of the associated CS invariants in table \ref{tab: ZGamma and CSGamma}. Note here that we do not include the three cases of $\mathfrak{e}_8, \mathfrak{f}_4$, and $\mathfrak{g}_2$ since these cases have trivial centers and so the corresponding intersections are trivial.

 \subsubsection{Reduction of the M-theory kinetic action}
Let us now look at the reduction of the kinetic term of the M-theory action \eqref{S M kin}. For this, let us first expand the differential character $\br{dG}_7$:
\begin{multline}\label{dG7 decomp}
        \br{dG}_7 = \br{f}_8\star \br{1} + \br{\mathcal{B}}_{6}^{\rm b} \star t_{2}^{\rm b}+\sum_{a} \br{\mathcal{B}}_{6,a}^{\rm f} \star t_{2,a}^{\rm f} + \sum_{\alpha = {\rm f, b}} \br{f}_{5}^\alpha \star \br{\rm vol}_3^\alpha + \sum_{a} \br{\mathcal{B}}_{5, a}\star \br{t}_{3,a}\\ + \sum_{a} \br{\mathcal{B}}_{4,a} \star \br{t}_{4, a}+\br{\mathcal{B}}_{3}^{\rm b} \star (\br{t}_{2}^{\rm b} \times \br{\rm vol}_3^{\rm f})+\sum_{ a} \br{\mathcal{B}}_{3, a}^{\rm f} \star (\br{t}_{2,a}^{\rm f} \times \br{\rm vol}_3^{\rm b}) + \br{f}_2 \star (\br{\rm vol}_3^{\rm f} \times \br{\rm vol}_3^{\rm b})~.
\end{multline}
The exactness of the differential form $dG_7$ imposes the following exactness constraints on each one of $Y_5$ differential characters:
\begin{align}\label{top class for dG7 modes}
    c(\br{f}^\bullet_{p+1}) = \frac{1}{2\pi} dh_p^\bullet~, \qquad c(\br{\mathcal{B}}_{p+1, \bullet}^{\bullet}) = \delta A_{p,\bullet}^\bullet~.
\end{align}

Plugging this decomposition for $\br{dG}_7$ and \eqref{G4 decomp} for $\br{G}_4$ in the kinetic term \eqref{S M kin}, one obtains the following action for the 5d SymTFT:\footnote{In this full expansion, one also encounters terms of the form $\int^{\br{H}}_{Y_5} \br{B}_{1,\bullet} \star \br{\mathcal{B}}_{4, \bullet}$. But for this case, the integral over the link space $L_6$ is trivial since it is a linking between two torsional 3-cycles $\br{t}_{3, \bullet}$.}
\begin{align}\label{S L6 BF}
\begin{split}
    S^{L_6}_{\rm BF} = &- \sum_{a, b} \mathsf{K}_{a, b} \left(\int_{Y_5}^{\br{H}} \br{B}_{0,a}\star \br{\mathcal{B}}_{5,b} +\int^{\br{H}}_{Y_5} \br{B}_{1,a}\star\br{\mathcal{B}}_{4,b}\right)\\
    &- 2 {\rm CS}_{B} \int^{\br{H}}_{Y_5} \br{B}_{2}^{\rm b} \star \br{\mathcal{B}}_{3}^{\rm b}- 2 \sum_{ a, b} \left[{\rm CS}_{\widetilde{\Gamma}_{\rm ext}}\right]_{a, b} \int^{\br{H}}_{Y_5} \br{B}_{2, a}^{\rm f} \star \br{\mathcal{B}}_{3, b}^{\rm f}\\
   & -\sum_{\alpha = {\rm f,b}} \int_{Y_5}^{\br{H}} \br{F}_1^{\alpha} \star \br{f}_5^{\neq \alpha} +\int_{Y_5}^{\br{H}} \br{F}_{4}\star \br{f}_{2}~.
\end{split}
\end{align}
Here, the coefficients $\mathsf{K}_{a,b}$ are defined in \eqref{mathsf K}.

Using \eqref{top class for dG4 modes} and \eqref{top class for dG7 modes}, we can write the BF action above as follows:

\begin{align}\label{S L6 BF SymTFT}
\begin{split}
    S^{L_6}_{\rm BF} = &- \sum_{a, b} \mathsf{K}_{a, b} \left(\int_{Y_5} {B}_{0,a}\smile \delta A_{4,b} +\int_{Y_5} {B}_{1,a}\smile \delta{A}_{3,b}\right)\\
    &- 2 {\rm CS}_{B} \int_{Y_5} {B}_{2}^{\rm b} \smile \delta {A}_{2}^{\rm b}- 2 \sum_{ a, b} \left[{\rm CS}_{\widetilde{\Gamma}_{\rm ext}}\right]_{a, b} \int_{Y_5} {B}_{2, a}^{\rm f} \smile \delta {A}_{2, b}^{\rm f}\\
   & -\sum_{\alpha = {\rm f,b}} \int_{Y_5} \frac{{F}_1^{\alpha}}{2\pi} \wedge \frac{{h}_4^{\neq \alpha}}{2\pi} +\int_{Y_5}\frac{{F}_{4}}{2\pi}\wedge \frac{{h}_{1}}{2\pi}~.
\end{split}
\end{align}

The full action of the 5d SymTFT obtained from the reduction along the link space $L_6$ is given summing the two contributions \eqref{S L6 twist SymTFT} and \eqref{S L6 BF SymTFT}, namely:
\begin{equation}\label{full S symtft}
        S_{\rm SymTFT}^{L_6}  = S_{\rm twist}^{L_6} + S_{\rm BF}^{L_6}~.
\end{equation}

\medskip
\noindent
\textbf{The associated HFSs.} Similar to our discussion above concerning the HFSs coming from decomposing $G_4$, in table \ref{tab: HFS from G7} we exhibit the possible HFSs coming from the decomposition of the $G_7$ field \eqref{dG7 decomp} and appearing in the BF action \eqref{S L6 BF}. Note here that, since in \eqref{dG7 decomp} we are decomposing $\br{dG}_7$ rather than $\br{G}_7$, then the cohomology class $h_p$ is to be interpreted as the field strength of the associated HFS rather than its background gauge field. We will discuss the details of these symmetries in the following subsections.

Furthermore, we note that the field strengths $h_{p}$ are not closed, a property inherited from the behavior of the topological flux $G_{7}$, which satisfies
\begin{equation}\label{dG7}
dG_{7} \,=\, -\frac{1}{4\pi}\,G_{4}\, \wedge \, G_{4}\,.
\end{equation}
However, one can refine $h_{p}$ by incorporating terms arising from the SymTFT action (the twist action). These contributions originate from reducing the topological M-theory action (\ref{S M CS}) along the free cycles of the link space.

An alternative to this refinement procedure is to directly start with the following action:
\begin{equation}
    S_{\text{kin}}^{\text{M}} + S_{\text{CS}}^{\text{M}} = \frac{1}{(2\pi)^2} \int_{M_{11}} G_4 \wedge \mathsf{P}_7\, ,
\end{equation}
where $\mathsf{P}_{7}$ is a locally defined 7-form in 11d supergravity, given by \cite[section 5]{Moore:2004jv}:
\begin{equation}\label{eq:P7in11d}
    \mathsf{P}_7 \,:=\, G_7 - \frac{1}{12\pi} \,C_3 \,\wedge\, G_4\, .
\end{equation}

While this alternative formulation is well-suited for capturing continuous symmetries, it does not adequately account for discrete (finite) symmetries. The refinement procedure, on the other hand, provides an intermediate framework that accommodates both continuous and discrete symmetries simultaneously. For further details, the reader is referred to \cite{Najjar:2024vmm}.

For the case at hand, the significance of the final term in (\ref{S L6 twist SymTFT}),
\begin{equation}\label{eq:F4F1F1}
  S_{\rm twist}^{L_6} \,\supset\,  \int^{\br H}_{Y_5} \br{F}_4\star \br{F}_1^{\rm f} \star\br{F}_1^{\rm b}\,,
\end{equation}
becomes evident, as it provides the necessary structure to refine the field strengths $h_{p}$ obtained from dimensional reduction of the BF-type terms. 

Let us consider a concrete example in which the term $\br{F}_{1}^{\rm{b}} \,\star\, \br{f}_{5}^{\rm{f}}$ is refined by incorporating the cubic term $\br{F}_4\star \br{F}_1^{\rm f} \star\br{F}_1^{\rm b}$ in the twisted action. Specifically, we take:
\begin{equation}
 \begin{split}
      S_{\rm{SymTFT}}^{L_{6}}\,&\,\supset\,   \int_{Y_5}^{\br{H}}\,\left[\,-\, \br{F}_{1}^{\rm{b}} \,\star\, \br{f}_{5}^{\rm{f}} \,+\, \br{F}_4\star \br{F}_1^{\rm f} \star\br{F}_1^{\rm b}\,\right]\,
      \\
      \,=&\, -\, \int_{Y_5}^{\br{H}}\,\br{F}_{1}^{\rm{b}}\,\star\,\left[\, \br{f}_{5}^{\rm{f}} \,+\, \br{F}_4\star \br{F}_1^{\rm f} \,\right] \,.
 \end{split}
\end{equation}
We now apply the $c$-map to the differential characters over $Y_{5}$, as given in (\ref{top class for dG7 modes}), by identifying:
\begin{equation}
    c(\br{f}_{5})\,=\,\frac{1}{2\pi}dh_{4}\,,\qquad c(\br{F}_{4})\,=\,\frac{1}{2\pi}dC_{3}\,,\qquad c(\br{F}_{0})\,=\,\frac{1}{2\pi}dC_{0}~.
\end{equation}
Substituting these into the SymTFT term yields:
\begin{equation}\label{refine-h4-def-tilda-h4}
    \begin{split}
        -\,\frac{1}{2\pi}\, \int_{Y_5}\,F_{1}^{\rm{b}}\,\wedge\,\left[\, h_{4}^{\rm{f}} \,+\,\frac{1}{4\pi}\,\left(\, dC_{3}\wedge F_1^{\rm f} \,+\, F_{4}\,\wedge\, dC_{0} \,\right)\,\right] \,\equiv\,  -\,\frac{1}{2\pi}\, \int_{Y_5}\,F_{1}^{\rm{b}}\,\wedge\, \widetilde{h}_{4}^{\rm{f}}~.
    \end{split}
\end{equation}
In other words, the quantity $\widetilde{h}_{4}^{\rm{f}}$ defined above is a refined, closed version of $h_{4}^{\rm{f}}$, and can be understood as a particular dimensional reduction of the local 7-form $\mathsf{P}_{7}$ along the link space $L_{6}$.

Analogous refinement procedures appear in (\ref{eq:g-4-i-refinment}) and (\ref{eq:refinment-g1}) below in the context of symmetry topological operators associated with continuous $m$-form symmetries $U(1)^{[m]}$.

\begin{table}[t]
    \renewcommand{\arraystretch}{1.7}
    \centering
    \begin{equation*}
        \begin{array}{|c|c|c|c|c|}
        \hline
        \br{A}_{2, \bullet}^\bullet & \br{A}_{3, \bullet} &\br{A}_{4, \bullet} & \br{f}_2 & \br{f}_5^\bullet\\
        \hline
        \Z_n^{[1]} &\Z^{[2]}_{m} & \Z_k^{[3]} & U(1)^{[-1]} & U(1)^{[2]} \\
        \hline
        \end{array}
    \end{equation*}   
    \caption{\textsc{First row:} all possible differential characters appearing in twist action \eqref{S L6 BF}. \textsc{Second row:} the corresponding HFS associated with these background gauge fields. Here $n, m$, and $k$ are the torsional degrees of the corresponding background gauge fields. The explicit form of these degrees depends on $C_0$ and $B$ of each one of the six cases \eqref{eq:the-6-cases}.}
    \label{tab: HFS from G7}
\end{table}

\subsection{5d SymTFT: local observer perspective and decomposition}\label{sec:local-observer}
Following the multiverse picture we proposed in the previous section, let us point out here that the derivation above of the 5d SymTFT \eqref{full S symtft} is done from the `global observer' perspective. In this final part of the section, we would like to refine our proposal by discussing what happens from the frame of reference of a local observer living in the $j$-th universe. 

We start by recalling the decomposition of the group $\Gamma_{\rm ns}$ acting on the link space $L_6$. From this, one can decompose the group $\widetilde{\Gamma}_{\rm ext}$ acting on the fibered 3-sphere $\mathbb{S}^{3}_{\rm f}$ as follows:
\begin{equation}\label{Gamma ext decompose}
    \widetilde{\Gamma}_{\rm{ext}}\,=\,C_{0}\, \sqcup\, \widetilde{\varphi}(b_{1})\,\cdot\,C_{0} \,\sqcup\,\cdots\,\sqcup\, \widetilde{\varphi}(b_{m-1})\,\cdot\,C_{0}~.
\end{equation}
As we discussed around \eqref{eq:ns-theory-decompose}, we choose the frame of the local observer in the $j$-th universe to be such that it only sees the $C_0$ singularity--i.e. it does not see anything about the action of $B$ on the base $\mathbb{S}^3_{\rm b}$. Said differently, in each one of the $m$ universes, the observer sees the M-theory reduction on the link space:
\begin{equation}\label{local L6}
        L_6^{\rm local} := \frac{\mathbb{S}_{\rm f}^{3}}{C_0}\, \times\,\mathbb{S}^3_{\rm b}~,
\end{equation}
rather than the `global' one \eqref{eq:def-L6-with-Gamma-ext}.

\medskip
\noindent
\textbf{Revisiting the homology ring calculation.} Let us, at this stage, revisit our proposal for calculating the homology ring of the link space $L_6$. From the above discussion, the local observer measures the torsional part of the first homology to be $\mathrm{Tor}H_{1}(L^{\rm local}_{6},\Z) \cong\Z_n$. To simplify our discussion, here we took $C_0^{\rm ab} \cong \Z_n$.

Collectively, since the $m$ distinct observers--each associated with one of the $m$ disjoint universes--are not simultaneously accessible within a single local frame, there exist $m$ distinct $C_0^{\rm ab}$ torsional 1-cycles that can be independently measured. These distinct 1-cycles differ by a ``dressing'' with the element $\widetilde{\varphi}(b_{j})$, which characterizes the action of the frame of the $j$-th observer. Based on this structure, the total space of torsional 1-cycles measured from the perspective of all observers is given by:
\begin{equation}
    \mathrm{Tor}H_{1}^{\rm{observers}}\,=\, \bigsqcup_{j=0}^{m-1}\,\Z_{n}^{(b_{j})}\,=\,\Z_{nm}\,.
\end{equation}
This coincides with the contribution of the fibered factor of the link space \eqref{eq:def-L6-with-Gamma-ext} to (\ref{eq:TorH1-Znm-Zm}).

\medskip
\noindent
\textbf{Projecting the 5d SymTFT.} Inspired by the decomposition of the group $\widetilde{\Gamma}_{\rm ext}$ in \eqref{Gamma ext decompose}, one can schematically define a set of projection maps $\pi^{(j)}$ that project the full 5d SymTFT \eqref{full S symtft} to the corresponding universe:
\begin{equation}
        {\rm SymTFT}^{(j)} := \pi^{(j)} ({\rm SymTFT}_{\rm full})~, 
\end{equation}
where by SymTFT$_{\rm full}$ we mean that whose action is \eqref{full S symtft}. To write down the explicit form for the action of SymTFT$^{(j)}$, let us first make some further comments on how it should project the different background gauge fields that come from the decompositions \eqref{G4 decomp} and \eqref{dG7 decomp} and enter in the final form of \eqref{full S symtft}. 

From the form of the link space $L_6^{\rm local}$ \eqref{local L6} that the local observer sees and the original one \eqref{eq:def-L6-with-Gamma-ext}, we make the following claims:\footnote{These claims are based drawn from our physical multiverse picture. At the time of writing this work, we are lacking the proper mathematical formulation through which one can make these claims more concrete.}
\begin{itemize}
    \item For the torsional background gauge fields $B_{2,a}^{\rm f}$ and $A_{2,b}^{\rm f}$ which are associated with the fibered 3-sphere, we have:
    \begin{equation}
                B_{2,a}^{(j)} \,\sim\, \pi^{(j)}_{*}(B_{2,a}^{\rm f})~,\qquad  A_{2,b}^{(j)} \,\sim\, \pi^{(j)}_{*}(A_{2,b}^{\rm f})~,
    \end{equation}
    and these both are in the cohomology ring of $\mathbb{S}^3_{\rm f}/C_0$. Here, we used `$\sim$' just to insist that we are being heuristic about the projection map $\pi^{(j)}$. This follows from the fact that the local observer sees \eqref{local L6} only.
    \item For the free background gauge fields $F_{\bullet}^{\bullet}$ and $f^{\bullet}_{\bullet}$ coming from both, the fibered and the base 3-spheres, the projection map acts on them as follows:
    \begin{equation}
        \pi^{(j)}_{*} (F_{\bullet}^{\bullet}) \,\sim\, F_{\bullet}^{\bullet}~, \qquad \pi^{(j)}_{*} (f_{\bullet}^{\bullet}) \,\sim\, f_{\bullet}^{\bullet}~.
    \end{equation}
    This follows from the fact that the local observer sees the two spheres in $L_6^{\rm local}$--albeit quotiented by different groups than in $L_6$.  
    \item For the rest, which are the torsional background gauge fields coming from the base in $L_6$, we claim that they are forgotten by the projection map. 
\end{itemize}

Using the claim above, we can decompose the full 5d SymTFT as follows:
\begin{equation}\label{full SymTFT decompose}
    {\rm SymTFT_{\rm full}}\, = \,\bigoplus_{j=0}^{m-1} {\rm SymTFT}^{(j)}\,\oplus\, {\rm SymTFT}_{\rm global}~.
\end{equation}
The action of the $j$-th SymTFT is given by:
\begin{equation}\label{S symtft j}
        S_{{\rm SymTFT}^{(j)}}\, = \,S_{\rm twist}^{(j)}\, +\, S_{\rm BF}^{(j)}~,
\end{equation}
where, 
\begin{equation}\label{S twist j}
\begin{split}
    S_{\rm twist}^{(j)} &=
\sum_{a,b} \left[{\rm CS}_{C_0}\right]_{a,b} \int_{Y_5} {B}_{2,a}^{(j)} \smile {B}^{(j)}_{2,b} \smile \frac{{F}_1^{\rm b}}{2\pi}+\int^{\br H}_{Y_5} \br{F}_4\star \br{F}_1^{\rm f} \star\br{F}_1^{\rm b}~.
    \end{split}
\end{equation}
and, 
\begin{equation}\label{S BF j}
     \begin{split}
        S_{\rm BF}^{(j)} &=
    - 2 \sum_{ a, b} \left[{\rm CS}_{C_0}\right]_{a, b} \int_{Y_5} {B}_{2, a}^{(j)} \smile \delta {A}_{2, b}^{(j)}
   -\sum_{\alpha = {\rm f,b}} \int_{Y_5} \frac{{F}_1^{\alpha}}{2\pi} \wedge \frac{{h}_4^{\neq \alpha}}{2\pi} +\int_{Y_5}\frac{{F}_{4}}{2\pi}\wedge \frac{{h}_{1}}{2\pi}~.
\end{split}
\end{equation}
As for the `global' part of the SymTFT decomposition \eqref{full SymTFT decompose}, its action contains the remaining terms that are not included in \eqref{S symtft j} for all values of $j$. This includes all the terms involving the torsional background gauge fields coming from the base. Additionally, it includes terms involving the CS coupling of $\mathbb{S}^3/\widetilde{\Gamma}_{\rm ext}$ which are not accounted for in CS$_{C_0}$.

\section{Charged defects and symmetry operators}\label{sec:charged top defects}
In the previous section, we pointed out that the differential characters on $Y_5$ appearing in the decomposition of the M-theory fields $\br{G}_4$ \eqref{G4 decomp} and $\br{dG}_7$ \eqref{dG7 decomp} are interpreted as background gauge fields for higher form symmetries of the resulting 4d $\mathcal{N}=1$ gauge theory. In this section, we construct the associated defects charged under these symmetries and their generating operators.

\subsection{General aspects}
From the M-theory perspective, the charged defects and symmetry operators--for both discrete and continuous symmetries--are constructed from M$2$ and M$5$ branes wrapping different homology cycles on $M_{11} = Y_5\times L_6$. As we will point out later in the discussion, the nature of the symmetry, being discrete or continuous, depends on the type of homology cycle, torsional or free, that the brane is wrapping in the link space $L_6$. So, let us start here by reviewing some aspects of this construction. Key contributions which have significantly advanced our understanding of these realizations include \cite{Apruzzi:2021phx,Apruzzi:2022rei,Heckman:2022muc,Cvetic:2023plv,Apruzzi:2023uma,Garcia-Valdecasas:2023mis,Najjar:2024vmm,Najjar:2025rgt,Najjar:2025htp}. Our discussion in this subsection closely follows the framework developed in subsection 2.2 of \cite{Najjar:2024vmm}.

\subsubsection{Charged defects}
The charged defects of generalized symmetry are non-dynamical extended objects. Therefore, the brane configurations that generate them must wrap non-compact cycles in the resolved or deformed cone space $\widetilde{X}_{7}$. These non-compact cycles can effectively be taken as cycles of the link space $L_{6}$ along with the radial direction of $ \widetilde{X}_{7}$. Hence, we define the set of spacetime supersymmetric $m$-dimensional defects, denoted by $\mathbb{D}^{(m)}$, as BPS branes wrapping such non-compact cycles. Specifically, we have: 
\begin{equation}\label{def:defect1}
    \mathbb{D}^{(m)} := \bigcup_{p=2,5}\, \left\{\,\text{M$p$-branes on } H_{p-m}(L_{6},\Z) \times [0, \infty) \right\}~.
\end{equation}

From 4d perspective, these are $m$-dimensional defects that are charged under an $m$-form symmetry $\mathsf{G}^{[m]}$. Later in this subsection, we will revisit these defects and discuss how to measure their charges under the HFS in terms of the linking with the symmetry topological operators, which we now discuss. 

\subsubsection{Symmetry topological operators}

The topological symmetry operators are constructed through (not necessarily BPS) branes wrapping cycles in $L_{6}$, and extending transversely to the radial direction of the cone $\widetilde{X}_7$. In particular, the set of all  $(m'+1)$-dimensional (for $m+m'=d-2 = 2$) symmetry operators can be defined as:
\begin{equation}\label{def:symmetryoperator1}
    \mathbb{U}^{(m'+1)} =\bigcup_{p=2,5}  \,\{\,\text{$p$-branes wrapping $H_{p-m'}(L_{6},\Z)$ and transverse to $[0,\infty)$}\}\,.
\end{equation}

From the 4d gauge theory perspective, these codimension-$(m+1)$ topological operators generate the $m$-form symmetry $\mathsf{G}^{[m]}$ under which the defects in $\mathbb{D}^{(m)}$ are charged. As we will discuss in more detail momentarily, the charges of these defects is measured by their linking pairing with the operators in $\mathbb{U}^{m'+1}$ \cite{Gaiotto:2014kfa}.

But before going into the details of this point, let us first review how these topological operators are constructed for both types of symmetries, discrete and continuous. In what follows, we will denote the elements of $\mathbb{U}^{(m'+1)}$ by $\mathcal{U}(\Sigma_{m'+1})$. Here, $\Sigma_{m'+1}\subset Y_5\setminus [0,\infty)$ is the $(m'+1)$-dimensional submanifold supporting the topological operator. 

\paragraph{Discrete symmetries.} In this case, the operators $\mathcal{U}(\Sigma_{m'+1})$ come from BPS M2 and M5 branes wrapping torsional cycles $\gamma_k$ of the link space $L_{6}$. The form of the symmetry operators $\mathcal{U}(\Sigma_{m'+1})$ can be expressed as:
\begin{equation}\label{eq:symmetry-operator-discrete(1)}
    \begin{split}
        \mathcal{U}^{\text{M2 on }\gamma_{k}} (\Sigma_{3-k}) &=  \exp{ 2\pi i \int^{\br{H}}_{\Sigma_{3-k} \times \gamma_{k}} \br{G}_4 } 
        \\
        &=  \exp{ 2\pi i \int^{\br{H}}_{\Sigma_{3-k} \times L_{6}} \br{\mathsf{PD} } (\gamma_{k}) \star \br{G}_4 }~,
    \end{split}
\end{equation}
and,
\begin{equation}\label{eq:symmetry-operator-discrete(2)}
    \begin{split}
        \mathcal{U}^{\text{M5 on }\gamma_{k}} (\Sigma_{6-k}) &= \exp{ 2\pi i \int^{\br{H}}_{\widehat{\Sigma}_{7-k} \times \gamma_{k}} \br{\dd G}_7 } 
        \\
        &=  \exp{ 2\pi i \int^{\br{H}}_{\widehat{\Sigma}_{7-k} \times L_{6}} \br{\mathsf{PD} } (\gamma_{k}) \star \br{\dd G}_7 }~.
    \end{split}
\end{equation}
Here, $\mathsf{PD}(\cdot)$ is the Poincar\'e duality isomorphism. In the second line above, $\widehat{\Sigma}_{7-k}$ is any chain with $\partial \widehat{\Sigma}_{7-k} = \Sigma_{6-k}$. By the exactness of $\dd G_7$, the result only depends on $\Sigma_{6-k}$, and thus, the operator is indeed topological. Similarly, from the exactness of the $G_4$ field-strength, the corresponding operator above is topological.

\begin{table}[t]
\centering
\begin{tabular}{|l|c|c|c|c|}
\hline
& M2 & & M5 & \\
\hline
\hline
Tor$H_{1}(L_{6},\Z)\times [0,\infty)$ \hspace{2pt} & electric line defects & {\color{red}$\diamondsuit$} & --- &   \\
Tor$H_{2}(L_{6},\Z)\times [0,\infty)$ \hspace{2pt} & Local operator & \begin{color}{blue}$\circ$\end{color} & Domain wall & ${\spadesuit}$ \\
Tor$H_{3}(L_{6},\Z)\times [0,\infty)$ \hspace{2pt} & --- & $\clubsuit$ & Surface defect & \begin{color}{blue}$\triangle$\end{color} \\
Tor$H_{4}(L_{6},\Z)\times [0,\infty)$ \hspace{2pt} & --- &  & magnetic line defects & {\color{red}$\heartsuit$}\\
\hline
Tor$H_{1}(L_{6},\Z)$ \hspace{4pt} & 1-form sym. generator &  {\color{red}$\heartsuit$}  & &  \\
Tor$H_{2}(L_{6},\Z)$ \hspace{4pt} & 2-form sym. generator & \begin{color}{blue}$\triangle$\end{color} & $(-1)$-form sym. generator & $\clubsuit$ \\
Tor$H_{3}(L_{6},\Z)$ \hspace{4pt} & 3-form sym. generator & $\spadesuit$ & 0-form sym. generator & \begin{color}{blue}$\circ$\end{color}  \\
Tor$H_{4}(L_{6},\Z)$ \hspace{4pt} & --- &   & 1-form sym. generator & {\color{red}$\diamondsuit$} \\
\hline
\end{tabular}
\caption{Branes wrapping torsional cycles in $L_6$ give rise to finite symmetries. We mark with an equal symbol the charged defect and the corresponding symmetry generators. This table also appeared as table 10 in \cite{Najjar:2024vmm}.}
\label{tab:M2M5-G2}
\end{table}

Pick $\gamma_k$ to be a generating torsional cycle in Tor$H_k(L_6, \Z)$ of degree $n$ ($n\cdot \gamma_k = 0$). From the engineered 4d gauge theory perspective, and upon some dimensional counting, one can see that the topological operator $\mathcal{U}^{\text{M2 on }\gamma_{k}} (\Sigma_{3-k}) $ generates a $k$-form symmetry $\mathbb{Z}_n^{[k]}$. In a similar fashion, the topological operator $\mathcal{U}^{\text{M5 on }\gamma_{k}} (\Sigma_{6-k})$ generates a $(k-3)$-form symmetry $\mathbb{Z}_n^{[k-3]}$. Since there is no meaning for a $(-2)$-form symmetry, for this case we take $k\geq 2$. We summarize these observations in table \ref{tab:M2M5-G2}. These results match with the observations made in tables \ref{tab: HFS from G4} and \ref{tab: HFS from G7}.

Now, let us simplify the expressions for the topological operators above even further. Let us recall the decomposition of $\br{G}_4$ in \eqref{G4 decomp} and $\br{dG}_7$ in \eqref{dG7 decomp}. We have:
\begin{equation}
  \br{G}_4= \br{B}_{3-k} \star \br{t}_{k+1} + \cdots\,,\qquad\mathrm{and} \qquad  \br{\dd G}_{7}= \br{\delta A}_{6-k} \star \br{t}_{k+1} + \cdots \,,
\end{equation}
where by ``$\cdots$" we are ignoring the other terms which are irrelevant for the following discussion. Then, the symmetry operators associated with M2- \eqref{eq:symmetry-operator-discrete(1)} and M5-branes \eqref{eq:symmetry-operator-discrete(2)} wrapping a torsional cycle $\gamma_{k}\in\mathrm{Tor}H_{k}(L_{6},\Z)$ take the form:
\begin{equation}\label{top operators discrete general}
\begin{aligned}
    \mathcal{U}^{\text{M2 on }\gamma_{k}} (\Sigma_{3-k}) &\,=\,  \exp{ 2\pi i \left( \int^{\br{H}}_{L_{6}} \br{\mathsf{PD} } (\gamma_{k}) \star \br{t}_{k+1} \right) \int_{\Sigma_{3-k}} B_{3-k} } ~, \\
    \mathcal{U}^{\text{M5 on }\gamma_{k}} (\Sigma_{6-k}) &\,= \,\exp{ 2\pi i \left( \int^{\br{H}}_{L_{6}} \br{\mathsf{PD} } (\gamma_{k}) \star \br{t}_{k+1} \right) \int_{\Sigma_{6-k}} A_{6-k} }~.
\end{aligned}
\end{equation}
Here, $B_{3-k}$ and $A_{6-k}$ are $\Z_n$ gauge field with $\Z$-periodicity. We will revisit these expressions momentarily for the link space $L_6$ whose homology is given by \eqref{CoH of L6}. We will show how to explicitly work out the integral over $L_6$ in terms of the CS invariants \eqref{CS invariants: def} and the coefficients \eqref{mathsf K} introduced earlier.

Let us make here the following two comments concerning these discrete symmetries:
\begin{itemize}
    \item The discrete symmetries come in dual pairs. The $\Z_n^{[2-k]}( = \widehat{\Z_n}^{[k]})$ HFS coming from M5 wrapping $\widehat\gamma_k\in {\rm Tor}H_{5-k}(L_6, \Z)$ is the magnetic (a.k.a. Pontryagin) dual of the $\Z_n^{[k]}$ HFS generated by an M2 brane wrapping the $k$-torsional cycle $\gamma_k\in{\rm Tor} H_k(L_6,\Z)$. Recall the duality \eqref{torsion duality} where here we take $\widehat{\gamma}_k$ to be the dual cycle of $\gamma_k$. From the M-theory perspective, this is none other than the EM duality which exchanges the roles of M2 with M5 and hence $G_4$ with $G_7$ (see, e.g., \cite{Apruzzi:2021nmk,Najjar:2024vmm}).
    \item The charges of the defects in $\mathbb{D}^{[k]}$ under the $k$-form symmetry $\Z_n^{(k)}$ are controlled by the non-trivial linking pairing in with the associated topological operators in $L_6$ via\footnote{Besides the usual linking between the two submanifolds in $Y_5$ on which the defect and the symmetry operator are supported.}:
    \begin{equation}\label{eq:pairing-Tor-Tor}
        \mathrm{Tor}H_{k}(L_{6},\Z) \,\times\, \mathrm{Tor}H_{5-k}(L_{6},\Z)  \ \longrightarrow \ \Q/\Z~.
\end{equation}
In table \ref{tab:M2M5-G2}, we list all types of charged defects that we can get and associate them with the symmetry group under which they are charged.
\end{itemize}

\paragraph{Continuous symmetries.} In this case, continuous $m$-form symmetries are generated by symmetry operators $\mathcal{U}(\Sigma_{m'+1})$, constructed from the Page charges $P_{4}$ and $P_{7}$, also referred to as fluxbranes \cite{Najjar:2024vmm}. The $P_{4} = G_4$ Page charge measures M5-brane charge, while $P_{7}$ is associated with M2-brane charge \cite{Page:1983mke}. The definition of the $P_7$ charge will be given momentarily, from which it will be clear why we should use it, rather than $G_7$, to define the topological operators.

To define symmetry operators from Page charges, the charges must be supported on free homology cycles of the link space $L_{6}$, rather than torsional ones as in the discrete symmetries above. The topological symmetry operator arising from the $P_{4}$ Page charge (i.e., the $G_{4}$-fluxbrane) takes the form:
\begin{align}\label{symmetry-operator-fluxbrane-G4-0}
\begin{split}
    \mathcal{U}^{G_4\text{-flux along }\gamma_{k}} (\Sigma_{4-k}) &= \exp{ \frac{i \varphi}{2\pi} \int_{\Sigma_{4-k} \times \gamma_{k}} {G_4} }\\
    &=  \exp{ i \frac{\varphi}{2\pi} \int_{\Sigma_{4-k} \times L_{6}} \mathsf{PD} (\gamma_{k}) \wedge G_4 }~ .
\end{split}
\end{align}
where $\gamma_{k}\in H_{k}(L_{6},\Z)_{\mathrm{free}}$ and $\mathsf{PD}(\gamma_k)\in H^{6-k}(L_6,\Z)_{\rm free}$ is the corresponding Poincar\'e dual.
Similarly, the symmetry operator associated with the $P_{7}$ Page charge is given by:
\begin{align}\label{eq:P7-Page-symmetry-operator}
\begin{split}
   \mathcal{U}^{P_7\text{-flux along }\gamma_{k}} (\Sigma_{7-k}) &= \exp{ i \frac{\varphi}{2\pi} \int_{\Sigma_{7-k} \times \gamma_{k}} P_7 }\\
   &=  \exp{ i \frac{\varphi}{2\pi} \int_{\Sigma_{7-k} \times L_{6}} \mathsf{PD} (\gamma_{k}) \wedge P_7 }~.
\end{split}
\end{align}

Note, as in the case of the discrete symmetries \eqref{eq:symmetry-operator-discrete(1)}, the operators coming from the $P_4$-fluxbranes \eqref{symmetry-operator-fluxbrane-G4-0} are topological since $dG_4 = 0$. Meanwhile, to show that those coming from the $P_7$-fluxbrane \eqref{eq:P7-Page-symmetry-operator} are topological, let us recall some aspects of these $P_7$ Page charges. 

The 7-form field $P_{7}$, is defined via the Hopf--Wess--Zumino (HWZ) action as \cite{Bandos:1997ui,Intriligator:2000eq}: 
\begin{equation}\label{eq:Page-P7-charge}
  P_{7}\, := \restr{\left(G_{7} \,+\,\frac{1}{4\pi}  H_{3}\wedge G_{4}\,\right)}{\Sigma_{7}}\, ,
\end{equation}
where $\Sigma_{7}$ satisfies $\partial\Sigma_{7}=\Sigma_6^{\text{M5}}$, with $\Sigma_6^{\text{M5}}$ supports the M5-brane worldvolume. The 3-form $H_{3}$ is a self-dual field on the M5-brane worldvolume satisfying \cite{Howe:1997vn}:\footnote{In this context, $H_{3}$ is the field strength of the 2-form gauge field $B_{2}$, which couples to strings on $\Sigma^{\mathrm{M5}}_{6}$. These strings originate from the intersection of M2-branes with the M5-brane.}
\begin{equation}\label{eq:dH3isG4}
    \dd H_{3} \,=\, \restr{G_{4}}{\Sigma^{\mathrm{M5}}_{6}}~.
\end{equation}
From this we can see that the Page charge $P_{7}$ is both closed and quantized, thereby ensuring that \eqref{eq:P7-Page-symmetry-operator} is a well-defined topological operator\footnote{This can be viewed as a motivation for using the Page charges to define the topological operators rather than the naive choice $G_7$ which is not closed \protect\eqref{dG7}.}

\begin{table}[t]
\centering
\begin{tabular}{|l|c|c|c|c|}
\hline
& M2 & & M5 & \\
\hline
\hline
$H_{0}(L_{6},\Z)\times [0,\infty)$ \hspace{2pt} & electric 2d defect & {\color{red}$\diamondsuit$} & --- &   \\
$H_{3}(L_{6},\Z)\times [0,\infty)$ \hspace{2pt} & electric $(-1)$d ``defect" & $\clubsuit$ & magnetic 2d defect & ${\spadesuit}$ \\
$H_{6}(L_{6},\Z)\times [0,\infty)$ \hspace{2pt} & --- & $ $ & magnetic $(-1)$d ``defect" & {\color{red}$\heartsuit$} \\
\hline
\hline 
& $P_{4}$ & & $P_{7}$ & \\
\hline
\hline
$H_{0}(L_{6},\Z)$ \hspace{4pt} & $(-1)$-form sym. generator &  {\color{red}$\heartsuit$}  & --- &  \\
$H_{3}(L_{6},\Z)$ \hspace{4pt} & 2-form sym. generator & $\spadesuit$ & $(-1)$-form sym. generator & $\clubsuit$ \\
$H_{6}(L_{6},\Z)$ \hspace{4pt} & --- &  & 2-form sym. generator & {\color{red}$\diamondsuit$}  \\
\hline
\end{tabular}
\caption{Fluxbranes wrapping free cycles in $L_6$ give rise to continuous symmetries. We mark with an equal symbol the charged defect and the corresponding symmetry generators.}
\label{Table:continuous-symm-G2-M-theory}
\end{table}

From the perspective of the 4d $\mathcal{N}=1$ gauge theory, the symmetry operators \eqref{symmetry-operator-fluxbrane-G4-0}  and \eqref{eq:P7-Page-symmetry-operator} generate continuous $U(1)$ higher form symmetries. Taking $\gamma_k$ to be a generating free $k$-cycle, and after some dimensional counting, the $G_4$-flux operator \eqref{symmetry-operator-fluxbrane-G4-0} generates a $(k-1)$-form symmetry $U(1)^{[k-1]}$. Meanwhile, the $P_7$-fluxbrane symmetry operator \eqref{eq:P7-Page-symmetry-operator} generates a $(k-4)$-form symmetry $U(1)^{[k-4]}$. For instance, since the link space $L_6$ has free cycles of degrees $0, 3$ and $6$, then, one deduces that \eqref{symmetry-operator-fluxbrane-G4-0} give $U(1)^{[-1]}$ and two $U(1)^{[2]}$ symmetries. On the other hand, \eqref{eq:P7-Page-symmetry-operator} give two $U(1)^{[-1]}$ and one $U(1)^{[2]}$. We summarize this discussion in table \ref{Table:continuous-symm-G2-M-theory}. This matches with the observations we made earlier in tables \ref{tab: HFS from G4} and \ref{tab: HFS from G7}.

Schematically, let us consider a free co-cycle $v_{k} \in H^{k}(L_{10-d},\Z)_{\text{free}}$, and expand $P_{4}$ and $P_{7}$ as
\begin{equation}\label{P4 P7 decomp}
P_{4}\,=\, G_{4}\, =\, F_{4-k}\, \wedge\, v_{k} +\cdots~, \quad  \mathrm{and} \ \ P_{7}\,=\, \widetilde{h}_{7-k}\wedge v_{k} + \cdots~.
\end{equation}
Let us comment here that, due the fact that it comes from $P_7$, the field strength $\widetilde{h}_{7-k}$ is flat and have quantized periods. Therefore, we may refer to $\widetilde{h}_{7-k}$ as the effective lower-dimensional Page charge\footnote{A similar discussion from the type IIB perspective can also be found at \protect\cite[\S3.2]{Najjar:2025htp}.}.

Putting the above expansions back in \eqref{symmetry-operator-fluxbrane-G4-0} and \eqref{eq:P7-Page-symmetry-operator}, we get:

\begin{align}\label{eq:Page-charges-reduction}
\begin{split}
    \mathcal{U}^{P_{4}\text{-flux along }\gamma_{k}}(\Sigma_{4-k}) &= \exp{  i \frac{\varphi}{2\pi} \left( \int_{L_{6}} \mathsf{PD} (\gamma_{k}) \wedge v_{k}\right) \int_{\Sigma_{4-k}} F_{4-k} } \\
    &= \exp{ i \frac{\varphi}{2\pi}  \int_{\Sigma_{4-k}} F_{4-k} } ~,  \\
    \vspace{3cm}\\
    \mathcal{U}^{P_7\text{-flux along }\gamma_{k}} (\Sigma_{7-k}) &= \exp{ i \frac{\varphi}{2\pi} \left( \int_{L_{6}} \mathsf{PD} (\gamma_{k}) \wedge v_{k}\right) \int_{\Sigma_{7-k}} \widetilde{h}_{7-k} } \\
    &= \exp{ i  \frac{\varphi}{2\pi}  \int_{\Sigma_{7-k}} \widetilde{h}_{7-k} } ~.
    \end{split}
    \end{align}
Without loss of generality, here, we assumed that the integral over $L_{6}$ is normalized to unity.

Let us make the following comments on the continuous symmetries constructed above:
\begin{itemize}
\item For the defects in $\mathbb{D}^{(k)}$ \eqref{def:defect1} that are charged under $U(1)^{[k]}$ have to respect the following pairing:
\begin{equation}\label{eq:pairing-free-free}
    H_{k}(L_{6},\Z)_{\mathrm{free}} \,\times\,  H_{6-k}(L_{6},\Z)_{\mathrm{free}}  \ \longrightarrow \ \Z~.
\end{equation}

\item As previously noted, the HWZ action—governed by the Page charge $P_{7}$—serves as the topological action for an M5-brane. In particular, the insertion of a localized M5-brane modifies the equations of motion of the $G$-fields in 11-dimensional supergravity by introducing a source term. As a consequence, the supergravity action itself receives a correction precisely given by the HWZ coupling. This indicates that a localized M5-brane sources the $P_{7}$ Page charge. Therefore, in any analysis involving $P_{7}$, one must account for the presence of such localized M5-branes. For a recent overview and further references, see \cite[Appendix C]{Najjar:2024vmm}. Analogously, a localized M2-brane sources the $P_{4}$ Page charge.

\item A given electric defect charged under a $P_{7}$ fluxbrane originates from an M2-brane configuration, as discussed around (\ref{def:defect1}). In such a setup, both M2- and M5-branes are involved: the M2-brane gives rise to the defect, while the M5-brane sources the $P_{7}$ Page charge. This naturally leads to a dual perspective: one may exchange the roles of the M2- and M5-branes. In this dual picture, the M2-brane now sources a $P_{4}$ Page charge, while the M5-brane defines a magnetic defect.

This exchange of roles can be formalized as follows:
\begin{equation}\label{eq:exchang-role-M2-M5}
\begin{aligned}
   \begin{rcases}
       \text{M2 on $\mathbb{R}_{+}\times \gamma_{k}$}
       \\
       \text{$P_{7}$ on $\gamma_{6-k}$}
   \end{rcases}\quad & \xleftrightarrow[ \ \text{ Exchanging role \ }]{\text{ \  M2 -- M5 branes \ }} \  \quad & \begin{cases}
       \text{M5 on $\mathbb{R}_{+}\times \gamma_{6-k}$}
       \\
       \text{$P_{4}$ on $\gamma_{k}$}\,,
   \end{cases}
\end{aligned}
\end{equation}
with $\gamma_{\bullet}\in H_{\bullet}^{\text{free}}(L_{6},\Z)$.

This correspondence implies a relation between $m$-form and $(d-m-3)$-form $U(1)$ symmetries,
\begin{equation}\label{eq:U(1)-m-dual-U(1)-d-m-3}
\begin{aligned}
    U(1)^{\scriptscriptstyle [m]} \quad & \xleftrightarrow{ \ \text{ M2 $\leftrightarrow$ M5 } \ } \quad & U(1)^{\scriptscriptstyle [d-m-3]} \,.
\end{aligned}
\end{equation}
Here, we take $k=2-m$ to connect with the above equation. 

Note that the above (M2–M5 role-exchange) duality should not to be confused with electric–magnetic duality in spacetime.

Applying the M2–M5 role-exchange duality to the link space $L_{6}$, we find that the $(-1)$-form symmetry $U(1)^{[-1]}$ is dual to the 2-form symmetry $U(1)^{[2]}$. More generally, one identifies three distinct dual pairs of such continuous $U(1)$ symmetries in this setup. The complete set of brane configurations and corresponding Page fluxes responsible for these dual symmetries is summarized in Table \ref{Table:continuous-symm-G2-M-theory}. A detailed discussion of these continuous symmetries and their topological operators can be found in section \ref{sec:continuous-symmetry-operators}.

\end{itemize}

For further details on the construction of such symmetry operators from Page charges, we refer the reader to \cite{Najjar:2024vmm}. A more exhaustive treatment of the $P_{7}$ Page charge can be found in \cite{Moore:2004jv}. For additional discussions on the HWZ action and its physical implications, see also \cite{Bandos:1997ui,Intriligator:2000eq}

In the next two subsections, we apply the above discussion to the general link space $L_{6}$, as presented in (\ref{CoH of L6}). A similar analysis was carried out in \cite{Najjar:2024vmm}; thus, some overlap with that discussion is expected. However, our goal here is to summarize and generalize those results to the broader scenarios we are developing in this paper.

\subsection{Discrete symmetry operators from \texorpdfstring{$L_{6}$}{L6}}
So far, our discussion in this section has been general with its treatment of the singular homology ring of the link space $L_6$. Using our earlier result for $H_{\bullet}(L_6,\Z)$ derived in \eqref{CoH of L6}, we would like now to be more explicit in our construction of these defects for the six theories we are considering in this work \eqref{eq:the-6-cases}. Our discussion here overlaps with that presented in \cite[\S4.3.3]{Najjar:2024vmm}.

Recall that, the discrete symmetries of the 4d $\mathcal{N}=1$ gauge theory coming from the wrapping of the M2 and M5 branes come in Pontryagin dual pairs. In what follows, we will discuss each one of these pairs separately. For convenience here, let us recall the explicit form of the homology ring of the link space \eqref{CoH of L6}:
\begin{equation}\label{CoH of L6 repeated}
    \begin{split}
        H_{\bullet}(L_{6},\Z) \,&=\,  \Z \,\oplus\, \begin{matrix}B^{\rm ab}\\\oplus \\\widetilde{\Gamma}^{\mathrm{ab}}_{\mathrm{ext}}\end{matrix}\,\oplus\,  {\rm Tor}_1^{\Z}(B^{\rm ab}, \widetilde{\Gamma}^{\mathrm{ab}}_{\mathrm{ext}})  \oplus\, \begin{matrix}\Z^2\\\oplus\\ {\rm Tor}_1^{\Z}(B^{\rm ab}, \widetilde{\Gamma}^{\mathrm{ab}}_{\mathrm{ext}}) \end{matrix}\,\oplus\, \begin{matrix}B^{\rm ab}\\\oplus\\ \widetilde{\Gamma}^{\mathrm{ab}}_{\mathrm{ext}}\end{matrix}\, \oplus\, 0\,\oplus\, \Z~.
    \end{split}
\end{equation}

\subsubsection{Electric and magnetic 1-form symmetries}

The first pair of these discrete symmetries consists of the 1-form symmetry coming from the M2-branes wrapping torsional 1-cycles and that coming from the M5-branes wrapping torsional 4-cycles. We will refer to the earlier as the electric 1-form symmetry and the latter as the magnetic one.

Following the discussion around \eqref{def:defect1}, the sets of all possible electric and magnetic charged defects are given by:

\begin{align}   
\begin{split}
    &\mathbb{D}^{(1)}_{\mathrm{electric}}\,=\, \{\,\text{M2-branes \,on}\,\, \left(B^{\rm ab}\oplus \widetilde{\Gamma}^{\mathrm{ab}}_{\mathrm{ext}}\right) \text{1-cycles}\,\times\,[0,\infty)\,\}~,\\
    &\mathbb{D}^{(1)}_{\mathrm{magnetic}}\,=\, \{\,\text{M5-branes \,on}\,\, \left(B^{\rm ab}\oplus \widetilde{\Gamma}^{\mathrm{ab}}_{\mathrm{ext}}\right) \text{$4$-cycles}\,\times\,[0,\infty)\,\}~.
\end{split}
\end{align}
Stated more explicitly, the electric and magnetic 1-form symmetries are given by:
\begin{equation}\label{1-form discrete symmetry}
    \mathsf{G}^{[1]}_{\rm electric} \cong \mathsf{G}^{[1]}_{\rm magnetic} \cong (B^{\rm ab})^{[1]} \oplus (\widetilde{\Gamma}_{\rm ext}^{\rm ab})^{[1]}~.
\end{equation}

Following the discussion around \eqref{def:symmetryoperator1}, let us now construct the explicit form of the topological operators generating these symmetries. As we will discuss momentarily, we will consider only a subsector of the full symmetry. In what follows, we will denote these subsectors by $\widetilde{\mathsf{G}}^{[1]}_{\rm electric}\subset \mathsf{G}_{\rm electric}^{[1]}$ and $\widetilde{\mathsf{G}}^{[1]}_{\rm magnetic}\subset \mathsf{G}_{\rm magnetic}^{[1]}$.

\medskip
\noindent
\textbf{Symmetry operators for $\widetilde{\mathsf{G}}^{[1]}_{\rm electric}$.} For the electric 1-form symmetry, we have:
\begin{equation}
\mathbb{U}^{(2)}_{\mathrm{electric}}\,=\, \{\,\text{M5-branes \,on}\,\, \left(B^{\rm ab}\oplus \widetilde{\Gamma}^{\mathrm{ab}}_{\mathrm{ext}}\right) \text{4-cycles}\,\}\,.
\end{equation}
Using the second expression in \eqref{top operators discrete general}, one can work out the form of these topological operators. Let us start with those generating the $B^{\rm ab}$ 1-form symmetry. Recall that, for our cases of interest \eqref{eq:the-6-cases}, $B^{\rm ab} \cong\Z_m$ for some $m$. Therefore, we have only one generator $\gamma_4 = \mathsf{PD}(t_2^{\rm b})$:
\begin{align} \label{base 1-form discrete top operator}
\begin{split}
    \mathcal{U}^{\mathrm{M5\,on\,}\mathrm{PD}(t^{\rm b}_{2})}(\Sigma_{2})
    &= \exp\left(2\pi i \int_{L_6}^{\br{H}} \br{t}_{2}^{\rm b}\star \br{t}_2^{\rm b} \star\br{\rm vol}^{\rm f}_3 \int_{\Sigma_2}A_2^{\rm b}\right)= \exp\left(\frac{2\pi i }{m} \int_{\Sigma_2}A_2^{\rm b}\right)~.
\end{split}
\end{align}
Where, in the last equality, we used the explicit form of the 3d CS invariant for $B = \Z_m$ given in table \ref{tab: ZGamma and CSGamma}. Moreover, we used the fact that periods of $A_2^{\rm b}$ are integer-quantized.

Let us now look at the second component of the electric 1-form symmetry, namely the contribution coming from the fibered 3-sphere of $L_6$. 

Following the local vs. global perspectives discussed in section \ref{sec:local-observer}, we will focus here on the sub-sector perceived by the local observer, namely, 
we will study here only the topological operators generating the $1$-form symmetry $(C_0^{\rm ab})^{[1]}\subset (\widetilde{\Gamma}_{\rm ext}^{\rm ab})^{[1]}$.\footnote{We will revisit this point again in section \ref{subsec: symtft outer} below.} So, in total, we are focusing here on the following subgroup of the full 1-form symmetry:
\begin{equation}\label{tilde G electric 1-form}
    \widetilde{\mathsf{G}}_{\rm electric}^{[1]} = (B^{\rm ab})^{[1]}\oplus (C_0^{(\rm ab)})^{[1]}\subset \mathsf{G}_{\rm electric}^{[1]}~.
\end{equation}

Similar to the derivation \eqref{base 1-form discrete top operator}, one can work out the topological operator generating $(C_0^{\rm ab})^{[1]}$ 1-form symmetry:
\begin{align} \label{fiber 1-form discrete top operator}
\begin{split}
    \mathcal{U}^{\mathrm{M5\,on\,}\mathrm{PD}(t^{\rm f}_{2, a})}(\Sigma_{2}) 
    &= \exp\left(2\pi i \sum_{b}\int_{L_6}^{\br{H}} \br{t}_{2, a}^{\rm f}\star \br{t}_{2,b}^{\rm f} \star\br{\rm vol}^{\rm b}_3 \int_{\Sigma_2}A_{2,b}^{\rm f}\right)\\
    &= \exp\left(4\pi i \sum_{b} [{\rm CS}_{C_0}]_{a,b}  \int_{\Sigma_2}A_{2, b}^{\rm f}\right)~.
\end{split}
\end{align}

From these explicit formulas, we can see that the differential 2-forms $A_{2,\bullet}^{\bullet}$ are the background gauge fields for these 1-form symmetries. This matches our interpretation given in table \ref{tab: HFS from G7}. But this interpretation is a bit tricky for the operators \eqref{fiber 1-form discrete top operator} due to the extra sum appearing in the argument of the exponent. Let us make some comments on this point here.

As shown in table \ref{tab:H1 for S3 fiber}, for the majority of the cases we are studying in this paper, $C_0^{\rm ab}$ consists of one torsional component. In these cases, the above expression for the corresponding topological symmetry operator  simplifies as follows:
\begin{align} \label{fiber 1-form discrete top operator simplified}
\begin{split}
    \mathcal{U}^{\mathrm{M5\,on\,}\mathrm{PD}(t^{\rm f}_{2})}(\Sigma_{2}) 
    &= \exp\left(4\pi i {\rm CS}_{C_0}  \int_{\Sigma_2}A_{2}^{\rm f}\right)~.
\end{split}
\end{align}

Meanwhile, for the cases where $C_0 \cong 2\D_{4N}$, we have that $C_0^{\rm ab} = \Z_2\oplus\Z_2$. The matrix of the corresponding CS invariants is given in table \ref{tab: ZGamma and CSGamma}. For this case, we get two $\Z_{2,b}^{[1]}$ 1-form symmetries with background gauge fields $A_{2,b}^{\rm f}$. But note that, for $a=1, 2$, the topological operator \eqref{fiber 1-form discrete top operator} is not the operator generating $\Z_{2,a}^{[1]}$ rather it generates a linear combination of these $\Z_{2,1}^{[1]}$ and $\Z_{2,2}^{[1]}$. 

Taking $\mathcal{A}^{(a)}_{2}$ to be the 2-form background gauge field for the 1-form symmetry generated by \eqref{fiber 1-form discrete top operator}, we have:
\begin{equation}\label{linear comb of bg}
        \begin{split}
                \mathcal{A}^{(1)}_{2} := \frac{N}{2}\, A_{2,1}^{\rm f} + \frac{N-1}{2}\,A_{2,2}^{\rm f}~,\\
                \mathcal{A}_{2}^{(2)} := \frac{N-1}{2}\, A_{2,1}^{\rm f} + \frac{N}{2}\,A_{2,2}^{\rm f}~.
        \end{split}
\end{equation}

\medskip
\noindent
\textbf{Symmetry operators for $\widetilde{\mathsf{G}}^{[1]}_{\rm magnetic}$.} Similarly, the topological operators for the magnetic 1-form symmetry are given by:
\begin{equation}
\mathbb{U}^{(2)}_{\mathrm{magnetic}}\,=\, \{\,\text{M2-branes \,on}\,\, \left(B^{\rm ab}\oplus \widetilde{\Gamma}^{\mathrm{ab}}_{\mathrm{ext}}\right) \text{1-cycles}\,\}\,.
\end{equation}
As for the electric case above, here we will focus mainly on the subgroup:
\begin{equation}\label{tilde G magnetic 1-form}
    \widetilde{\mathsf{G}}_{\rm magnetic}^{[1]} \cong (B^{\rm ab})^{[1]}\oplus (C_0^{(\rm ab)})^{[1]}\subset \mathsf{G}_{\rm magnetic}^{[1]}~.
\end{equation}

From \eqref{top operators discrete general}, the symmetry operator generating the 1-form symmetry $(B^{\rm ab})^{[1]}$ is given by:
\begin{equation}
  \begin{split}
        \mathcal{U}^{\mathrm{M2\,on\,}\mathrm{PD}(t^{\rm b}_{2}\smile\vol_{3}^{\rm f})}(\widetilde{\Sigma}_{2}) &= \exp\left(2\pi i \int_{L_6}^{\br{H}}\br{t}_2^{\rm b}\star \br{\rm vol}_3^{\rm f}\star\br{t}_2^{\rm b}\int_{\widetilde{\Sigma}_2}B_2^{\rm b}\right)\\
        &= \exp\left(\frac{2\pi i}{m}\int_{\widetilde{\Sigma}_2}B_2^{\rm b}\right)~.
  \end{split}
\end{equation}
In the second equality we used the explicit form for the CS invariant for $B = \Z_m$, and the fact that $B_{2}^{\rm{b}}$ has integer quantized periods. 

The topological operators for the $(C_0^{\rm ab})^{[1]}\subset (\widetilde{\Gamma}_{\rm ext}^{\rm ab})^{[1]}$ can be constructed in a similar fashion. In this case we have:

\begin{equation}\label{fiber 1-from magentic}
\begin{split}
     \mathcal{U}^{\mathrm{M2\,on\,}\mathrm{PD}(t^{\rm f}_{2,a}\,\smile\,\vol_{3}^{\rm b})}(\widetilde{\Sigma}_{2}) &=\exp\left(2\pi i\sum_{b} \int_{L_6}^{\br{H}}\br{t}_{2,a}^{\rm f}\star \br{\rm vol}_3^{\rm b}\star\br{t}_{2,b}^{\rm f}\int_{\widetilde{\Sigma}_2}B_{2,b}^{\rm f}\right)\\
     &= \exp\left(4\pi i\sum_{b} [{\rm CS}_{C_0}]_{a,b}\int_{\widetilde{\Sigma}_2}B_{2,b}^{\rm f}\right)~.
\end{split}
\end{equation}
 
Similar to the observation made above concerning the background gauge fields $A_{2,\bullet}^{\bullet}$, note here that the differential forms $B_{2,\bullet}^{\bullet}$ are indeed background gauge fields for the magnetic 1-form symmetries. This also matches the interpretations given in table \ref{tab: HFS from G4} and \ref{tab: HFS from G7}. Note that, in \eqref{fiber 1-from magentic} above, one needs to consider linear combinations of the background fields of the form \eqref{linear comb of bg} for the cases with $C_0 \cong 2\D_{4N}$.

\subsubsection{Electric 0-form and magnetic 2-form symmetries}

Let us now move to the second pair of dual symmetries, which are the 0-form $\mathsf{G}^{[0]}_{\rm electric}$ and 2-form symmetries $\mathsf{G}^{[2]}_{\rm magnetic}$. In the same spirit as for the previous case, we will refer to the 0-form symmetry as the electric symmetry and the 2-form symmetry as the magnetic one.

Recall from the discussion around \eqref{def:defect1}, the defects charged under these symmetries are constructed from M2 and M5 branes as follows:
\begin{equation}
    \begin{split}
        &\mathbb{D}^{(0)}_{\rm electric}  = \{\,\text{M2-branes on \,} \left(B^{\rm ab}\otimes \widetilde{\Gamma}^{\rm ab}_{\rm ext}\right) \,\text{2-cycles}\times [0,\infty)\,\}~,\\
        &\mathbb{D}_{\rm magnetic}^{(2)} = \{\,\text{M5-branes on \,} \left(B^{\rm ab}\otimes \widetilde{\Gamma}^{\rm ab}_{\rm ext}\right) \,\text{3-cycles}\times [0,\infty)\,\}~.
    \end{split}
\end{equation}
From this, we deduce that, depending on the link space $L_6$ among the six cases \eqref{eq:the-6-cases}, the geometrically engineered 4d $\mathcal{N}=1$ gauge theory has the following 0-form and 2-form symmetries:
\begin{equation}\label{0/2 form full}
            \mathsf{G}^{[0]}_{\rm electric} \cong \mathsf{G}^{[2]}_{\rm magnetic} \cong B^{\rm ab} \otimes \widetilde{\Gamma}_{\rm ext}^{\rm ab}~.
\end{equation}
Next, we will explicitly construct the topological operators generating these symmetries. Note that, as we will see below, in these two symmetries, we do not run into the issue with calculating the CS invariants for $\widetilde{\Gamma}_{\rm ext}$; hence, we can calculate the topological operators generating the full symmetry.

\medskip
\noindent
\textbf{Symmetry operators for ${\mathsf{G}}_{\rm electric }^{[0]}$.}
For the electric 0-form symmetry, we have the following set of topological operators \eqref{def:symmetryoperator1}:
\begin{equation}
    \mathbb{U}_{\rm electric}^{(3)} := \{\,\text{M5-branes on\,} (B^{\rm ab}\otimes \widetilde{\Gamma}_{\rm ext}^{\rm ab})\, \text{3-cycles}\,\}~.
\end{equation}

More explicitly, the generators of this symmetry are of the form \eqref{top operators discrete general}:
\begin{equation}\label{0-form disc top op}
\begin{split}
        \mathcal{U}^{\mathrm{M5\,on\,}\mathrm{PD}(t_{3,a})}(\Sigma_3) &= \exp\left(2\pi i \sum_{b} \int_{L_6}^{\br{H}} \br{t}_{3,a} \star \br{t}_{4,b} \int_{\Sigma_3} A_{3,b}\right)\\
        &= \exp\left(-\frac{2\pi i}{n_a}  \int_{\Sigma_3} A_{3,a}\right)~,
\end{split}
\end{equation}
where in the second equality, we used \eqref{mathsf K} with $n_a$ being the torsional degree of the class $t_{3,a}\in {\rm Tor} H^3(L_6, \Z)$.

\medskip
\noindent
\textbf{Symmetry operators for ${\mathsf{G}}_{\rm magnetic }^{[2]}$.} 
As for the dual magnetic 2-form symmetry, we have the following set of topological operators \eqref{def:symmetryoperator1}:
\begin{equation}\label{2-form discrete generators}
    \mathbb{U}_{\rm magnetic}^{(1)} := \{\,\text{M2-branes on\,} (B^{\rm ab}\otimes {\widetilde \Gamma}_{\rm ext}^{\rm ab})\, \text{2-cycles}\,\}~.
\end{equation}

The topological operators generating this symmetry are given by \eqref{top operators discrete general}:
\begin{equation}
\begin{split}
        \mathcal{U}^{\mathrm{M2\,on\,}\mathrm{PD}(t_{4,a})}(\Sigma_1) &= \exp\left(2\pi i \sum_{b} \int_{L_6}^{\br{H}} \br{t}_{4,a} \star \br{t}_{3,b} \int_{\Sigma_1} B_{1,b}\right)\\
        &= \exp\left(-\frac{2\pi i}{n_a}  \int_{\Sigma_1} B_{1,a}\right)~.
\end{split}
\end{equation}
In the second line we followed the same steps as we did in \eqref{0-form disc top op}. 

Let us remark here that, from the explicit form of the topological operators for this pair of dual symmetries, we see that the differential 1-form $B_{1,a}$ is a background gauge field for the 0-form symmetry, hence it enters the definition of the topological operator of the dual 2-form symmetry. Similarly, for the differential 3-form $A_{3,a}$. This matches with our earlier interpretation in tables \ref{tab: HFS from G4} and \ref{tab: HFS from G7}.
\subsubsection{Electric \texorpdfstring{$(-1)$}{-1}-form and magnetic 3-form symmetries}

Now we come to the last pair of discrete symmetries for the resulting 4d $\mathcal{N}=1$ gauge theory. This pair consists of the electric $(-1)$-form symmetry $\mathsf{G}_{\rm electric}^{[-1]}$ and the magnetic $3$-form symmetry $\mathsf{G}_{\rm magnetic}^{[3]}$. The sets of defects charged under these symmetries are give by \eqref{def:defect1}:\footnote{In general, the defects $\mathbb{D}^{(-1)}$ are `morally' constructed as  such configurations are physically obstructed. Hence, no physical defects of this type exist, which is consistent with field theory analysis, e.g., \cite{Aloni:2024jpb,Santilli:2024dyz}.}
\begin{equation}
    \begin{split}
            &\mathbb{D}_{\rm electric}^{(-1)} = \{\,\text{M2-branes on\,} \left(B^{\rm ab}\otimes \widetilde{\Gamma}_{\rm ext}^{\rm ab}\right)\, \text{3-cycle}\times [0,\infty)\,\}~,\\
            &\mathbb{D}_{\rm magnetic}^{(3)} = \{\,\text{M5-branes on \,} \left(B^{\rm ab}\otimes \widetilde{\Gamma}^{\rm ab}_{\rm ext}\right)\,\text{2-cycles} \times [0, \infty)\,\}~.
    \end{split}
\end{equation}
Therefore, similar to the previous case \eqref{0/2 form full}, we have:
\begin{equation}
          \mathsf{G}^{[-1]}_{\rm electric}\cong \mathsf{G}_{\rm magnetic}^{[3]} \cong B^{\rm ab}\otimes \widetilde{\Gamma}_{\rm ext}^{\rm ab}~.  
\end{equation}

\medskip
\noindent
\textbf{Symmetry operators for ${\mathsf{G}}_{\rm electric}^{[-1]}$.} For this symmetry, we have the following set of topological operators \eqref{def:symmetryoperator1}:
\begin{equation}
        \mathbb{U}_{\rm electric}^{(4)} = \{\,\text{M5-branes on \,} (B^{\rm ab}\otimes {\widetilde \Gamma}_{\rm ext}^{\rm ab})\, \text{2-cycles}\,\}~.
\end{equation}
More explicitly, the topological operators generating this symmetry are of the form \eqref{top operators discrete general}:
\begin{equation}
\begin{split}
    \mathcal{U}^{\mathrm{M5\,on\,}\mathrm{PD}(t_{4,a})}(\Sigma_4) &= \exp\left(2\pi i\sum_{b} \int_{L_6}^{\br{H}} \br{t}_{4,a}\star \br{t}_{3,b}\int_{\Sigma_4} A_{4,b} \right)~\\
    &=\exp\left(-\frac{2\pi i}{n_a} \int_{\Sigma_4}A_{4,a} \right)~.
    \end{split}
\end{equation}
Where, in the second equality, we used \eqref{mathsf K} and $n_a$ is the torsional degree of the class $t_{4,a}\in {\rm Tor}H_4(L_6,\Z)$.

\medskip
\noindent
\textbf{Symmetry operators for ${\mathsf{G}}_{\rm magnetic }^{[3]}$.} 
As for this symmetry, we have the following set of topological operators \eqref{def:symmetryoperator1}:
\begin{equation}
        \mathbb{U}_{\rm magnetic}^{(0)} = \{\,\text{M2-branes on \,} (B^{\rm ab}\otimes {\widetilde \Gamma}_{\rm ext}^{\rm ab})\, \text{3-cycles}\,\}~.
\end{equation}
More explicitly, the topological operators generating this symmetry are of the form \eqref{top operators discrete general}:
\begin{equation}
\begin{split}
    \mathcal{U}^{\mathrm{M2\,on\,}\mathrm{PD}(t_{3,a})} ({\rm pt}) &:= \exp\left(2\pi i \sum_b\int_{L_6}^{\br{H}} \br{t}_{3,a}\star \br{t}_{4,b} \int_{\rm pt} B_{0,b}\right)~\\
    &=\exp\left(-\frac{2\pi i}{n_a} B_{0,a}\mid_{\rm pt}\right)~.
\end{split}
\end{equation}

Similar to the previous two dual pairs, one can indeed check that this construction matches with the interpretation in tables \ref{tab: HFS from G4} and \ref{tab: HFS from G7} of the differential forms $B_{0,a}$ and $A_{4,a}$ as background gauge fields for the $(-1)$-form and $3$-form symmetries, respectively.

\subsection{Continuous symmetry operators from \texorpdfstring{$L_{6}$}{L6}}\label{sec:continuous-symmetry-operators}

The continuous symmetries arising from the $P_{7}$-fluxbrane wrapping non-trivial free cycles of the link space $L_{6}=(\mathbb{S}^{3}_{\rm f}\times \mathbb{S}^{3}_{\rm b})/\Gamma$ have already been discussed in \cite[\S 4.3.4]{Najjar:2024vmm}. For completeness, we briefly review that discussion here. The various continuous symmetries that may arise in this context are summarized in table \ref{Table:continuous-symm-G2-M-theory}. 

\subsubsection{Electric continuous symmetries}

Let us begin by considering the symmetry topological operators arising from the various configurations of the $P_{7}$ Page charge wrapping non-trivial free cycles of the link space.

\paragraph{Continuous electric $(-1)$-form symmetries.} There are two continuous $U(1)^{\scriptscriptstyle [-1,\alpha]}$ $(-1)$-form symmetries given by $P_7$-fluxbranes wrapping either generator in $H_{3}(L_{6},\Z)_{\text{free}}=\Z\oplus\Z$. The symmetry topological operators can be constructed using the second operator in (\ref{eq:Page-charges-reduction}):
\begin{equation}\label{Sym-op-(-1)-form-U(1)}
    \begin{aligned}
      \mathcal{U}^{P_7\text{-flux along }\mathsf{PD}({\rm vol}_3^{\alpha})}(\Sigma_4) &= \exp \left\{ i \frac{\varphi}{2\pi} \int_{L_6 \times \Sigma_4} \vol_3^{\alpha} \wedge P_7 \right\} \\
        &= \exp \left\{ i \frac{\varphi}{2\pi} \int_{L_6 \times \Sigma_4} \vol_3^{\alpha} \wedge \left[ \sum_{\beta=\mathrm{f,b}} \widetilde{h}_4^{\beta} \wedge \vol_3^{\beta} + \cdots \right] \right\}  \\
        &= \exp \left\{ i \frac{\varphi }{2\pi}\int_{\Sigma_4 } \, \widetilde{h}_{4}^{\alpha} \right\} \,.
\end{aligned}
\end{equation}
In the second equality above, we used the expansion \eqref{P4 P7 decomp} for the $P_7$ Page charge. Recall here that $\mathsf{PD}({\rm vol}_3^{\rm f}) = [\mathbb{S}^3_{\rm b}/B]$ and $\mathsf{PD}({\rm vol}_3^{\rm b}) = [\mathbb{S}^3_{\rm f}/\widetilde{\Gamma}_{\rm ext}]$.

Recalling \eqref{eq:Page-P7-charge}, the field $\widetilde{h}_{4}^{\alpha}$ is defined as the sum
\begin{equation}
    \widetilde{h}_{4}^{\alpha}\,=\, h_{4}^{\alpha} \,+\, g_{4}^{\alpha}\,,
\end{equation}
where $h_{4}^{\alpha}$ arises from the expansion of the M-theory $G_{7}$ along the 3-form volume $\vol_{3}^{\alpha}$, and $g_{4}^{\alpha}$ is defined as:
\begin{equation}\label{eq:g-4-i-refinment}
\begin{aligned}
        g_4^{\alpha} \,&=\, \frac{1}{4\pi} \int_{\mathsf{PD}({\rm vol}_3^\alpha)} H_3 \wedge G_4 \\
        \,&=\, \frac{1}{4\pi} \int_{\mathsf{PD}({\rm vol}_3^{\alpha})} \left( \mathsf{h}_{3} \,+\, \sum_{\beta=\mathrm{f},\mathrm{b}}\mathsf{h}_0^{\beta} \vol_3^{\beta} \right) \wedge \left( F_4 \,+\, \sum_{\gamma=\mathrm{f},\mathrm{b}}F_1^{\gamma} \vol_3^{\gamma}  \right) \\
        \,&=\, \frac{1}{4\pi} \left( \mathsf{h}_{3} \wedge F_1^{\alpha} \,+\, \mathsf{h}_0^{\alpha} F_{4} \right) \,,
\end{aligned}
\end{equation}
with $\dd \mathsf{h}_3 = F_4$ and $\dd \mathsf{h}_0^{\alpha} = F_1^{\alpha}$.

Let us also comment here that, the continuous $(-1)$-form symmetry associated with the fiber, denoted by $U(1)^{ [-1,\mathrm{f}]}$, is identified with the Chern--Weil type $(-1)$-form symmetry \cite{Najjar:2024vmm}. We will discuss this point on more detail in section \ref{sec:phy-implications} below. 

\paragraph{Continuous universal electric 2-form symmetry.} 
For the universal $U(1)^{ [2]}$ electric 2-form symmetry, the symmetry topological operator is given by a $P_7$-fluxbrane wrapping the whole link space $L_{6}$, i.e., $H_{6}(L_{6},\Z)_{\rm free}\cong\Z$. Using (\ref{eq:Page-charges-reduction}), the symmetry generator can be expressed as:
\begin{equation}\label{gen:2-formContinuous}
    \begin{aligned}
       \mathcal{U}^{P_7\text{-flux along }L_6}(\Sigma_{1})&=\exp \left\{ i \frac{\varphi}{2\pi} \int_{L_6 \times \Sigma_1} P_7 \right\} \\
        &= \exp \left\{ i \frac{\varphi}{2\pi} \int_{L_6 \times \Sigma_1} \left[ \widetilde{h}_1 \wedge \vol_3^{\mathrm{f}} \wedge \vol_3^{\mathrm{b}} + \cdots \right] \right\}  \\
        &= \exp \left\{ i \frac{\varphi}{2\pi} \oint_{\Sigma_1 } \, \widetilde{h}_{1}  \right\} .
    \end{aligned}
    \end{equation}
Here, $\Sigma_{1}$ is loop defined over the spacetime, and $\widetilde{h}_1=h_1 + g_1$ with correction term 
   \begin{equation}\label{eq:refinment-g1}
       g_{1}\,=\, \frac{1}{4\pi} \,\int_{L_6}\, H_{3}\, \wedge \,G_{4} \,=\, \frac{1}{4\pi} \,\left(\, \mathsf{h}_{0}^{\mathrm{f}} F_{1}^{\mathrm{b}} \,+\, \mathsf{h}_{0}^{\mathrm{b}} F_{1}^{\mathrm{f}} \,\right) ~,
   \end{equation}
subject to $\dd \mathsf{h}_0^{\alpha} = F_1^{\beta}$, hence $2 \pi \dd g_1 = F_1^{\mathrm{f}} \wedge F_1^{\mathrm{b}}$.

In this case, the charged defects for these electric $U(1)^{\scriptscriptstyle[2]}$ $2$-form symmetries are realized through M2-branes  as follows:
and extend along the radial direction: \begin{equation}
\mathbb{D}^{(2)}_{\rm{electric}}\,=\,\{\, \text{M2-branes on }\, H_{0}(L_{6},\Z)\, \times[0,\infty)\}\,.
\end{equation}
This defines $\Z$-valued 2d non-dynamical defects charged under the universal 2-form symmetry.

\subsubsection{Magnetic continuous symmetries}

Having constructed the symmetry topological operators associated with electric defects originating from M2-branes, we now turn our attention to the magnetic sector, specifically, to the symmetry topological operators arising from M5-branes. As we discussed earlier around \eqref{symmetry-operator-fluxbrane-G4-0}, these come from $P_4 = G_4$ Page charges (fluxbranes).

\paragraph{Continuous magnetic $(-1)$-form symmetry.}

Consider an M5-brane wrapping the link space $L_{6}$ and extended along the radial direction of the SymTFT. 

The corresponding defect is charged under a Page charge $P_{4}$, with flux sourced by elements of  $H_{0}(L_{6},\Z)$. Using the first equation (\ref{eq:Page-charges-reduction}), 
\begin{equation}
    \mathcal{U}^{P_{4}\text{-charge along} \,{\rm pt}}(\Sigma_{4})\,=\, \exp{i\frac{\varphi}{2\pi}\,\int_{\Sigma_{4}}F_{4}}\,.
\end{equation}

Via the M2-M5-exchange duality discussed around (\ref{eq:exchang-role-M2-M5}), this $(-1)$-form symmetry is dual to the universal 2-form symmetry whose associated topological operator is given in (\ref{gen:2-formContinuous}). These dual symmetries constitute one of the pairs of dual continuous symmetries associated with the topology of the link space $L_{6}$ as discussed earlier.

\paragraph{Continuous magnetic 2-form symmetries.} There are two $U(1)^{\scriptscriptstyle[2]}$ 2-form symmetries which arise from $P_{4}$ Page charges wrapping either free 3-cycle in $L_6$. Following the general structure in (\ref{eq:Page-charges-reduction}), the  resulting symmetry operators can be written as:
\begin{equation}
\begin{split}
        \mathcal{U}^{P_{4}\text{-charge along} \,\mathsf{PD}({\rm vol}_3^\alpha)}(\widetilde{\Sigma}_{1})\,&=\,\exp{i\frac{\varphi}{2\pi}\,\int_{\widetilde{\Sigma}_{1}} \,\vol_{3}^{\alpha}\,\wedge\,\left[\sum_{\beta=\rm{f,b}}F_{1}^{\beta}\wedge\,\vol^{\beta}_3\right]}
        \\
        \,&=\, \exp{i\frac{\varphi}{2\pi}\,\int_{\widetilde{\Sigma}_{1}}F_{1}^{\alpha}}~.
\end{split}
\end{equation}

These 2-form symmetries are dual—in the sense described around (\ref{eq:exchang-role-M2-M5})—to the pair of electric $(-1)$-form symmetries whose symmetry operators are given in (\ref{Sym-op-(-1)-form-U(1)}). Together, these constitute two out of the three dual pairs of continuous symmetries associated with the topology of the link space $L_{6}$, thereby completing the expected three dual symmetry pairs.

The charged defects are constructed via M5-branes as
\begin{equation}
\mathbb{D}^{(2)}_{\rm{magnetic}}\,=\,\{\, \text{M5-branes on }\, H_{3}(L_{6},\Z)\, \times[0,\infty)\}\,.
\end{equation}
In particular, following our comment around \eqref{eq:pairing-Tor-Tor}, we see that the defect coming from wrapping the M5-brane around the free 3-cycle indexed by $\alpha$ is charged under the symmetry coming from the 3-cycle indexed by $\beta\neq \alpha$.


\section{SymTFT and outer automorphism}\label{subsec: symtft outer}

So far, we have analyzed the physics of the non-splittable quotients classified in (\ref{eq:the-6-cases}), their associated SymTFTs, and the origin of their symmetry topological operators and defects. Additionally, we derived non-simply laced gauge theories through the folding procedure discussed in section \ref{sec:outer-auto}. In this context, a natural question arises: What are the SymTFTs corresponding to these non-simply laced gauge theories? This section aims to address this question.

Our key observation is that the monodromy maps discussed around (\ref{Mon action as Aut}) extend naturally to a subgroup of the torsional cycles of the link space, which originate from the quotient of the fiber 3-sphere $\bbS^{3}_{\rm{f}}$. Consequently, the folding procedure applies directly to this subsector, retaining only the relevant subgroups of it \footnote{Except in the case of $\mk{su}(2N+1)$, which we will discuss separately.}. This, in turn, modifies the spectrum of discrete background gauge fields associated with the $p$-form symmetries, and hence their associated charged defects and topological generators. Importantly, the folding process preserves the universe decomposition of the theory according to $B\cong \Z_m$. As we will show, the induced folding on the link space reproduces precisely the expected spectrum of electric Wilson lines and 't Hooft lines (in each of the $m$ distinct universes), thereby providing strong evidence for the validity of our construction in this section.

In subsection \ref{sec:Tor-outer-folding}, we develop the argument establishing the presence of an induced monodromy action on the torsional cycles of the link space. Subsequently, in subsection \ref{sec:folding-center-torH1}, we describe how this induced folding manifests at the level of the torsional homology groups of the link space. In the last subsection, we will characterize the structure of the SymTFTs associated with the non-simply laced gauge theories and explicitly determine their corresponding coefficients.

\subsection{Torsional cycles and outer automorphisms}\label{sec:Tor-outer-folding}

In M-theory, the ’t Hooft screening argument \cite{DelZotto:2015isa,GarciaEtxebarria:2019caf,Morrison:2020ool,Albertini:2020mdx} asserts that unscreened $m$-dimensional non-dynamical defects arise from M$p$-branes wrapping non-compact $(p+1-m)$-cycles and can be characterized as:
\begin{equation}
\mathbb{D}^{(m)}\,=\, \bigcup_{p=2,5}\,\, \biggl\{\, \text{M$p$-branes on \,\,}\,\, \frac{H_{p+1-m}(\widetilde{X}_{7},L_{6},\Z)}{H_{p+1-m}(\widetilde{X}_{7},\Z)} \biggr\}~.
\end{equation}
Here, $\widetilde{X}_{7}$ denotes the resolved cone space over the link $L_{6}$. Moreover, $H_{p+1-m}(\widetilde{X}_{7},L_{6},\Z)$ is the relative homology which captures the data of cycles which are non-compact but have a boundary on $L_{6}$. This definition is equivalent to the earlier one given in (\ref{def:defect1}). This is simply because a non-compact relative $(p+1-m)$-chain can be viewed as a $(p-m)$-cycle in the link space extended along the radial direction.

In the following, we are interested in the set of line operators $\mathbb{D}^{(1)}$ characterized by the relative and absolute homology groups $H_{2}(\widetilde{X}_{7},L_{6},\Z)$ and $H_{2}(\widetilde{X}_{7},\Z)$. Equivalently, as described in (\ref{def:defect1}), these line defects are also classified by the torsional part of $H_{1}(L_{6},\Z)$. Assuming $H_1(\widetilde{X}_7,\mathbb{Z})=0$,  the consistency between these two characterizations follows from the long exact sequence in relative homology (see subsection 3.3 of \cite{Apruzzi:2021nmk} for instance): 
\begin{equation}\label{eq:exact-sequence}
H_{2}(\widetilde{X}_{7},\Z)\,\xrightarrow[]{g}\,H_{2}(\widetilde{X}_{7},L_{6},\Z)\,\xrightarrow[]{f}\, \mathrm{Tor}H_{1}(L_{6},\Z)\,\to\,0~.
\end{equation}
In particular, elements of $H_{2}(\widetilde{X}_{7},\Z)$ are the compact vanishing 2-cycles in $\widetilde{X}_{7}$.
 In contrast, $H_{2}(\widetilde{X}_{7},L_{6},\Z)$ describes non-compact 2-cycles in the cone $\widetilde{X}_{7}$. We write $\mathrm{Tor}H_{1}(L_{6},\Z)$ rather than the full $H_{1}(L_{6},\Z)$, since, as shown in (\ref{CoH of L6}), the 1-cycles of the link space $L_{6}$ are purely torsional.

Now, consider a torsional cycle $\gamma\in H_{1}(L_{6},\Z)$, of order $n\in\mathbb{N}$: $n\cdot\gamma=0$. The exactness of \eqref{eq:exact-sequence} implies the existence of a compact 2-cycle $S\in H_{2}(\widetilde{X}_{7},\Z)$, such that:
\begin{equation}
    g(S)\,=\,n\,\widetilde{\gamma}~,\quad\mathrm{with}\quad f(\widetilde{\gamma})\,=\,\gamma~,
\end{equation}
for some non-compact 2-chain $\widetilde{\gamma}\in H_{2}(\widetilde{X}_{7},L_{6},\Z)$.

This establishes a correspondence between vanishing compact 2-cycles and a subset of ${\rm Tor}H_{1}(L_{6},\Z)$, namely those torsional 1-cycles that originate from the fiber 3-sphere:
\begin{equation}
    \restr{\mathrm{Tor}H_{1}(L_{6},\Z)}{\bbS^{3}_{\rm{f}}}~.
\end{equation}
The associated line defects, denoted by $\restr{\mathbb{D}^{(1)}}{\scriptscriptstyle\bbS^{3}_{\rm{f}}}$, define a proper subset of the full electric line defect space. Furthermore, following the discussion around (\ref{tilde G electric 1-form}), we can focus on a further refined subset corresponding to Wilson lines, denoted by $\mathbb{W}^{(1)}$. These are identified with the line defects charged under the center of the gauge group obtained in the singular limit of $\widetilde{X}_{7}$, and determined by $C_{0}$; hence, the name Wilson lines. In other words, these Wilson lines satisfy the embedding:\footnote{Said differently, we focus on the Wilson lines within a given universe from the perspective of a local observer; see the discussion in section \ref{sec:local-observer} and around (\ref{tilde G electric 1-form}).}
\begin{equation}\label{eq:W1-embeeding}
    \mathbb{W}^{(1)}\subset \restr{\mathbb{D}^{(1)}}{\bbS^{3}_{\rm{f}}}\subset \mathbb{D}^{(1)}~,
\end{equation} 
and they come from wrapping M2-branes around $C_0^{\rm ab}$ 1-cycles in $L_6$. Their associated symmetry topological operators are given explicitly in (\ref{fiber 1-form discrete top operator}).

The field theory perspective associates Wilson lines $\mathbb{W}^{(1)}$ with the center $\mathcal{Z}(G)$ of the gauge group $G$ \cite{Kapustin:2005py,Gaiotto:2014kfa}, with $\mathcal{Z}(G)$ being determined by the short exact sequence:
\begin{equation}\label{eq:root-weight-center}
    0\,\to\,\Lambda_{\mathrm{root}}\,\to\,\Lambda_{\mathrm{weight}}\,\to\,\mathcal{Z}(G)\,\to\,0~.
\end{equation}
Here, $\Lambda_{\mathrm{weight}}$ denotes the weight lattice and $\Lambda_{\mathrm{root}}$ the root lattice of the Lie algebra $\mk{g}$ of the gauge group $G$. Consistency between geometric engineering and field theory requires the following isomorphisms (e.g., \cite{GarciaEtxebarria:2019caf,Morrison:2020ool}):
\begin{equation}\label{eq:top-algebra-correspondence}
\Lambda_{\rm root}\,\cong\, H_{2}(\widetilde{X}_{7},\Z) \,,\quad \Lambda_{\mathrm{weight}}\,\cong\,  H_{2}(\widetilde{X}_{7},L_{6},\Z) ~,
\end{equation}
and, 
\begin{equation}\label{eq:top-algebra-correspondence-2}
\mathcal{Z}(G) \,\cong\,C_{0}^{\rm{ab}}\subset \restr{{\rm Tor}H_{1}(L_{6},\Z)}{\bbS^{3}_{\rm{f}}}~.
\end{equation}
In particular, the final isomorphism reflects the fact that we are focusing on Wilson lines given by the embedding (\ref{eq:W1-embeeding}). See the discussion in subsection \ref{sec:co-homology} regarding the embedding of $C_{0}^{\rm{ab}}$ into $\restr{{\rm Tor}H_{1}(L_{6},\Z)}{\bbS^{3}_{\rm{f}}}$.

One might reasonably question the second isomorphism in \eqref{eq:top-algebra-correspondence}, since the weight lattice of the gauge algebra corresponds only to a particular sublattice within the relative homology group $H_{2}(\widetilde{X}_{7},L_{6},\Z)$. This arises because relative cycles can be viewed as chains extending from the link space to the zero-section, and our focus on $C_{0}^{\rm ab}$ (as appropriate for a single observer) should be reflected in this isomorphism. This concern is indeed justified. However, to avoid overloading the discussion with additional formalism and notation, we adopt the isomorphism as a schematic one.

We now argue that the outer automorphism discussed in subsection \ref{sec:outer-auto} extends to the homological cycles of the link space. To start, recall that the outer automorphism is defined through the monodromy action on the Dynkin diagrams in (\ref{eq:Mon-def-on-DD-ADE}): 
\begin{equation}\label{Mon action on DDADE}
    \mathrm{Mon}\,:\,\pi_{1}(\mathbb{S}^{3}_{\mathrm{b}}/B)\,\longrightarrow\,  \mathrm{Aut}(\mathrm{DD}_{\mathfrak{g}})\,.
\end{equation}
As we did earlier in subsection \ref{subsec: physics dictionary}, we denote the algebra automorphism associated with a loop $\gamma_{\rm b}\in \pi_1(\mathbb{S}^3_{\rm b}/B)$ by ${\rm Mon}_{\gamma_{\rm b}}$. Equivalently, the monodromy acts on the roots of the Dynkin diagrams and hence on the root lattice $\Lambda_{\mathrm{root}}$. Due to the short exact sequence in \eqref{eq:root-weight-center}, the outer automorphism action extends to the weight lattice $\Lambda_{\mathrm{weight}}$ and the center $\mathcal{Z}(G)$ via the commutative diagram:
\begin{equation}\label{eq:commut-diag-algebra}
\begin{tikzcd}
1 \arrow[r, ""] & \Lambda_{\mathrm{root}} \arrow[r, "g"] \arrow[d, "\mathrm{Mon}_{\gamma_{\rm b}}"] & \Lambda_{\mathrm{weight}} \arrow[r, "f"] \arrow[d, "\mathrm{Mon}'_{\gamma_{\rm b}}"] & \mathcal{Z}(G) \arrow[d, "\mathrm{Mon}''_{\gamma_{\rm b}}"] \arrow[r, ""] & 1
\\
1 \arrow[r, ""] &\Lambda_{\mathrm{root}} \arrow[r, "g"]  & \Lambda_{\mathrm{weight}} \arrow[r, "f"]& \mathcal{Z} (G) \arrow[r, ""] & 1 \,.
\end{tikzcd}
\end{equation}
The induced maps $\mathrm{Mon}'_{\gamma_{\rm b}}$ and $\mathrm{Mon}''_{\gamma_{\rm b}}$ are defined through the commutativity of the above diagram.

The outer automorphism action naturally extends to all components of the exact sequence in (\ref{eq:exact-sequence}). This follows from the McKay correspondence \cite{mckay}:
\begin{equation}\label{McKay Correspondence}
\mathrm{DD}_{\scriptscriptstyle ADE} \,\cong\, H_2(\widetilde{X}_{7},\mathbb{Z})~.
\end{equation}
which we reviewed earlier in subsection \ref{subsec: physics dictionary}. Recall that $H_{2}(\widetilde{X}_{7},\Z)$ is the space of vanishing 2-cycles arising from the resolution of the codimension-four singularity of the form $\R^{4}/C_0$.
 Via \eqref{McKay Correspondence}, automorphisms of the Dynkin diagram can be associated with automorphisms on $H_{2}(\widetilde{X}_{7},\Z)$:
\begin{equation}
   \mathrm{Aut}(\mathrm{DD}_{\scriptscriptstyle ADE})\, \cong\, \mathrm{Aut}(H_{2}(\widetilde{X}_{7},\Z))~.
\end{equation}

The isomorphism above induces a monodromy action on $H_2(\widetilde{X}_7,\Z)$ via the action \eqref{Mon action on DDADE}:\footnote{Here we are abusing notation by denoting this monodromy action using the same notation we used for the monodromy action on the Lie algebra. Things should be clear to the reader from context.}
\begin{equation}
     \mathrm{Mon}\,:\, \pi_{1}(\mathbb{S}^{3}_{\mathrm{b}}/B)\,\to\,  \mathrm{Aut}(H_{2}(\widetilde{X}_{7},\Z))\,.
\end{equation}
The action extends coherently to the entire exact sequence in (\ref{eq:exact-sequence})  via the commutative diagram:
\begin{equation}\label{eq:commut-diagram-geo}
\begin{tikzcd}
H_{2}(\widetilde{X}_{7},\Z) \arrow[r, "g"] \arrow[d, "\mathrm{Mon}_{\gamma_{\rm b}}"] & H_{2}(\widetilde{X}_{7},L_{6},\Z) \arrow[r, "f"] \arrow[d, "\mathrm{Mon}'_{\gamma_{\rm b}}"] & C_0^{\rm ab} \arrow[d, "\mathrm{Mon}''_{\gamma_{\rm b}}"] \arrow[r, ""]  & 0
\\
H_{2}(\widetilde{X}_{7},\Z) \arrow[r, "g"]  & H_{2}(\widetilde{X}_{7},L_{6},\Z) \arrow[r, "f"]& C_0^{\rm ab} \arrow[r, ""]  & 0 \,,
\end{tikzcd}
\end{equation}
with $\mathrm{Mon}_{\gamma_{\rm b}}\in\mathrm{Aut}(H_{2}(\widetilde{X}_{7},\Z))$ and $\mathrm{Mon}'_{\gamma_{\rm b}}$ and $\mathrm{Mon}''_{\gamma_{\rm b}}$ are induced maps on $ H_{2}(\widetilde{X}_{7},L_{6},\Z)$ and $C_0^{\rm ab}\subset\restr{\mathrm{Tor}H_{1}(L_{6},\Z)}{\bbS^{3}_{\rm{f}}}$, respectively. The similarity between (\ref{eq:commut-diag-algebra}) and (\ref{eq:commut-diagram-geo}) is due to the consistency requirement in (\ref{eq:top-algebra-correspondence}). The map $\mathrm{Mon}''_{\gamma_{\rm b}}$ specifically acts on $C_0^{\rm ab} \subset \restr{\mathrm{Tor}H_{1}(L_{6},\Z)}{\bbS^{3}_{\rm{f}}}$, meanwhile leaving the rest of Tor$H_1(L_6, \Z)$ invariant.

As established in Section \ref{sec:co-homology}, this induced automorphism affects not just $\mathrm{Tor}H_{1}(L_{6},\Z)|_{\mathbb{S}^{3}_{\mathrm{f}}}$ but also extends to other torsional cycles of $L_{6}$. In particular, the induced maps influence:
\begin{itemize}
\item \textbf{Torsional 4-cycles}: Recall that, an element of $\restr{\mathrm{Tor}H_{4}(L_{6},\Z)}{{\mathbb{S}^{3}_{\mathrm{f}}}}$ take the form $\gamma \times [\mathbb{S}^3_{\rm b}/B]$, where $\gamma$ represents a torsional 1-cycle coming from the fiber factor as above. 
Put differently, the action of $\mathrm{Mon}''_{\gamma_{\rm b}}$ must necessarily extend to torsional $C_0^{\rm ab}$ 4-cycles in $L_{6}$ to maintain consistency with the duality between torsional cycles \eqref{torsion duality}.

\item \textbf{Torsional 2-cycles:} From K\"unneth formula, the torsional part of the second homology group (and hence the third homology group) is constructed from the torsional part of the first homology ring. From the above discussion, we established that the monodromy acts on Tor$H_1(L_6,\Z)$. Therefore, we expect this action to extend to the second homology as well. See table \ref{Table:tor-cycles-after-foldings} below for instance.
\end{itemize}

\subsection{Foldings, centers, and torsional 1-cycles}\label{sec:folding-center-torH1}

Recall that folding (or twisting) a simply-laced $\mk{g}$ gauge theory (i.e. $\mk{g}$ is of $ADE$ type) by the set of outer automorphisms $\mathrm{Aut}(\mathrm{DD}_{\mk g})$ 
gives a non-simply laced gauge algebra $\widetilde{\mk{g}}$ as discussed in subsection \ref{subsec: physics dictionary} (see the discussion in and around table \ref{Table:outer-automorphisms} for instance). Accordingly, the folding alters the center $\mathcal{Z}(G_{\scriptscriptstyle ADE})$ of the gauge theory. In the context of Lie algebra, one way to find the center of the corresponding $\widetilde{\mk{g}}$ is by the following steps:
\begin{itemize}
\item First, the center of a simple Lie group can be read off from its extended (affine) Dynkin diagram--See section 2 of \cite{Larouche_2011}.
  Expanding the highest root $\alpha_{0}$ in terms of the simple roots:\footnote{See footnote around \eqref{E6 example inner auto} for the definition of the highest root.}
\begin{align}
    \alpha_{0} = \sum_{k=1 }^{{\rm rank}(\mk{g})} m_k \, \alpha_k~,
\end{align}
we refer to $m_k$ as the mark of the root $\alpha_k$. From this perspective, the center $\mathcal{Z}(G)$ is written as:
\begin{equation}\label{eq:center-mark-one-nodes}
 \mathcal{Z}(G)\,=\,   \Z_{\#(\mathrm{nodes\,with\,}m=1)}\,.
\end{equation}
The above formula can be applied to all Lie algebras apart from $\mk{d}_{2n}$. For the affine $\Hat{\mk{d}}_{2n}$, the number of nodes with $m=1$ is equal to 4; however, the center of the associated group is given as $\Z_{2}\oplus\Z_{2}$ rather than $\Z_4$.

The extended Dynkin diagrams along with their corresponding marks are shown in figure \ref{fig:dynkin+marks}.

\item Second, we extend the folding procedure to the affine $ADE$ Dynkin diagrams. See figure \ref{Figure:ADE-folding}. Upon this, the center $\mathcal{Z}(G)$ is reduced to the center $\mathcal{Z}(\widetilde{G})$ (with $\widetilde{G}$ is the Lie group of the resulting non-simply laced Lie algebra $\widetilde{\mk{g}}$) of the corresponding affine Dynkin diagrams of $\widetilde{\mk{g}}$. The new center $\mathcal{Z}(\widetilde{G})$ can be read off as explained in the first point.
\end{itemize}

The reduction of the center via an outer automorphism twist:
\begin{equation}
    \mathcal{Z}(G)\,\to\,\mathcal{Z}(\widetilde{G})~, 
\end{equation}
fundamentally alters the spectrum of admissible Wilson lines $\mathbb{W}^{(1)}$. This is because, in field theory, Wilson lines are classified by the center of the gauge group \cite{Gaiotto:2014kfa}; thus, the modified center $\mathcal{Z}(\widetilde{G})$ generically admits a different set of topological line operators.

These observations provide the physical intuition for why the automorphism $A_{\mathrm{Mon}}$ naturally extends to the (torsional) homological cycles of the link space, as encoded in the commutative diagram (\ref{eq:commut-diagram-geo}). Furthermore, the folding also affects other torsional $p$-cycles, since the outer automorphism acts non-trivially on such cycles in the link space, as discussed following equation (\ref{eq:commut-diagram-geo}).

Assuming that the topological–algebraic correspondence (\ref{eq:top-algebra-correspondence}) remains valid after folding, we propose that the action of folding by outer automorphisms on torsional 1-cycles is precisely captured by the isomorphism\footnote{Apart from the $\mk{su}(2n+1)$ cases, which will be treated separately below.}:
\begin{equation}\label{eq:center-tor-1-folded}
   \mathcal{Z}(\widetilde{G})\,\cong\, \left(\mathrm{Tor}H_{1}(L_{6},\Z)|_{\mathbb{S}^{3}_{\mathrm{f}}}\right)^{{\rm Aut}({\rm DD}_{\mk{g}})}
\end{equation}
Here,  the RHS denotes the invariant subgroup under the outer automorphism action, i.e., the torsional 1-cycles that survive the folding. Explicitly, we should take the invariant subgroup of the $C_{0}^{\rm{ab}}$. To clarify what we mean, let us take $C_{0}^{\rm{ab}}=\Z_{n}$ and $B=\Z_{m}$. The group $\mathrm{Tor}H_{1}(L_{6},\Z)|_{\mathbb{S}^{3}_{\mathrm{f}}}$ can be expressed through (\ref{eq:Znm-disjoint-unioun}) and the invariant subgroup then is given as:
\begin{equation}\label{universe decomp after fold}
         \left(\Z_{nm}\right)^{{\rm Aut}({\rm DD}_{\mk{g}})}\,=\, \bigsqcup_{b_{j}\,\in\,\Z_{m}}\,\,\left(\Z_{n}^{(b_{j})}\right)^{{\rm Aut}({\rm DD}_{\mk{g}})}\,~,
\end{equation}
where, 
\begin{equation}
 \left(\Z_{n}^{(b_{j})}\right)^{{\rm Aut}({\rm DD}_{\mk{g}})}\,:=\, \widetilde{\varphi}(b_{j})_{\rm{ab}}\,\cdot\,\left(\Z_{n}\right)^{{\rm Aut}({\rm DD}_{\mk{g}})}~.
\end{equation}
In the context of geometric engineering, this demonstrates how folding alters the spectrum of admissible Wilson lines.

\subsubsection{Explicit folding of the center: testing the proposal for \texorpdfstring{$\mathcal{Z}(\widetilde{G})$}{ZG}}

In the following, we examine our assumption \eqref{eq:center-tor-1-folded} in the cases of interest 
and demonstrate that it holds. Our strategy can be summarized as follows:
\begin{itemize}
    \item \textbf{Folding the center 1-form symmetry.} The divisors correspond to the center 1-form symmetry of gauge theory $G$ with Lie algebra $\mk{g}$ takes the following generic form \cite[section 3]{Bhardwaj:2020phs} (see also table \ref{tab: ZGamma and CSGamma}):
\begin{equation}
D_{\mathcal{Z}(G)}\,=\, \sum_{j}\,n_{j}\,S_{j}~.
\end{equation}

Since the divisors $S_{\bullet}$ are in one-to-one correspondence with the simple roots of the Dynkin diagram, the action of the outer automorphism on $S_{\bullet}$ can be determined directly. A similar discussion, excluding the step of folding, can be found in section 4 of \cite{DeMarco:2025pza}. Upon twisting by the outer automorphism, we indeed recover $D_{\mathcal{Z}(\widetilde{G})}$, as expected.

\item \textbf{Folding the torsional 1-cycles.} We apply the outer automorphism folding to the space of torsional 1-cycles $\mathrm{Tor}H_{1}(L_{6},\Z)|_{\mathbb{S}^{3}_{\mathrm{f}}}$ by retaining only the subspace invariant under the automorphism action, as expressed in equation (\ref{eq:center-tor-1-folded}). The resulting set of torsional 1-cycles must satisfy two conditions:
\begin{itemize}
    \item From the perspective of geometric engineering, it must yield the correct spectrum of Wilson lines for the gauge theory with gauge group $\widetilde{G}$.
    \item From the field-theoretic viewpoint, it must reproduce the center $\mathcal{Z}(\widetilde{G})$.
\end{itemize}
For our examples of interest, both conditions are satisfied as we show explicitly now.
\end{itemize}

The above procedure provides evidence in support of the proposal in equation (\ref{eq:center-tor-1-folded}), justifying its application in full generality. This includes, for instance, the $\mk{su}(2N+1)$ case, which we will discuss at the end. 

We now proceed with the examples, following the strategy outlined above. The final results for these examples are summarized in table 

\begin{table}[t]
    \centering
    \begin{tabular}{|c|c|c||c|c||c|c|}
    \hline
       {\rm Case} &$B$ & $C_0$&${\mk g}$ &$\widetilde{\mk{g}}$ &$ H_1(L_6,\Z)^{\scriptscriptstyle{\rm Aut}({\rm DD}_{\mk{g}})}$&$H_2(L_6,\Z)^{\scriptscriptstyle{\rm Aut}({\rm DD}_{\mk{g}})}$\\
        \hline
        \hline
         1&$\Z_{2}$& $\Z_{N}$&$\mk{su}(N)$& $\mk{sp}(N)$ & $\Z_{4}\oplus \Z_{2}$ & $\Z_{2}$ \\
         \hline
          2&$\Z_4$ & $\Z_{2N+1}$&${\mk su}(2N+1)$ & $\mk{so}(2N+1)$   &$\Z_{8} \oplus \Z_{4}$& $\Z_{4} $\\
          \hline
        3&$\Z_{2}$ & $\Z_{2N}$ & ${\mk su}(2N)$&$\mk{sp}(2N)$ & $\Z_{4} \oplus \Z_2$ & $\Z_2$\\
           \hline
         4&$\Z_3$& $2\D_{4}$ & ${\mk so}(8)$&$\mk{g}_{2}$  & $   \Z_{3}\oplus \Z_{3}$& $\Z_3$\\
         \hline
        5(a)&$\Z_2$ & $2\D_{4N}$& ${\mk so}(4N)$&$\mk{so}({4N-1})$ & $\Z_{4} \oplus \Z_2$ & $\Z_{2}$\\
         \hline
          5(b)&$\Z_2$ & $2\D_{4N+2}$ &$\mathfrak{so}(4N+2)$& $\mk{so}({4N+1})$  & $\Z_{4} \oplus \Z_2$ &$\Z_2$ \\
         \hline
         6&$\Z_2$ & $2\T$ & $\mathfrak{e}_6$&$\mk{f}_{4}$ & $ \Z_2\oplus\Z_2$ & $\Z_2$ \\
         \hline
    \end{tabular}
    \caption{For each one of the six cases in \eqref{eq:the-6-cases}, this table contains the corresponding non-simply laced Lie algebra obtained after folding. The last two columns are for the first and second homology rings associated with the resulting theory. Compare this with those of the original theory in table \ref{tab:H1 H2 for L6}. The third and fourth homology rings can be worked out using the duality \eqref{torsion duality}. For the first case, recall from the discussion around \eqref{eq:condition-on-gamma-Npr} that $N$ is even since $p=2$. Moreover, for the cases $\mathfrak{g}_2$ and $\mathfrak{f}_4$, the first component in the first homology comes from the extension of the trivial centre by $B$.}
    \label{Table:tor-cycles-after-foldings}
\end{table}

\medskip
\noindent
 \paragraph{\underline{The $\mk{su}(2N)$ case:}}

In this case, the generator of the $\mathbb{Z}_{2N}$ center 1-form symmetry is given by:
\begin{equation}\label{eq:D-su-center-1-form}
\begin{split}
    D_{\mathcal{Z}(SU(2N))}\,&=\,\sum_{j=1}^{2N-1}\,j\,S_{i}=\,S_{1}\,+\,2S_{2}\,+\,\cdots\,+\, (2N-1) S_{2N-1}\,.
\end{split}
\end{equation}
Since the center is $\Z_{2N}$, we can take the integers $n_{j} \mod 2N$. The complete set of $2N$ center symmetry elements are: 
\begin{equation}
\begin{split}
 S_{1}\,&+\,2S_{2}\,+\,\cdots\,+\, (2N-1) S_{2N-1}~,
\\
 2S_{1}\, &+\, 4  S_{2}\,+ \,\cdots\, +\, 2(2N-1)S_{2N-1}~,
\\
&\qquad\qquad \quad  \vdots
\\
  NS_{1} \, &+\, 2N  S_{2}\,+ \,\cdots\, +\, N(2N-1)S_{2N-1}~,
\\
&\qquad \qquad \quad  \vdots
\\
  2NS_{1} \, &+\, 2N  S_{2}\,+ \,\cdots\, +\, 2NS_{2N-1}~.
\end{split}
\end{equation}
The last element is the identity. We note that the middle element is represented as:
\begin{equation}\label{eq:mid-element-Z-su(2n)}
    NS_{1}\, +\, NS_{3}\,+\,\cdots\, +\,NS_{2N-3}\,+\, N S_{2N-1}~,
\end{equation}
which generates a $\mathbb{Z}_2$ subgroup of $\Z_{2N}$. Note that only divisors with odd index survive.

        Following the discussion in section \ref{sec:outer-auto}, the $\Z_{2}$ outer automorphism of the $\mk{su}(2N)$ algebra does the following exchange:

        \begin{equation}\label{fold su(2n)}
          S_{j} ~ {\longleftrightarrow} ~ S_{2N-j}~, 
        \end{equation}
      with $S_{n}$ corresponds to the $\Z_2$-invariant node of the $\mk{su}(2N)$ Dynkin diagram. 
Folding by the above action, we define the new set of divisors $\widetilde{S}_{j}$ as:\footnote{Keeping in mind that, for $N$ odd, the divisor $S_N$ does not appear in the find expression \eqref{eq:mid-element-Z-su(2n)}.}
\begin{equation}\label{redfn su2n divisors}
    \left\{ \, \widetilde{S}_{j} \, =\, \frac{1}{2}\, (S_{j}\,+\,S_{2N-j})\,\,,\,\, \widetilde{S}_{N}\, =\, S_{N}\,,\quad \mathrm{for}\,\, j=1,2,\cdots\,N-1\,\right\}.
\end{equation}
The divisors $\widetilde{S}_{j}$ and $\widetilde{S}_{N}$ are related to the Dynkin nodes of the $\mk{sp}(2N)$, which is the resulting Lie algebra after folding--See table \ref{Table:outer-automorphisms}. Applying \eqref{fold su(2n)} on the element \eqref{eq:mid-element-Z-su(2n)} generating $\Z_2\subseteq \Z_{2N}$, we find:\footnote{The over all factor of $N$ here is very essential to take care of the $1/N$ factor when we define the symmetry topological operators in terms of the exponential of these divisors.}
\begin{equation}\label{gen for sp2N}
    N\,\sum_{j=1}^N \frac{1-(-1)^j}{2}\,\widetilde{S}_j~.
\end{equation}
 This precisely matches the generator of the $\Z_2$ center structure of $\mathfrak{sp}(2N)$ \cite[section 3]{Bhardwaj:2020phs}. 

The induced action of the $\Z_{2}$ outer automorphism on the space $\Z_{2N}\subset \restr{\mathrm{Tor}H_{1}(L_{6},\Z)}{\mathbb{S}^3_{\rm f}}$ is given by the following involution:
\begin{equation}
    \gamma_{1}\, \longmapsto\, -\,\gamma_{1} \,,\qquad \gamma_{1}\,\in\, \Z_{2N}~.
\end{equation}
The subgroup of $\Z_{2N}$-valued torsional cycles invariant under this action contains the $\Z_{2}$-valued torsional cycles satisfying:
\begin{equation}
  2\gamma_{1}\,=\,0~.
\end{equation}
In other words, we may consider:
\begin{equation}
    \Z_{2N}\,\subset\, \restr{\mathrm{Tor}H_{1}(L_6, \Z)}{\mathbb{S}^3_{\rm f}}
  \,\,\xrightarrow{\,\,\Z_{2}\,\text{outer}\,\,} 
  \,\,(\Z_{2N})^{\Z_2} \,\cong\, \Z_{2}\,.
\end{equation}
Here, $\Z_{2N}$ is equivalent to $C_{0}^{\rm{ab}}$ of the current case of interest. Therefore, the full first homology of the link space coming from the fiber factor reduces from $\Z_{4n}$ to $\Z_4$, which is the extension of the invariant subsector by $B = \Z_2$. This matches exactly the center reduction under folding given above. Hence, the result is consistent with the assumption in  (\ref{eq:center-tor-1-folded}).

\medskip
\noindent
\paragraph{\underline{The $\mk{so}(8)$ case:}}

The generators of the $\Z_{2}\oplus\Z_{2}$ center 1-form symmetry are:
\begin{equation}
 D_{\mathcal{Z}(SO(8))}^{(1)}\,=\, S_{1}\,+\,S_{3}~,\qquad D_{\mathcal{Z}(SO(8))}^{(2)}\,=\, S_{4}\, +\, S_{1}\,+\, S_{3} ~.
\end{equation}
The action of the $\Z_{3}$ outer automorphism is given as:
\begin{equation}
   S_1\,\longmapsto\,S_3~, \quad S_3\,\longmapsto\,S_4~, \quad S_4\,\longmapsto\,S_1~.
  \end{equation}
Upon folding, the $\mk{so}(8)$ turns into the $\mk{g}_{2}$ Lie algebra. Only the second generator survives the folding and gives:
\begin{equation}
   3 \widetilde{S}_{1} := (S_1\,+\,S_3\,+\,S_4)~.
\end{equation}
However, one can show that it gives a trivial center as expected. 

On torsional 1-cycles, the $\Z_{3}$ outer automorphism gives:
\begin{equation}\label{eq:torh1-so8-folding-trivial}
\Z_{2}\,\oplus\,\Z_{2}\,\subset\, \restr{\mathrm{Tor}H_{1}(L_6,\Z)}{\mathbb{S}^3_{\rm f}}\,  \, 
  \,\,\xrightarrow{\,\,\Z_{3}\,\text{outer}\,\,} 
  \,\, (\Z_{2}\,\oplus\,\Z_{2})^{\Z_{3}} \,\cong\, \mathrm{trivial}\,.
\end{equation}
The result agrees with the assumption in (\ref{eq:center-tor-1-folded}).

\medskip
\noindent
\paragraph{\underline{The $\mk{so}(4N)$ case:}}

The generators of the $\Z_{2}\oplus\Z_{2}$ center $1$-form symmetry are given by:
\begin{equation}\label{so4n generators}
\begin{split}
 &D_{\mathcal{Z}(SO(4N))}^{(1)}\,=\, S_{2N-1}\,+\, \sum_{j=1}^{2N-2}\,\frac{1-(-1)^{j}}{2} \,S_{j}~,\\
 & D_{\mathcal{Z}(SO(4N))}^{(2)}\,=\, S_{2N}\, +\, \sum_{j=1}^{2N-2}\,\frac{1-(-1)^{j}}{2} \,S_{j} ~.
 \end{split}
\end{equation}
The action of the $\Z_{2}$ outer automorphism on the divisors amounts to the following exchange:
\begin{equation}
    S_{2N}\,\longleftrightarrow\, S_{2N-1}~.
\end{equation}
Meanwhile, the other divisors are left invariant. 

At the level of the generators above, this amounts to the following exchange:
\begin{equation}
    D_{\mathcal{Z}(SO(4N))}^{(1)} \, \longleftrightarrow\, D_{\mathcal{Z}(SO(4N))}^{(2)}\,,
\end{equation}
Therefore, only one $\Z_2$ generator survives the folding, namely, the diagonal one (i.e., the sum of the two generators in \eqref{so4n generators}): 
\begin{equation}\label{gen for so4N-1}
		2\,\widetilde{S}_{2N-1} := (S_{2N-1} + S_{2N})~.
\end{equation}

The induced $\Z_{2}$ outer automorphism  action on torsional 1-cycles $\Z_{2}\oplus\Z_{2}$ is given as:
\begin{equation}
    \alpha\, \longleftrightarrow\, \beta
\end{equation}
with $\alpha$ and $\beta$ are the generators of the two $\Z_{2}$ factors. The effect of folding is to choose the invariant $\Z_{2}$ subgroup of 
\begin{equation}
 \Z_{2}\oplus\Z_{2}\,\cong\,  \{\,(1,1)\,,\,(\alpha,1)\,,\,(1,\beta)\,,\,(\alpha,\beta) \,\}~,
\end{equation}
which is given by:
\begin{equation}
  \Z_{2}\,\cong\,  \{\,(1,1)\,,\,(\alpha,\beta) \,\}~.
\end{equation}
Therefore, we have:
\begin{equation}
 \Z_{2}\,\oplus\,\Z_{2}\,\subset\,\restr{\mathrm{Tor}H_{1}(L_6, \Z)}{\mathbb{S}^3_{\rm f}}
  \,\,\xrightarrow{\,\,\Z_{2}\,\text{outer}\,\,} 
  \,\, (\Z_{2}\,\oplus\,\Z_{2})^{\Z_{2}} \,\cong\, \Z_{2}~.
\end{equation}
The above result agrees with the action of the automorphism on the center of $SO(4n)$ and supports \eqref{eq:center-tor-1-folded}. 

\medskip
\noindent
 \paragraph{\underline{The $\mk{so}(4N+2)$ case:}}

For this case, the generator of the $\Z_{4}$ center 1-form symmetry is given as:
\begin{equation}
    D_{\mathcal{Z}(SO(4N+2))}\,=\, 3S_{2N+1}\,+\,S_{2N}\,+\, \sum_{j=1}^{2N-1}\,(1\,-\,(-1)^{j})\,S_{j}~.
\end{equation}
The complete set of the 4 operators is given by:
\begin{equation}
\begin{split}
     &3S_{2N+1}\,+\,S_{2N}\,+\, 2\sum_{j=1}^{2N-1}\,\frac{1}{2}\,(1\,-\,(-1)^{j})\,S_{j}~,
     \\
     & 2S_{2N+1}\,+\,2S_{2N}\,+\, 0\sum_{j=1}^{2N-1}\,\frac{1}{2}\,(1\,-\,(-1)^{j})\,S_{j}~,
     \\
      &5S_{2N+1}\,+\,3S_{2N}\,+\, 2\sum_{j=1}^{2N-1}\,\frac{1}{2}\,(1\,-\,(-1)^{j})\,S_{j}~,
      \\
      & 0\,S_{2N+1}\,+\,0\,S_{2N}\,+\,0\, \sum_{j=1}^{2N-1}\,\frac{1}{2}\,(1\,-\,(-1)^{j})\,S_{j}~,
\end{split}
\end{equation}
with the coefficient of the divisors $S_{\bullet}$ are defined mod $4$. Note the second element defines the $\Z_{2}$ subgroup of $\Z_{4}$.

The $\Z_{2}$ outer automorphism acts as:
\begin{equation}
     S_{2N+1} \,\longleftrightarrow \,  S_{2N}~. 
\end{equation}
Folding by the $\Z_{2}$ automorphism sends $\mk{so}(4N+2)$ to $\mk{so}(4N+1)$ with a new generator:
\begin{equation}\label{gen for so4N+1}
    2 \,\widetilde{S}_{2N}\,:=\,\,\left(S_{2N+1}+S_{2N}\right)\,.
\end{equation}
This generator defines a $\Z_{2}$ center 1-form symmetry for the Lie group $SO(4N+1)$ as discussed in section 3 of \cite{Bhardwaj:2020phs}.

At the level of the torsional 1-cycles, we need to check the invariant subgroup of $\Z_{4}$ under the $\Z_{2}$ outer automorphism. This indeed gives us a $\Z_{2}$ subgroup:
\begin{equation}
 \Z_{4}\,\subset \,\restr{\mathrm{Tor}H_{1} (L_6, \Z)}{\mathbb{S}^3_{\rm f}} \,\,\xrightarrow{\,\,\Z_{2}\,\text{outer}\,\,} 
  \,\, (\Z_{4})^{\Z_{2}} \,\cong\, \Z_{2}~.
\end{equation}
Again, the two sides (torsional and center) agree and are in favor of \eqref{eq:center-tor-1-folded}.

\medskip
\noindent
\paragraph{\underline{The $\mk{e}_{6}$ case:}}

The generator of the $\Z_{3}$ center 1-form symmetry in the case is given as:
\begin{equation}
    D_{\mathcal{Z}(E_6)}\,=\, S_{1}\,+\,2S_{2}\,+\,3S_{3}\,+\,4S_{4}\,+\,5S_{5}~.
\end{equation}
The coefficient in front of the divisors $S_{\bullet}$ is taken to be defined mod 3. Thus, we can rewrite the generator as:
\begin{equation}
    D_{\mathcal{Z}(E_6)}\,=\, S_{1}\,+\,2 S_{2}\,+\,0S_{3}\,+\,S_{4}\,+\,2 S_{5}~.
\end{equation}
 The action of the $\Z_{2}$ outer automorphism is given by:
\begin{equation}
    S_{1}\,\longleftrightarrow\, S_{5}\,,\qquad  S_{2}\,\longleftrightarrow\, S_{4}\,.
\end{equation}
Hence, the folding gives the following operator
\begin{equation}
     3\left( \widetilde{S}_{1}\,+\, \widetilde{S}_{2} \right)~,
\end{equation}
where, $\widetilde{S}_1:=\frac{1}{2}(S_1 + S_5)$ and $\widetilde{S}_2:= \frac{1}{2}(S_2+S_4)$. One can show that the above generator gives a trivial 1-form symmetry. This matches the expectations, as folding sends $\mk{e}_{6}$ to $\mk{f}_{4}$ which have trivial centers.

From a geometric engineering perspective, we consider the invariant subgroup under $\Z_{2}$ outer automorphism: 
\begin{equation}\label{eq:torh1-e6-folding-trivial}
 \Z_{3}\,\subset\, \restr{\mathrm{Tor}H_{1}(L_6,\Z)}{\mathbb{S}^3_{\rm f}}\,\,
  \,\,\xrightarrow{\,\,\Z_{2}\,\text{outer}\,\,} 
  \,\, (\Z_{3})^{\Z_{2}} \,\cong\, \mathrm{trivial}~.
\end{equation}
Hence, we arrive at the expected results.

\medskip
\noindent
\paragraph{\underline{The $\mk{su}(2N+1)$ case.}}

One can observe that the folding method discussed above does not directly apply to the Lie algebra $\mk{su}(2N+1)$. In particular, there is no $\Z_{2}$ subgroup in the center $\Z_{2N+1}$, and the subgroup of $\Z_{2N+1}\subset\restr{\mathrm{Tor}H_{1}(L_6, \Z)}{\mathbb{S}^3_{\rm f}}$ that is invariant under the induced $\Z_{2}$ outer automorphism is trivial.

Nevertheless, one can still define the correct 1-form symmetry of the resulting $\mk{so}(2N+1)$ gauge theory, as we now explain. First, note that there is a one-to-one correspondence between the simple roots $\alpha_{j}$ of $\mk{su}(2N+1)$ and the set of divisors $S_{j}$. The $\Z_{2}$ outer automorphism acts on these divisors by exchanging: 
\begin{equation}\label{eq:outer-action-Sj-su-odd}
    S_{j}\,\,\longleftrightarrow \,\, S_{2N+1-j}\,.
\end{equation}

Upon folding, we define the divisors $\widetilde{S}_{j}$ corresponding to the simple roots of $\mk{so}(2N+1)$ analogously to (\ref{eq:long-short-su-odd}). Explicitly, we have:
\begin{equation}
    \widetilde{S}_{j}\,=\, \frac{1}{2}\,(S_{j} + S_{2N+1-j} )\,, \quad \mathrm{with} \ \ \  j = 1,\cdots,N\,.
\end{equation}
Among these, only the $\widetilde{S}_{j}$ corresponding to the outermost node, generates the $\Z_{2}$ 1-form symmetry of the $\mk{so}(2N+1)$ gauge theory.

This can be justified by inspecting the expression for the highest root of $\mk{so}(2N+1)$:
\begin{equation}
    \widetilde{\alpha}_{0}\,:=\, \widetilde{\alpha}_{1}\,+\,2\widetilde{\alpha}_{2}\,+\,\cdots\,2\widetilde{\alpha}_{N}~. 
\end{equation}
which shows that only $\widetilde{\alpha}_{1}$ appears with unit coefficient. Hence, according to the argument given around (\ref{eq:center-mark-one-nodes}), only $\widetilde{S}_{1}$ contributes non-trivially to the 1-form symmetry group. This matches the generator given in section 3 of \cite{Bhardwaj:2020phs}.

From the link space perspective, one can similarly define the $\Z_{2}$ torsional cycles. Begin with the generating torsional 1-cycle $\gamma$ in $\Z_{2N+1}$ for the $\mk{su}(2N+1)$ theory, satisfying:
\begin{equation}
(2N+1)\,\gamma \,=\, 0 ~.
\end{equation}
Any other torsional 1-cycle $\gamma_j$ can be written as:
\begin{equation}
    \gamma_{j}\,=\, j\,\gamma\,, \quad\mathrm{with} \quad j\,=\,0,\,1,\,\cdots,\, 2N ~.
\end{equation}
The induced $\Z_{2}$ outer automorphism acts on the set $\{\gamma_{j}\}$ as
\begin{equation}
    \gamma_{j}\,\,\longleftrightarrow\,\,\gamma_{2N+1-j}\,,
\end{equation}
mirroring the action on the divisors (\ref{eq:outer-action-Sj-su-odd}). By matching $S_{j}\leftrightarrow\gamma_{j}$, one finds that the folded torsional 1-cycle representative can be taken as
\begin{equation}
\widetilde{\gamma}\, = \,\frac{1}{2}(\gamma_1 + \gamma_{2N}) \,, \qquad \text{with} \quad 2\,\widetilde{\gamma} \,=\, 0 \,,\,
\end{equation}
identifying a $\Z_{2}$-torsional 1-cycle in the link space corresponding to the 1-form symmetry of the folded $\mk{so}(2N+1)$ theory.


\subsection{New 5d SymTFTs due to foldings}\label{subsec: new 5d SymTFT}
Using the results we obtained so far in this section, let us now discuss aspects of the 5d SymTFTs associated with the 4d gauge theories after folding. Recall that, from our earlier discussion in subsection \ref{sec:reduce-action-SymTFT}, the explicit form of the full action \eqref{full S symtft} for each one of the six cases depends on the topological information that we worked out in table \ref{Table:tor-cycles-after-foldings}. So, in the following, we will make several comments on different aspects of these SymTFTs and the corresponding higher-form symmetries mimicking our discussion in the previous sections \ref{sec:SymTFT} and \ref{sec:charged top defects} above.

\medskip
\noindent
\textbf{Universe decomposition after folding.} From our discussion around \eqref{eq:center-tor-1-folded}, we observe that the universe decomposition of the 4d gauge theory is preserved upon folding. For instance, from \eqref{universe decomp after fold} we can see that we still have an extension of the resulting center $\mathcal{Z}(\widetilde{G})$ by the group $B \cong\Z_m$.

Following the discussion in subsection \ref{sec:local-observer}, this means that, after folding, we get $m$ universes where each local observer perceives a 4d $\mathcal{N}=1$ gauge theory with gauge group $\widetilde{{G}}$ and measures the Wilson lines that are classified by the center $\mathcal{Z}(\widetilde{G})$.

\begin{table}[t]
    \renewcommand{\arraystretch}{1.7}
    \centering
    \begin{equation*}
        \begin{array}{|c|c|c|}
        \hline
         \widetilde{g} & Z_{\widetilde{g}}  & -{\rm CS}_{\widetilde{g}}^{(3)}  \\
        \hline
        \hline
        \mathfrak{b}_N  \equiv \mathfrak{so}(2N+1)& \widetilde{S}_{2N} & \frac{1}{2}\\
        \hline
        \mathfrak{c}_N \equiv \mathfrak{sp}(2N) &\sum_{j=1}^N\frac{1-(-1)^j}{2}\widetilde{S}_{j} & \frac{N}{4}\\
        \hline
        \end{array}
    \end{equation*}   
    \caption{\textsc{Second column:} the central divisors associated with the non-simply laced Lie algebras obtained upon folding. \textsc{Third column:} the Chern--Simons invariants associated with these Lie algebras. These are obtained by applying the divisors in the second column to \eqref{CS-integral-intersection}.}
    \label{tab: CS invariants for folded g}
\end{table}

\medskip
\noindent
\textbf{The topological couplings.} Let us start with the topological invariants $\mathsf{K}_{a,b}$ \eqref{mathsf K} and the Chern--Simons invariants \eqref{CS invariants: def} appearing in the explicit twist and BF actions \eqref{S L6 twist SymTFT} and \eqref{S L6 BF SymTFT}. For the first one, note from its definition that it depends explicitly on the torsional degrees of the differential forms $t_{3,a}$ and $t_{4,b}$. After folding, the form of these invariants is still given by the far RHS of \eqref{mathsf K} keeping in mind that the torsional order $n_a$ depends on the final result we got in table \ref{Table:tor-cycles-after-foldings}.

As for the CS invariants after the folding, similar to what we have been doing so far, let us focus on those associated with the center of the gauge group $\mathcal{Z}(\widetilde{G})$ before the extension by $B$. Looking at the six cases in table \eqref{Table:tor-cycles-after-foldings}, we note that there are only two cases for which we need to compute these invariants: namely, ${\mk b}_N = \mathfrak{so}(2N+1)$ and $\mathfrak{c}_N = \mathfrak{sp}(2N)$. Keep in mind that, as we also found explicitly in the previous subsection, for the cases $F_4$ and $G_2$, the center is trivial, and so the associated CS invariants are trivial as well in the subsector we are interested in.

To compute the CS invariants for the non-trivial cases ${\mk b}_N$ and ${\mk c}_N$, we will follow the algorithm reviewed around \eqref{CS-integral-intersection}. We take $\widetilde{X}'_{4}$ to be the resolved auxiliary 4-manifold obtained after folding $\widetilde{X}_{4}$, where $\widetilde{X}_{4}$ is the auxiliary space arising from the resolution of the original $C_{0}$ singularity. As for the divisors $Z_a$, we take them to be the ones we obtained in \eqref{gen for sp2N} for $\mathfrak{c}_N$ case and in $\eqref{gen for so4N-1}$ and $\eqref{gen for so4N+1}$ for $\mathfrak{b}_N$ case. In table \ref{tab: CS invariants for folded g} above, we summarize the forms of these generators and the corresponding CS invariants. The corresponding Cartan matrices for these Lie algebras can be found in table 6 in \cite{SLANSKY19811}.

\medskip
\noindent
\textbf{Higher-form symmetries and universe decomposition.} Similar to our discussion in section \ref{sec:SymTFT}, the 4d gauge theories with non-simply laced gauge algebra resulting from the folding also exhibit higher-form symmetries. The list of these symmetries can be worked out in the exact same manner as we did for those of $ADE$ type. The explicit form of the discrete symmetries can be read off directly from table \ref{Table:tor-cycles-after-foldings}. As for the continuous $U(1)^{[\bullet]}$, these are preserved after the folding since they depend on the free part of the homology groups. Moreover, it is straightforward to write down the explicit form of the topological generators of these symmetries (both, discrete and continuous) by mimicking the formulas in section \ref{sec:charged top defects}.

\section{Physical implications of the 5d SymTFTs}\label{sec:phy-implications}

The purpose of this section is to investigate the physical aspects of the SymTFTs derived in section \ref{sec:SymTFT}. The central idea (as also explained in section 4 of \cite{Najjar:2024vmm}) is that SymTFTs encode not only the generalized $p$-form symmetries exhibited by the theory, but also potential TQFTs that may couple to the dynamical QFTs residing on the physical boundary. This extends the geometric engineering program by enabling the detection and characterization of TQFTs that may appear arbitrary or unexpected from a purely field-theoretic perspective.

The following construction draws on subsection 4.4 of \cite{Najjar:2024vmm} and extends it to encompass gauge theories beyond the case of $\mk{su}(N)$ theory. In particular, we extend it to the 6 different gauge theories |with $\mk{su}(n)$, $\mk{so}(2n)$, and $\mk{e}_{6}$ Lie algebras| studied in subsection \ref{sec:ADE-gauge-theories} which are associated with the non-splittable quotients given in (\ref{eq:the-6-cases}). Further, we extend our analysis to gauge theories obtained via folding constructions, discussed in subsection \ref{subsec: physics dictionary}, leading to gauge algebras of types $\mk{so}(2n+1)$, $\mk{sp}(2n)$, $\mk{f}_{4}$, and $\mk{g}_{2}$.

This section begins by reviewing the mixed ’t Hooft anomaly between the periodicity of the Yang–Mills theta angle $\theta_{\rm{YM}}$ and the electric 1-form symmetry $\Z_{n}^{[1]}$ for general gauge theories\footnote{As in the previous sections, we set $C_{0}^{\rm{ab}}=\Z_{n}$, which determines the electric 1-form symmetry $\Z_{n}^{\scriptscriptstyle[\rm{e},1]}$. This correspond to the center of the gauge group $G$, $\mathcal{Z}(G)$, with $G$ is specified by the $ADE$ finite group $C_{0}$.}, following \cite{Cordova:2019uob} and references therein. This anomaly inflow is naturally encoded within our SymTFT. Recognizing this structure allows us to identify the presence of a CW $(-1)$-form symmetry.

Following the discussion presented in section \ref{sec:local-observer}, we now focus on the SymTFT as perceived by a particular local observer—corresponding to a specific universe—along with certain contributions inherited from the global SymTFT. Explicitly, our primary objective is to investigate the physical implications of the following SymTFT action:
\begin{equation}\label{5dSymTFT terms}
    S_{\rm SymTFT}^{\rm 5d} = \int_{Y_{5}}\left[\frac{F_{4}}{2\pi}\wedge\frac{\widetilde{h}_{1}}{2\pi}+\frac{F_{1}^{\rm{b}}}{2\pi}\wedge\frac{\widetilde{h}_{4}^{\rm{f}}}{2\pi}+\frac{1}{m} B_{0}\smile\delta A_{4}+\mathrm{CS}_{C_{0}}\,\frac{F_{1}^{\rm{b}}}{2\pi}\smile B_{2}^{\rm{e}}\smile B_{2}^{\rm{e}} \right]\,.
\end{equation}
In particular, as we will see, the last term captures the anomaly inflow discussed earlier. Leading to identifying $F_{1}^{\rm{b}}$ with the carvature of $\theta_{\rm{YM}}$ in the 5d SymTFTs. This identification is key to the construction presented in this section.

Before proceeding with our analysis, we introduce a change of conventions for the gauge fields associated with the discrete sector of the SymTFT presented above.

\paragraph{Conventions for discrete gauge fields.} A discrete $\Z_{n}$ $p$-form gauge field $B_{p}$ can be modeled using a pair of $U(1)$-valued gauge fields $(\widetilde{B}_{p},\widetilde{B}_{p-1})$ subject to the constraint:
\begin{equation}\label{eq:nB=F}
  n\,\widetilde{B}_{p} \,=\, \widetilde{\rm{F}}_{p} \,. 
\end{equation}
Here, $\widetilde{F}_{p}$ denotes the field strength of $\widetilde{B}_{p-1}$, which locally takes the form $\dd\widetilde{B}_{p-1}$. Taking the periods of $\widetilde{F}_{p}$ are $2\pi \Z$-quantized, it follows from the above constraint that: 
\begin{equation}
   \int_{\Sigma_{p}}\, \widetilde{B}_{p}\,\in\, \frac{2\pi\Z}{n}~,
\end{equation}
for any $p$-cycle $\Sigma_{p}$.

In the SymTFT action, we implement this redefinition by setting
\begin{equation}
    B_{p}\, \mapsto\, \frac{n}{2\pi} \widetilde{B}_{p}\,,
\end{equation}
and subsequently omit the tilde to avoid notational clutter.

With this new convention, the SymTFT action \eqref{5dSymTFT terms} takes the following form:
\begin{equation}\label{5dSymTFT terms-to-use}
    S_{\rm SymTFT}^{\rm 5d} = \frac{1}{(2\pi)^{2}} \int_{Y_{5}}\left[ \,F_{4}\wedge\widetilde{h}_{1}+F_{1}^{\rm{b}}\wedge\widetilde{h}_{4}^{\rm{f}}\,+\,m\, B_{0}\wedge\dd A_{4}+\frac{n^{2}\mathrm{CS}_{C_{0}}}{2\pi}\,F_{1}^{\rm{b}}\wedge B_{2}^{\rm{e}}\wedge B_{2}^{\rm{e}} \right]
\end{equation}

\paragraph{SymTFTs, outer automorphism, and anomaly inflow.}

It is important to note that the above discussion extends naturally to the class of gauge theories obtained via folding, as discussed in section \ref{subsec: symtft outer}. In particular, we find that gauge theories with gauge algebra of type $\mk{b}_{n}$ and $\mk{c}_{n}$ admit, structurally, the same SymTFT presented above. The corresponding $\mathrm{CS}_{C_{0}}$ invariants for these cases can be found in Table \ref{tab: CS invariants for folded g}.

In contrast, for Lie algebras of type $\mathfrak{g}_2$ and $\mathfrak{f}_4$, the absence of torsional 1-cycles in the fiber, as shown in (\ref{eq:torh1-so8-folding-trivial}) and (\ref{eq:torh1-e6-folding-trivial}), implies the absence of both electric and magnetic 1-form symmetries. Consequently, the anomaly inflow contributions vanish in these cases, consistent with field-theoretic expectations.

We now proceed to project the above bulk terms onto the physical boundary and investigate the implications in two folds:
\begin{itemize}
    \item In section \ref{sec:mod-inst-sum},  we gauge a finite subgroup $\Z_{K}^{[2]}\subset U(1)^{[2]}$ of the universal continuous 2-form symmetry, which corresponds to the first term in the above SymTFT. That subsection demonstrates that this strategy enables us to modify the sum over instanton sectors for all gauge theories (simply laced and non-simply laced) encountered in this paper.

    \item In section \ref{subsec:4group}, we aim to further gauge the electric 1-form symmetry $\Z_{n}^{[1]}$. However, the anomaly-inflow encoded in the last term of the above SymTFT induces a fractional instanton number, which obstructs the consistent gauging of $\Z_{n}^{[1]}$ unless one simultaneously gauges a finite 3-form symmetry $\Z_{m}^{[3]}$. This ultimately leads to a nontrivial 4-group structure as observed in \cite{Tanizaki:2019rbk,Najjar:2024vmm,Najjar:2025htp}. In this section, we extend the analysis of the 4-group structure to a broader class of gauge theories whose gauge algebras are of types $\mk{a}_{n}$, $\mk{b}_{n}$, $\mk{c}_{n}$, $\mk{d}_{n}$, and $\mk{e}_{6}$.
\end{itemize}

\subsection{Yang--Mills theta angle and a mixed 't Hooft anomaly} \label{subsec: mixed t hooft}
In this subsection, we explore the first physical aspect of the 4d $\mathcal{N}=1$ gauge theories that we obtain from geometric engineering in M-theory. This aspect concerns the interaction between the periodicity of the theta angle and the electric 1-form symmetry, which comes in the form of a mixed 't Hooft anomaly.

\subsubsection{$\theta_{\rm YM}$ and instanton counting} 
Let us first look at the Chern--Weil (CW) $U(1)^{[-1]}$ $(-1)$-form symmetry. As we discussed around \eqref{Sym-op-(-1)-form-U(1)} (see also \cite{Najjar:2024vmm} for details), this is the $U(1)^{[-1,\rm f]}$ that we get from wrapping the $P_7$ fluxbrane around the class $[\mathbb{S}^3_{\rm b}/B]  = \mathsf{PD}(\vol_3^{\rm f})$. The 0-form background gauge field for this symmetry is none other than the Yang--Mills theta angle $\theta_{\rm YM}$, which, at the level of the action of the 4d theory, appears as the following topological term:
\begin{equation}\label{S theta YM}
 S_{\theta_{\mathrm{YM}}} \,=\,   \,\frac{\theta_{\mathrm{YM}}}{8\pi^{2}}\,\int_{M_{4}} \, \tr(\mathsf{F}\wedge \mathsf{F})~.
\end{equation}
Here, $\mathsf{F}$ is the $G$ field strength and $M_4$ is the 4-manifold where the 4d gauge theory lives. We will assume here that it is closed. Therefore, 
\begin{equation}\label{instanton G}
       k\,\equiv \, \frac{1}{8\pi^{2}}\,\int_{M_{4}} \, \tr(\mathsf{F}\wedge \mathsf{F}) \,\in\,\Z~,
\end{equation}
is the instanton number, which is classified by $\pi_3(G)$. Since we are taking the gauge group $G$ to be the simply-connected cover for the gauge algebra $\mathfrak{g}$, then we have $\pi_3(G)\cong \Z$.

\medskip
\noindent
\textbf{Periodicity of $\theta_{\rm YM}$.}
The partition function of the full 4d theory decomposes into contributions from the different instanton sectors:\label{Z G}
\begin{equation}
    Z^G[\theta_{\rm YM}] \,=\, \sum_{k\in\Z} e^{i\,k\,\theta_{\rm YM}}\,Z_k~.
\end{equation}
From this explicit expression, we can see that, for instance, shifting the theta angle by $2\pi$ renders the partition function of the theory invariant. This is the $2\pi$ periodicity property of $\theta_{\rm YM}$. Said differently, this is the gauge transformation of $U(1)^{[-1,\rm f]}$ CW symmetry.

\subsubsection{A mixed 't Hooft anomaly}
Let us look at a local observer in one of the universes. This observer detects the Wilson lines of the $G$ gauge theory charged under the center 1-form symmetry $\mathcal{Z}(G)^{[1]}$. What we would like to do in the rest of this subsection is to explore the connection between this electric 1-form symmetry and the periodicity of the theta angle (and hence, on the instanton number counting) that we reviewed above.

We will discuss this connection for all possible cases for the gauge group $G$ that we are interested in in this work. But let us start with summarizing the main observation. Take $B_2^{\rm e}\in H^2(M_4,\mathcal{Z}(G))$ to be a 2-form background gauge field for the 1-form symmetry $\mathcal{Z}(G)^{[1]}$ with a Stiefel--Whitney (SW) class $w_2\sim B_2^{\rm e}$. Turning on this field amounts to twisting the principal $G$-bundle to a $G/\mathcal{Z}(G)$-bundle. As we will review in examples below, this results in the following anomaly in the periodicity of $\theta_{\rm YM}$ \cite{Kapustin:2014gua,Gaiotto:2014kfa,Cordova:2019uob}:
\begin{equation}\label{eq:Zdifferby2pi-Anomaly}
    \frac{Z^{G}[\theta_{\text{YM}}+2\pi,B_{2}^{\mathrm{e}}]}{Z^{G}[\theta_{\text{YM}},B_{2}^{\mathrm{e}}]}   \, = \, \exp(2\pi i \,{\Phi_{G}}\,\int_{M_4}\frac{\mathcal{P}(w_2)}{2})~.
\end{equation}
Here, $\Phi_G$ is the anomaly factor and $\mathcal{P}$ is the Pontryagin square (see appendix in \cite{Kapustin:2013qsa} and appendix C in \cite{Closset:2024sle}):
\begin{equation}\label{pontryagin}
    \mathcal{P} \ : \ H^{2}(M_{4}, \Z_N) \, \longrightarrow \, H^{4}(M_{4}, \widehat{\mathcal{A}}(\Z_N))~,    
\end{equation}  
such that:
\begin{equation}
    \mathcal{P}(B)\,=\,
    \begin{cases}
         B\,\cup\,B\,,\qquad &N = 2n+1~,
        \\
        B\,\cup\, B\quad \mathrm{mod}\  n \,,\qquad  &N = 2n~.
    \end{cases}
\end{equation}
In \eqref{pontryagin}, we introduced the notation $\widehat{\mathcal{A}}(\Z_N)$ which is defined as:
\begin{equation}
        \widehat{\mathcal{A}}(\Z_N) \, = \, \begin{cases}
                    \Z_N~,\qquad &N\in 2\Z +1~,\\
                    \Z_{2N}~, \qquad &N\in2\Z~. 
        \end{cases}
\end{equation}  

The anomaly (\ref{eq:Zdifferby2pi-Anomaly}) is a $U(1)^{[-1,\rm f]}-\mathcal{Z}(G)^{[1]}$ mixed 't Hooft anomaly which can be described by having the following 5d anomaly inflow term \cite{Cordova:2019uob}:
\begin{equation}\label{S inflow}
 S_{\mathrm{inflow}} \,=\,  \Phi_{G}\,\int_{Y_{5}}\, \frac{\dd\theta_{\mathrm{YM}}}{2\pi}\,\cup\,\mathcal{P}(B_{2}^{\mathrm{e}})~.
\end{equation}
The $\theta_{\mathrm{YM}}$ is promoted to a compact scalar with background field strength $F^{\rm b}_{1}\equiv \dd\theta_{\mathrm{YM}}$ \cite{Najjar:2024vmm,Cordova:2019uob,Heidenreich:2020pkc}. Upon the identification:
\begin{equation}\label{phi G CS C0}
        \Phi_G  \equiv \, -\, {\rm CS}_{C_0}~,
\end{equation}
the inflow term  \eqref{S inflow} is none other than the last term appearing in \eqref{5dSymTFT terms} of the 5d SymTFT that we obtained from M-theory reduction. 

\paragraph{Towards the sandwich construction.} Before we go into more details of these anomaly factors, let us make the following comment, which will be essential for us in the following subsections when discussing the sandwich construction of the 5d SymTFT. Interpreting $F_{1}^{\rm{b}}$ as the curvature associated with the $\theta_{\rm{YM}}$-angle allows us to establish the identification:
\begin{equation}\label{eq:h4-identified-with-TrFF}
    \frac{\widetilde{h}_4^{\mathrm{f}}}{2\pi}  \quad  \xlongleftrightarrow{\ \rm{identify}\ }\quad -\frac{1}{8\pi^{2}}\,\tr\,\mathsf{F}\,\wedge\, \mathsf{F}~.
\end{equation}
The explicit meaning of this identification will be clear later in the next subsection. This identification is supported by two key properties:
\begin{itemize}
    \item[--] \textbf{Flatness:} The flatness condition $\dd\widetilde{h}_4^{\rm{f}}=0$ mirrors the Bianchi identity $\dd\tr\mathsf{F}\,\wedge\, \mathsf{F}=0$ for the second Chern class.
    
    \item[--] \textbf{Quantization:} The integrals of both sides of (\ref{eq:h4-identified-with-TrFF}) over spacetime lie in $\Z$, reflecting topological charge quantization.
\end{itemize}

.  

We will start our discussion below by reviewing the case $G=SU(N)$. This will be the basic case from which we will extract $\Phi_G$ for the other groups $G$. We will do this by using the trick of embedding a product of $SU(N)$ groups in $G$ following \cite{Witten:2000nv, Cordova:2019uob}. From the identification \eqref{phi G CS C0} above, one can read off the form of $\Phi_G$ for the cases with $G$ is of $ADE$ type from table \ref{tab: ZGamma and CSGamma}. Therefore, the calculations we will perform now can be viewed as a test of this identification. But our main purpose will be to discuss the embedding trick here, which we will use in subsection \ref{subsec:4group} below.

\medskip
\noindent
\textbf{Mixed 't Hooft anomaly in $SU(N)$ gauge theory.} Take $G = SU(N)$. In this case, the gauge group of the 4d $\mathcal{N}=1$ SYM theory we get from the first three cases of \eqref{eq:the-6-cases}. The electric 1-form symmetry in this case is $\Z_N^{[1]}$. Let us turn on a 2-form background gauge field $B_2^{\rm e}$. This amounts to twisting the $SU(N)$ bundle to an $PSU(N)$ bundle with SW class given by $w_2 = \frac{N}{2\pi}B_2^{\rm e}$.

As we reviewed at the beginning of this section, one can turn $B_2^{{\rm e}}$ into a $U(1)$ gauge field by introducing the pair $(B_2^{\rm e},B_1^{\rm e})$ such that $N B_2^{\rm e} = \dd B_1^{\rm e}$.\footnote{Here we are abusing notation by denoting both, the discrete and continuous 2-forms, by $B_2^{\rm e}$.} Following \cite{Gaiotto:2014kfa, Kapustin:2014gua} (see also section 3 of \cite{Brennan:2023mmt} for a more pedagogical introduction), we will introduce this background gauge field into the action of the 4d SYM theory by replacing $\mathsf{F}$ by $\widetilde{\mathsf{F}}$:
\begin{equation}\label{UN field strength}
        \widetilde{\mathsf{F}} := \mathsf{F} + B_2^{\rm e}~.
\end{equation}
The $U(N)$ field strength $\widetilde{\mathsf{F}}$ is invariant under the $\Z_N^{[1]}$ gauge transformation:
\begin{equation}\label{1-form sym transform}
\begin{split}
        &B_2^{\rm e} \,\longmapsto\, B_2^{\rm e} \,+ \, {\rm d}\Lambda_1~,\qquad\widetilde{\mathsf{F}} \,\longmapsto\,\widetilde{\mathsf{F}} \,+\,{\rm d}\Lambda_1~.
\end{split}
\end{equation}
We refer to $\Lambda_{1}$ as the electric gauge parameter. 

Focusing on topological term \eqref{S theta YM}, in terms of the $U(N)$ field strength, it is of the following form:
\begin{equation}
       S^{SU(N)}_{\theta_{\rm YM}}= \frac{\theta_{\rm YM}}{8\pi^2}\int_{M_4} \tr\,(\widetilde{\mathsf{F}}-B_2^{\rm e})\wedge(\widetilde{\mathsf{F}}-B_2^{\rm e}) = \frac{\theta_{\rm YM}}{8\pi^2} \int_{M_4}\left(\tr\widetilde{\mathsf{F}}\,\wedge\,\widetilde{\mathsf{F}}\, - \,N\,B_2^{\rm e}\,\wedge\,B_{2}^{\rm e}\right)~.
\end{equation}
Shifting $\theta_{\rm YM}$ by $2\pi$, the first term remains invariant. But the second term gives rise to a new contribution proportional to the integral of $B_2^{\rm e}\wedge B_2^{\rm e}$. Since the background 2-form is identified with the SW class $w_2$, then we have:
\begin{equation}\label{quantisation condition}
       \frac{1}{8\pi^2} \int_{M_4} B_2^{\rm e}\,\wedge\,B_2^{\rm e} 
       \,\in\,\frac{1}{N^2}\,\Z~.
\end{equation}

Putting all these observations together, we find that the mixed 't Hooft anomaly is of the following form:\footnote{In deriving this anomaly form, one needs to keep in mind that the instanton density of $U(N)$ gauge theory is given by:
\protect\begin{equation*}
    \frac{1}{8\pi^2}\,\left(\tr\,\widetilde{\mathsf{F}}\,\wedge\,\widetilde{\mathsf{F}}\,-\,\,\tr\,\widetilde{\mathsf{F}}\wedge\,\tr\,\widetilde{\mathsf{F}}\right)
\end{equation*}
rather than from $\frac{1}{8\pi^2}\,\tr\,\widetilde{\mathsf{F}}\,\wedge\,\widetilde{\mathsf{F}}$ as for the $SU(N)$ case \protect\eqref{instanton G}. The extra contribution comes from the added $U(1)$ factor.}

\begin{equation}\label{SUN anomlay}
        \frac{Z^{SU(N)}[\theta_{\rm YM}+2\pi,\,B_2^{\rm e}]}{Z^{SU(N)}[\theta_{\rm YM},\,B_2^{\rm e}]}\,=\, \exp(2\pi i\,
        \frac{ N-1}{2N} \int_{M_4}\mathcal{P}\left(\frac{N}{2\pi}B_2^{\rm e}\right)) ~.
\end{equation}
Comparing this with \eqref{eq:Zdifferby2pi-Anomaly}, we find that:
\begin{equation}
        \Phi_{SU(N)} = \frac{N-1}{2N}~,
\end{equation}
which matches with the corresponding CS$_{\Z_N}$ in table \ref{tab: ZGamma and CSGamma}.

\begin{table}[t]
\centering
\renewcommand{\arraystretch}{1.5}
\begin{tabular}{|c|c|c|c|c|}
\hline
{$G$} & $H$& $G/\mathcal{Z}(G)$-bundle with SW class $B_2^{\rm e}$
\\
\hline
\hline
{$SU(N)$}& $SU(N)$ & $(PSU(N),\,B_2^{\rm e})$ 
\\
\hline
{$Sp(2N)$} & $SU(2)^N$ & $(PSU(2), B_2^{\rm e})^{\otimes N}$
\\
\hline
{$E_6$}& $SU(3)^3$& $SU(3)\otimes (SU(3), B_2^{\rm e})\otimes(SU(3), B_2^{\rm e}) $
\\
\hline
{$E_7$}   &$SU(2)\times SU(4)^2$& $SU(4)\otimes (PSU(2),B_{2}^{\rm e})\otimes (PSU(4),2B_2^{\rm e})$
\\
\hline
$Spin(2N+1)$ & $ Spin(2N-3)\times SU(2)^2$& $Spin(2N-3)\otimes (PSU(2),B_2^{\rm e})^{\otimes 2}$
\\
\hline
$Spin(4N+2)$ &$Spin(6)\times Spin(4)^{N-1}$& $SU(2)^{N-1}\otimes (PSU(2),B_2^{\rm e})^{N-1} \otimes (PSU(4),B_2^{\rm e})$ 
\\
\hline
$Spin(4N)$  & $SU(2)^{2N}$& $(PSU(2),\,B_{2}^{{\rm e},\,l})^{N}\otimes(PSU(2),\,B_{2}^{{\rm e},\,r})^{N}$ 
\\
\hline
\end{tabular}
\caption{For each gauge group $G$, in the second column, we consider a particular $H$ embedded in $G$. In the third column, we realize the twisted $G/\mathcal{Z}(G)$ bundle with a SW class $B_2^{\rm e}$ as a tensoring of twisted bundles coming from the components of $H$. For the $Spin(4N+2)$ case note that $Spin(6)\cong SU(4)$ and $SU(2)^2\subset Spin(4)$. Moreover, for the last case, recall that $\mathcal{Z}(G) = \Z_2^{\ell}\times \Z_2^{r}$.
}
\label{Table:1-form-alpha-G}
\end{table}

\medskip
\noindent
\textbf{Generalization to other gauge group.} Let us now make some comments about the generalization of the discussion above for gauge groups $G$ other than $SU(N)$. Following \cite{Witten:2000nv, Cordova:2019uob}, the $G/\mathcal{Z}(G)$ bundle can be realized as a tensoring of several $PSU(N)$ bundles that are obtained by considering a particular embedding of the original $SU(N)$'s inside $G$. Denoting this subgroup by $H\subseteq G$, In table \ref{Table:1-form-alpha-G}, we list the choice of each $H$ for each $G$ and the SW class for each $SU(N)$ component following \cite{Cordova:2019uob}. The constraint on the embedded subgroup $H$ is that ${\rm rank}\,H = {\rm rank}\,G$. 

To see how this one can use this embedding to work out the anomaly factor $\Phi_G$, let us consider the example of $G = E_6$ in detail. As shown in table \ref{Table:1-form-alpha-G} above, we consider the embedding of $H =SU(3)_1\times SU(3)_2\times SU(3)_3\subset E_6$. Let us denote the field strength associated with the $SU(3)_i$-bundle by $\mathsf{F}_i$, for $i=1, 2, 3$.  

As prescribed in the last column of table \ref{Table:1-form-alpha-G}, the $E_6/\Z_3$-bundle with a background 2-form $B_2^{\rm e}$ is equivalent to the tensor product:
\begin{equation}
        E_6/\Z_3 ~{\rm with }~B_2^{\rm e} \,=\, SU(3)_1~\otimes~ (PSU(3)_2~{\rm with}~ B_2^{\rm e})~\otimes~(PSU(3)_3~{\rm with}~ B_2^{\rm e})~.
\end{equation}
Similar to what we did earlier for the $SU(N)$ case in \eqref{UN field strength}, to include these background gauge fields into the action of the 4d $E_6$ SYM theory in a gauge invariant way, we consider the following field strengths:
\begin{equation}\label{tilde F s for E6}
\begin{split}
        &\widetilde{\mathsf{F}}_1 \,:=\,{\mathsf{F}}_1~,\\
        &\widetilde{\mathsf{F}}_2 \,:=\,{\mathsf{F}}_2 + B_2^{\rm e}~,\\
        &\widetilde{\mathsf{F}}_3 \,:=\,{\mathsf{F}}_3 + B_2^{\rm e}~.
\end{split}
\end{equation}
The $Z_3^{[1]}$ gauge transformations are the same as in \eqref{1-form sym transform}, namely:
\begin{equation}
B_2^{\rm e} \,\longmapsto\, B_2^{\rm e} \,+ \, {\rm d}\Lambda_1~,\qquad\widetilde{\mathsf{F}}_{2} \,\longmapsto\,\widetilde{\mathsf{F}}_{2} \,+\,{\rm d}\Lambda_1~,\qquad\widetilde{\mathsf{F}}_{3} \,\longmapsto\,\widetilde{\mathsf{F}}_{3} \,+\,{\rm d}\Lambda_1~,
\end{equation}
and $\widetilde{\mathsf{F}}_1$ remains invariant.

Now, let us plug the field strengths \eqref{tilde F s for E6} in the topological action \eqref{S theta YM} for the $E_6$ theory. We get:
\begin{equation}
        S_{\theta_{\rm YM}}^{E_6} = \frac{\theta_{\rm YM}}{8\pi^2} \int_{M_4}\left[\tr\,\widetilde{\mathsf{F}}_1\,\wedge\,\widetilde{\mathsf{F}}_1+\sum_{i=2,3}\tr\,(\widetilde{\mathsf{F}}_i-B_2^{\rm e})\wedge(\widetilde{\mathsf{F}}_i-B_2^{\rm e})\right]~.
\end{equation}
Following the footnote around \eqref{SUN anomlay}, we rewrite the last two terms as:
\begin{equation}
\begin{split}
        &\tr\,(\widetilde{\mathsf{F}}_i-B_2^{\rm e})\wedge(\widetilde{\mathsf{F}}_i-B_2^{\rm e}) = \tr\, \widetilde{\mathsf{F}}_i\,\wedge\,\widetilde{\mathsf{F}}_i\,-\,\tr\, \widetilde{\mathsf{F}}_i\,\wedge\,\tr\, \widetilde{\mathsf{F}}_i\,+6\,B_2^{\rm e}\,\wedge\,B_2^{\rm e}~,\\
\end{split}
\end{equation}
for $i=2,3$.

Putting these observations together, one can see that the anomaly in the periodicity of the YM theta angle will come from the term:
\begin{equation}
            \frac{12\,\theta_{\rm YM}}{8\pi^2}\int_{M_4}B_2^{\rm e}\,\wedge\,B_2^{\rm e}~.
\end{equation}
Taking $\theta_{\rm YM} \mapsto \theta_{\rm YM} + 2\pi$, we find that the partition function of the $E_6$ theory transforms as:
\begin{equation}
        \frac{Z^{E_6}[\theta_{\rm YM}+2\pi,\, B_2^{\rm e}]}{Z^{E_6}[\theta_{\rm YM},\, B_2^{\rm e}]} \,= \,\exp(2\pi i \,\frac{2}{3}\int_{M_4} \mathcal{P}\left(\frac{3}{2\pi}B_2^{\rm e}\right))~.
\end{equation}
Here we used the quantisation condition \eqref{quantisation condition}. So, we find that $\Phi_{E_6} = \frac{2}{3}$ which indeed matches the Chern--Simons invariant $-$CS$_{2\T}$ in table \ref{tab: ZGamma and CSGamma}. One can follow similar steps for the other gauge groups using table \ref{Table:1-form-alpha-G}. For the anomaly factor, one indeed gets the corresponding CS invariants in table \ref{tab: ZGamma and CSGamma} for the ADE cases and table \ref{tab: CS invariants for folded g} for $Sp(2N)$ and $Spin(2N+1)$.

\subsection{Modified instanton sum}\label{sec:mod-inst-sum}

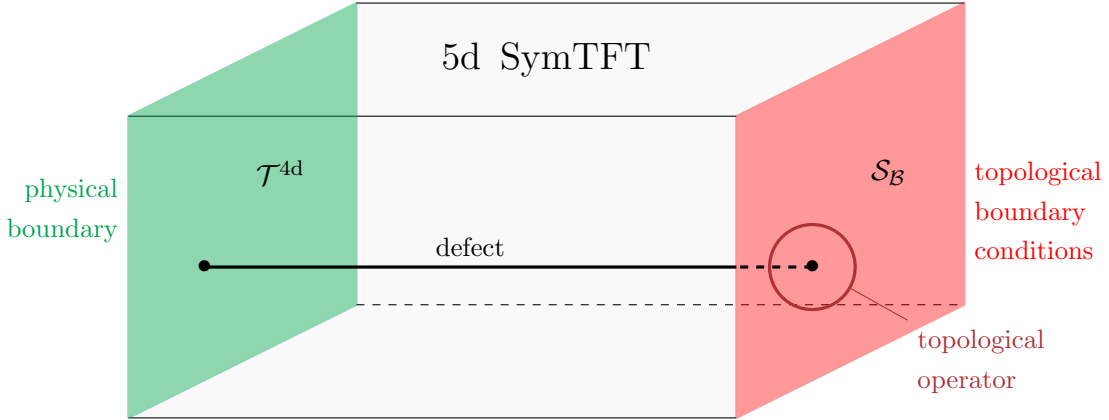
\begin{figure}[t]
\centering
\begin{tikzpicture}
    \draw[gray,thin,fill,opacity=0.05] (0,0) -- (3,1.5) -- (3,5.5) -- (11,5.5) -- (8,4) -- (8,0) -- (0,0);
    \draw[black,thin] (0,0) -- (8,0);
    \draw[black,thin,dashed] (3,1.5) -- (11,1.5);
    \draw[black,thin] (0,4) -- (8,4);
    \draw[black,thin] (3,5.5) -- (11,5.5);
    \draw[Green,thick,fill,opacity=0.45] (0,0) -- (3,1.5) -- (3,5.5) -- (0,4) -- (0,0); 
    \draw[red,thick,fill,opacity=0.4] (8,0) -- (11,1.5) -- (11,5.5) -- (8,4) -- (8,0);

    \node at (5.5,4.75) {\Large 5d \,SymTFT};
    \node at (2,3.25) {$\mathcal{T}^{\rm 4d}$};
    \node at (10,3.25) {$\mathcal{S}_{\mathcal{B}}$};
    \node[Green,anchor=east,align=right] at (0,2.75) {\small physical\\ \small boundary};
    \node[red,anchor=west,align=left] at (11,2.75) {\small topological\\ \small boundary\\ \small  conditions};

    \path[black,very thick] (1,2) edge node[anchor=south] {\small defect} (8,2);
    \draw[black,very thick,dashed] (8.05,2) -- (9,2);

    \node[] at (1,2) { $\bullet$ };
     \node[] at (9,2) { $\bullet$ };
    \node[Maroon,draw,circle,very thick] (to) at (9,2) { \hspace{24pt} };    
    \node[Maroon,anchor=west,align=left] (tt) at (10.25,.75) {\small topological\\ \small operator};
    \draw[Maroon,thin] (to) -- (tt);
\end{tikzpicture}
\caption{Visual representation of the 5d SymTFT. The physical theory is located at the \textcolor{Green}{green} boundary. The symmetry boundary is shown in \textcolor{red}{red}. The black line, stretched in the radial direction, corresponds to a defect in the physical theory. A topological symmetry operator acting on the defect is drawn in \textcolor{Maroon}{maroon}.}
\label{fig:symth}
\end{figure}


As a second physical implication for the 4d theories we explored in this work, in this subsection, we show how the range of instanton numbers gets modified in these theories upon gauging a discrete 2-form symmetry building on and generalizing the earlier works \cite{Seiberg:2010qd, Tanizaki:2019rbk, Najjar:2024vmm, Najjar:2025htp}. 

First, we start with reviewing the sandwich construction of the 5d SymTFT and how to obtain the 4d gauge theory action from it. Then, in this second part of this subsection, we go through the details of deriving a general formula for the modified instanton sum that applies to the six cases we study in this work.

\subsubsection{Sandwich construction of SymTFT: lightning review}

Let us start this subsection by giving a brief review of some essential tools of the sandwich construction of the 5d SymTFT. For more details on these constructions, see appendix B of \cite{Najjar:2024vmm} for a more recent review and references therein. 

From the perspective of our work here, the main idea is that the 5d SymTFT \eqref{5dSymTFT terms} realizes all possible (generalized) global symmetries that the geometrically engineered 4d $\mathcal{N}=1$ gauge theory can exhibit. But, these symmetries cannot all be realized simultaneously in the 4d theory--In particular, one cannot have the $p$-form symmetry $\Z_n^{[p]}$ and its Pontryagin dual  $(2-p)$-form symmetry  $\widehat{\Z_n}^{[p]}$ at the same time. Rather, one symmetry is obtained from gauging the other.

Compactifying the 5d theory on an interval $I := [0,1]$, besides the 5d SymTFT, the sandwich construction also consists of the two boundary theories at the end points of $I$. On one side, one has \textit{the physical boundary} $\mathcal{T}^{\rm 4d}$, which is none other than the actual physical 4d $\mathcal{N}=1$ gauge theory. Meanwhile, on the other side, we have \textit{the topological boundary} that we denote by $\mathcal{S}_{\mathcal{B}}$. This side consists of a 4d TQFT (e.g., BF terms) containing information about the symmetries of the 4d physical theory, and it says nothing about the actual dynamics. The full construction is summarized in figure \ref{fig:symth}. 

For any background gauge field appearing in the 5d SymTFT, there are two possible topological boundary conditions for it to choose from. Our choice is correlated to the type of generalized symmetry that we would like the 4d gauge theory to have. To make this point more clear, let us first recall the definition of the projection map $\widetilde{\delta}$.

\medskip
\noindent
\textbf{The projection map $\widetilde{\delta}$.} Depending on the topological boundary conditions chosen, this map projects the background gauge fields of the 5d theory into 4d gauge fields that enter in the definition of $\mathcal{T}^{\rm 4d}$.

To see how this works, let us consider a $p$-form symmetry $\Z_n^{[p]}$ whose background gauge field is $A_{p+1}$.\footnote{Similar discussion can be done for continuous symmetries. For more details, see appendix B of \cite{Najjar:2024vmm}.} As pointed out above, there are two possible choices for the topological boundary conditions that one can impose on $A_{p+1}$, namely, Neumann or Dirichlet boundary conditions. Schematically, the action of $\widetilde{\delta}$ on $A_{p+1}$ is given by:
\begin{equation}
        \widetilde{\delta}(A_{p+1}) = \begin{cases}
                    a_{p+1}|_{M_4}~,\qquad &{\rm Neumann}~,\\
                    A_{p+1}|_{M_4}~, \qquad &{\rm Dirichlet}~.
        \end{cases}
\end{equation}
That is:
\begin{itemize}
    \item The Neumann boundary condition gauges the background field $A_{p+1}$ (hence what we mean by $a_{p+1}$) and so the associated symmetry $\Z_n^{[p]}$ is gauged in the 4d physical theory. Therefore, the 4d gauge theory exhibits the corresponding Pontryagin dual $(d-p-2)$-form symmetry $\Z_n^{[d-p-2]} = \widehat{\Z_n}^{[p]}$.
    \item The Dirichlet boundary condition keeps the gauge field as a background field in the 4d theory. Said differently, the 4d physical theory will have a global $p$-form $\Z_n^{[p]}$.
\end{itemize}
Note from these two cases that choosing a Dirichlet boundary condition for $A_{p+1}$ is equivalent to choosing a Neumann boundary condition for the background gauge field $B_{d-p-1}$ of the dual symmetry $\Z_{n}^{[d-p-2]}$, and vice versa.

\subsubsection{Modifying the instanton number}
Based on our discussion above concerning the sandwich construction, let us now focus on the Chern--Weil $U(1)^{[-1, {\rm f}]}$ $(-1)$-form symmetry. As we discussed earlier, this is the symmetry whose $0$-form background gauge field is the theta angle $\theta_{\rm YM}$ of the 4d gauge theory. To obtain this in the 4d theory, one has to gauge the dual magnetic $U(1)^{[2,{\rm f}]}$ 2-form symmetry by choosing its topological boundary condition to be Neumann.  What we would like to do now is to study how the instanton number counting can be modified in the 4d gauge theory. Following \cite{Najjar:2024vmm}, we will do this by gauging a finite subgroup $\Z_K$ of the universal 2-form symmetry $U(1)^{[2]}$ by projecting it to the 4d physical theory with a Neumann topological boundary condition.

To this end, let us focus on the following two terms of the 5d SymTFT coming from the first two terms of \eqref{5dSymTFT terms-to-use}: 
\begin{equation}\label{terms for  modified instantons}
        S_{\rm SymTFT} \supset -\,\frac{1}{(2\pi)^{2}}\,\int_{Y_5}\,\left[\, F_1^{\rm b}\wedge \widetilde{h}_4^{\rm f} \,+\,F_4\wedge \widetilde{h}_1\,\right]~.
\end{equation}
Recall that the $F_4$ here is the field strength of the background gauge field of universal $U(1)^{[2]}$ and $h_1$ is that of the dual $(-1)$-form symmetry $U(1)^{[-1]}$. Following \cite{Najjar:2024vmm}, to gauge a finite $\Z_K$ subgroup of the 2-form symmetry, we take the following projection map actions on $F_4$ and $h_1$:
\begin{equation}\label{delta on -1 ZK}
        \widetilde{\delta}(F_4) = \dd c_3~, \qquad \widetilde{\delta}(\widetilde{h}_1) = \theta^{(2)} + K\chi^{(2)}~,  
\end{equation}
where here $c_3$ is the $2\pi$-periodic 3-form gauged field of $\Z_K^{[2]}$,\footnote{Keep in mind that, due to our conventions for the $2\pi$ factors appearing in the continuous symmetry case, the field $c_3$ has quantisation condition on its periods $\int_{M_4}\dd c_3\in 2\pi\Z$ rather than being integer quantized.} $\theta^{(2)}$ is the $0$-form background gauge field for the dual $(-1)$-form $\Z_K$ symmetry. Moreover, we shifted $\theta^{(2)}$ by the Lagrange multiplier $\chi^{(2)}$ to impose the constraint:
\begin{equation}
K\dd c_3 = 0~.
\end{equation}
This is none other than the statement that we are gauging the subgroup $\Z_K^{[2]}\subset U(1)^{[2]}$ rather than the full continuous symmetry.

To write down the contribution of the terms \eqref{terms for  modified instantons} to the full 4d topological action, let us also consider the following projections of the remaining two fields:
\begin{equation}\label{delta on -1 CW}
    \widetilde{\delta}(F_1^{\rm b})\, = \,\theta_{\rm YM} + \chi^{(1)}~, \qquad \widetilde{\delta}\left( \frac{\widetilde{h}_{4}^{\rm{f}}}{2\pi} \right) \,= \,-\frac{1}{8\pi^{2}}\tr \mathsf{F}\wedge \mathsf{F}~. 
\end{equation}
Here, in the first action, we used the fact that the Chern--Weil $(-1)$-form symmetry is associated with the gauge $\theta$-angle. The role of the shift by $\chi^{(1)}$ will be clear in the following. As for the second action, here we used the earlier identification we made in \eqref{eq:h4-identified-with-TrFF}. 

Applying the map $\widetilde{\delta}$ on \eqref{terms for  modified instantons} and using the explicit actions in \eqref{delta on -1 ZK} and \eqref{delta on -1 CW}, we obtain:

\begin{equation}\label{4d / ZK2 TQFT action}
 S_{\text{TQFT}}^{G/\Z_{K}^{\scriptscriptstyle[2]}}\,=\, \int_{M_{4}}\,\,\left[  \frac{\theta_{\text{YM}} + \chi^{(1)}}{8\pi^{2}}\,\tr(\mathsf{F}\,\wedge\,\mathsf{F})\,+\, \frac{\theta^{(2)}+\,K\,\chi^{(2)}}{2\pi}\,\dd c_{3}\,\right]\,.
\end{equation}
Remark that the equation of motion for $\chi^{(1)}$ implies:
\begin{equation}
    \frac{1}{8\pi^{2}}\int_{M_{4}}\, \tr(\mathsf{F}\,\wedge\,\mathsf{F})\, =\, 0~.
\end{equation}
That is, it sets the instanton number to zero. But, to access the sectors of the 4d theory with non-trivial instanton numbers, we will take the case:
\begin{equation}
    \chi^{(1)} \,= \, \chi^{(2)} \,\equiv\, \chi \,.
\end{equation}
This choice will be justified further in the next subsection in terms of the gauge invariance of the above action.

With this choice, the action \eqref{4d / ZK2 TQFT action} becomes:
\begin{equation}\label{4d TQFT /ZK2 final}
\begin{split}
S_{\text{TQFT}}^{G/\Z_{K}^{\scriptscriptstyle[2]}} \,&=\, \int_{M_{4}}\,\,\left[ \frac{\theta_{\text{YM}}}{8\pi^{2}}\,\tr(\mathsf{F}\,\wedge\,\mathsf{F})\,+\, \frac{\theta^{(2)}}{2\pi}\,\dd c_{3}\right]\\
&-\int_{M_4}\chi\,\wedge\,\left(\,\frac{1}{8\pi^{2}}\tr(\mathsf{F}\,\wedge \,\mathsf{F})- \frac{K}{2\pi}\,\dd c_{3}\,\right)~.
\end{split}
\end{equation}
Now, the EOM for the field $\chi$ imposes:
\begin{equation}\label{modified instanton}
      \frac{1}{8\pi^{2}} \tr(\mathsf{F}\,\wedge\,\mathsf{F}) \,=\, \frac{K}{2\pi}\,\dd c_{3}~.
\end{equation}
From the $2\pi\Z$ quantisation condition on the periods of the gauge field $c_3$, we see that the EOM for the field $\chi$ constrains the possible instanton sectors of the 4d gauge theory to be a multiple of $K$ \cite{Seiberg:2010qd, Tanizaki:2019rbk, Najjar:2024vmm,Najjar:2025htp}. 

\medskip
\noindent
\textbf{Modifying $\theta_{\rm YM}$ periodicity.} For instance, when computing the partition function of this 4d theory, one finds that it decomposes into the following form:
\begin{equation}
    Z^{G/\Z_K^{\scriptscriptstyle{[2]}}} = \sum_{\nu\in K\Z} e^{2\pi i\, \nu \,\theta_{\rm YM}}\, Z_{\nu}~.
\end{equation}
Here, $Z_\nu$ is the contribution from each sector. From the phase factor, one expects that the $2\pi$-periodicity of the theta angle that we discussed around \eqref{Z G} gets modified as well.

At the level of the action \eqref{4d TQFT /ZK2 final}, this can be seen by substituting the constraint \eqref{modified instanton} and combining the two theta terms. In the end, we get the term:
\begin{equation}
    (\theta^{(2)} + K\,\theta_{\rm YM}) \,\int_{M_4}\, \dd c_{3} ~.
\end{equation}
Since $\theta^{(2)}$ is $2\pi$-periodic, one can deduce that the YM theta angle will have the following periodicity condition:
\begin{equation}
        \theta_{\rm YM} \,\sim\, \theta_{\rm YM} \,+\, \frac{2\pi }{K}~.
\end{equation}

\medskip
\noindent
\textbf{Universality of $U(1)^{[2]}$.} Note that, from the multiverse perspective, each one of the local observers can perform the steps above to get a modification in the counting of instanton numbers--and hence, in the periodicity of the YM theta angle. This follows from our choice of the universal 2-form symmetry $U(1)^{[2]}$. Since this symmetry does not depend on the explicit form of (the quotients of) the fiber and base 3-spheres, it can be detected by all the local observers, hence the `universality' nature of this symmetry.

\medskip
\noindent
\textbf{Modified instantons for non-simply laced gauge theories.} Before concluding this section, we note that the above procedure applies not only to the six cases of interest listed in (\ref{eq:the-6-cases}), but also to gauge theories associated with non-simply laced Lie algebras obtained via folding. This follows from the fact that the folding procedure does not affect the free cycles of the link. Consequently, the resulting folded theories inherit the same spectrum of continuous $U(1)^{\scriptscriptstyle[m]}$ $m$-form symmetries as their simply laced counterparts. Thus, this section generalizes the notion of modified instanton sums to gauge theories with Lie algebras of type $\mk{su}(N)$, $\mk{so}(2N+1)$, $\mk{sp}(2N)$, $\mk{so}(2N)$, $\mk{e}_{6}$, $\mk{f}_{4}$, and $\mk{g}_{2}$.

\subsection{4-group structures} \label{subsec:4group}
The previous subsection demonstrated that gauging a finite $\Z_{K}^{[2]}\subset U(1)^{[2]}$ symmetry and projecting it along with the $U(1)^{[-1]}$ CW $(-1)$-form symmetry onto the physical boundary yields a TQFT that modifies the sum over instanton sectors as presented in (\ref{4d TQFT /ZK2 final}).

Connecting this discussion with the mixed 't Hooft anomaly we discussed in subsection \ref{subsec: mixed t hooft}, we go one step further where we investigate the enlarged 4d TQFT we get from gauging the electric 1-form symmetry $\mathcal{Z}(G)^{[1]}$. As emphasized in \cite{Tanizaki:2019rbk} for the $G = SU(N)$ case, the consistent gauging of $\Z_{N}^{[1]}$ requires the simultaneous gauging of a finite 3-form symmetry. Fortunately, for the class of non-splittable quotients considered in this work, the resulting $\mk{su}(N)$, $\mk{so}(2N)$, and $\mk{e}_{6}$ gauge theories, as well as the $\mk{so}(2N+1)$ and $\mk{sp}(2N)$ theories obtained via the folding procedure, naturally admit such a finite 3-form symmetry, denoted by $\Z_{m}^{[3]}$. The simultaneous gauging of these symmetries induces a non-trivial higher-group structure, which we seek to uncover in the remainder of this subsection.

\subsubsection{Gauging the electric 1-form symmetry} 

Here, we are interested in gauging the electric $\mathcal{Z}(G)^{ [1]}$ 1-form symmetry.  As can be seen from the examples in (\ref{UN field strength}) and (\ref{tilde F s for E6}), turning on a non-trivial 2-form gauge field $B_{2}^{\rm{e}}$ requires modifying the gauge theory field strength $\sf{F}$ to a twisted field strength $\widetilde{\sf{F}}$, satisfying the constraint:
\begin{equation}\label{eq:tr-tilde-F=nB2e}
     \tr(\widetilde{\mathsf{F}})\,=\, n\,B_{2}^{\mathrm{e}}~.
\end{equation}
To simplify the discussion, here we are taking the 1-form symmetry to be $\mathcal{Z}(G)^{[1]} \cong \Z_n$. The generalization to the case $G = Spin(4N)$ where $\mathcal{Z}(G) =\Z_2\times\Z_2 $ should be straightforward. The above constraint is understood via the embedding trick due to \cite{Cordova:2019uob} as reviewed around table \ref{Table:1-form-alpha-G}.

With this in mind, the projection of the $(-1)$-form CW-symmetry operator $\widetilde{h}_{4}^{\rm{f}}$ \eqref{delta on -1 CW} should now involve the twisted field strength $\widetilde{\sf{F}}$. Explicitly, on the physical boundary, we have:
\begin{equation}
 \widetilde{\delta}\left( \frac{\widetilde{h}_{4}^{\rm{f}}}{2\pi} \right) \,= \,-\,\frac{1}{8\pi^{2}}\,\tr(\widetilde{\mathsf{F}}\wedge \widetilde{\mathsf{F}})~,
\end{equation}
while the projection of $F_{1}^{\rm{b}}$ is kept the same as in (\ref{delta on -1 CW}).

Additionally, gauging $\Z_{n}^{[1]}$ requires the projection of the anomaly inflow given in (\ref{S inflow}) as:
\begin{equation}\label{eq:projec-anomaly-inflow}
\widetilde{\delta}\left(\frac{n^{2}\Phi_{\scriptscriptstyle G}}{2\pi} \  F_{1}^{\rm{b}} \wedge B_{2}^{\rm{e}}\wedge B_{2}^{\rm{e}} \right)\,=\, -\,\frac{1}{8\pi^{2}}\,\widetilde{\Phi}_{\scriptscriptstyle G}\  (\theta_{\rm YM} + \chi)\,\wedge\,  b_{2}\,\wedge\, b_{2}\,.
\end{equation}
Here, we imposed free boundary conditions on the 2-form gauge field $B_{2}^{\rm{e}}$: $\widetilde{\delta}(B_2^{\rm e}) = b_2$. Moreover, we  defined:
\begin{equation}
    \frac{n^{2}\,\Phi_{\scriptscriptstyle G}}{2\pi}\,:=\,-\,\frac{1}{8\pi^{2}}\,\widetilde{\Phi}_{\scriptscriptstyle G}~.
\end{equation}

At this stage, the TQFT we obtained via the projection procedure can be displayed as
\begin{equation}\label{eq:TQFT-theta-1form-2form}
\begin{split}
S_{\text{TQFT}}^{G/\Z_n^{[1]}\times\Z_K^{[2]}} \,&=\, \int_{M_{4}}\,\,\left[ \frac{\theta_{\text{YM}}}{8\pi^{2}}\,\left(\tr(\widetilde{\mathsf{F}}\,\wedge\,\widetilde{\mathsf{F}})-\widetilde{\Phi}_{\scriptscriptstyle G} \, b_{2}\wedge b_{2}\right)\,+\, \frac{\theta^{(2)}}{2\pi}\,\dd c_{3}\right]\\
&-\int_{M_4}\chi\,\wedge\,\left[\,\frac{1}{8\pi^{2}}\,\left(\tr(\widetilde{\mathsf{F}}\,\wedge\,\widetilde{\mathsf{F}})-\widetilde{\Phi}_{\scriptscriptstyle G} \, b_{2}\wedge b_{2}\right)- \frac{K}{2\pi}\,\dd c_{3}\,\right]~.
\end{split}
\end{equation}
Apart from the terms (\ref{eq:projec-anomaly-inflow}), this is the TQFT in (\ref{4d TQFT /ZK2 final}) from the previous subsection.

Furthermore, one can verify, using the embedding trick that we reviewed in subsection \ref{subsec: mixed t hooft}, that the following combination:
\begin{equation}
    \tr(\widetilde{\mathsf{F}}\,\wedge\,\widetilde{\mathsf{F}})-\widetilde{\Phi}_{\scriptscriptstyle G} \, b_{2}\wedge b_{2}
\end{equation}
is invariant under the small gauge transformation of the 2-form gauge field $b_{2}$ \eqref{1-form sym transform}, which transforms as $b_{2}\to b_{2}+\dd\Lambda_{1}$.

Let us consider the equation of motion for the Lagrange multiplier $\chi$ in \eqref{eq:TQFT-theta-1form-2form}:
\begin{equation}\label{eq:EOM-chi-puzzel}
   \,\frac{1}{8\pi^{2}}\,\left(\tr(\widetilde{\mathsf{F}}\,\wedge\,\widetilde{\mathsf{F}})-\widetilde{\Phi}_{\scriptscriptstyle G} \, b_{2}\wedge b_{2}\right)- \frac{K}{2\pi}\,\dd c_{3}\,=\,0~.
\end{equation}
Upon integrating this equation over spacetime $M_{4}$, one finds that both
\begin{equation}
   \frac{1}{8\pi^{2}} \int_{M_{4}}\, \tr(\widetilde{\mathsf{F}}\,\wedge\,\widetilde{\mathsf{F}})\,,\qquad \mathrm{and}\,\qquad \frac{K}{2\pi}\int_{M_{4}}\dd c_{3}~,
\end{equation}
are integers, since the first is the second Chern number and the second term is a properly quantized 4-form flux. However, the remaining term:
\begin{equation}\label{1/n quantisation}
   \frac{\widetilde{\Phi}_{\scriptscriptstyle G}}{8\pi^{2}}\, \int_{M_{4}}\, \, b_{2}\wedge b_{2} \ \in\  \frac{1}{n}\,\Z~,
\end{equation}
which is fractional and cannot cancel the integer contributions from the other two terms. This leads to a contradiction, signaling that the equation of motion is inconsistent unless further refinement is considered.

We have arrived precisely at the contradiction (or puzzle) discussed in section 3.2 of \cite{Tanizaki:2019rbk}. The resolution proposed there involves gauging a discrete 3-form symmetry, a mechanism we will now implement in our setup.

\subsubsection{Gauging the $\Z_{m}^{[3]}$ 3-form symmetry} 

This procedure amounts to imposing a free (Neumann-type) boundary condition on the 4-form gauge field $A_{4}$:
    \begin{equation}
    \widetilde{\delta}\left(A_{4}\right)\,=\, a_{4}~.
    \end{equation}
Consistently, one must impose Dirichlet boundary conditions on the scalar field $B_{0}$. Explicitly, we set  
\begin{equation}
\widetilde{\delta}\left(mB_{0}\right)\,=\,\theta^{(3)}\,+\,m\chi^{(3)}~.
\end{equation}
Here, $\theta^{(3)}\in\R/2\pi\Z$ is a $2\pi$ periodic theta-angle associated with the dynamical top-form $a_{4}$, and $\chi^{(3)}$ is a Lagrange multiplier shift.

With this prescription, the TQFT action in (\ref{eq:TQFT-theta-1form-2form}) is now refined to:
\begin{equation}\label{eq:TQFT-(1)}
\begin{split}
S_{\text{TQFT}}^{G/\Z_K^{[2]}\times\Z_n^{[1]}{\times}\Z_m^{[3]}} \,&=\, \int_{M_{4}}\,\,\left[ \frac{\theta_{\text{YM}}}{8\pi^{2}}\,\tr((\widetilde{\mathsf{F}}-b_{2})^{2})+\, \frac{\theta^{(2)}}{2\pi}\,\dd c_{3}\,+\,\frac{\theta^{(3)}}{2\pi}\,a_{4}\right]\\
&-\int_{M_4}\chi\,\wedge\,\left[\,\frac{1}{8\pi^{2}}\left(\tr(\widetilde{\mathsf{F}}\,\wedge \,\widetilde{\mathsf{F}}) -\widetilde{\Phi}_{\scriptscriptstyle G}\, b_{2}\wedge b_{2} \right)- \frac{K}{2\pi}\,\dd c_{3}\,\right]
\\
\,&+\,\int_{M_{4}}\,\frac{m}{2\pi}\chi^{(3)}\wedge\, a_{4}~.
\end{split}
\end{equation}
Here, the term $-\widetilde{\Phi}_{\scriptscriptstyle G}\, b_{2}\wedge b_{2}$ in has been absorbed into the $\theta_{\rm{YM}}$-term by implementing the embedding trick reviewed in subsection \ref{subsec: mixed t hooft} above.

The above TQFT action is not yet in its final form. One would ideally reduce the number of independent parameters and Lagrange multipliers appearing in the action. A systematic approach to achieve this involves imposing invariance under large gauge transformations of the finite gauge fields, which in turn imposes non-trivial constraints among the theta angles $\theta$ and the shift parameters $\chi$ appearing in the action.

To illustrate this, consider a large gauge transformation of the 3-form gauge field $c_{3}$:
\begin{equation}
    c_{3}\,\to \,c_{3}\,+\,\Lambda_{3}~,
\end{equation}
where $\Lambda_{3}$ is the gauge parameter. Requiring gauge invariance of the TQFT action under this transformation implies the following conditions:
\begin{equation}
    \delta_{\Lambda_{3}}a_{4}\,=\,-\,\dd\Lambda_{3}  \,,\qquad\, \chi\,=\,\chi^{(3)} \,,\qquad K\,=\,m~.
\end{equation}

Substituting these constraints into the TQFT action given in equation (\ref{eq:TQFT-(1)}), one obtains:
\begin{equation}
\begin{split}
S_{\text{TQFT}}^{G/\Z_m^{[2]}\times\Z_n^{[1]}\times \Z_m^{[3]}} \,&=\, \int_{M_{4}}\,\,\left[\, \frac{\theta_{\text{YM}}}{8\pi^{2}}\,\tr((\widetilde{\mathsf{F}}-b_{2})^{2})\,+\, \frac{\theta^{(2)}}{2\pi}\,(\dd c_{3}\,-\,a_{4})\,\right]\\
&-\int_{M_4}\chi\,\wedge\,\left[\,\frac{1}{8\pi^{2}}\left(\tr(\widetilde{\mathsf{F}}\,\wedge \,\widetilde{\mathsf{F}}) -\widetilde{\Phi}_{\scriptscriptstyle G}\, b_{2}\wedge b_{2} \right)\,+\, \frac{m}{2\pi}(\,-\dd c_{3}\, +\,a_{4})\,\right]\,.
\end{split}
\end{equation}

\subsubsection{Solving the puzzle: a 4-group structure} 
We are now equipped with all the necessary ingredients to resolve the puzzle discussed around \eqref{1/n quantisation}. The key step involves refining the definition of the field strength associated with $m\,a_{4}$, as introduced in (\ref{eq:nB=F}), to consistently absorb the anomalous term $-\widetilde{\Phi}_{\scriptscriptstyle G} b_{2}\wedge b_{2}$ term. Concretely, we redefine the 4-form field strength as: 
\begin{equation}\label{eq:ma4=da3-b2b2}
    \frac{m}{2\pi}\,a_{4}\,\equiv\,\mathrm{F}_{4}\,:=\, \frac{1}{2\pi}\,\dd a_{3} \,-\,\frac{\widetilde{\Phi}_{\scriptscriptstyle G}}{8\pi^{2}}\  b_{2}\wedge b_{2} ~.
\end{equation}
Without this refinement, the naive expression for the field strength would simply be $\mathrm{F}_{4}=\dd a_{3}$.

Let us now revisit the terms coupled to the scalar field $\chi$. The corresponding equation of motion now reads:
\begin{equation}
    \frac{1}{8\pi^{2}}\tr(\widetilde{\mathsf{F}}\,\wedge \,\widetilde{\mathsf{F}}) \,+\, \frac{1}{2\pi}\dd a_{3} \,-\,\frac{m}{2\pi}\dd c_{3}\,=\,0~.
\end{equation}
Since $a_{3}$ is a $U(1)$-valued gauge field, it satisfies the quantization condition 
\begin{equation}
    \int_{M_{4}} \dd a_{3}\ \in \Z\,,
\end{equation}
which resolve the issue we encountered in (\ref{eq:EOM-chi-puzzel}).

However, to guarantee full consistency, we must also check gauge invariance under the 1-form electric symmetry gauge parameter $\Lambda_{1}$, especially since the dual field strength $\widetilde{\mathsf{F}}$ transforms non-trivially. Taking $a_{4}$ to be gauge invariant under $\Lambda_{1}$, we require that $a_{3}$ itself transforms non-trivially as:
\begin{equation}
    \delta_{\Lambda_{1}}a_{3}\,=\, \frac{\widetilde{\Phi}_{\scriptscriptstyle G}}{2} \ b_{2}\,\wedge\,\Lambda_{1} \,+\, \widetilde{\Phi}_{\scriptscriptstyle G}\  \Lambda_{1}\,\wedge\,\dd \Lambda_{1}\,.  
\end{equation}
Further, one can verify that the integral of $\delta_{\Lambda_{1}}a_{3}$ gives an integer, confirming that the above procedure resolves the puzzle.    

The transformation rule for $a_{3}$ reveals the presence of a higher symmetry structure, specifically a 4-group. The resulting non-trivial extension takes the form:
\begin{equation}
    \Z_{n}^{[1]}\ \widetilde{\times}\ \Z_{m}^{[3]}\,.
\end{equation}
Signaling a non-trivial coupling between the electric 1-form symmetry $\Z_{n}^{[1]}$ and the finite 3-form symmetry $\Z_{m}^{ [3]}$. 

The above discussion extends the realization of 4-group symmetry—originally demonstrated in the context of $\mk{su}(n)$ gauge theories from a field-theoretic perspective in \cite{Tanizaki:2019rbk}, from M-theory geometric engineering in \cite{Najjar:2024vmm}, and from holography in \cite{Najjar:2025htp}—to encompass $\mk{so}(2n)$ and $\mk{e}_{6}$ gauge theories.

\medskip
\noindent
\textbf{Foldings and 4-group structure.} We emphasize that the higher 4-group structure discussed above also extends to non-simply laced gauge theories with non-trivial centers, obtained via the folding procedure outlined in section \ref{sec:outer-auto}. In particular, our analysis demonstrates that gauge theories with Lie algebras $\mk{so}(2N+1)$ and $\mk{sp}(2N)$ naturally exhibit a 4-group structure arising from the simultaneous gauging of the electric 1-form symmetry and the finite 3-form symmetry.  


\section{Conclusions and outlook}

In this work, we study novel quotients of the Bryant–Salamon space of the form $(\R^4\times \mathbb{S}^3_{\rm b})/\Gamma_{\rm ns}$, where part of the quotient group acts simultaneously on both the fiber and the base. Such quotients are closely related to extensions of finite $ADE$ subgroups of $SU(2)$, of which six cases were classified by Reid in \cite{Reid:1985}. We reviewed how these quotients can be systematically described using Goursat’s theorem and, following Cortés and Vázquez \cite{cortés2014locallyhomogeneousnearlykahler}, ensured that the resulting group actions are free on the link space $\bbS^{3}_{\rm{f}}\times\bbS^{3}_{\rm{b}}$.

For each of the six cases, we analyzed the 4d $\mathcal{N}=1$ gauge theories geometrically engineered from M-theory on $M_4\times (\R^4\times\mathbb{S}^3_{\rm b})/\Gamma_{\rm ns}$ and studied their associated symmetries via the construction of the corresponding SymTFTs. Our analysis revealed the following key implications of these quotients:
\begin{itemize}
	\item \textbf{Decomposition into universes.} We interpreted the simultaneous action of the quotient group in terms of decomposing the resulting 4d gauge theory into multiple universes. Each one of these universes contains the same gauge algebra $\mathfrak{g}$. Furthermore, these universes are separated by domain walls which are defects charged under a discrete 3-form symmetry that results also results from the nature of the quotient group $\Gamma_{\rm ns}$. From M-theory, this interpretation follows earlier ideas and observations of \cite{Witten:1997kz,Acharya:2020vmg,Najjar:2022eci}.
	 \item \textbf{Inner and outer automorphisms.} The second implication of these types of quotients that we studied in this work concerns the inner and outer automorphisms that act on the resulting gauge algebra $\mathfrak{g}$. We showed how the action of these automorphisms gives rise to a series of symmetry-breaking of $\mathfrak{g}$. For instance, we demonstrated how our construction parallels previously studied GUT breaking patterns, such as those discussed in \cite{Pati:1974yy,Georgi:1974my,Georgi:1974sy,Fritzsch:1974nn,Georgi:1974yf,Gursey:1975ki,Dimopoulos:1981yj}. Moreover, we discussed how one can use the outer automorphism to perform the folding on the simply-laced algebra $\mathfrak{g}$ to obtain a non-simply laced one. Later on in the work, this trick enabled us to study many properties of non-simply laced 4d gauge theories.
        \item \textbf{5d SymTFTs.}  At the level the homology of the link space $(\mathbb{S}^3_{\rm f}\times \mathbb{S}^3_{\rm b})/\Gamma_{\rm ns}$, inspired by the decomposition property, we proposed an algorithm that computes these groups. We found very rich structure of torsional cycles that later on we used to construct the topological operators generating the different higher form symmetries that can be exhibited by the 4d gauge theory. This opened the doors for us to many physical implications, including mixed 't Hooft anomalies, modified instanton sums, and higher group structures. We studied these aspects for theories with simply-laced gauge algebras as well as non-simply-laced ones. For the latter type, we employed the folding trick by the set of outer automorphisms. We believe that studying these latter cases is one of the novelties of our work.
\end{itemize}

Our systematic analysis opens up new possibilities for future research work:
\begin{itemize}
    \item As noted in section \ref{sec:geo-eng-4d}, our analysis focused on a specific component of the quotient space $(\R^{4}\times\mathbb{S}^3_{\rm b})/\Gamma_{\rm ns}$. A natural extension of this work would be to study the full moduli space associated with such quotients and to explore the interpretation and role of generalized symmetries across its different components.

    \item It would be interesting to investigate QFTs in other dimensions that can be geometrically engineered using variants of the non-splittable groups considered in this work.

    \item In this work, we analyzed the effects of outer automorphisms on SymTFTs and derived novel SymTFTs (via M-theory) for non-simply laced gauge theories. A natural next step is to investigate the role of inner automorphisms on the SymTFTs and derive some physical implications. 

    \item In general, higher-group structures imply non-trivial correlations between the operators of the constituent $p$-form symmetries—or equivalently, between their associated defects—as they cross each other. It would be natural to investigate how such effects are realized from their M-theoretic origins. 
\end{itemize}


\acknowledgments
We would like to thank Leonardo Foscolo, Yi-Nan Wang, Leonardo Santilli, Lakshya Bhardwaj, and Haynes Miller for discussions and correspondence. MN is grateful to Leonardo Santilli and Yi-Nan Wang for collaboration on a related project, and to SIMIS for hospitality during part of this work. Special thanks go to MN’s son, Tameem, for his patience during the summer holidays while this project was in progress. MN is supported by National Natural Science Foundation of China under Grant No. 12175004, No. 12422503, and by Young Elite Scientists Sponsorship Program by CAST (2023QNRC001, 2024QNRC001).


\appendix

\section{Supplementary figures}

\begin{figure}[H]
\centering
\includegraphics[scale = 0.82]{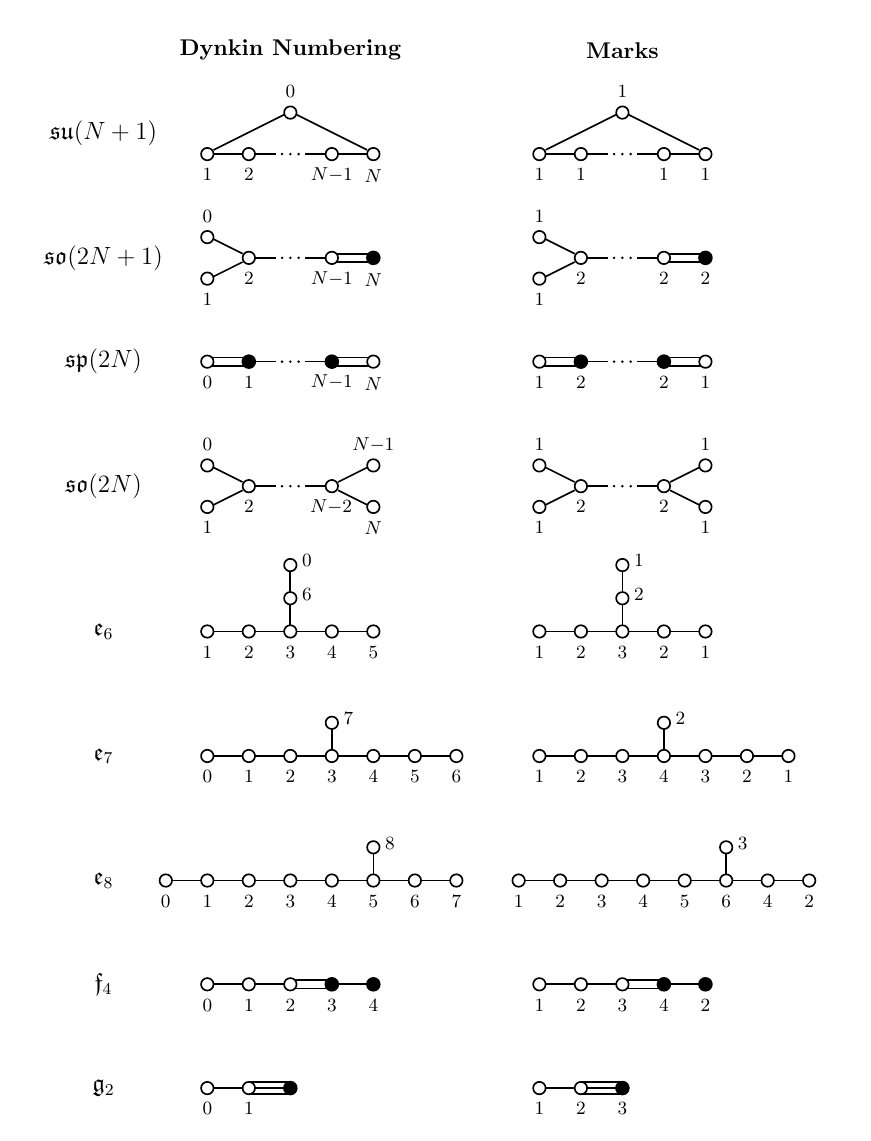}
\caption{The affine Dynkin diagrams along with the marked affine Dynkin diagrams. The figure above is borrowed from \cite{Larouche_2011}. The center $\mathcal{Z}(G)$ of the Lie groups associated with the Lie algebra above is given by the finite cyclic groups with order equal to the number of nodes with marks equal $1$.}
\label{fig:dynkin+marks}
\end{figure}


\bibliographystyle{JHEP}
\bibliography{ref}

\end{document}